\documentclass[twocolumn,aps,prb,notitlepage]{revtex4-2}
\pdfoutput=1
\usepackage{bm}
\usepackage{graphicx}
\usepackage{amssymb}
\usepackage{amsmath}
\usepackage{dcolumn}
\usepackage{amsfonts}
\usepackage{upgreek}
\usepackage{comment}
\usepackage[margin=.7in]{geometry}
\usepackage[euler]{textgreek}
\usepackage{lineno}
\usepackage{braket}
\usepackage[caption=false]{subfig}
\usepackage{floatrow}
\usepackage{gensymb}
\usepackage{multirow}
\usepackage{wasysym}
\usepackage{mathtools}
\usepackage[usenames,dvipsnames]{xcolor}
\usepackage[normalem]{ulem}
\usepackage[colorlinks]{hyperref}
\hypersetup{
    colorlinks=true,
    citecolor=blue,
    linkcolor=blue,
    filecolor=magenta,      
    urlcolor=blue,
    }
\usepackage{lineno}
\usepackage{tikz} 
\allowdisplaybreaks
\newcommand{\stkout}[1]{\ifmmode\text{\sout{\ensuremath{#1}}}\else\sout{#1}\fi}
\newcommand{\bs}{\boldsymbol}
\newcommand{\pd}{\partial}
\newcommand{\pr}{^{\prime}}
\newcommand{\dpr}{^{\prime\prime}}

\newcommand{\mbk}{\mathbf{k}}
\newcommand{\mbkpr}{\mathbf{k}^{\prime}}
\newcommand{\mbkdpr}{\mathbf{k}^{\prime \prime}}

\newcommand{\veps}{\varepsilon}
\newcommand{\dprime}{\prime\prime}

\begin{document}

\title{Quantum kinetic theory of quadratic responses}
\author{M. Mehraeen}
\email{mxm1289@case.edu}
\affiliation{Department of Physics, Case Western Reserve University, Cleveland, Ohio 44106, USA}
\date{\today}
\begin{abstract}
Recent work has revealed a general procedure for incorporating disorder into the semiclassical model of carrier transport, whereby the predictions of quantum linear response theory can be recovered within a quantum kinetic approach based on a disorder-averaged density-matrix formalism. Here, we present a comprehensive generalization of this framework to the nonlinear response regime. In the presence  of an electrostatic potential and random impurities, we solve the quantum Liouville equation to second order in an applied electric field and derive the carrier densities and equations of motion. In addition to the anomalous velocity arising from the Berry curvature and the Levi-Civita connection of the quantum metric tensor, a host of extrinsic velocities emerge in the equations of motion, reflecting the various possibilities for random interband walks of the carriers in this transport regime. Furthermore, several scattering and conduction channels arise, which can be classified in terms of distinct physical processes, revealing numerous unexplored mechanisms for generating nonlinear responses in disordered condensed matter systems. 
\end{abstract}
\maketitle

\section{Introduction}

Owing to its fundamental interest and practical significance, the study of carrier transport in disordered condensed matter systems has a rich and multifaceted history, which has inspired a variety of complementary theoretical approaches, including Kubo formulas~\cite{kubo1957statistical}, Keldysh theory~\cite{keldysh2024diagram} and density-matrix methods~\cite{karplus1954hall, kohn1957quantum, luttinger1958quantum, luttinger1958theory}. Despite their seemingly different approaches to obtaining response functions, which range from evaluating dressed propagators and vertex functions via diagrammatic methods to ensemble-averaging observables weighted by carrier distribution functions, the underlying commonality of all these methods may be regarded as the attempt to quantify the evolution of the density operator in the presence of disorder and other interactions.

In light of this shared connection, and with the goal of deriving a more complete description of carrier conduction on macroscopic scales, a recurring research theme in the linear response regime has been the attempt to incorporate the results of diagrammatic methods into quantum kinetic theory~\cite{sinitsyn2005disorder, sinitsyn2007anomalous, sinitsyn2007semiclassical, culcer2017interband, sekine2017quantum, xiao2017semiclassical, xiao2019temperature, atencia2022semiclassical}. The former approach allows for a robust treatment of perturbations and is readily generalizable to many-body problems~\cite{mahan2000many, bruus2004many}. The kinetic approach, while also providing a systematic perturbative analysis~\cite{vasko2006quantum}, deals directly with carrier distribution functions and thus often allows for a more transparent transition to the physically intuitive and relatively simple semiclassical limit~\cite{xiao2010berry}.

Through a series of recent works, a proposal has been put forward of a modified semiclassical framework that is based on solving the quantum Liouville equation for the disorder-averaged density matrix~\cite{culcer2017interband, sekine2017quantum, atencia2022semiclassical}. This is a versatile approach that allows for a systematic analysis of the effects of disorder scattering on physical observables and has recently been applied to a variety of problems in response theory, including the study of resonance structures~\cite{bhalla2020resonant, bhalla2022resonant}, probing quantum geometry~\cite{cullen2021generating, bhalla2022resonant} and exploring nonlinear electric and thermal responses~\cite{bhalla2023quantum, atencia2023disorder, varshney2023quantum}. In this approach, the effect of disorder naturally arises not only in the carrier distributions--which is traditionally the case~\cite{ashcroft1976solid}--but also in the equations of motion governing the carrier dynamics. Thus, in the linear response regime, in addition to the familiar group and (intrinsic) anomalous velocities~\cite{karplus1954hall, kohn1957quantum, adams1959energy, chang1995berry, chang1996berry, sundaram1999wave}, the semiclassical equations of motion are modified by an additional disorder-dependent term dubbed the extrinsic velocity~\cite{atencia2022semiclassical}. In this picture, just as the intrinsic anomalous velocity is related to interband coherence effects mediated by an electric field, the extrinsic velocity can similarly be thought of as arising from extrinsic interband coherence effects due to the Berry connection and disorder. 

As a consequence of introducing the disorder average at the level of the quantum mechanical density matrix, the modified semiclassical equations can now be regarded as describing the averaged motion of carriers after many scattering events. The extrinsic velocity thus quantifies the average change in the carrier velocity after multiple random interband walks due to disorder scattering~\cite{atencia2022semiclassical} and is related to the side-jump velocity associated with the coordinate shift at scattering events~\cite{sinitsyn2006coordinate}. 

In addition to the equations of motion, the extrinsic velocity also modifies the linear response functions. For the specific case of the anomalous Hall effect~\cite{karplus1954hall, luttinger1958theory, smit1958spontaneous, berger1970side, nozieres1973simple}, it has been shown that including the current density associated with the extrinsic velocity results in a total conductivity that agrees with that obtained via diagrammatic methods in the noncrossing approximation in the presence of disorder~\cite{atencia2022semiclassical}. This is a notable result, as it allows one to obtain diagrammatically equivalent results by a relatively simple and physically transparent semiclassical approach. In addition, it can improve the accuracy of computational methods, which are commonly performed using the semiclassical method~\cite{wang2006abinitio, wang2007fermi, gradhand2012first, he2012berry, chen2013weyl, bianco2014how, chen2014anomalous, olsen2015valley, feng2016first, dai2017negative, martiny2019tunable, wuttke2019berry, du2020berry, he2020giant, he2021superconducting}.

Going beyond linear responses, the past decade has witnessed a surge of interest in nonlinear spin and charge transport phenomena in a variety of materials systems with differing magnetic order and band topology~\cite{du2021nonlinear, ideue2021symmetry, ortix2021nonlinear, nagaosa2024nonreciprocal, shim2024spin}. Several novel transport phenomena have emerged, including unidirectional~\cite{avci2015magnetoresistance, avci2015unidirectional, olejnik2015electrical, zhang2016theory, yasuda2016large, avci2018origins, lv2018unidirectional, duy2019giant, guillet2020observation, zelezny2021unidirectional, guillet2021large, hasegawa2021enhanced, liu2021magnonic, liu2021chirality, chang2021large, shim2022unidirectional, mehraeen2022spin, ding2022unidirectional, lou2022large, mehraeen2023quantum, cheng2023unidirectional, fan2023observation, zheng2023coexistence,  mehraeen2024proximity, zou2024nonreciprocal, zhao2024large, aoki2024evaluation, huang2024spin, kao2024unconventional} and bilinear~\cite{rikken2001electrical, he2018bilinear, dyrdal2020spin, zhang2022large, wang2022large, fu2022bilinear, golub2023electrical,  marx2024nonlinear, boboshko2024bilinear, kim2024spin} magnetoresistance effects, as well as a variety of nonlinear Hall effects~\cite{sodemann2015quantum, low2015topological, yasuda2017current, du2018band, facio2018strongly, you2018berry, zhang2018electrically, zhang2018berry, he2019nonlinear, ma2019observation, kang2019nonlinear, du2019disorder, wang2021intrinsic, li2021nonlinear, zeng2021nonlinear, wang2022observation, gao2023quantum, wang2023quantum, kaplan2023general, das2023intrinsic, ma2023anomalous, zhuang2024intrinsic, wang2024intrinsic}. From a fundamental point of view, nonlinear responses provide an opportunity to probe the nontrivial quantum geometry of Bloch states when linear responses are prohibited by symmetry~\cite{liu2024quantum}, thereby allowing for an extension of this analysis to a more diverse range of quantum materials with different lattice and space-time symmetries. Recent observations of the nonlinear Hall effect in time-reversal-symmetric conditions~\cite{ma2019observation, kang2019nonlinear} and intrinsic quadratic responses in topological antiferromagnets~\cite{gao2023quantum, wang2023quantum} with combined parity-inversion time-reversal symmetry not only yield valuable insights into the quantum geometry of these materials systems through the Berry curvature dipole or quantum metric tensor, but can also be harnessed for useful applications, such as Neel vector detection in parity-inversion time-reversal symmetric antiferromagnets~\cite{wang2021intrinsic, liu2021intrinsic}.

Furthermore, even when linear effects dominate the response of the system, the unidirectional nature of quadratic responses with respect to the electric field implies that they can rather straightforwardly be distinguished from the linear-response signal through a simple reversal in the direction of the applied electric field. In addition, as has recently been demonstrated in a variety of magnetic heterostructures~\cite{avci2015unidirectional, olejnik2015electrical, yasuda2016large, lv2018unidirectional, duy2019giant, zelezny2021unidirectional, guillet2021large, hasegawa2021enhanced, liu2021magnonic, chang2021large, shim2022unidirectional, ding2022unidirectional, lou2022large, cheng2023unidirectional, fan2023observation, zheng2023coexistence,  mehraeen2024proximity, zhao2024large, aoki2024evaluation, huang2024spin}, nonlinear transport effects such as unidirectional magnetoresistance are typically also odd in the in-plane magnetization, and are thus suitable for magnetic order detection, regardless of the presence of linear responses. Given the complementary applications and different origins and symmetry properties that nonlinear responses have with respect to their linear-response counterparts and the growing number of nonlinear transport effects being discovered, it is highly desirable to develop a unifying theoretical framework to capture these effects.

Physically, several mechanisms have been identified for generating nonlinear responses, which can generally be attributed to the combined effects of disorder scattering, band topology and the quantum geometry of Bloch states--realized through the Berry curvature and quantum metric tensors. And a number of theoretical extensions of linear-response methods have been proposed to accommodate nonlinear effects within the modern perspective of transport, including Kubo formulas~\cite{parker2019diagrammatic, du2021quantum, rostami2021gauge, kaplan2023unifying, mckay2024charge}, modified kinetic frameworks~\cite{xiao2019theory, nandy2019symmetry, bhalla2023quantum, atencia2023disorder, ba2023nonlinear, huang2023scaling}, drift-diffusion models~\cite{zhang2016theory, mehraeen2022spin} and nonequilibrium Keldysh theory~\cite{freimuth2021theory}.

\begin{figure*}[hpt]
\vspace{-.7cm}
\captionsetup[subfigure]{labelformat=empty}
    \sidesubfloat[]{\includegraphics[width=0.7\linewidth,trim={3cm -1cm 3cm -1cm}]{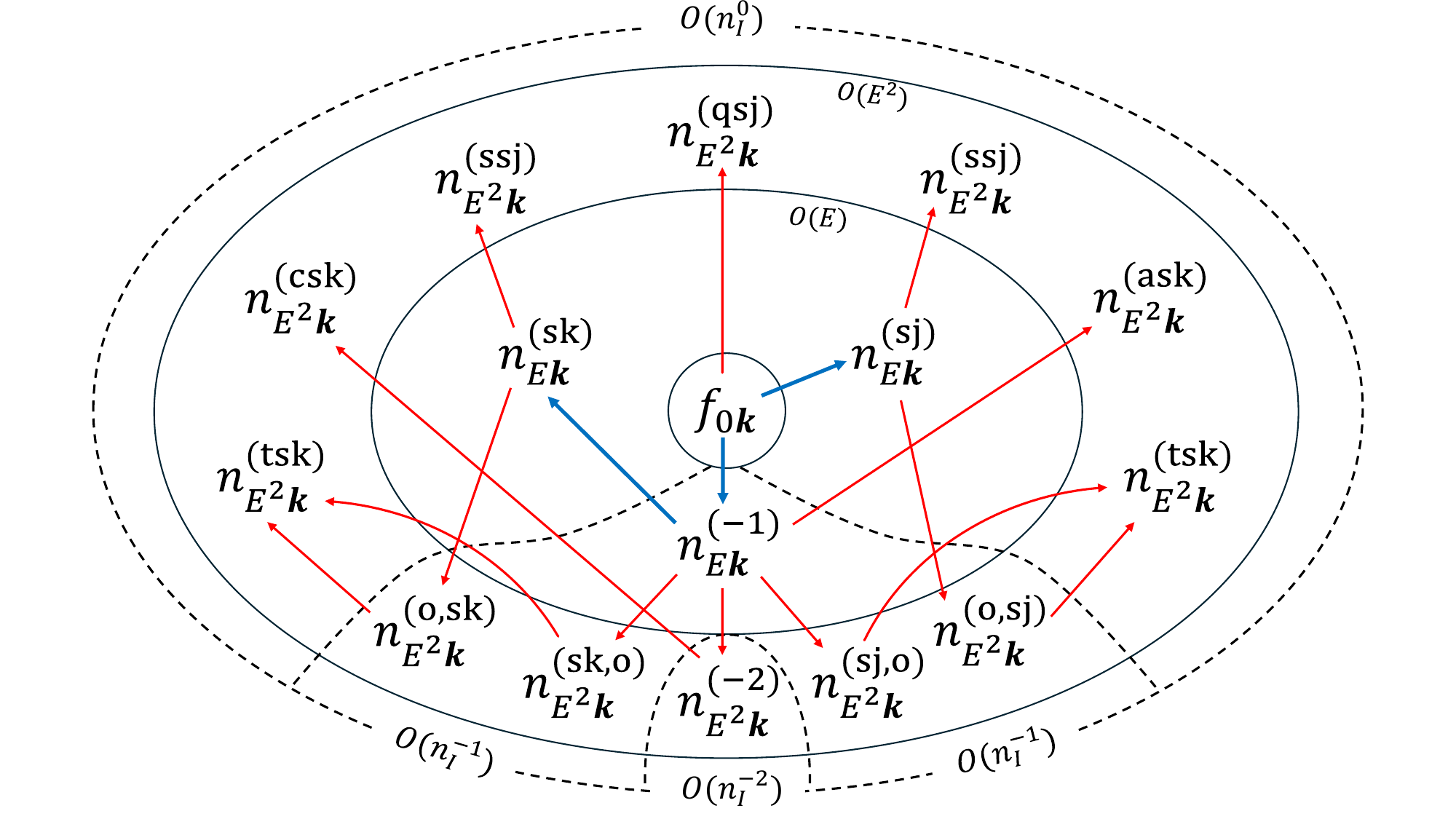}\label{fig1a}}
    \caption{Schematic map of carrier densities at first and second order in the applied electric field and their physical origins, starting from the equilibrium (Fermi-Dirac) distribution $f_{0\mbk}$. The three blue central arrows represent, ordinary, side-jump and skew scattering in linear response theory, while the surrounding red arrows arise in quadratic response theory. The dashed lines separate different orders in the impurity density.}
    \label{fig1}
\end{figure*}

In this work, we present a generalization of the linear response theory proposed in recent works~\cite{culcer2017interband, sekine2017quantum, atencia2022semiclassical} to the nonlinear response regime. In the presence of the electrostatic interaction and in the weak-disorder limit, we systematically solve the quantum Liouville equation to second order in the applied electric field and to three orders in the impurity density, revealing an extensive pattern of scattering processes. As illustrated in Fig.~\ref{fig1}, disorder scattering in linear response theory is represented by the triality of ordinary, side-jump and skew scattering~\cite{smit1958spontaneous, berger1970side, nagaosa2010anomalous}. In the quadratic response regime, we show that this triality underpins a relatively large number of scattering processes that contribute to the carrier density at different orders in the impurity density.

Furthermore, we uncover several field-independent and -dependent extrinsic velocities, which, in conjunction with the linear-response extrinsic velocity, account for disorder corrections to the equations of motion up to second order in the applied field. Specifically, we find that the equation of motion for the carrier position in the presence of disorder reads
\begin{equation}
\label{eom}
\begin{split}
\dot{x}_{\mu}^a
&=
v_{\mu \mbk}^a
-
\varepsilon_{\mu \nu \rho} \dot{k}_{\nu}^a \Omega_{\rho \mbk}^a
+
\dot{k}_{\nu}^a \dot{k}_{\rho}^a
\sum_{b} M_{\mbk}^{ab}
\Gamma_{\nu \rho \mu \mbk}^{ba}
\\
&+
\alpha_{\mu \mbk}^{a} 
+
\beta_{\mu \mbk}^{a}
+
\gamma_{\mu \mbk}^{a}
+
\kappa_{\mu \mbk}^{a}
+
\chi_{\mu \mbk}^{a},
\end{split}
\end{equation}
where $a$ is the band index and Greek letters represent spatial indices. Here
$\bs{v}_{\mbk}^a 
=
\bs{\pd}_{\mbk} \veps_{\mbk}^a /\hbar$ is the group velocity of carriers in the energy band $\veps_{\mbk}^a$, $\bs{\Omega}_{\mbk}^{ab}
=
\bs{\pd}_{\mbk} 
\times
\bs{\mathcal{A}}_{\mbk}^{ab}$ is the Berry curvature, with 
$\bs{\Omega}_{\mbk}^{a}
\equiv
\bs{\Omega}_{\mbk}^{aa}$ and $\bs{\mathcal{A}}_{\mbk}^{ab}
 =
 i \braket{u_{\mbk a} | \bs{\pd}_{\mbk} u_{\mbk b}}$ the Berry connection, with $\ket{u_{\mbk a}}$ the periodic part of the Bloch state. And
 $\Gamma_{\mu \nu \rho \mbk}^{a b}
=
\frac{1}{2} 
( \pd^{\mbk}_{\rho} g_{\mu \nu \mbk}^{ab}
+
\pd^{\mbk}_{\nu} g_{\mu \rho \mbk}^{ab}
-
\pd^{\mbk}_{\mu} g_{\nu \rho \mbk}^{ab})$ is a Bloch-space Levi-Civita connection component of the band-resolved quantum metric~\cite{provost1980riemannian, cheng2010quantum},
$
g_{\mu \nu \mbk}^{ab}
=
\text{Re}(\mathcal{A}_{\mu \mbk}^{ab} \mathcal{A}_{\nu \mbk}^{ba})
$.
The remaining terms in Eq.~(\ref{eom}), which we present in Sec.~\ref{sec_emergent}, are  the extrinsic velocities that emerge from the solution of the quantum Liouville equation and constitute the disorder corrections to the carrier motion.

Physically, the intrinsic terms can be understood as representing the effects of purely field-induced corrections to the Bloch states as a result of interband mixing, while the extrinsic velocities can be interpreted as encapsulating the averaged effects of random walks between energy bands due to impurity scattering, which may or may not depend on the electric field. In this sense, the approach presented here places the effects of electrostatic and disorder interactions on a more equal footing as far as the carrier dynamics is concerned and includes both their effects in the equations of motion. As we show below, the various carrier densities and velocities that emerge allow for numerous physically distinct conductivity channels to exist, reflecting the multitude of mechanisms that have partially been explored in the literature.

The remainder of the paper is organized as follows. In Sec.~\ref{sec_qke}, we introduce the model and derive the quantum kinetic equation. In Sec.~\ref{sec_linear}, we briefly review the derivation of the linear-response densities. In Sec.~\ref{sec_quad}, we discuss in detail the derivation of the quadratic density matrix. In Sec.~\ref{application}, we apply the theory to study the nonlinear transport in a model of 2D Dirac fermions. Finally, we present concluding remarks and offer an outlook.  

\section{Quantum kinetic equation}
\label{sec_qke}

We begin our analysis with the Hamiltonian
\begin{equation}
H
=
H^0 + V(\mathbf{r}) + U(\mathbf{r}),
\end{equation}
where $H^0$ describes the unperturbed system, $V(\mathbf{r})=e \mathbf{E} \cdot \mathbf{r}$ is the electrostatic potential and $U(\mathbf{r})$ is the disorder potential, which we assume describes scalar point scatterers with a white noise distribution, $\braket{U(\mathbf{r})}=0$,  
$\braket{U(\mathbf{r}) U(\mathbf{r\pr})}=n_I U_0^2 \delta(\mathbf{r} - \mathbf{r}\pr)$, and neglect higher-order non-Gaussian disorder correlations. Here, $n_I$ is the impurity density, $U_0$ measures the strength of the impurity interaction and $\braket{\cdots}$ indicates the disorder average~\footnote{It should be stressed that the choice of scalar disorder here is merely one of convenience and the theory is readily applicable to spin-dependent disorder profiles as well.}. We work within the crystal momentum representation, $\ket{\psi_{\mathbf{k}a}}
=
e^{i \mathbf{k} \cdot \mathbf{r}}
\ket{u_{\mathbf{k}a}}$, where $\ket{\psi_{\mathbf{k}a}}$ is an eigenstate of the unperturbed system, with 
$H^0 \ket{\psi_{\mathbf{k}a}} = \veps_{\mbk}^a\ket{\psi_{\mathbf{k}a}}$. The disorder averages then read $\braket{U_{\mbk \mbkpr}^{a a\pr}} = 0$ and 
$\braket{U_{\mbk \mbkpr}^{a a\pr} 
U_{\mbkpr \mbk}^{b\pr b}} 
=
n_I U_0^2 \braket{u_{\mbk a} | u_{\mbkpr a\pr}}
\braket{u_{\mbkpr b\pr} | u_{\mbk b}}$, where
$U_{\mbk \mbkpr}^{a a\pr}
\equiv
\Braket{\psi_{\mbk a} | U | \psi_{\mbkpr a\pr}}$.

Consider the quantum Liouville equation for the single-particle density operator 
\begin{equation}
\label{qke}
\frac{\pd \rho}{\pd t} + \frac{i}{\hbar} \left[ H, \rho \right]
=
0.
\end{equation}
We express the density operator as the sum of a disorder-averaged part and a fluctuating part, 
$\rho = \braket{\rho} + \delta \rho$. Then, within the Born approximation, Eq.~(\ref{qke}) decomposes as
\begin{subequations}
\label{qke_decomp}
\begin{gather}
\label{qke_avg}
\pd_t \braket{\rho}
+
\frac{i}{\hbar} \left[ H^0, \braket{\rho} \right]
+
J \left( \braket{\rho}\right)
=
- \frac{i}{\hbar} \left[ V, \braket{\rho} \right],
\\
\label{qke_fluc}
\pd_t \delta \rho
+
\frac{i}{\hbar} \left[ H^0, \delta \rho \right]
=
- \frac{i}{\hbar} \left[ U, \braket{\rho} \right]
-
\frac{i}{\hbar} \left[ V, \delta \rho \right],
\end{gather}
\end{subequations}
where $J \left( \braket{\rho}\right) \equiv \frac{i}{\hbar} \braket{ \left[ U, \delta \rho \right]}$ is the collision integral due to disorder scattering. For the purpose of studying nonlinear responses, we consider the field expansion of the density operator,
$\braket{\rho}
=
\sum_{n=0} \braket{\rho_{E^n}}$
and 
$\delta \rho
=
\sum_{n=0} \delta \rho_{E^n}$ (with the convention 
$A_0 \equiv A_{E^0}$), which yields the solution
\begin{equation}
\label{delta_rho_0}
\delta \rho_0 (t)
=
\int_{-\infty}^{\infty} \frac{d\veps}{2 \pi i}
G^R(\veps) \left[ U, \braket{\rho_0 (t)}\right] G^A(\veps),
\end{equation}
for the field-independent part and the recursive solution
\begin{equation}
\label{delta_rho_n}
\begin{split}
\delta \rho_{E^n} (t)
=&
\int_{-\infty}^{\infty} \frac{d\veps}{2\pi i}
G^R(\veps) \mathcal{R}_{E^n}(t) G^A(\veps),
\end{split}
\end{equation}
for the higher-order fluctuations, where
$\mathcal{R}_{E^n}(t)
=
[ U, \braket{\rho_{E^n} (t)} ] + [ V, \delta \rho_{E^{n-1}} (t)]$
and
$G^{R/A}(\veps)
=
\left( \veps - H^0 \pm i \delta \right)^{-1}$ 
is the retarded (advanced) Green's function of the unperturbed system~\footnote{Here, as in Ref.~\cite{atencia2022semiclassical}, we assume the Markovian approximation, whereby time derivatives of the fluctuations can be neglected.}. To exploit the sum separability of the recursive solution, it is convenient to henceforth introduce the notation with square brackets as 
$\delta \rho_{E^n}(\braket{\rho_{0}}, \ldots, \braket{\rho_{E^n}} )
\equiv
\sum_{l=0}^n
\delta \rho_{E^n} \left[ \braket{\rho_{E^l}} \right]$
to specify individual contributions to the fluctuations.

In order to solve for the disorder-averaged part of the density operator, we first express Eq.~(\ref{qke_avg}) in the crystal momentum representation, which gives rise to the quantum kinetic equation
\begin{equation}
\label{qke_avgk}
\pd_t f_{\mbk}
+
\frac{i}{\hbar} \left[ H_{\mbk}^0, f_{\mbk} \right]
+
J_{\mbk} \left( f_{\mbk}\right)
=
\frac{e}{\hbar} \mathbf{E} \cdot \bs{\mathcal{D}}_{\mbk} f_{\mbk},
\end{equation}
where $f_{\mbk}^{ab} \equiv
\braket{\psi_{\mbk a} | \braket{\rho} | \psi_{\mbk b}}$ is the distribution function and $\bs{\mathcal{D}}_{\mbk}$ is the Berry covariant derivative, which is obtained from the the k-space representation of the position operator~\cite{blount1962formalisms}, 
$\mathbf{r}_{\mbk \mbk\pr}
=
-i \bs{\pd}_{\mbk\pr} \delta_{\mbk\pr \mbk}
+
\delta_{\mbk\pr \mbk} \bs{\mathcal{A}}_{\mbk}$, resulting in the familiar form 
\begin{equation}
\bs{\mathcal{D}}_{\mbk} \mathcal{O}_{\mbk}
=
 \bs{\pd}_{\mbk} \mathcal{O}_{\mbk}
 -
 i \left[
 \bs{\mathcal{A}}_{\mbk}, \mathcal{O}_{\mbk} \right],
\end{equation}
for a generic k-diagonal operator $ \mathcal{O}_{\mbk}$. To second order in the electric field, the momentum-space density operator reads
$f_{\mbk}
=
f_{0\mbk} + f_{E\mbk} + f_{E^2\mbk}$, with $f_{0\mbk}$ the equilibrium distribution, which--for most practical purposes--is the Fermi-Dirac distribution, and we further decompose the field-dependent parts to diagonal and off-diagonal components in band space as 
$f_{E\mbk} = n_{E\mbk} + S_{E\mbk}$ and $f_{E^2\mbk} = n_{E^2\mbk} + S_{E^2\mbk}$. Then, through a perturbative analysis in powers of the electric field and disorder parameter, we can solve for both the diagonal part $n_{\mbk}$ and the off-diagonal components $S_{\mbk}$. To achieve this, the collision integral is also decomposed using the square-bracket notation,  
$J_{\lambda \mbk} \left( f_{\eta \mbk} \right)
\equiv
\frac{i}{\hbar}
\Braket{
\left[ U , \delta \rho_{\lambda} \left[ \braket{\rho_{\eta} } \right] \right]
}_{\mbk}$, with $\lambda, \eta = 0,E,E^2$, the explicit form of which is presented in Appendix~\ref{App_collision_int}.

\section{Linear density matrix}
\label{sec_linear}

First, we revisit the quantum kinetic equation at linear order in the applied electric field. This transport regime has been extensively studied in the literature and the discussion presented in this section is not new. However, since the contributions to the linear-response densities and their physical interpretations are required for obtaining and classifying quadratic-response quantities, for completeness--as well as for notational uniformity--we briefly review here the main results. A detailed treatment of this can be found in Ref.~\cite{atencia2022semiclassical}.

As $n_{E\mbk}$ is at most linear the scattering time, we start at $O(n_I^{-1})$ in the perturbative analysis and take 
$n_{E\mbk} = n_{E\mbk}^{(-1)} + n_{E\mbk}^{(0)}$. Assuming a steady-state solution for the band-diagonal carrier density, inspection of Eq.~(\ref{qke_avgk}) reveals that the lowest-order contribution from $S_{E\mbk}$ is at $O(n_I^{0})$. Thus, the field-linear contribution to the quantum kinetic equation is decomposed as
\begin{subequations}
\label{qke_avgk_E}
\begin{gather}
\label{qke_avgk_E_a}
J_{0\mbk}^a (n_{E\mbk}^{(-1)})
=
\frac{e}{\hbar} \mathbf{E} \cdot \bs{\pd}_{\mbk} f_{0\mbk}^a,
\\
\label{qke_avgk_E_b}
\pd_t S_{E \mbk}^{(0)}
+
\frac{i}{\hbar} \left[ H^0, S_{E \mbk}^{(0)} \right]
=
D_{E \mbk}^{(0)}
+
I_{E \mbk}^{(0)},
\\
\label{qke_avgk_E_c}
J_{0\mbk} ( f_{E\mbk}^{(0)})
+
J_{E\mbk} ( f_{0\mbk})
= 0,
\end{gather}
\end{subequations}
where Eqs.~(\ref{qke_avgk_E_a}) and (\ref{qke_avgk_E_b}) are--respectively--the diagonal and off-diagonal elements of Eq.~(\ref{qke_avgk}) at $O(n_I^0)$, while Eq.~(\ref{qke_avgk_E_c}) represents the $O(n_I)$ terms. In Eq.~(\ref{qke_avgk_E_b}), the evolution of the off-diagonal density matrix $S_{E \mbk}^{(0)}$ is governed by a term originating from the covariant derivative,
$D_{E \mbk}^{(0) ab}
=
\frac{ie}{\hbar} \mathbf{E} \cdot \mathbf{\mathcal{A}}_{\mbk}^{ab}
\left( f_{0\mbk}^a - f_{0 \mbk}^b \right)$, as well as a collision-integral-type term 
$I_{E \mbk}^{(0) ab}
=
- J_{0 \mbk}^{ab} ( n_{E \mbk}^{(-1)} )$. 

Assuming a transport time $\tau_{\mbk}^a$ that takes into account both self-energy and vertex corrections in the diagrammatic language, Eq.~(\ref{qke_avgk_E_a}) is readily solved as
\begin{equation}
\label{n_Ek_-1}
n_{E \mbk a}^{(-1)}
=
\frac{e}{\hbar} 
\tau_{\mbk}^a \mathbf{E} \cdot \bs{\pd}_{\mbk}
f_{0 \mbk}^{a}.
\end{equation}
For a general anisotropic system, extracting the carrier density from the collision integral is often challenging, which translates to difficulty in obtaining a closed-form solution for the transport time. However, in the isotropic limit, $\tau_{\mbk} \rightarrow \tau_{0\mbk}$, one obtains the familiar form with the angular weighting factor arising from vertex corrections~\cite{mahan2000many}
\begin{equation}
\label{tau_0k}
\frac{1}{\tau_{0 \mbk}^a}
=
\frac{2\pi}{\hbar} 
\sum_{\mbkpr b}
\delta ( \veps_{0 k}^a - \veps_{0 k^{\prime}}^b)
\Braket{
U_{\mbk \mbkpr}^{ab} U_{\mbkpr \mbk} ^{ba} 
}
\left[ 1 - \cos (\phi - \phi\pr) \right],
\end{equation}
with $\veps_{0 k}^a$ an energy eigenvalue of the isotropic subsystem and $\phi$ the azimuthal angle in momentum space. This solution for the transport time can then be used as a perturbative basis for obtaining the full transport time in weakly anisotropic systems.  

The solution to Eq.~(\ref{qke_avgk_E_b}) is also straightforward, leading to the following form for the band-off-diagonal density matrix
\begin{equation}
\label{S_Ek_0}
S_{E \mbk}^{(0)} 
=
\hbar \int_{-\infty}^{\infty} \frac{d\veps}{2 \pi}
G^R (\veps) \left[ D_{E \mbk}^{(0)}
+
I_{E \mbk}^{(0)} \right] G^A (\veps),
\end{equation}
which we express here as 
$S_{E \mbk}^{(0)}
=
S_{E \mbk}^{(0)} \left[D_{E \mbk}^{(0)} \right]
+
S_{E \mbk}^{(0)} \left[ I_{E \mbk}^{(0)} \right]$ for brevity and relegate the full forms to Appendix~\ref{App_off-diagonal}. The off-diagonal nature of $S_{E \mbk}^{(0)} \left[ I_{E \mbk}^{(0)} \right]$ given by Eq.~(\ref{S_Ek_0I}) implies that the contribution of this term to the equilibrium collision integral consists of antisymmetric parts of the Gaussian disorder distribution, which correspond to skew scattering in the semiclassical theory. $D_{E \mbk}^{(0)}$, however, is a function of the equilibrium distribution function. Therefore, its contribution is included in the electric-field correction to $f_{0\mbk}$, which semiclassically describes a side-jump process. This then reveals the physical nature of Eq.~(\ref{qke_avgk_E_c}); introducing the decomposition 
$n_{E\mbk}^{(0)}
=
n_{E\mbk}^{(\text{sk})} + n_{E\mbk}^{(\text{sj})}$, we arrive at the two equations
\begin{subequations}
\label{linear_carrier_eq}
\begin{align}
\label{J0k_nEk_a}
&J_{0\mbk}^a \left( n_{E\mbk}^{(\text{sk})} \right)
+
J_{0\mbk}^a \left( S_{E \mbk}^{(0)} \left[ I_{E \mbk}^{(0)} \right] \right)
=
0,
\\
\label{J0k_nEk_b}
&J_{0\mbk}^a \left( n_{E\mbk}^{(\text{sj})} \right)
+
J_{E\mbk}^a \left( f_{0\mbk} \right)
+
J_{0\mbk}^a \left( S_{E \mbk}^{(0)} \left[ D_{E \mbk}^{(0)} \right] \right)
=
0,
\end{align}
\end{subequations}
which are then readily solved for the skew-scattering and side-jump carrier densities (see Appendix~\ref{App_carrier_densities} for the explicit forms of the carrier densities). This completes the derivation of the field-linear density matrix.

\section{Quadratic density matrix}
\label{sec_quad}

We now extend this framework to nonlinear responses. In the quadratic response regime, wherein one is also interested in the second-order deviation of an electron distribution from equilibrium, there are also $O(\tau^2)$ contributions to the density operator. Thus, the appropriate expansion to consider is 
$n_{E^2\mbk} = n_{E^2\mbk}^{(-2)} + n_{E^2\mbk}^{(-1)} + n_{E^2\mbk}^{(0)}$, where we retain sub-subleading disorder corrections as well. This is an essential formal inclusion, partly due to the general expectation that intrinsic effects should appear at any order in the field expansion in the absence of disorder. Similar to the linear response regime--where band-off-diagonal elements do not appear at leading order in the disorder--for steady-state solutions, one can readily verify that $S_{E^2\mbk}^{(-2)}$ will not contribute to the density operator. Thus, the field-quadratic decomposition of the quantum kinetic equation results in the five general equations
\begin{subequations}
\label{qke_avgk_E2}
\begin{gather}
\label{qke_avgk_E2_a}
J_{0\mbk}^a \left(n_{E^2\mbk}^{(-2)}\right)
=
\frac{e}{\hbar} \mathbf{E} \cdot \bs{\pd}_{\mbk} n_{E\mbk a}^{(-1)},
\\
\label{qke_avgk_E2_b}
\pd_t S_{E^2 \mbk}^{(-1)}
+
\frac{i}{\hbar} \left[ H^0, S_{E^2 \mbk}^{(-1)} \right]
=
D_{E^2 \mbk}^{(-1)}
+
I_{E^2 \mbk}^{(-1)},
\\
\label{qke_avgk_E2_c}
J_{0 \mbk}^a \left( f_{E^2\mbk}^{(-1)}\right)
+
J_{E\mbk}^a \left( f_{E\mbk}^{(-1)}\right)
=
\frac{e}{\hbar} \mathbf{E} \cdot
\left( \bs{\mathcal{D}}_{\mbk} f_{E\mbk}^{(0)} \right)^a,
\\
\label{qke_avgk_E2_d}
\begin{split}
\pd_t S_{E^2 \mbk}^{(0)}
+
\frac{i}{\hbar} \left[ H^0, S_{E^2 \mbk}^{(0)} \right]
&=
D_{E^2 \mbk}^{(0)}
+
D_{E^2 \mbk}^{\prime (0)}
+
I_{E^2 \mbk}^{(0)}
\\
&+
I_{E^2 \mbk}^{\prime (0)}
+
I_{E^2 \mbk}^{\prime \prime (0)},
\end{split}
\\
\label{qke_avgk_E2_e}
J_{0\mbk} \left( f_{E^2\mbk}^{(0)} \right)
+
J_{E\mbk} \left( f_{E\mbk}^{(0)} \right)
+
J_{E^2\mbk} \left( f_{0\mbk}\right) = 0.
\end{gather}
\end{subequations}
The solution to Eq.~(\ref{qke_avgk_E2_a}) for the nonlinear density is simply
\begin{equation}
\label{n_E2_-2}
n_{E^2 \mbk a}^{(-2)}
=
\left( \frac{e}{\hbar}\right)^2 E_{\mu} E_{\nu} \tau_{\mbk}^a \pd_{\mbk}^{\mu}
( \tau_{\mbk}^a \pd_{\mbk}^{\nu} f_{0 \mbk}^a ),
\end{equation}
which quantifies the ordinary scattering of electrons at second order in the field expansion. Below, we discuss in detail the solutions of the remaining four equations and their associated physical processes. 

\subsection{Off-diagonal elements}

Consider first the off-diagonal elements. The evolution equation for $S_{E^2 \mbk}^{(-1)}$, Eq.~(\ref{qke_avgk_E2_b}), is quite similar to that of $S_{E \mbk}^{(0)}$, given by Eq.~(\ref{qke_avgk_E_b}), and may be regarded as its quadratic-response counterpart, with the analogous higher-order driving terms
$D_{E^2 \mbk}^{(-1) ab}
=
\frac{ie}{\hbar} \mathbf{E} \cdot \mathbf{\mathcal{A}}_{\mbk}^{ab}
( n_{E \mbk a}^{(-1)} - n_{E \mbk b}^{(-1)})$ and
$I_{E^2 \mbk}^{(-1) ab}
=
- J_{0 \mbk}^{ab} ( n_{E^2 \mbk}^{(-2)} )$ and the solution
\begin{equation}
\label{S_E2k_-1}
S_{E^2 \mbk}^{(-1)} 
=
\hbar \int_{-\infty}^{\infty} \frac{d\veps}{2 \pi}
G^R (\veps) \left[ D_{E^2 \mbk}^{(-1)}
+
I_{E^2 \mbk}^{(-1)} \right] G^A (\veps),
\end{equation}

A quick comparison between Eq.~(\ref{S_Ek_0_decomp}) and Eq.~(\ref{S_E2k_-1_decomp}) reveals that $S_{E^2 \mbk}^{(-1)}$ simply corresponds to $S_{E \mbk}^{(0)}$ with the densities replaced by their higher-order counterparts in the electric field. The situation is different for the other off-diagonal component, $S_{E^2 \mbk}^{(0)}$, which includes terms that have no linear-response counterparts. As indicated in Eq.~(\ref{qke_avgk_E2_d}), its evolution is determined by five distinct driving terms, which consist of two covariant-derivative-type terms and three collision integral-type terms, given by 
\begin{subequations}
\label{S_E2k_0_terms}
\begin{align}
\label{S_E2k_0_terms_a}
D_{E^2 \mbk}^{(0) ab}
&=
\frac{ie}{\hbar} \mathbf{E} \cdot \mathbf{\mathcal{A}}_{\mbk}^{ab}
\left[ n_{E \mbk a}^{(0)} - n_{E \mbk b}^{(0)} \right],
\\
\label{S_E2k_0_terms_b}
D_{E^2 \mbk}^{\prime (0) ab}
&=
\frac{e}{\hbar} \mathbf{E} \cdot
\left[
\bs{\mathcal{D}}_{\mbk} S_{E \mbk}^{(0)}
\right]^{ab},
\\
\label{S_E2k_0_terms_c}
I_{E^2 \mbk}^{(0) ab}
&=
- J_{0 \mbk}^{ab} \left( n_{E^2 \mbk}^{(-1)} \right),
\\
\label{S_E2k_0_terms_d}
I_{E^2 \mbk}^{\prime (0) ab}
&=
- J_{E \mbk}^{ab} \left( n_{E \mbk}^{(-1)} \right),
\\
\label{S_E2k_0_terms_e}
I_{E^2 \mbk}^{\prime \prime (0) ab}
&=
- J_{0 \mbk}^{ab} \left( S_{E^2 \mbk}^{(-1)} \right),
\end{align}
\end{subequations}
with the familiar formal solution
\begin{equation}
\label{S_E2k_0}
\begin{split}
S_{E^2 \mbk}^{(0)} 
&=
\hbar \int_{-\infty}^{\infty} \frac{d\veps}{2 \pi}
G^R (\veps)
\left[ D_{E^2 \mbk}^{(0)}
+
D_{E^2 \mbk}^{\prime (0)}
+
I_{E^2 \mbk}^{(0)}
\right.
\\
&\left.+
I_{E^2 \mbk}^{\prime (0)}
+
I_{E^2 \mbk}^{\prime \prime (0)} \right]
G^A (\veps).
\end{split}
\end{equation}
It is worth noting that the two driving terms $D_{E^2 \mbk}^{\prime (0)}$ and $I_{E^2 \mbk}^{\dprime (0)}$ are themselves functions of off-diagonal density matrix elements and can therefore be further decomposed as
\begin{subequations}
\label{S_E2k_Dpr_Idpr}
\begin{align}
S_{E^2 \mbk}^{(0)} \left[D_{E^2 \mbk}^{\prime (0)} \right]
&=
S_{E^2 \mbk}^{(0)} \left[D_{E^2 \mbk}^{\prime (0)} [ D_{E \mbk}^{(0)}]
\right]
+
S_{E^2 \mbk}^{(0)} \left[D_{E^2 \mbk}^{\prime (0)} [ I_{E \mbk}^{(0)}]
\right],
\\
S_{E^2 \mbk}^{(0)} \left[I_{E^2 \mbk}^{\dprime (0)} \right]
&=
S_{E^2 \mbk}^{(0)} \left[I_{E^2 \mbk}^{\dprime (0)} [ D_{E^2 \mbk}^{(-1)}]
\right]
+
S_{E^2 \mbk}^{(0)} \left[I_{E^2 \mbk}^{\dprime (0)} [ I_{E^2 \mbk}^{(-1)}]
\right],
\end{align}
\end{subequations}

We thus observe that, in comparison to the relatively simple picture in the linear response regime, there are a large number of terms which contribute to the off-diagonal density matrix elements in the quadratic response regime, reflecting a wealth of interconnected scattering processes that are illustrated schematically in Fig.~\ref{fig1}. We next discuss the derivation of the carrier densities and the ensuing physical classification of these processes.
\vspace{-.2cm}
\subsection{Mixed scattering}

Consider Eq.~(\ref{qke_avgk_E2_c}), which must be solved for $n_{E^2 \mbk}^{(-1)}$. Based on the previously stated correspondence between $S_{E^2 \mbk}^{(-1)}$ and $S_{E \mbk}^{(0)}$, it is straightforward to conclude that this equation is essentially the field-quadratic counterpart to Eq.~(\ref{qke_avgk_E_c}) and thus describes the special (skew, side-jump) scattering of ordinary-scattered electrons. A difference, however, with the linear response is the additional covariant-derivative term on the right-hand side of Eq.~(\ref{qke_avgk_E2_c}), which includes terms that physically describe the ordinary scattering of electrons having previously undergone side jump or skew scattering at linear order in the electric field [see Eqs.~(\ref{n_Ek^sk}) and (\ref{n_Ek^sj})]. The conclusion, therefore, is that $n_{E^2 \mbk}^{(-1)}$ measures the density of electrons which undergo a mixture of consecutive special and ordinary scatterings and thus consists of the four distinct contributions
\begin{equation}
\label{n_E^2ka^-1}
n_{E^2 \mbk a}^{(-1)}
=
n_{E^2 \mbk a}^{(\text{sj,o})}
+
n_{E^2 \mbk a}^{(\text{sk,o})}
+
n_{E^2 \mbk a}^{(\text{o,sj})}
+
n_{E^2 \mbk a}^{(\text{o,sk})},
\end{equation}
where
$n_{E^2 \mbk}^{(\text{sp,o})}$ is the density of ordinary-scattered electrons which subsequently experience special scattering (sp=sj,sk), and $n_{E^2 \mbk}^{(\text{o,sp})}$ is the density of special-scattered electrons which then undergo ordinary scattering. As a result, Eq.~(\ref{qke_avgk_E2_c}) is also decomposed as
\begin{subequations}
\begin{align}
\begin{split}
&J_{0\mbk}^a \left( n_{E^2\mbk}^{(\text{sj,o})} \right)
+
J_{0\mbk}^a \left( S_{E^2 \mbk}^{(-1)} \left[ D_{E^2 \mbk}^{(-1)} \right] \right)
+
J_{E \mbk}^a \left( n_{E \mbk}^{(-1)}\right)
\\
&=
- \frac{i e}{\hbar} \mathbf{E} \cdot
\left[ \bs{\mathcal{A}}_{\mbk} ,
S_{E \mbk}^{(0)} \right]^a,
\end{split}
\\
&J_{0\mbk}^a \left( n_{E^2\mbk}^{(\text{sk,o})} \right)
+
J_{0\mbk}^a \left( S_{E^2 \mbk}^{(-1)} ]\left[ I_{E^2 \mbk}^{(-1)} \right] \right)
=
0,
\\
&J_{0\mbk}^a \left( n_{E^2 \mbk}^{(\text{o,sj})} \right)
=
\frac{e}{\hbar} \mathbf{E} \cdot 
\bs{\pd}_{\mbk} n_{E \mbk a}^{(\text{sj})},
\\
&J_{0\mbk}^a \left( n_{E^2 \mbk}^{(\text{o,sk})} \right)
=
\frac{e}{\hbar} \mathbf{E} \cdot 
\bs{\pd}_{\mbk} n_{E \mbk a}^{(\text{sk})},
\end{align}
\end{subequations}
the solutions of which--presented in Appendix~\ref{App_carrier_densities}--yield the mixed-scattering carrier densities.

\subsection{Zeroth-order scattering}

We next turn to Eq.~(\ref{qke_avgk_E2_e}), the last of the five main equations arising from the quantum kinetic equation in the quadratic response regime, which must be solved for $n_{E^2 \mbk}^{(0)}$. Similar to the special scattering processes in the linear regime, this includes the density matrix elements which are formally of $O(n_I^0)$ in the disorder expansion, but can nevertheless arise from the concerted actions of disorder and the geometry of Bloch states. A relatively large number of collision integrals contribute to the density matrix at this order, leading to a multitude of scattering processes. Upon inspection, these are classified as five physically distinct processes, which we label as
\begin{equation}
n_{E^2 \mbk}^{(0)}
=
n_{E^2 \mbk}^{\text{(ssj)}}
+
n_{E^2 \mbk}^{\text{(qsj)}}
+
n_{E^2 \mbk}^{\text{(tsk)}}
+
n_{E^2 \mbk}^{\text{(csk)}}
+
n_{E^2 \mbk}^{\text{(ask)}},
\end{equation}
resulting in the decomposition
\begin{subequations}
\label{zeroth_collision}
\begin{align}
\label{J0k_nE^2k_c}
&J_{0\mbk}^a \left( n_{E^2 \mbk}^{(\text{ssj})} \right)
+
J_{0\mbk}^a \left( S_{E^2 \mbk}^{(0)}\left[ D_{E^2 \mbk}^{(0)} \right] \right)
+
J_{E\mbk}^a \left( n_{E \mbk}^{(0)} \right)
=
0,
\\
\label{J0k_nE^2k_d}
&J_{0\mbk}^a \left( n_{E^2 \mbk}^{(\text{csk})} \right)
+
J_{0\mbk}^a 
\left( S_{E^2 \mbk}^{(0)} \left[I_{E^2 \mbk}^{\dprime (0)} [ I_{E^2 \mbk}^{(-1)}]
\right] \right)
=
0,
\\
\label{J0k_nE^2k_e}
&J_{0\mbk}^a \left( n_{E^2 \mbk}^{(\text{tsk})} \right)
+
J_{0\mbk}^a \left( S_{E^2 \mbk}^{(0)}\left[ I_{E^2 \mbk}^{(0)} \right] \right)
=0,
\\
\label{J0k_nE^2k_a}
&J_{0\mbk}^a \left( n_{E^2\mbk}^{(\text{ask})} \right)
+
\mathcal{I}_1
= 0,
\\
\label{J0k_nE^2k_b}
&J_{0\mbk}^a \left( n_{E^2\mbk}^{(\text{qsj})} \right)
+
\mathcal{I}_2
= 0,
\end{align}
\end{subequations}
with the lengthier expressions $\mathcal{I}_1$ and $\mathcal{I}_2$ presented in Appendix~\ref{app_I_1}. Below, we elucidate this classification and elaborate on the physical significance of each term.
\vspace{-.2cm}
\subsubsection{secondary side jump}

Consider first Eq.~(\ref{J0k_nE^2k_c}). Note that this equation is structurally analogous to Eq.~(\ref{J0k_nEk_a}) and thus describes a side-jump process, with the difference that it acts on $n_{E\mbk}^{(0)}$. It is then clear that $n_{E^2 \mbk}^{(\text{ssj})}$ measures the density the electrons that are first subjected to a side-jump or skew scattering process and then undergo an additional side jump. This secondary side-jump density is thus itself comprised of two terms
\begin{equation}
n_{E^2 \mbk a}^{\text{(ssj)}}
=
n_{E^2 \mbk a}^{\text{(sj,sj)}}
+
n_{E^2 \mbk a}^{\text{(sj,sk)}},
\end{equation}
each of which is obtained by solving the relevant contribution to Eq.~(\ref{J0k_nE^2k_c}).
\vspace{-.2cm}
\subsubsection{tertiary skew scattering}

Given the presence of secondary side-jump scatterings, it is natural to ask whether a special scattering event could be proceeded by a skew-scattering process as well. To answer this in the affirmative, consider next Eq.~(\ref{J0k_nE^2k_e}). Based on the form of the collision integral
$J_{0\mbk}^a \left( S_{E^2 \mbk}^{(0)}\left[ I_{E^2 \mbk}^{(0)} \right] \right)$ given in Appendix~\ref{App_collision_int}, it is evident that this measures the skew scattering of electrons which have initially experienced mixed scattering. Thus, as illustrated in Fig.~\ref{fig1}, overall, this describes a three-stage scattering process at least, which we refer to as tertiary skew scattering. From the fourfold decomposition of $n_{E^2\mbk}^{(-1)}$ given by Eq.~(\ref{n_E^2ka^-1}), it is clear that $n_{E^2 \mbk a}^{\text{(tsk)}}$ also decomposes as
\begin{equation}
n_{E^2 \mbk a}^{\text{(tsk)}}
=
n_{E^2 \mbk a}^{\text{(sk,sj,o)}}
+
n_{E^2 \mbk a}^{\text{(sk,o,sj)}}
+
n_{E^2 \mbk a}^{\text{(sk,sk,o)}}
+
n_{E^2 \mbk a}^{\text{(sk,o,sk)}},
\end{equation}
yielding four distinct contributions.
\vspace{-.5cm}

\subsubsection{quadratic side jump}

In the quadratic response regime, there is yet another source of side-jump scattering. In addition to the secondary side jump, which arises from collision integrals containing corrections to $n_{E\mbk}^{(0)}$ that are linear in the electric field, one must also take into consideration field-quadratic corrections to the equilibrium distribution function itself. These quadratic side-jump contributions are obtained by solving Eq.~(\ref{J0k_nE^2k_b}) for $n_{E^2 \mbk}^{\text{(qsj)}}$, which reads
\begin{equation}
\label{n_E2k_qsj}
\begin{split}
n_{E^2\mbk a}^{(\text{qsj})}
&=
- \tau_{\mbk}^a
\left\{
J_{0\mbk}^a 
\left( S_{E^2 \mbk}^{(0)} \left[D_{E^2 \mbk}^{\prime (0)} [ D_{E \mbk}^{(0)}]
\right] \right)
\right.
\\
&\left.+
J_{E\mbk}^a \left( S_{E \mbk}^{(0)} \left[ D_{E \mbk}^{(0)} \right] \right)
+
J_{E^2 \mbk}^a 
\left( f_{0 \mbk} \right)
\right\}.
\end{split}
\end{equation}
The explicit form of $n_{E^2 \mbk}^{\text{(qsj)}}$ is rather lengthy and can be found in Appendix~\ref{App_carrier_densities}. It is interesting to note, however, the appearance of the quantum metric in Eq.~(\ref{N_qsj_8}), which then compels one to interpret the quadratic side jump as being partly driven by the quantum metric tensor. This is in line with the findings of Ref.~\cite{gao2014field}, which relates the quantum metric to the field-induced positional shift of the carriers.
\vspace{-.55cm}
\subsubsection{cubic skew scattering}

In addition to the tertiary skew scattering, the carrier density of which contains $\braket{UU}^2$-type terms, there is another source of skew scattering, which consists of terms that are cubic in $\braket{UU}$ and acts on the density of ordinary-scattered electrons, $n_{E^2 \mbk}^{(-2)}$, guaranteeing that the resultant density is of O($n_I^0$). The carrier density from this cubic skew scattering is obtained by solving Eq.~(\ref{J0k_nE^2k_d}), yielding
\begin{equation}
n_{E^2 \mbk a}^{(\text{csk})}
=
- \tau_{\mbk}^a
J_{0\mbk}^a 
\left( S_{E^2 \mbk}^{(0)} \left[I_{E^2 \mbk}^{\dprime (0)} [ I_{E^2 \mbk}^{(-1)}]
\right] \right),
\end{equation}
where, once again, we relegate the lengthy explicit form to Appendix~\ref{App_carrier_densities}. 

\subsubsection{anomalous skew scattering}

Lastly, we consider Eq.~(\ref{J0k_nE^2k_a}), the solution of which is
\begin{equation}
\begin{split}
n_{E^2\mbk a}^{(\text{ask})}
&=
-\tau_{\mbk}^a
\left\{
J_{0\mbk}^a 
\left( S_{E^2 \mbk}^{(0)} \left[D_{E^2 \mbk}^{\prime (0)} [ I_{E \mbk}^{(0)}]
\right] \right)
\right.
\\
&\left.+
J_{0\mbk}^a 
\left( S_{E^2 \mbk}^{(0)} \left[I_{E^2 \mbk}^{\dprime (0)} [ D_{E^2 \mbk}^{(-1)}]
\right] \right)
\right.
\\
&\left.+
J_{E\mbk}^a 
\left( S_{E^2 \mbk}^{(0)} \left[I_{E^2 \mbk}^{\prime (0)} \right] \right)
+
J_{E\mbk}^a 
\left( S_{E \mbk}^{(0)} \left[I_{E \mbk}^{(0)} \right] \right)
\right\}.
\end{split}
\end{equation}
This describes an anomalous skew scattering process, whereby electrons which scatter ordinarily at linear order in the electric field, $n_{E \mbk}^{(-1)}$, subsequently undergo a field-dependent scattering with a skew scattering disorder profile. As is clear from the relevant expressions in Appendix~\ref{App_carrier_densities}, the resultant carrier density is thus comprised of terms proportional to the product of the covariant derivative and the antisymmetric parts of $\braket{UU} \braket{UU}$-type terms.
\vspace{-.5cm}

\subsection{General remarks on the formalism}

Having presented the derivation of the density matrix, we note that a feature of the density-matrix formalism that becomes evident in the kinetic approach is the ability to provide a detailed mapping of the scattering channels that arise through the various processes and the interconnections between the various carrier densities. One could argue that this feature is slightly obscured in other methods, such as diagrammatic considerations, which typically do not deal directly with carrier densities. An alternative distinction scheme, which complements the viewpoint presented here, is to classify all non-ordinary scatterings as either side jump or skew scattering, where the former process would involve contributions from the conduction band only, while the latter receives contributions from both the valence and conduction bands. Despite this, as we illustrate for a specific model in Section~\ref{application}, one could argue for the merits of a more detailed distinction between the various scatterings, as processes which would otherwise fall within the same classification subtype turn out to have noticeably different scaling signatures in the response of the system.

In closing this section, it is also useful to briefly compare the formalism presented in this work to another recent quantum kinetic approach discussed in Ref.~\cite{bhalla2023quantum}. There, the collision integral is approximated as being proportional to the density matrix via a constant relaxation time such that scatterings are accounted for by introducing intraband and interband relaxation times for the diagonal and off-diagonal components of the quantum kinetic equation, respectively. In the present approach, we assume the diagonal elements of the collision integral are proportional to a general momentum-dependent transport time, the solution of which is obtained from extracting the carrier density from the collision integral. This results in a transport time that is physically equivalent to including self-energy and vertex corrections in its derivation~\cite{mahan2000many}. Furthermore, the solution of the off-diagonal quantum kinetic equation at different field and impurity orders emerges once the relevant carrier densities are known. Therefore, the transport time manifests in the off-diagonal density matrix as well through the carrier densities. Nevertheless, as is evident in both approaches, at higher orders beyond linear responses, intraband and interband processes are intricately connected, leading to a variety of physical phenomena in the nonlinear response regime.
\vspace{.5cm}

\section{Emergent Velocities and Modified Semiclassics}
\label{sec_emergent}

Having delved into the solution of the quantum Liouville equation, we may now proceed to evaluate the modified semiclassical equations of motion and, subsequently, the disorder-dressed quadratic current density. The former is found by adopting the proposed prescription outlined in Ref.~\cite{atencia2022semiclassical}, the essential idea of which is to reexpress the ensemble average of the velocity operator entirely in terms of the diagonal elements of the density matrix, \textit{i.e.}, the carrier densities. That is, we solve the equation
\begin{equation}
\label{rdot_v}
\sum_{\mbk a} \dot{\mathbf{r}}^a n_{\mbk a}
=
\sum_{\mbk a b} \mathbf{v}_{\mbk}^{b a} f_{\mbk}^{ab},
\end{equation}
for the time evolution of the carrier position 
$\dot{\mathbf{r}}^a$, where 
$\mathbf{v}_{\mbk} \equiv
(1/ \hbar) \bs{\mathcal{D}}_{\mbk} H_{\mbk}^0$ is the velocity operator. Using the results from the previous section, the right-hand side of Eq.~(\ref{rdot_v}) reads
\begin{equation}
\label{rdot_v_expanded}
\sum_{\mbk a b} \mathbf{v}_{\mbk}^{b a} f_{\mbk}^{ab}
=
\sum_{\mbk a}
\bs{v}_{\mbk}^a n_{\mbk a}
-
\frac{i}{\hbar} \sum_{\mbk a b}  \left(\veps_{\mbk}^a - \veps_{\mbk}^b \right)
\bs{\mathcal{A}}_{\mbk}^{b a} 
S_{\mbk}^{ab}.
\end{equation}
The first term on the right, which is already diagonal, describes the ordinary group velocity experienced by all the various carriers. Our goal then is to band-diagonalize the second term. For the linear-response term, inserting Eq.~(\ref{S_Ek_0}) leads to contributions from the anomalous and extrinsic velocities
\begin{equation}
\begin{split}
\label{S_Ek_contrib}
-\frac{i}{\hbar} \sum_{\mbk a b}
\left(\veps_{\mbk}^a - \veps_{\mbk}^b \right) \mathcal{A}_{\mbk}^{b a}
S_{E \mbk}^{(0)ab}
&=
\sum_{\mbk a}
\left[
\frac{e}{\hbar} \mathbf{E} \times \bs{\Omega}_{\mbk}^a f_{0 \mbk}^a
\right.
\\
&\left.+
\bs{\alpha}_{\mbk a} n_{E \mbk a}^{(-1)}
\right],
\end{split}
\end{equation}
where the extrinsic velocity is defined as~\cite{atencia2022semiclassical}
\begin{equation}
\bs{\alpha}_{\mbk}^a
=
\frac{1}{\hbar} \int_{-\infty}^{\infty} \frac{d\veps}{2\pi}
\Braket{
\left[ U, G^A(\veps) \left[ U, \bs{\mathcal{A}}\pr \right] G^R(\veps)\right]
}_{\mbk}^a,
\end{equation}
with the prime indicating that only off-diagonal elements of the Berry connection are evaluated--reflecting the gauge invariance of the extrinsic velocity. Explicitly, this reads
\begin{equation}
\label{alpha_ka}
\bs{\alpha}_{\mbk}^a
=
\frac{2}{\hbar} \sum_{\mbkpr b} \text{Im}
\left(
\frac{
\Braket{U_{\mbk \mbkpr}^{ab}
\left[ U, \bs{\mathcal{A}}^{\prime} \right]_{\mbkpr \mbk}^{ba}}}
{\veps_{\mbk}^a - \veps_{\mbkpr}^b - i \delta}
\right).
\end{equation}

We now extend this analysis to the quadratic response contributions. The first term to consider is
$S_{E^2 \mbk}^{(-1)}$. Based on the previously stated analogy with its linear response counterpart $S_{E \mbk}^{(0)}$, it is straightforward to verify that
\begin{equation}
\begin{split}
-\frac{i}{\hbar} \sum_{\mbk a b}
\left(\veps_{\mbk}^a - \veps_{\mbk}^b \right) \mathcal{A}_{\mbk}^{ba}
S_{E^2 \mbk}^{(-1)ab}
&=
\sum_{\mbk a}
\left[
\frac{e}{\hbar} \mathbf{E} \times \bs{\Omega}_{\mbk}^a n_{E \mbk a}^{(-1)}
\right.
\\
&\left.+
\bs{\alpha}_{\mbk}^a n_{E^2 \mbk a}^{(-2)}
\right].
\end{split}
\end{equation}
Thus the effect of $S_{E^2 \mbk}^{(-1)}$ is to simply describe the contributions of the anomalous and extrinsic velocities in the quadratic response regime, where we note that the anomalous contribution is that of the familar Berry curvature dipole~\cite{sodemann2015quantum}, albeit with the full disorder dependence of the transport time. Similarly, for the $S_{E^2 \mbk}^{(-1)}$ contributions from 
$D_{E^2 \mbk}^{(0)}$ and
$I_{E^2 \mbk}^{(0)}$, inserting the relevant terms of Eq.~(\ref{S_E2k_0}) into Eq.~(\ref{rdot_v_expanded}) yields   
\begin{equation}
\begin{split}
- \sum_{\mbk a b}
\mathcal{A}_{\mbk}^{b a}
\left[ D_{E^2 \mbk}^{(0)} + I_{E^2 \mbk}^{(0)} \right]^{ab}
&=
\sum_{\mbk a}
\left[
\frac{e}{\hbar} \mathbf{E} \times \bs{\Omega}_{\mbk}^a n_{E \mbk a}^{(0)}
\right.
\\
&\left.+
\bs{\alpha}_{\mbk}^a n_{E^2 \mbk a}^{(-1)}
\right],
\end{split}
\end{equation}
showing that the higher-order carriers also experience these two velocities. 

We next turn to the contributions from Eq.~(\ref{S_E2k_0}) which have no counterparts in the linear response regime. The first term to consider is 
$D_{E^2 \mbk}^{\prime (0)}$, given by Eq.~(\ref{S_E2k_0_terms_b}). After a little algebra, it can be shown that the contribution of this term to Eq.~(\ref{rdot_v_expanded}) is also comprised of an intrinsic part and an extrinsic part as
\begin{equation}
\begin{split}
\label{eom_metric_beta}
- \sum_{\mbk a b}
\mathcal{A}_{\mu \mbk}^{b a}
D_{E^2 \mbk}^{\prime (0) ab}
&=
\sum_{\mbk a}
\left[
\frac{e^2}{\hbar^2} E^{\nu} E^{\rho}
\sum_{b}
M_{\mbk}^{ab}
\Gamma_{\nu \rho \mu \mbk}^{b a} 
f_{0\mbk}^a
\right.
\\
&\left.+
\beta_{\mu \mbk}^{a} n_{E \mbk a}^{(-1)},
\right],
\end{split}
\end{equation}
where 
$M_{\mbk}^{ab}
\equiv
2 \hbar / (\varepsilon_{\mbk}^a - \varepsilon_{\mbk}^b)
$ with $a \neq b$. We thus see that from the off-diagonal density matrix, the geodesic term naturally emerges through this approach in the equations of motion and carrier transport, which is quite reasonable, given the Riemannian structure on the underlying manifold of quantum states~\cite{provost1980riemannian}.

The remaining term in Eq.~(\ref{eom_metric_beta}) arises from a new extrinsic velocity, $\bs{\beta}_{\mbk}$, given by
\begin{equation}
\begin{split}
\bs{\beta}_{\mbk}^a
&=
\frac{i}{\hbar}
\int_{-\infty}^{\infty} \frac{d\veps}{2 \pi}
\int_{-\infty}^{\infty} \frac{d\veps\pr}{2 \pi}
\\
&\times
\Braket{
\left[U, G^A(\veps\pr)
\left[U, G^A(\veps)
\left[V, \bs{\mathcal{A}}\pr \right] G^R(\veps)
\right]
G^R(\veps\pr)
\right]
}_{\mbk}^a,
\end{split}
\end{equation}
or explicitly as
\begin{equation}
\bs{\beta}_{\mbk}^a
=
\frac{2e}{\hbar} \sum_{\mbkpr b} \text{Re}
\left(
\frac{
\Braket{U_{\mbk \mbkpr}^{ab}
\left[ U, \bs{\mathcal{F}} \right]_{\mbkpr \mbk}^{ba}}}
{\veps_{\mbk}^a - \veps_{\mbkpr}^b - i \delta}
\right),
\end{equation}
with 
$\bs{\mathcal{F}}_{\mbk}^{ab}
=
- \left[ \left( 
\mathbf{E} \cdot \bs{\mathcal{D}} \right) \bs{\mathcal{A}}^{\prime} \right]_{\mbk}^{ab} 
/
( \veps_{\mbk}^a - \veps_{\mbk}^b + i \delta )$, 
which is linear in the applied electric field and thus appears only in the nonlinear transport regime, in contrast to the field-independent $\bs{\alpha}_{\mbk}$, which manifests in both linear and nonlinear responses. This then suggests a general pattern, in which extrinsic velocities that are at most of $O(E^{n-1})$ contribute to the $n$-th order response, in agreement with a recent study based on a semiclassical Boltzmann transport analysis~\cite{huang2023scaling}. This is indeed the case for the remaining quadratic-response velocities, as we show below.

Moving on to Eq.~(\ref{S_E2k_0_terms_d}), the contribution from 
$I_{E^2 \mbk}^{\prime(0)}$ is also that of an extrinsic velocity, which reads
\begin{equation}
- \sum_{\mbk a b}
\mathcal{A}_{\mu \mbk}^{b a}
I_{E^2 \mbk}^{\prime(0) ab} 
=
\sum_{\mbk a}
\gamma_{\mu \mbk}^{a} n_{E \mbk a}^{(-1)},
\end{equation}
with
\begin{equation}
\begin{split}
\bs{\gamma}_{\mbk}^a
&=
\frac{i}{\hbar}
\int_{-\infty}^{\infty} \frac{d\veps}{2 \pi}
\int_{-\infty}^{\infty} \frac{d\veps\pr}{2 \pi}
\\
&\times
\Braket{
\left[U, G^A(\veps\pr)
\left[V, G^A(\veps)
\left[ U, \bs{\mathcal{A}}\pr \right] G^R(\veps)
\right]
G^R(\veps\pr)
\right]
}_{\mbk}^a,
\end{split}
\end{equation}
or
\begin{equation}
\bs{\gamma}_{\mbk}^a
=
\frac{2e}{\hbar} \sum_{\mbkpr b} \text{Re}
\left(
\frac{
\Braket{U_{\mbk \mbkpr}^{ab}
\left[ \left( 
\mathbf{E} \cdot \bs{\mathcal{D}} \right) \bs{\mathcal{G}} \right]_{\mbkpr \mbk}^{ba}}
}
{\veps_{\mbk}^a - \veps_{\mbkpr}^b - i \delta}
\right),
\end{equation}
with 
$\bs{\mathcal{G}}_{\mbk \mbkpr}^{ab}
=
- \left[ U, \bs{\mathcal{A}}^{\prime} \right]_{\mbk \mbkpr}^{ab} 
/
( \veps_{\mbk}^a - \veps_{\mbkpr}^b + i \delta )$. As can be seen, this differs from $\bs{\beta}_{\mbk}$ by an exchange in the orders of the electrostatic and disorder potential in the formal definition.

Lastly, the contribution arising from $I_{E^2 \mbk}^{\prime \prime (0)}$ can be expressed as two separate extrinsic velocities acting on the linear and quadratic densities
\begin{equation}
- \sum_{\mbk a a\pr}
\mathcal{A}_{\mu \mbk}^{a\pr a}
I_{E^2 \mbk}^{\prime \prime (0) aa\pr}
=
\sum_{\mbk a}
\left[ \kappa_{\mu \mbk}^{a} n_{E \mbk a}^{(-1)} + \chi_{\mu \mbk}^{a} n_{E^2 \mbk a}^{(-2)} \right],
\end{equation}
with
\begin{equation}
\bs{\kappa}_{\mbk}^a
=
i e \int_{-\infty}^{\infty} \frac{d\veps}{2\pi} 
\left[
G^R(\veps)  \mathbf{E} \cdot \bs{\mathcal{A}}\pr G^A(\veps) , \bs{\alpha} \right]_{\mbk}^a,
\end{equation}
\begin{equation}
\begin{split}
\bs{\chi}_{\mbk}^a
&=
\int_{-\infty}^{\infty} \frac{d\veps}{2 \pi}
\int_{-\infty}^{\infty} \frac{d\veps\pr}{2 \pi}
\\
&\times
\Braket{
\left[U G^A(\veps\pr) G^A(\veps)
,
\bs{\alpha} G^R(\veps) U G^R(\veps\pr)
\right] + \text{h.c.} }_{\mbk}^a,
\end{split}
\end{equation}
which explicitly read
\begin{equation}
\bs{\kappa}_{\mbk}^a
=
2e \sum_b \text{Re}
\left(
\frac{\mathbf{E} \cdot \bs{\mathcal{A}}_{\mbk}^{\prime ab} \bs{\alpha}_{\mbk}^{\prime ba} 
}
{\veps_{\mbk}^a - \veps_{\mbk}^b - i \delta }
\right),
\end{equation}
and
\begin{equation}
\begin{split}
\bs{\chi}_{\mbk}^a
&=
2 \sum_{bc} \text{Re}
\left[
\frac{
\braket{U_{\mbk \mbkpr}^{ab} U_{\mbkpr \mbk}^{ca}} \bs{\alpha}_{\mbkpr}^{bc}}
{(\veps_{\mbk}^a - \veps_{\mbkpr}^b - i \delta)
(\veps_{\mbkpr}^b - \veps_{\mbkpr}^c + i \delta)}
\right.
\\
&\left.-
\frac{
\braket{U_{\mbk \mbkpr}^{bc} U_{\mbkpr \mbk}^{ca}} \bs{\alpha}_{\mbk}^{ab}}
{(\veps_{\mbkpr}^c - \veps_{\mbk}^a - i \delta)
(\veps_{\mbk}^a - \veps_{\mbk}^b + i \delta)}
\right].
\end{split}
\end{equation}
It is perhaps interesting to note from these terms that the off-diagonal elements of the linear-response velocity $\bs{\alpha}_{\mbk}$ also contribute to the transport, albeit as higher-order extrinsic terms.

Gathering all the above terms, and using the equation of motion for the carrier momentum, 
$\dot{\mathbf{k}}^a = -e \mathbf{E}/\hbar$,
we ultimately arrive at the equation of motion for the carrier position given by Eq.~(\ref{eom}). And at second order in the electric field, the total current density--defined as 
$\mathbf{j}
\equiv
-e \text{Tr} (\mathbf{v} \braket{\rho})$--is expressed as
\begin{widetext}
\begin{equation}
\label{j_E2_tot}
\begin{split}
j_{\mu E^2}
&=
-e \sum_{\mbk a} \left[ 
v_{\mu \mbk}^a  n_{E^2 \mbk a}
+
\frac{e}{\hbar} \varepsilon_{\mu \nu \rho} E_{\nu} \Omega_{\rho \mbk}^a n_{E \mbk a}
+
\left( \frac{e}{\hbar} \right)^2 E_{\nu} E_{\rho}
\sum_{b} M_{\mbk}^{ab}
\Gamma_{\nu \rho \mu \mbk}^{b a} f_{0 \mbk}^a
+
\alpha_{\mu \mbk}^{a} \left( 
n_{E^2 \mbk a}^{(-2)} + n_{E^2 \mbk a}^{(-1)}\right)
\right.
\\
&\left. 
\hspace{0.55\linewidth}
+
\left( \beta_{\mu \mbk}^{a}
+
\gamma_{\mu \mbk}^{a}
+
\kappa_{\mu \mbk}^{a} \right) n_{E \mbk a}^{(-1)}
+
\chi_{\mu \mbk}^{a} n_{E^2 \mbk a}^{(-2)},
\right].
\end{split}
\end{equation}
\vspace{-.5cm}
\end{widetext}
Several remarks regarding these results are in order. First, as one would expect, the leading-order disorder contribution to the quadratic current is
\begin{equation}
\label{j_E2_gr}
\bs{j}_{E^2}^{(-2)}
=
-e \sum_{\mbk a} \bs{v}_{\mbk}^a  n_{E^2 \mbk a}^{(-2)},
\end{equation}
which arises from the group velocity of ordinary-scattered carriers. At next order in the impurity density, three terms contribute, which are related to the group velocity of mixed-scattered carriers, the Berry curvature dipole~\cite{sodemann2015quantum} and the leading order contribution from the (linear-response) extrinsic velocity as
\begin{equation}
\label{j_E2_mixexbc}
\bs{j}_{E^2}^{(-1)}
=
\bs{j}_{E^2}^{(\text{mix})}
+
\bs{j}_{E^2}^{(\text{ex})}
+
\bs{j}_{E^2}^{(\text{BC})},
\end{equation}
with
\begin{subequations}
\begin{align}
\label{j_E2_mix}
&\bs{j}_{E^2}^{(\text{mix})}
=
-e \sum_{\mbk a} \bs{v}_{\mbk}^a  n_{E^2 \mbk a}^{(-1)},
\\
\label{j_E2_ex}
&\bs{j}_{E^2}^{(\text{ex})}
=
-e \sum_{\mbk a}  \bs{\alpha}_{\mbk}^{a} n_{E^2 \mbk a}^{(-2)},
\\
\label{j_E2_BC}
&\bs{j}_{E^2}^{(\text{BC})}
=
- \frac{e^2}{\hbar} \sum_{\mbk a} \mathbf{E} \times \bs{\Omega}_{\mbk}^a  n_{E \mbk a}^{(-1)}.
\end{align}
\end{subequations}
The remaining terms are then of $O(n_I^0)$. One such term is the anomalous velocity experienced by $n_{E\mbk}^{(0)}$ carriers, leading to anomalous nonlinear currents arising from side-jump and skew scattering processes. Interestingly, such nonlinear responses have also been recently proposed in Refs.~\cite{ma2023anomalous, atencia2023disorder}. Another $O(n_I^0)$ term in Eq.~(\ref{j_E2_tot}) is that of the quantum metric, which has recently garnered significant interest as an intrinsic nonlinear response~\cite{wang2021intrinsic, bhalla2022resonant, gao2023quantum, wang2023quantum, das2023intrinsic, hetenyi2023fluctuations, zhuang2024intrinsic, wang2024intrinsic} and makes a natural appearance in the density-matrix formalism. The remaining $O(n_I^0)$ terms are related to the group velocity of the carriers due to zeroth-order scattering and the various extrinsic velocities experienced by the carriers, revealing the intricate interplay between disorder and quantum geometry at this order.

In order to gain better insight into the emergence of extrinsic velocities in the carrier dynamics and transport in the nonlinear response regime, it is helpful to first revisit the linear response of the system, in which the nonequilibrium Fermi-surface carriers $n_{E\mbk}^{(-1)}$ experience the extrinsic velocity $\bs{\alpha}_{\mbk}$. In the presence of disorder, the electrons undergo random scatterings off the impurities, which leads to fluctuations in their velocities as they scatter between different energy bands. The net effect of this process is an overall change in the average velocity of the electrons within each energy band, which is captured by $\bs{\alpha}_{\mbk}^a$ and contributes to the transport at the Fermi level through the nonequilibrium carriers.

Extending this picture to the quadratic responses, it is clear that Eq.~(\ref{j_E2_ex}) describes a similar effect of $\bs{\alpha}_{\mbk}$ on the field-corrected distribution $n_{E^2\mbk}^{(-2)}$. The effect at the Fermi surface can be understood through an integration by parts
\begin{equation}
\label{ell}
-e \sum_{\mbk a}  \bs{\alpha}_{\mbk}^{a} n_{E^2 \mbk a}^{(-2)}
=
\frac{e^2}{\hbar} \sum_{\mbk a} n_{E \mbk a}^{(-1)} \mathbf{E} \cdot \bs{\pd}_{\mbk}   \bs{\ell}_{\alpha \mbk}^a,
\end{equation}
where we have introduced the extrinsic transport length vector
$\bs{\ell}_{\alpha \mbk}^a
=
\tau_{\mbk}^a \bs{\alpha}_{\mbk}^a$, which measures the average displacement the carriers experience over the transport time by the random interband walks on the Fermi surface. From Eq.~(\ref{ell}), we see that this average displacement receives a correction from the electric field--absent in the linear response--which results in a nonlinear current contribution at $O(n_I^{-1})$. At next order in the impurity density, a similar argument can be applied for the contribution of $\bs{\chi}_{\mbk}$ as well as for that of $\bs{\alpha}_{\mbk}$, with the difference that in the latter case, the mixed-scattering carriers $n_{E^2 \mbk}^{(-1)}$ come into play, indicating additional intermediate side-jump and skew-scattering processes.

Furthermore, the emergence of $\bs{\beta}_{\mbk}$, $\bs{\gamma}_{\mbk}$ and $\bs{\kappa}_{\mbk}$ can be attributed to the fact that the electric field itself can also induce interband coherence effects through its coupling with the position operator in the electrostatic interaction. In the absence of disorder, this field-mediated interband mixing can be understood as giving rise to the intrinsic Berry curvature and quantum metric contributions in Eq.~(\ref{eom}). And when disorder is present, the combined effect of the two interactions can give rise to new means of interband mixing such that the resulting extrinsic velocities acquire field dependencies. This is in contrast to the linear response regime, where--within the weak-disorder limit--the electrostatic and scattering corrections to the Bloch states can be separated, resulting in only independent corrections to physical observables~\cite{xiao2017semiclassical}.

\section{Application to massive Dirac fermions}
\label{application}

\subsection{Broken time reversal symmetry}

To illustrate the leading and subleading responses that emerge from the theory, consider a minimal model of 2D tilted massive Dirac fermions
\begin{equation}
\label{H_Dirac}
H_{\mbk}^0
=
v \bs{\sigma} \cdot \mbk + \mbk \cdot \mathbf{t} + m \sigma_z,
\end{equation}
where $m$ is the mass gap energy, while $v$ and $\mathbf{t} = (t_x, t_y)$ measure the strengths of the spin-orbit interaction and tilting of the Dirac cones, respectively. This is a fairly ubiquitous model used to describe the transport-relevant band structure in a host of quantum materials and may be used to develop more detailed band structures~\cite{goerbig2008tilted, du2018band, ma2019observation}. The eigenenergies read 
$\veps_{\mbk}^a
=
\veps_{0 k}^a + \mbk \cdot \mathbf{t}$ where 
$\veps_{0 k}^a = a h_k$ is the energy of the isotropic subsystem in the absence of tilting, with 
$h_k = \sqrt{k^2v^2+m^2}$ and $a=\pm$. Note that the tilting breaks the time reversal symmetry of the energy bands as well as the Hamiltonian, as opposed to the mass, which only breaks the time reversal symmetry at the level of the Hamiltonian. The Bloch eigenstates are expressed as
\vspace{-.2cm}
\begin{equation}
\ket{u_{\mathbf{k}a}}
=
\begin{pmatrix}
\left( \frac{1+ a}{2}\right) 
\cos \frac{\theta}{2}
+
\left( \frac{1-a}{2}\right) 
\sin \frac{\theta}{2}
\\
e^{i\phi}
\left[
\left( \frac{1+a}{2}\right) 
\sin \frac{\theta}{2}
-
\left( \frac{1-a}{2}\right) 
\cos \frac{\theta}{2}
\right]
\end{pmatrix},
\vspace{-.2cm}
\end{equation}
with $\theta = \cos^{-1}(m/h_k)$. For simplicity, we assume the weakly anisotropic case $t \ll v$, such that the eccentricity of the Fermi surface is small. We also assume the Fermi level lies in the upper band. The transport time is hence expressed as
\begin{equation}
\vspace{0cm}
\label{tau_k}
\frac{1}{\tau_{\mbk}^a}
=
\frac{2\pi}{\hbar} 
\sum_{\mbkpr b}
\delta ( \veps_{\mbk}^a - \veps_{\mbkpr}^b)
\Braket{
U_{\mbk \mbkpr}^{ab} U_{\mbkpr \mbk} ^{ba} 
}
\left( 1
- 
\frac{\mathbf{e}_E \cdot \bs{v}_{\mbkpr}^b}
{\mathbf{e}_E \cdot \bs{v}_{\mbk}^a}
\right),
\vspace{-0cm}
\end{equation}
with $\mathbf{e}_E$ the unit vector along the electric field, where we note that in the isotropic limit, Eq.~(\ref{tau_k}) reduces to Eq.~(\ref{tau_0k}). To first order in tilting, this yields
\begin{widetext}
\begin{equation}
\frac{1}{\tau_{\mbk}^a}
=
\frac{1}{\tau_{0 k}}
\left( 1 
-
\frac{a h_k}{v^2}
\frac{\mathbf{e}_E \cdot \mathbf{t}}
{\mathbf{e}_E \cdot \mathbf{k}}
\right)
-
\frac{a n_I U_0^2}{32 \hbar}
\left( \frac{h_k}{v^2} \right)^2
\left[
\left(
11 + 4 \cos 2\theta + \cos 4\theta \right)
\frac{\mathbf{e}_E \cdot \mathbf{t}}
{\mathbf{e}_E \cdot \mathbf{k}}
+
\left(
21 + 12 \cos 2\theta - \cos 4\theta \right)
\mathbf{t} \cdot 
\frac{\mathbf{e}_{\mbk}}{k}
\right],
\end{equation}
\end{widetext}
with $\mathbf{e}_{\mbk} = \mbk/k$ and the isotropic transport time given by
\begin{equation}
\frac{1}{\tau_{0 k}}
=
\frac{n_I U_0^2}{4 \hbar} \frac{h_k}{v^2}
\left( 1 + 3 \cos^2 \theta \right).
\end{equation}
To evaluate the nonlinear conductivities, we start with the leading-order contribution in disorder, Eq.~(\ref{j_E2_gr}), which arises from ordinary scattering. Without loss of generality, let us set the electric field along the $x$ direction, so that fewer elements of the conductivity tensor need to be evaluated. Inserting Eq.~(\ref{n_E2_-2}) into Eq.~(\ref{j_E2_gr}) and using the approximation
$\pd f_{0\mbk}^a / \pd \veps_{\mbk}^a
\simeq
- \delta(\veps_{\mbk}^a - \veps_F)
$, the quadratic conductivities associated with the (longitudinal) nonlinear response and nonlinear Hall effect are obtained as
\begin{subequations}
\label{sigma_ord}
\begin{align}
\sigma_{xxx}^{(\text{o})}
&=
- \frac{e^3}{8 \pi \hbar} (n_I U_0^2)^{-2}
\frac{v^4}{h_{k_F}^2} t_x
\mathcal{S}_{xxx}^{(\text{o})}(\theta),
\\
\sigma_{yxx}^{(\text{o})}
&=
- \frac{e^3}{8 \pi \hbar} (n_I U_0^2)^{-2}
\frac{v^4}{h_{k_F}^2} t_y
\mathcal{S}_{yxx}^{(\text{o})}(\theta),
\end{align}
\end{subequations}
respectively, where the superscript (o) indicates the ordinary-scattering origin of the conductivity. Here, the functions
$\mathcal{S}_{\mu xx}(\theta)$ encapsulate the mass dependence and are presented in Appendix~\ref{app_dirac}. 

To understand the dependence of the nonlinear current associated with Eq.~(\ref{sigma_ord}) on the tilting vector, it suffices to consider the symmetries of the terms in Eq.~(\ref{j_E2_gr}), namely the quadratic carrier density $n_{E^2 \mbk a}^{(-2)}$ and the group velocity of the Dirac fermions, $\bs{v}_{\mbk}^a = (a v^2/ \hbar h_k) \mbk + \mathbf{t}/\hbar$. In the limit $t_{\mu} = 0$, the group velocity is odd under the mirror symmetry transformation $k_{\mu} \rightarrow - k_{\mu}$, while the carrier density term is even. Thus the corresponding integral in Eq.~(\ref{j_E2_gr}) vanishes, which leads to the conclusion $\sigma_{\mu xx}^{(\text{o})} \propto t_{\mu}$. For the conductivities that are discussed below, a similar symmetry analysis can be applied, from which one can infer the angular dependence on the tilting vector for each term.

\begin{figure*}[hpt]
\captionsetup[subfigure]{labelformat=empty}
    \sidesubfloat[]{\includegraphics[width=0.32\linewidth,trim={0cm 0cm 0cm 0cm}]{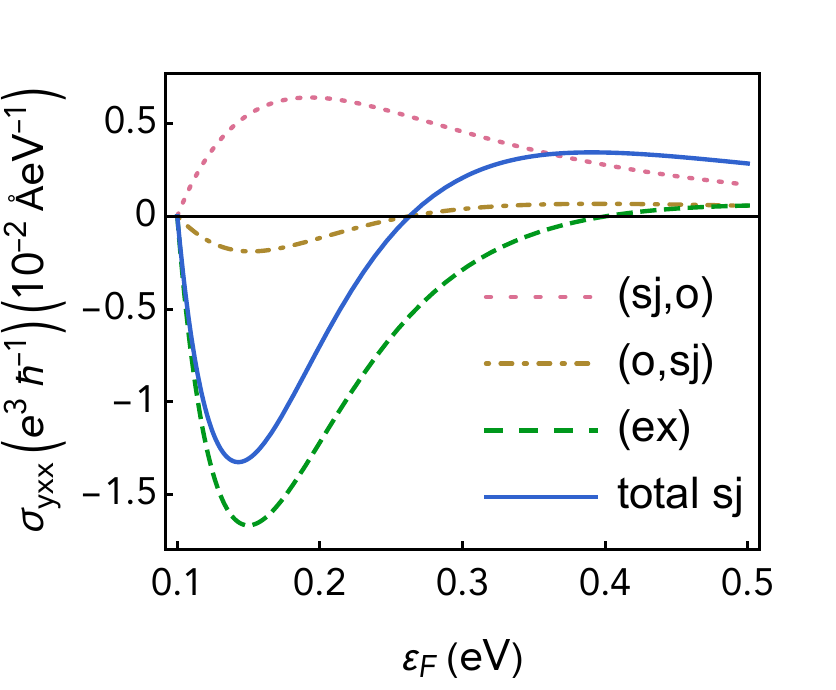}\label{fig2a}}
    \sidesubfloat[]{\includegraphics[width=0.32\linewidth,trim={0cm 0cm 0cm 0cm}]{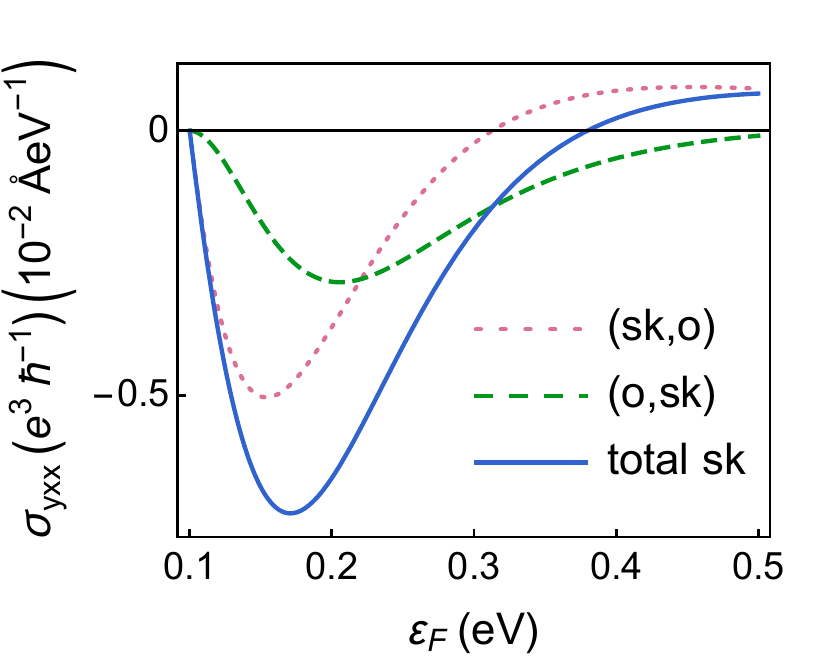}\label{fig2b}}
    \sidesubfloat[]{\includegraphics[width=0.32\linewidth,trim={0cm 0cm 0cm 0cm}]{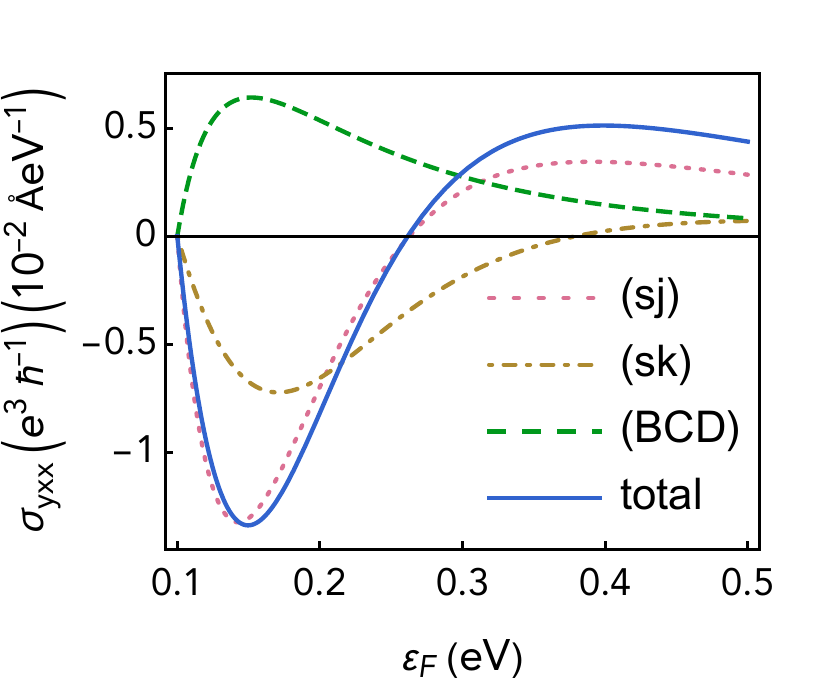}\label{fig2c}}
    \caption{Scalings of leading-order contributions to the nonlinear Hall conductivity in the presence of time reversal symmetry as a function of the Fermi energy, starting from the bottom of the conduction band. Parameters used: $m=0.1$ eV, $t_x = 0.1$ eV  \AA, $t_y = 0$, $v=1$ eV \AA, and $n_I U_0^2=10^2$ eV$^2$ \AA$^2$.}
    \label{fig2}
\end{figure*}

We next look at the subleading contributions to the nonlinear current, which are those of the mixed scattering, Berry curvature dipole and extrinsic velocity. For simplicity, we do not consider the sub-subleading processes in this work, as they are higher order effects and can be neglected in the weak disorder limit. And to obtain analytical results in the weak tilting limit, we use the isotropic transport time. The current density arising from mixed scattering is given by Eq.~(\ref{j_E2_mix}), where the explicit forms of the relevant carrier densities for the tilted Dirac model are presented in full in Eqs.~(\ref{n_Ek_sj_dirac} - \ref{N_3k_sko_dirac}). Inserting these into Eq.~(\ref{j_E2_mix}), we obtain the side-jump conductivities
\begin{subequations}
\begin{align}
\begin{split}
\sigma_{xxx}^{(\text{sj,o})}
&=
- \frac{e^3}{8 \pi \hbar} (n_I U_0^2)^{-1}
\frac{v^2}{h_{k_F}^2} t_y 
\mathcal{S}_{xxx}^{(\text{sj,o})}(\theta),
\end{split}
\\
\sigma_{yxx}^{(\text{sj,o})}
&=
\frac{e^3}{8 \pi \hbar} (n_I U_0^2)^{-1}
\frac{v^2}{h_{k_F}^2} t_x
\mathcal{S}_{yxx}^{(\text{sj,o})}(\theta),
\end{align}
\end{subequations}
and
\begin{subequations}
\begin{align}
\sigma_{xxx}^{(\text{o,sj})}
&=
\frac{e^3}{8 \pi \hbar} (n_I U_0^2)^{-1}
\frac{v^2}{h_{k_F}^2} t_y
\mathcal{S}_{xxx}^{(\text{o,sj})}(\theta),
\\
\sigma_{yxx}^{(\text{o,sj})}
&=
- \frac{e^3}{2 \pi \hbar} (n_I U_0^2)^{-1}
\frac{v^2}{h_{k_F}^2} t_x
\mathcal{S}_{yxx}^{(\text{o,sj})}(\theta),
\end{align}
\end{subequations}
as well as the skew-scattering contributions
\begin{subequations}
\begin{align}
\sigma_{xxx}^{(\text{sk,o})}
&=
-\frac{e^3}{64 \pi \hbar} (n_I U_0^2)^{-1}
\frac{v^2}{h_{k_F}^2} t_y
\mathcal{S}_{xxx}^{(\text{sk,o})}(\theta),
\\
\sigma_{yxx}^{(\text{sk,o})}
&=
- \frac{e^3}{32 \pi \hbar} (n_I U_0^2)^{-1}
\frac{v^2}{h_{k_F}^2} t_x
\mathcal{S}_{yxx}^{(\text{sk,o})}(\theta),
\end{align}
\end{subequations}
and
\begin{subequations}
\begin{align}
\sigma_{xxx}^{(\text{o,sk})}
&=
\frac{e^3}{4 \pi \hbar} (n_I U_0^2)^{-1}
\frac{v^2}{h_{k_F}^2} t_y
\mathcal{S}_{xxx}^{(\text{o,sk})}(\theta),
\\
\sigma_{yxx}^{(\text{o,sk})}
&=
-\frac{e^3}{\pi \hbar} (n_I U_0^2)^{-1}
\frac{v^2}{h_{k_F}^2} t_x
\mathcal{S}_{yxx}^{(\text{o,sk})}(\theta).\end{align}
\end{subequations}
Next, we consider the leading-order extrinsic velocity contribution, Eq.~(\ref{j_E2_ex}). Evaluating the energy-conserving part of the extrinsic velocity in Eq.~(\ref{alpha_ka}) for the massive Dirac model, we obtain
\begin{widetext}
\begin{equation}
\label{alpha_k+}
\bs{\alpha}_{\mbk}^+
=
\frac{n_I U_0^2}{4 \hbar v}
\left[
\sin 2\theta \mathbf{e}_{\phi} 
-
\frac{1}{2 v}
\cos \theta \left(3 + \cos 2\theta \right) \mathbf{e}_{z} \times \mathbf{t}
+
\frac{1}{4 v}
\sin \theta \sin 2\theta
\left\{
\left(
\mathbf{e}_{2\phi}  \cdot \mathbf{t} \right) \mathbf{e}_{E}
+
\left[
\mathbf{e}_{z} \cdot \left( \mathbf{e}_{2\phi} \times \mathbf{t} \right) \right]
\mathbf{e}_{z} \times \mathbf{e}_{E}
\right\}
\right],
\end{equation}
\end{widetext}
which generalizes the expression obtained in Ref.~\cite{atencia2022semiclassical} to the tilted case. Here, $\mathbf{e}_{\phi}$ and $\mathbf{e}_{2 \phi}$ are the azimuthal unit vectors corresponding to the angles $\phi$ and $2\phi$, respectively, and $\mathbf{e}_{z}$ is the unit vector in the $z$ direction, which is perpendicular to the system plane. Note this implies that, in the presence of tilting, the extrinsic velocity is no longer limited to the transverse direction and can, in principle, also give rise to a longitudinal current. Indeed,  inserting Eq.~(\ref{alpha_k+}) into the current expression given by Eq.~(\ref{j_E2_ex}), we arrive at the conductivities
\begin{subequations}
\begin{align}
\sigma_{xxx}^{(\text{ex})}
&=
\frac{e^3}{8 \pi \hbar} (n_I U_0^2)^{-1}
\frac{v^2}{h_{k_F}^2} t_y
\mathcal{S}_{xxx}^{(\text{ex})}(\theta),
\\
\sigma_{yxx}^{(\text{ex})}
&=
- \frac{e^3}{8 \pi \hbar} (n_I U_0^2)^{-1}
\frac{v^2}{h_{k_F}^2} t_x
\mathcal{S}_{yxx}^{(\text{ex})}(\theta).
\end{align}
\end{subequations}
The last term to consider at this order in the disorder expansion is that of the familiar Berry curvature dipole, the leading order contribution of which is due to the anomalous velocity felt by the ordinary-scattered electrons
\begin{equation}
\label{j_E2_BC}
\delta \bs{j}_{E^2}^{(\text{BC})}
=
- \frac{e^2}{\hbar} \sum_{\mbk a} \mathbf{E} \times \bs{\Omega}_{\mbk}^a  n_{E \mbk a}^{(-1)}.
\end{equation}
The out-of-plane component of the Berry curvature vector is given by 
$\Omega_{z k}^a
=
- a \cos \theta v^2 / 2 h_k^2$, which yields the nonlinear Hall conductivity
\begin{equation}
\sigma_{yxx}^{(\text{BC})}
=
\frac{e^3}{\pi \hbar} (n_I U_0^2)^{-1}
\frac{v^2}{h_{k_F}^2} t_x
\mathcal{S}_{yxx}^{(\text{BC})}(\theta).
\end{equation}
Gathering all the above terms, the total nonlinear conductivities are obtained, which are relatively complicated expressions. It is thus useful to consider two physically relevant limits. The first is the low-energy limit whereby the Fermi level lies close to the mass gap, \textit{i.e.}, $k_Fv \ll m$. In this case, we find
\begin{subequations}
\begin{align}
\sigma_{xxx}
&
\overset{k_Fv \ll m}{\simeq}
- \frac{3 e^3}{8 \pi \hbar} (n_I U_0^2)^{-2}
\frac{v^4}{m^2} t_x,
\\
\sigma_{yxx}
&
\overset{k_Fv \ll m}{\simeq}
- \frac{e^3}{8 \pi \hbar} (n_I U_0^2)^{-2}
\frac{v^4}{m^2} t_y.
\end{align}
\end{subequations}
where we assume the disorder is weak enough to neglect subleading processes. Interestingly, this results in the following ratio between the the nonlinear Hall and longitudinal conductivities
\begin{equation}
\frac{\sigma_{yxx}}{\sigma_{xxx}}
\overset{k_Fv \ll m}{=}
\frac{1}{3} \cot \phi_t,
\end{equation}
with $\phi_t= \cot^{-1}(t_x/t_y)$, in agreement with results obtained through diagrammatic~\cite{mehraeen2023quantum, mehraeen2024proximity} and Boltzmann transport~\cite{marinescu2023magnetochiral} approaches for various systems, which suggests a general relationship between the two conductivity components in the weak-disorder limit. And at the other extreme, when the Fermi level lies much higher than the gap, $k_Fv \gg m$, we find
\begin{subequations}
\begin{align}
\sigma_{xxx}
&
\overset{k_Fv \gg m}{\simeq}
\frac{11 e^3}{\pi \hbar} (n_I U_0^2)^{-2}
\frac{v^2}{k_F^2} t_x,
\\
\sigma_{yxx}
&
\overset{k_Fv \gg m}{\simeq}
\frac{e^3}{\pi \hbar} (n_I U_0^2)^{-2}
\frac{v^2}{k_F^2} t_y,
\end{align}
\end{subequations}
and the ratio
\begin{equation}
\frac{\sigma_{yxx}}{\sigma_{xxx}}
\overset{k_Fv \gg m}{=}
\frac{1}{11} \cot \phi_t.
\end{equation}
This suggests that, in addition to the individual dependencies of the conductivities on the material and disorder parameters, these differing ratios may prove useful in exploring the band topology and materials parameter dependencies of various systems via nonlinear transport measurements.
\vspace{-0cm}
\subsection{Restoring time reversal symmetry}
\vspace{-0cm}
In order to better explore the contributions of the subleading terms and their different scalings with the system parameters, it is useful to also consider the time-reversal symmetric counterpart of the model given by Eq.~(\ref{H_Dirac}), which can be obtained by extending the system to include the transformed Hamiltonian with negative tilt and mass, $(\mathbf{t}, m) \rightarrow - ( \mathbf{t} ,  m)$. Note that negating the mass corresponds to the transformation $\theta \rightarrow \pi - \theta$ in the angular functions $\mathcal{S}_{\mu xx} (\theta)$ given by Eqs.~(\ref{S_muxx^o}-\ref{S_yxx^BC}) in Appendix \ref{app_dirac}. Upon doing so, it straightforward to verify that the ordinary-scattering conductivity does not contribute to the nonlinear transport in the presence of time reversal symmetry, as
\begin{equation}
\sigma_{\mu xx}^{(\text{o})} (- \mathbf{t} , - m)
=
- \sigma_{\mu xx}^{(\text{o})} (\mathbf{t} , m).
\end{equation}
Therefore, the subleading terms in the broken time reversal system are the leading-order terms in the extended time-reversal-invariant system and will dominate the transport. And their contributions to the transport are obtained by simply doubling the broken time reversal expressions of the conductivities presented above.

In Fig.~\ref{fig2}, we plot the leading-order contributions to the nonlinear Hall effect for the time-reversal-symmetric case as a function of the Fermi energy, which reveals several notable features: as shown in Fig.~\ref{fig2a}, three terms comprise the conventional side-jump contribution to the nonlinear transport, including the extrinsic velocity. As can be seen, there are noticeable qualitative and quantitative differences between the contributions, such that the sum of the side-jump terms alone can give rise to a sign change of the conductivity (see the solid blue curve). A similar situation holds for the skew-scattering conductivities, plotted in Fig.~\ref{fig2b}. Interestingly, while the contributions from $n_{E^2\mbk}^{(\text{sk,o})}$ and $n_{E^2\mbk}^{(\text{o,sk})}$ act constructively near the band bottom, at higher Fermi energies, they contribute with opposite signs. Finally, in Fig.~\ref{fig2c}, we plot the total nonlinear Hall conductivities involving the Berry curvature dipole and disorder contributions, which again exhibit Fermi-energy-dependent sign changes from the competition between intrinsic and extrinsic effects, in relatively good agreement with previous studies of the nonlinear Hall effect based on a Boltzmann transport formalism with the coordinate shift~\cite{du2019disorder, du2021nonlinear}.
\vspace{-.2cm}
\section{Conclusion and Outlook}

In this work, we have presented a nonlinear response theory based on a recently-developed density-matrix formalism in the linear response regime~\cite{culcer2017interband, sekine2017quantum, atencia2022semiclassical}, whereby the density matrix is decomposed into disorder-averaged and fluctuating parts. By treating the diagonal and off-diagonal elements of the density matrix in band space on an equal footing, the full contribution of the disorder interaction can be taken into account in a systematic manner. And through the perturbative solution of the quantum Liouville equation in the presence of electrostatic and weak disorder potentials, we have shown how the combined actions of the electric field and random impurities give rise to a multitude of scattering channels in the quadratic response regime, resulting in a classification of carrier densities in terms of distinct physical processes.
\vspace{-.4cm}

\vspace{-.4cm}
Moreover, in addition to the previously known linear-response extrinsic velocity, several extrinsic velocities have been identified, which only appear in the nonlinear response of the system, revealing the rich structure of interband coherence effects in this transport regime. As an application of this theory, we have studied the quadratic response in a prototypical model of tilted Dirac fermions, and have derived the leading and subleading disorder contributions to the nonlinear conductivities.
\vspace{.1cm}

\vspace{-.3cm}
This work suggests several follow-up studies. One possible direction would be to extend the analysis to nonlinear spin currents, which would allow for a detailed classification of spin-orbit torques in the nonlinear response regime. Another direction would be to apply the theory to more complicated disorder profiles, which have recently been shown to play a role in generating nonlinear responses~\cite{dyrdal2020spin, mehraeen2024proximity, boboshko2024bilinear}. Yet another possible direction would be to explore the zeroth-order contributions in specific materials--as has recently been initiated through other approaches as well~\cite{huang2023scaling, gong2024nonlinear}--in order to develop a better understanding of the contributions of the quadratic-response extrinsic velocities and their interplay with quantum geometry in this transport regime.
\vspace{-.1cm}

\vspace{-.1cm}
In closing, we note some physically relevant limits that can be relaxed in future studies based on the present approach. First, we have focussed our analysis on DC responses. Generalizing this to the AC regime, which is particularly significant in the study of nonlinear optical responses, will allow for the continued exploration of the resonance structures of nonlinear responses. This can have interesting signatures in quantum materials~\cite{bhalla2020resonant} and can be utilized as useful probes of quantum geometry~\cite{bhalla2022resonant}. In addition, in this initial study, we have limited our analysis to low temperatures. Relaxing this enables further analysis of finite-temperature effects via the quantum kinetic approach, including the effects of thermal magnons and nonlinear thermoelectric responses~\cite{varshney2023quantum}. Finally, while it is customary to include the effect of a magnetic field in semiclassical transport studies of spin-orbit-coupled materials via the Zeeman interaction--similar to the exchange coupling in magnetic systems, a thorough analysis would also require the inclusion of the magnetic field in the derivation of the equations of motion. This leads to additional contributions, such as the coupling to the Berry curvature and modifying the density of states, which can then manifest in the nonlinear response of the system~\cite{gao2014field}.
\vspace{-.5cm}

\section*{Acknowledgments}

We thank Niloufar Dadkhah, Hai-Zhou Lu, Cong Xiao and Shulei Zhang for helpful discussions and comments on the manuscript.

\appendix

\section{Collision integrals}
\label{App_collision_int}

Inserting Eqs.~(\ref{delta_rho_0}) and (\ref{delta_rho_n}) into the general definition of the collision integral, to second order in the electric field, we obtain
\begin{widetext}
\vspace{-.5cm}
\begin{subequations}
\begin{align}
\label{J_0k_fk}
J_{0\mbk} ( f_{\lambda \mbk} )
&=
\frac{1}{\hbar}
\int_{-\infty}^{\infty} \frac{d\veps}{2\pi}
\Braket{
\left[ U, G^R(\veps) \left[ U, \braket{\rho_{\lambda}} \right] G^A(\veps)
\right]
}_{\mbk},
\\
\label{J_Ek_fk}
J_{E \mbk} ( f_{\lambda \mbk} )
&=
- \frac{i}{\hbar}
\int_{-\infty}^{\infty} \frac{d\veps}{2\pi}
\int_{-\infty}^{\infty} \frac{d\veps\pr}{2\pi}
\Braket{
\left[ U, G^R(\veps)
\left[ V, G^R(\veps\pr)
\left[ U, \braket{\rho_{\lambda}} \right] G^A(\veps\pr) \right]
G^A(\veps)
\right]
}_{\mbk},
\\
\label{J_E2k_fk}
J_{E^2 \mbk} ( f_{\lambda \mbk} )
&=
- \frac{1}{\hbar}
\int_{-\infty}^{\infty} \frac{d\veps}{2\pi}
\int_{-\infty}^{\infty} \frac{d\veps\pr}{2\pi}
\int_{-\infty}^{\infty} \frac{d\veps\dpr}{2\pi}
\Braket{
\left[ U, G^R(\veps)
\left[ V, G^R(\veps\pr)
\left[ V, G^R(\veps\dpr)
\left[ U, \braket{\rho_{\lambda}} \right] G^A(\veps\dpr) 
\right] G^A(\veps\pr)
\right] G^A(\veps)
\right]
}_{\mbk},
\end{align}
\end{subequations}
where the integrals are evaluated with the relation
\begin{equation}
\int_{\infty}^{\infty}\frac{d\veps}{2 \pi i} 
G^A_{\mbk a}(\veps)
G^R_{\mbk^{\prime} b}(\veps)
=
\frac{1}{\veps_{\mbk}^a - \veps_{\mbkpr}^b + i \delta},
\end{equation}
and the useful identity
\begin{equation}
\label{dirac_derivative}
\frac{\delta (x)}{x^{n}}
=
\frac{(-1)^n}{n!} \frac{d^n}{dx^n}\delta (x).
\end{equation}
The explicit forms of the collision integrals are then derived once the off-diagonal density matrix elements are known, which are presented in the next section.

\section{Off-diagonal density matrix}
\label{App_off-diagonal}

Inserting the various driving terms into the formal solution of the off-diagonal density matrix, one finds
\begin{subequations}
\label{S_Ek_0_decomp}
\begin{gather}
\label{S_Ek_0D}
S_{E \mbk}^{(0) ab} \left[D_{E \mbk}^{(0)} \right]
=
e \mathbf{E} \cdot 
\bs{\mathcal{A}}_{\mbk}^{\prime ab}
\frac{f_{0 \mbk}^a - f_{0 \mbk}^{b}}{\veps_{\mbk}^a - \veps_{\mbk}^b},
\\
\label{S_Ek_0I}
S_{E \mbk}^{(0) ab} \left[ I_{E \mbk}^{(0)} \right]
=
i \pi
\sum_{\mbkpr c}
\frac{\Braket{U_{\mbk \mbk\pr}^{ac} U_{\mbk\pr \mbk}^{cb}}}
{\veps_{\mbk}^a - \veps_{\mbk}^b}
\left[
\left(n_{E \mbk a}^{(-1)} - n_{E \mbkpr c}^{(-1)}\right) \delta(\veps_{\mbk}^a 
-
\veps_{\mbkpr}^c)
+
\left(n_{E \mbk b}^{(-1)} - n_{E \mbkpr c}^{(-1)}\right) \delta(\veps_{\mbk}^b 
-
\veps_{\mbkpr}^c)
\right],
\end{gather}
\end{subequations}
at linear order in the electric field. And, similarly, for the analogous quadratic-response  counterparts, we obtain
\begin{subequations}
\label{S_E2k_-1_decomp}
\begin{gather}
\label{S_E2k_-1D}
S_{E^2 \mbk}^{(-1) ab} \left[D_{E^2 \mbk}^{(-1)} \right]
=
e \mathbf{E} \cdot 
\bs{\mathcal{A}}_{\mbk}^{\prime ab}
\frac{n_{E \mbk a}^{(-1)} - n_{E \mbk b}^{(-1)}}{\veps_{\mbk}^a - \veps_{\mbk}^b },
\\
\label{S_E2k_-1I}
S_{E^2 \mbk}^{(-1) ab} \left[ I_{E^2 \mbk}^{(-1)} \right]
=
i \pi
\sum_{\mbkpr c}
\frac{\Braket{U_{\mbk \mbk\pr}^{ac} U_{\mbk\pr \mbk}^{cb}}}
{\veps_{\mbk}^a - \veps_{\mbk}^b}
\left[
(n_{E^2 \mbk a}^{(-2)} - n_{E^2 \mbkpr c}^{(-2)} ) \delta(\veps_{\mbk}^a 
-
\veps_{\mbkpr}^c)
+
(n_{E^2 \mbk b}^{(-2)} - n_{E^2 \mbkpr c}^{(-2)}) \delta(\veps_{\mbk}^b 
-
\veps_{\mbkpr}^c)
\right],
\end{gather}
\end{subequations}
as well as
\begin{subequations}
\label{S_E2k_0_decomp}
\begin{gather}
\label{S_E2k_0D}
S_{E^2 \mbk}^{(0) ab} \left[D_{E^2 \mbk}^{(0)} \right]
=
e \mathbf{E} \cdot 
\bs{\mathcal{A}}_{\mbk}^{\prime ab}
\frac{n_{E \mbk a}^{(0)} - n_{E \mbk b}^{(0)}}{\veps_{\mbk}^a - \veps_{\mbk}^b },
\\
\label{S_E2k_0I}
S_{E^2 \mbk}^{(0) ab} \left[ I_{E^2 \mbk}^{(0)} \right]
=
i \pi
\sum_{\mbkpr c}
\frac{\Braket{U_{\mbk \mbk\pr}^{ac} U_{\mbk\pr \mbk}^{cb}}}
{\veps_{\mbk}^a - \veps_{\mbk}^b}
\left[
(n_{E^2 \mbk a}^{(-1)} - n_{E^2 \mbkpr c}^{(-1)} ) \delta(\veps_{\mbk}^a 
-
\veps_{\mbkpr}^c)
+
( n_{E^2 \mbk b}^{(-1)} - n_{E^2 \mbkpr c}^{(-1)} ) \delta(\veps_{\mbk}^b 
-
\veps_{\mbkpr}^c)
\right].
\end{gather}
\end{subequations}
The contribution from 
$I_{E^2 \mbk}^{\prime (0)}$ is obtained by inserting Eq.~(\ref{J_Ek_fk}) into the relevant term in Eq.~(\ref{S_E2k_0}), which yields
\begin{equation}
\label{S_E2k_0Ipr}
\begin{split}
S_{E^2 \mbk}^{(0) ab} \left[ I_{E^2 \mbk}^{\prime (0)} \right]
&=
- \frac{ e}
{\veps_{\mbk}^a - \veps_{\mbk}^b }
\mathbf{E} \cdot
\sum_{\mbkpr c}
\frac{1}
{\veps_{\mbk}^b - \veps_{\mbkpr}^c + i \delta}
\Biggl\{
\Braket{i U_{\mbk \mbk\pr}^{ac} 
(\bs{\pd}_{\mbk} + \bs{\pd}_{\mbkpr})
\left(
U_{\mbk\pr \mbk}^{cb}
\frac{n_{E \mbk b}^{(-1)} - n_{E \mbkpr c}^{(-1)}}
{\veps_{\mbk}^b - \veps_{\mbkpr}^c + i \delta}
\right)}
\\
&+
\sum_d
\left[
\Braket{U_{\mbk \mbk\pr}^{ac} U_{\mbk\pr \mbk}^{db}} \bs{\mathcal{A}}_{\mbkpr}^{cd}
\frac{n_{E \mbk b}^{(-1)} - n_{E \mbkpr d}^{(-1)}}
{\veps_{\mbk}^b - \veps_{\mbkpr}^d + i \delta}
-
\Braket{U_{\mbk \mbk\pr}^{ac} U_{\mbk\pr \mbk}^{cd}} \bs{\mathcal{A}}_{\mbk}^{db}
\frac{n_{E \mbk d}^{(-1)} - n_{E \mbkpr c}^{(-1)}}
{\veps_{\mbk}^d - \veps_{\mbkpr}^c + i \delta}
\right]
\Biggr\}
\\
&-
\frac{e}
{\veps_{\mbk}^a - \veps_{\mbk}^b}
\mathbf{E} \cdot
\sum_{\mbkpr c}
\frac{1}
{\veps_{\mbkpr}^c - \veps_{\mbk}^a + i \delta}
\Biggl\{
\Braket{i U_{\mbkpr \mbk}^{cb} 
(\bs{\pd}_{\mbk} + \bs{\pd}_{\mbkpr})
\left(
U_{\mbk \mbkpr}^{ac}
\frac{n_{E \mbk a}^{(-1)} - n_{E \mbkpr c}^{(-1)}}
{\veps_{\mbkpr}^c - \veps_{\mbk}^a + i \delta}
\right)}
\\
&+
\sum_d
\left[
\Braket{U_{\mbk \mbk\pr}^{dc} U_{\mbkpr \mbk}^{cb}} \bs{\mathcal{A}}_{\mbk}^{ad}
\frac{n_{E \mbk d}^{(-1)} - n_{E \mbkpr c}^{(-1)}}
{\veps_{\mbkpr}^c - \veps_{\mbk}^d + i \delta}
-
\Braket{U_{\mbk \mbk\pr}^{ad} U_{\mbk\pr \mbk}^{cb}} \bs{\mathcal{A}}_{\mbkpr}^{dc}
\frac{n_{E \mbk a}^{(-1)} - n_{E \mbkpr d}^{(-1)}}
{\veps_{\mbkpr}^d - \veps_{\mbk}^a + i \delta}
\right]
\Biggr\}.
\end{split}
\end{equation}
The find the remaining contributions from $D_{E^2 \mbk}^{\prime (0)}$ and $I_{E^2 \mbk}^{\dprime (0)}$, recalling Eq.~(\ref{S_E2k_Dpr_Idpr}), we arrive at the following terms
\begin{equation}
\begin{split}
S_{E^2 \mbk}^{(0) ab} \left[D_{E^2 \mbk}^{\prime (0)} [ D_{E \mbk}^{(0)}]
\right]
&=
- \frac{i e \mathbf{E} \cdot \bs{\pd}_{\mbk}
S_{E \mbk}^{(0) ab} \left[D_{E \mbk}^{(0)} \right]}
{\veps_{\mbk}^a - \veps_{\mbk}^b}
-
\frac{e^2 E^{\mu} E^{\nu}}
{\veps_{\mbk}^a - \veps_{\mbk}^b}
\Big\{
\mathcal{A}_{\mu \mbk}^{\prime ab}
(\mathcal{A}_{\nu \mbk}^{a}
-
\mathcal{A}_{\nu \mbk}^{b})
\frac{f_{0\mbk}^a - f_{0\mbk}^b}
{\veps_{\mbk}^a - \veps_{\mbk}^b}
\\
&-
\sum_{c}
\mathcal{A}_{\mu \mbk}^{\prime ac}
\mathcal{A}_{\nu \mbk}^{\prime cb}
\left(
\frac{f_{0\mbk}^a - f_{0\mbk}^c}
{\veps_{\mbk}^a - \veps_{\mbk}^c}
-
\frac{f_{0\mbk}^b - f_{0\mbk}^c}
{\veps_{\mbk}^b - \veps_{\mbk}^c}
\right)
\Bigg\},
\end{split}
\end{equation}
\begin{equation}
\begin{split}
S_{E^2 \mbk}^{(0) ab} \left[D_{E^2 \mbk}^{\prime (0)} [ I_{E \mbk}^{(0)}]
\right]
&=
- \frac{i e \mathbf{E} \cdot}
{\veps_{\mbk}^a - \veps_{\mbk}^b }
\Big\{
\bs{\pd}_{\mbk}
S_{E \mbk}^{(0) ab} \left[I_{E \mbk}^{(0)} \right]
\\
&+
\pi \sum_{\mbkpr cd}
\frac{
\bs{\mathcal{A}}_{\mbk}^{ac}
\Braket{U_{\mbk \mbk\pr}^{cd} U_{\mbk\pr \mbk}^{db}}
}
{\veps_{\mbk}^c - \veps_{\mbk}^b }
\left[
( n_{E \mbk b}^{(-1)} - n_{E \mbkpr d}^{(-1)} ) 
\delta(\veps_{\mbk}^b - \veps_{\mbkpr}^d)
+
( n_{E \mbk c}^{(-1)} - n_{E \mbkpr d}^{(-1)} )
\delta(\veps_{\mbk}^c - \veps_{\mbkpr}^d)
\right]_{c \neq b}
\\
&-
\pi \sum_{\mbkpr cd}
\frac{
\bs{\mathcal{A}}_{\mbk}^{cb}
\Braket{U_{\mbk \mbk\pr}^{ad} U_{\mbk\pr \mbk}^{dc}}
}
{\veps_{\mbk}^a - \veps_{\mbk}^c}
\left[
( n_{E \mbk a}^{(-1)} - n_{E \mbkpr d}^{(-1)} ) 
\delta(\veps_{\mbk}^a - \veps_{\mbkpr}^d)
+
( n_{E \mbk c}^{(-1)} - n_{E \mbkpr d}^{(-1)} )
\delta(\veps_{\mbk}^c - \veps_{\mbkpr}^d)
\right]_{c \neq a}
\Big\},
\end{split}
\end{equation}
\begin{equation}
\begin{split}
S_{E^2 \mbk}^{(0) ab} \left[I_{E^2 \mbk}^{\dprime (0)} [ D_{E^2 \mbk}^{(-1)}]
\right]
&=
i \pi e \mathbf{E} \cdot
\sum_{\mbkpr cd}
\Bigg\{
\left[
\frac{
\Braket{U_{\mbk \mbk\pr}^{ac} U_{\mbkpr \mbk}^{cd}} 
\bs{\mathcal{A}}_{\mbk}^{\prime db}
}
{\veps_{\mbk}^a - \veps_{\mbk}^b }
\frac{n_{E \mbk d}^{(-1)} - n_{E \mbk b}^{(-1)}}
{\veps_{\mbk}^d - \veps_{\mbk}^b }
-
\frac{
\Braket{U_{\mbk \mbk\pr}^{ac} U_{\mbk\pr \mbk}^{db}}
\bs{\mathcal{A}}_{\mbkpr}^{\prime cd}
}
{\veps_{\mbk}^a - \veps_{\mbk}^b }
\frac{n_{E \mbkpr c}^{(-1)} - n_{E \mbkpr d}^{(-1)}}
{\veps_{\mbkpr}^c - \veps_{\mbkpr}^d }
\right]
\delta(\veps_{\mbk}^b - \veps_{\mbkpr}^c)
\\
&-
\left[
\frac{
\Braket{U_{\mbk \mbk\pr}^{ad} U_{\mbkpr \mbk}^{cb}}
\bs{\mathcal{A}}_{\mbkpr}^{\prime dc}
}
{\veps_{\mbk}^a - \veps_{\mbk}^b }
\frac{n_{E \mbkpr d}^{(-1)} - n_{E \mbkpr c}^{(-1)}}
{\veps_{\mbkpr}^d - \veps_{\mbkpr}^c }
-
\frac{
\Braket{U_{\mbk \mbk\pr}^{dc} U_{\mbk\pr \mbk}^{cb}}
\bs{\mathcal{A}}_{\mbk}^{\prime ad}
}
{\veps_{\mbk}^a - \veps_{\mbk}^b }
\frac{n_{E \mbk a}^{(-1)} - n_{E \mbk d}^{(-1)}}
{\veps_{\mbk}^a - \veps_{\mbk}^d }
\right]
\delta(\veps_{\mbk}^a - \veps_{\mbkpr}^c)
\bigg\},
\end{split}
\end{equation}

\begin{equation}
\label{S_E2_Ipr_I-1}
\begin{split}
S_{E^2 \mbk}^{(0) ab} \left[I_{E^2 \mbk}^{\dprime (0)} [ I_{E^2 \mbk}^{(-1)}]
\right]
&=
- \frac{\pi^2}
{\veps_{\mbk}^a - \veps_{\mbk}^b}
\sum_{\mbkpr \mbkdpr cde}
\delta(\veps_{\mbk}^b - \veps_{\mbkpr}^c)
\\
&\times
\bigg\{
\frac{
\Braket{U_{\mbk \mbk\pr}^{ac} U_{\mbk\pr \mbk}^{cd}}
\Braket{U_{\mbk \mbkdpr}^{de} U_{\mbkdpr \mbk}^{eb}}
}
{\veps_{\mbk}^d - \veps_{\mbk}^b }
\left[
( n_{E^2 \mbk b}^{(-2)} - n_{E^2 \mbkdpr e}^{(-2)}) 
\delta(\veps_{\mbk}^b - \veps_{\mbkdpr}^e )
+
( n_{E^2 \mbk d}^{(-2)} - n_{E^2 \mbkdpr e}^{(-2)} )
\delta(\veps_{\mbk}^d - \veps_{\mbkdpr}^e)
\right]_{d \neq b}
\\
&-
\frac{
\Braket{U_{\mbk \mbk\pr}^{ac} U_{\mbk\pr \mbk}^{db}}
\Braket{U_{\mbkpr \mbkdpr}^{ce} U_{\mbkdpr \mbkpr}^{ed}}
}
{\veps_{\mbkpr}^c - \veps_{\mbkpr}^d }
\left[
( n_{E^2 \mbkpr d}^{(-2)} - n_{E^2 \mbkdpr e}^{(-2)}) 
\delta(\veps_{\mbkpr}^d - \veps_{\mbkdpr}^e )
+
( n_{E^2 \mbkpr c}^{(-2)} - n_{E^2 \mbkdpr e}^{(-2)} )
\delta(\veps_{\mbkpr}^c - \veps_{\mbkdpr}^e)
\right]_{d \neq c}
\Big\}
\\
&-
\frac{\pi^2}
{\veps_{\mbk}^a - \veps_{\mbk}^b}
\sum_{\mbkpr \mbkdpr cde}
\delta(\veps_{\mbk}^a - \veps_{\mbkpr}^c)
\\
&\times
\bigg\{
\frac{
\Braket{U_{\mbk \mbk\pr}^{dc} U_{\mbk\pr \mbk}^{cb}}
\Braket{U_{\mbk \mbkdpr}^{ae} U_{\mbkdpr \mbk}^{ed}}
}
{\veps_{\mbk}^a - \veps_{\mbk}^d}
\left[
( n_{E^2 \mbk d}^{(-2)} - n_{E^2 \mbkdpr e}^{(-2)}) 
\delta(\veps_{\mbk}^d - \veps_{\mbkdpr}^e )
+
( n_{E^2 \mbk a}^{(-2)} - n_{E^2 \mbkdpr e}^{(-2)} )
\delta(\veps_{\mbk}^a - \veps_{\mbkdpr}^e)
\right]_{d \neq a}
\\
&-
\frac{
\Braket{U_{\mbk \mbk\pr}^{ad} U_{\mbk\pr \mbk}^{cb}}
\Braket{U_{\mbkpr \mbkdpr}^{de} U_{\mbkdpr \mbkpr}^{ec}}
}
{\veps_{\mbkpr}^d - \veps_{\mbkpr}^c}
\left[
( n_{E^2 \mbkpr c}^{(-2)} - n_{E^2 \mbkdpr e}^{(-2)}) 
\delta(\veps_{\mbkpr}^c - \veps_{\mbkdpr}^e )
+
( n_{E^2 \mbkpr d}^{(-2)} - n_{E^2 \mbkdpr e}^{(-2)} )
\delta(\veps_{\mbkpr}^d - \veps_{\mbkdpr}^e)
\right]_{d \neq c}
\Big\}.
\end{split}
\end{equation}
Thus, all the off-diagonal density matrix elements are expressed now in terms of diagonal terms. When inserted into the various collision integrals, the carrier densities are derived, the forms of which are presented in the next section.

\section{$\mathcal{I}_1$ and $\mathcal{I}_2$ in Eq.~(\ref{zeroth_collision})}
\label{app_I_1}

\begin{equation}
\mathcal{I}_1
=
J_{0\mbk}^a 
\left( S_{E^2 \mbk}^{(0)} \left[D_{E^2 \mbk}^{\prime (0)} [ I_{E \mbk}^{(0)}]
\right] \right)
+
J_{0\mbk}^a 
\left( S_{E^2 \mbk}^{(0)} \left[I_{E^2 \mbk}^{\prime (0)} \right] \right)
+
J_{E\mbk}^a 
\left( S_{E \mbk}^{(0)} \left[I_{E \mbk}^{(0)} \right] \right)
+
J_{0\mbk}^a 
\left( S_{E^2 \mbk}^{(0)} \left[I_{E^2 \mbk}^{\dprime (0)} [ D_{E^2 \mbk}^{(-1)}]
\right] \right),
\end{equation}
\begin{equation}
\mathcal{I}_2
=
J_{0\mbk}^a 
\left( S_{E^2 \mbk}^{(0)} \left[D_{E^2 \mbk}^{\prime (0)} [ D_{E \mbk}^{(0)}]
\right] \right)
+
J_{E\mbk}^a \left( S_{E \mbk}^{(0)} \left[ D_{E \mbk}^{(0)} \right] \right)
+
J_{E^2 \mbk}^a 
\left( f_{0 \mbk} \right).
\end{equation}

\section{Carrier densities}
\label{App_carrier_densities}

\subsection{Linear-response densities}

The linear-response carrier densities are obtained by solving Eq.~(\ref{linear_carrier_eq}). As outlined above, this yields~\cite{atencia2022semiclassical}
\begin{equation}
\label{n_Ek^sk}
\begin{split}
n_{E\mbk a}^{(\text{sk})}
&=
-\frac{2\pi^2}{\hbar} \tau_{\mbk}^a
\sum_{\mbkpr \mbkdpr bcd}
\delta(\veps_{\mbk}^a 
-
\veps_{\mbkpr}^b)
\\
&\times
\text{Im} \Biggl\{
\frac{
\Braket{U_{\mbk \mbk\pr}^{ab} U_{\mbkpr \mbk}^{bc}}
\Braket{U_{\mbk \mbkdpr}^{cd} U_{\mbkdpr \mbk}^{da}}} 
{\veps_{\mbk}^a - \veps_{\mbk}^c} 
\left[
\left( n_{E\mbk a}^{(-1)} - n_{E\mbkdpr d}^{(-1)}\right) \delta(\veps_{\mbk}^a 
-
\veps_{\mbkdpr}^d)
+
\left( n_{E\mbk c}^{(-1)} - n_{E\mbkdpr d}^{(-1)}\right) \delta(\veps_{\mbk}^c 
-
\veps_{\mbkdpr}^d)
\right]_{c \neq a}
\\
&+
\frac{
\Braket{U_{\mbk \mbk\pr}^{ab} U_{\mbkpr \mbk}^{ca}}
\Braket{U_{\mbkpr \mbkdpr}^{bd} U_{\mbkdpr \mbkpr}^{dc}}} 
{\veps_{\mbkpr}^b - \veps_{\mbkpr}^c}
\left[
\left( n_{E\mbkpr b}^{(-1)} - n_{E\mbkdpr d}^{(-1)}\right) \delta(\veps_{\mbkpr}^b 
-
\veps_{\mbkdpr}^d)
+
\left( n_{E\mbkpr c}^{(-1)} - n_{E\mbkdpr d}^{(-1)}\right) \delta(\veps_{\mbkpr}^c 
-
\veps_{\mbkdpr}^d)
\right]_{c \neq b} \biggr\} ,
\end{split}
\end{equation}
and
\begin{equation}
\label{n_Ek^sj}
n_{E\mbk a}^{(\text{sj})}
=
- \frac{2\pi}{\hbar} \tau_{\mbk}^a
\frac{\pd f_{0\mbk}^a}{\pd \veps_{\mbk}^a}
e \mathbf{E} \cdot \sum_{\mbkpr b}
\delta(\veps_{\mbk}^a 
-
\veps_{\mbkpr}^b)
\Bigl\{
\text{Im}
\Braket{
\left[
\left( \bs{\pd}_{\mbk} + \bs{\pd}_{\mbkpr} \right)
U_{\mbk \mbk\pr}^{ab}
\right] U_{\mbk\pr \mbk}^{ba}}
-
\Braket{
U_{\mbk \mbk\pr}^{ab} U_{\mbk\pr \mbk}^{ba}}
\left( \bs{\mathcal{A}}_{\mbk}^a
-
\bs{\mathcal{A}}_{\mbkpr}^b\right)
\Bigr\}.
\end{equation}

The various quadratic-response carrier densities are considerably more numerous. However, they are evaluated in a fairly similar manner, the explicit forms of which are presented below.

\subsection{Mixed scattering}

The contribution from $n_{E^2 \mbk a}^{\text{(sj,o)}}$ is expressed as
\begin{equation}
\label{n_E2k_sjo}
n_{E^2 \mbk a}^{\text{(sj,o)}}
=
\mathcal{N}^{(\text{sj,o})}_{1,\mbk a}
+
\mathcal{N}^{(\text{sj,o})}_{2,\mbk a},
\end{equation}
with
\begin{equation}
\begin{split}
\mathcal{N}^{(\text{sj,o})}_{1,\mbk a}=&
- \frac{2 \pi}{\hbar^2}  e E^{\mu} \tau_{\mbk}^a
\sum_{\mbkpr b}
\delta ( \veps_{\mbk}^a - \veps_{\mbkpr}^b )
\pd_{\mbkpr}^{\rho}
\left[
\frac{v_{\rho \mbkpr}^b}{\lvert \bs{v}_{\mbkpr}^b \rvert^2}
\text{Im}
\Braket{
U_{\mbk \mbk\pr}^{ab} 
\left[
\left( \pd_{\mbk}^{\mu} + \pd_{\mbkpr}^{\mu} \right) U_{\mbk\pr \mbk} ^{ba} 
\right]
}
\left(
n_{E \mbk a}^{(-1)} - n_{E \mbkpr b}^{(-1)} 
\right)
\right]
\\
&-
\frac{2 \pi}{\hbar^2}  e E^{\mu} \tau_{\mbk}^a
\sum_{\mbkpr b}
\delta ( \veps_{\mbk}^a - \veps_{\mbkpr}^b )
\pd_{\mbkpr}^{\rho}
\left[
\frac{v_{\rho \mbkpr}^b}{\lvert \bs{v}_{\mbkpr}^b \rvert^2}
\left(
\mathcal{A}_{\mu \mbk}^a - \mathcal{A}_{\mu \mbkpr}^b 
\right)
\Braket{
U_{\mbk \mbk\pr}^{ab} U_{\mbk\pr \mbk} ^{ba} 
}
\left(
n_{E \mbk a}^{(-1)} - n_{E \mbkpr b}^{(-1)} 
\right)
\right],
\end{split}
\end{equation}
\begin{equation}
\begin{split}
\mathcal{N}^{(\text{sj,o})}_{2,\mbk a}
=&
- \frac{4 \pi}{\hbar}  e E^{\mu} \tau_{\mbk}^a
\sum_{\mbkpr bc}
\delta ( \veps_{\mbk}^a - \veps_{\mbkpr}^b )
\left(
\text{Re}
\left[
\Braket{
U_{\mbk \mbk\pr}^{ab} U_{\mbk\pr \mbk} ^{bc}
}
\mathcal{A}_{\mu \mbk}^{\prime ca}
\right]
\frac{
n_{E \mbk a}^{(-1)} - n_{E \mbkpr b}^{(-1)}
}
{\veps_{\mbk}^a - \veps_{\mbk}^c}
+
\text{Re}
\left[
\Braket{
U_{\mbk \mbk\pr}^{ab} U_{\mbk\pr \mbk} ^{ca}
}
\mathcal{A}_{\mu \mbkpr}^{\prime bc}
\right]
\frac{
n_{E \mbk a}^{(-1)} - n_{E \mbkpr b}^{(-1)}
}
{\veps_{\mbk}^a - \veps_{\mbkpr}^c}
\right),
\end{split}
\end{equation}
where the velocity factors here and elsewhere arise from application of Eq.~(\ref{dirac_derivative}). The remaining mixed-scattering terms read
\begin{equation}
\label{n_E2_sko}
\begin{split}
n_{E^2 \mbk a}^{(\text{sk,o})}
&=
- \frac{2\pi^2}{\hbar} \tau_{\mbk}^a
\sum_{\mbkpr \mbkdpr bcd}
\delta(\veps_{\mbk}^a 
-
\veps_{\mbkpr}^b)
\\
&\times
\Biggl\{
\frac{
\text{Im}
\left[
\Braket{U_{\mbk \mbk\pr}^{ab} U_{\mbkpr \mbk}^{bc}}
\Braket{U_{\mbk \mbkdpr}^{cd} U_{\mbkdpr \mbk}^{da}}
\right]
} 
{\veps_{\mbk}^a - \veps_{\mbk}^c}
\left[
( n_{E^2 \mbk a}^{(-2)} - n_{E^2 \mbkdpr d}^{(-2)}) 
\delta(\veps_{\mbk}^a -\veps_{\mbkdpr}^d)
+
( n_{E^2\mbk c}^{(-2)} - n_{E^2\mbkdpr d}^{(-2)}) 
\delta(\veps_{\mbk}^c -\veps_{\mbkdpr}^d)
\right]_{c \neq a}
\\
&+ 
\frac{
\text{Im}
\left[
\Braket{U_{\mbk \mbk\pr}^{ab} U_{\mbkpr \mbk}^{ca}}
\Braket{U_{\mbkpr \mbkdpr}^{bd} U_{\mbkdpr \mbkpr}^{dc}}
\right]
} 
{\veps_{\mbk}^a - \veps_{\mbkpr}^c}
\left[
( n_{E^2\mbkpr b}^{(-2)} - n_{E^2\mbkdpr d}^{(-2)})
\delta(\veps_{\mbkpr}^b -\veps_{\mbkdpr}^d)
+
( n_{E^2\mbkpr c}^{(-2)} - n_{E^2\mbkdpr d}^{(-2)}) 
\delta(\veps_{\mbkpr}^c - \veps_{\mbkdpr}^d)
\right]_{c \neq b} \biggr\} ,
\end{split}
\end{equation}

\begin{subequations}
\begin{align}
n_{E^2 \mbk a}^{(\text{o,sj})}
&=
\frac{e}{\hbar} \tau_{\mbk}^a
\mathbf{E} \cdot 
\bs{\pd}_{\mbk} n_{E \mbk a}^{(\text{sj})},
\\
n_{E^2 \mbk a}^{(\text{o,sk})}
&=
\frac{e}{\hbar} \tau_{\mbk}^a 
\mathbf{E} \cdot 
\bs{\pd}_{\mbk} n_{E \mbk a}^{(\text{sk})}.
\end{align}
\end{subequations}

\subsection{Secondary side jump}

The secondary side jump density is given by
\begin{equation}
n_{E^2 \mbk a}^{\text{(ssj)}}
=
\mathcal{N}^{(\text{ssj})}_{1,\mbk a}
+
\mathcal{N}^{(\text{ssj})}_{2,\mbk a},
\end{equation}
where

\begin{equation}
\begin{split}
\mathcal{N}^{(\text{ssj})}_{1,\mbk a}
=&
- \frac{2 \pi}{\hbar^2}  e E^{\mu} \tau_{\mbk}^a
\sum_{\mbkpr b}
\delta ( \veps_{\mbk}^a - \veps_{\mbkpr}^b )
\pd_{\mbkpr}^{\rho}
\left[
\frac{v_{\rho \mbkpr}^b}{\lvert \bs{v}_{\mbkpr}^b \rvert^2}
\text{Im}
\Braket{
U_{\mbk \mbk\pr}^{ab} 
\left[
\left( \pd_{\mbk}^{\mu} + \pd_{\mbkpr}^{\mu} \right) U_{\mbk\pr \mbk} ^{ba} 
\right]
}
\left(
n_{E \mbk a}^{(0)} - n_{E \mbkpr b}^{(0)} 
\right)
\right]
\\
&-
\frac{2 \pi}{\hbar^2}  e E^{\mu} \tau_{\mbk}^a
\sum_{\mbkpr b}
\delta ( \veps_{\mbk}^a - \veps_{\mbkpr}^b )
\pd_{\mbkpr}^{\rho}
\left[
\frac{v_{\rho \mbkpr}^b}{\lvert \bs{v}_{\mbkpr}^b \rvert^2}
\left(
\mathcal{A}_{\mu \mbk}^a - \mathcal{A}_{\mu \mbkpr}^b 
\right)
\Braket{
U_{\mbk \mbk\pr}^{ab} U_{\mbk\pr \mbk} ^{ba} 
}
\left(
n_{E \mbk a}^{(0)} - n_{E \mbkpr b}^{(0)} 
\right)
\right],
\end{split}
\end{equation}

\begin{equation}
\begin{split}
\mathcal{N}^{(\text{ssj})}_{2,\mbk a}
=&
- \frac{2 \pi}{\hbar}  e E^{\mu} \tau_{\mbk}^a
\sum_{\mbkpr bc}
\frac{
\text{Re}
\left[
\Braket{
U_{\mbk \mbk\pr}^{ab} U_{\mbk\pr \mbk} ^{bc}
}
\mathcal{A}_{\mu \mbk}^{\prime ca}
\right]
}
{\veps_{\mbk}^a - \veps_{\mbk}^c}
\left[
\left(
n_{E \mbk a}^{(0)} - n_{E \mbkpr b}^{(0)}
\right)
\delta ( \veps_{\mbk}^a - \veps_{\mbkpr}^b ) 
-
\left(
n_{E \mbk c}^{(0)} - n_{E \mbkpr b}^{(0)}
\right)
\delta ( \veps_{\mbk}^c - \veps_{\mbkpr}^b ) 
\right]
\\
&-
\frac{2 \pi}{\hbar}  e E^{\mu} \tau_{\mbk}^a
\sum_{\mbkpr bc}
\frac{
\text{Re}
\left[
\Braket{
U_{\mbk \mbk\pr}^{ab} U_{\mbk\pr \mbk} ^{ca}
}
\mathcal{A}_{\mu \mbkpr}^{\prime bc}
\right]
}
{\veps_{\mbkpr}^b - \veps_{\mbkpr}^c}
\left[
\left(
n_{E \mbk a}^{(0)} - n_{E \mbkpr b}^{(0)}
\right)
\delta ( \veps_{\mbk}^a - \veps_{\mbkpr}^b ) 
-
\left(
n_{E \mbk a}^{(0)} - n_{E \mbkpr c}^{(0)}
\right)
\delta ( \veps_{\mbk}^a - \veps_{\mbkpr}^c ) 
\right].
\end{split}
\end{equation}

\subsection{Tertiary skew scattering}

Similar to the previous skew scattering expressions, the tertiary skew scattering density reads
\begin{equation}
\begin{split}
n_{E^2 \mbk a}^{(\text{tsk})}
&=
- \frac{2\pi^2}{\hbar} \tau_{\mbk}^a
\sum_{\mbkpr \mbkdpr bcd}
\delta(\veps_{\mbk}^a 
-
\veps_{\mbkpr}^b)
\\
&\times
\Biggl\{
\frac{
\text{Im}
\left[
\Braket{U_{\mbk \mbk\pr}^{ab} U_{\mbkpr \mbk}^{bc}}
\Braket{U_{\mbk \mbkdpr}^{cd} U_{\mbkdpr \mbk}^{da}}
\right]
} 
{\veps_{\mbk}^a - \veps_{\mbk}^c}
\left[
( n_{E^2 \mbk a}^{(-1)} - n_{E^2 \mbkdpr d}^{(-1)}) 
\delta(\veps_{\mbk}^a -\veps_{\mbkdpr}^d)
+
( n_{E^2\mbk c}^{(-1)} - n_{E^2\mbkdpr d}^{(-1)}) 
\delta(\veps_{\mbk}^c -\veps_{\mbkdpr}^d)
\right]_{c \neq a}
\\
&+ 
\frac{
\text{Im}
\left[
\Braket{U_{\mbk \mbk\pr}^{ab} U_{\mbkpr \mbk}^{ca}}
\Braket{U_{\mbkpr \mbkdpr}^{bd} U_{\mbkdpr \mbkpr}^{dc}}
\right]
} 
{\veps_{\mbk}^a - \veps_{\mbkpr}^c}
\left[
( n_{E^2\mbkpr b}^{(-1)} - n_{E^2\mbkdpr d}^{(-1)})
\delta(\veps_{\mbkpr}^b -\veps_{\mbkdpr}^d)
+
( n_{E^2\mbkpr c}^{(-1)} - n_{E^2\mbkdpr d}^{(-1)}) 
\delta(\veps_{\mbkpr}^c - \veps_{\mbkdpr}^d)
\right]_{c \neq b} \biggr\}.
\end{split}
\end{equation}

\subsection{Quadratic side jump}

The quadratic side jump density is rather lengthy, as it is comprised of the three collision integrals given by Eq.~(\ref{n_E2k_qsj}). After a number of cancellations, the remaining terms are expressed as
\begin{equation}
n_{E^2\mbk a}^{(\text{qsj})}
=
\sum_{i=1}^{8}
\mathcal{N}^{(\text{qsj})}_{i,\mbk a},
\end{equation}
where
\begin{equation}
\begin{split}
\mathcal{N}^{(\text{qsj})}_{1,\mbk a}
=&
- \frac{\pi}{4} \hbar e^2 \tau_{\mbk}^a
\frac{\pd^4 f_{0\mbk}^a}{\pd(\veps_{\mbk}^a)^4}
\sum_{\mbkpr b}
\delta ( \veps_{\mbk}^a - \veps_{\mbkpr}^b )
\Braket{U_{\mbk \mbk\pr}^{ab} U_{\mbk\pr \mbk}^{ba}}
\left[
\mathbf{E} \cdot \left(\bs{v}_{\mbk}^a + \bs{v}_{\mbkpr}^b \right)
\right]^2
\\
&-
\pi \hbar e^2  E^{\mu} E^{\nu} \tau_{\mbk}^a
\frac{\pd^3 f_{0\mbk}^a}{\pd(\veps_{\mbk}^a)^4}
\sum_{\mbkpr b}
\delta ( \veps_{\mbk}^a - \veps_{\mbkpr}^b )
\biggl\{
\pd_{\mbkpr}^{\rho} 
\left[ \frac{v_{\rho \mbkpr}^b}{\lvert \bs{v}_{\mbkpr}^b \rvert^2} 
\Braket{U_{\mbk \mbk\pr}^{ab} U_{\mbk\pr \mbk}^{ba}}
v_{\mu \mbk}^a \left(v_{\nu \mbk}^a + v_{\nu \mbkpr}^b \right)
\right]
\\
&+ 
\frac{1}{3} \left( \pd_{\mbk}^{\mu} v_{\nu \mbk}^a
+
2 \pd_{\mbkpr}^{\mu} v_{\nu \mbkpr}^b \right)
\Braket{U_{\mbk \mbk\pr}^{ab} U_{\mbk\pr \mbk}^{ba}}
+
\left(v_{\mu \mbk}^a + v_{\mu \mbkpr}^b \right)
\text{Re}
\Braket{
U_{\mbk \mbk\pr}^{ab} 
\left[
\left( \pd_{\mbk}^{\nu} + \pd_{\mbkpr}^{\nu} \right) U_{\mbk\pr \mbk} ^{ba} 
\right]
}
\biggr\},
\end{split}
\end{equation}
\begin{equation}
\begin{split}
\mathcal{N}^{(\text{qsj})}_{2,\mbk a}
=&
- \frac{\pi}{\hbar}  e^2 E^{\mu} E^{\nu}\tau_{\mbk}^a
\frac{\pd^2 f_{0\mbk}^a}{\pd(\veps_{\mbk}^a)^2}
\sum_{\mbkpr b}
\delta ( \veps_{\mbk}^a - \veps_{\mbkpr}^b )
\biggl\{
\frac{1}{2} \pd_{\mbkpr}^{\rho}
\left(
\frac{v_{\rho \mbkpr}^b}{\lvert \bs{v}_{\mbkpr}^b \rvert^2}
\pd_{\mbkpr}^{\sigma}
\left[
\frac{v_{\sigma \mbkpr}^b}{\lvert \bs{v}_{\mbkpr}^b \rvert^2}
\left(
v_{\mu \mbk}^a v_{\nu \mbk}^a
-
v_{\mu \mbkpr}^b v_{\nu \mbkpr}^b
\right)
\Braket{U_{\mbk \mbk\pr}^{ab} U_{\mbk\pr \mbk}^{ba}}
\right]
\right)
\\
&+
\pd_{\mbkpr}^{\rho}
\left[
\frac{v_{\rho \mbkpr}^b}{\lvert \bs{v}_{\mbkpr}^b \rvert^2}
\left(
\pd_{\mbk}^{\mu} v_{\nu \mbk}^a
+
\pd_{\mbkpr}^{\mu} v_{\nu \mbkpr}^b
\right)
\Braket{U_{\mbk \mbk\pr}^{ab} U_{\mbk\pr \mbk}^{ba}}
\right]
+
\pd_{\mbkpr}^{\rho}
\left[
\frac{v_{\rho \mbkpr}^b}{\lvert \bs{v}_{\mbkpr}^b \rvert^2}
\left(
3 v_{\mu \mbk}^a
+
v_{\mu \mbkpr}^b
\right)
\text{Re}
\Braket{
U_{\mbk \mbk\pr}^{ab} 
\left[
\left( \pd_{\mbk}^{\nu} + \pd_{\mbkpr}^{\nu} \right) U_{\mbk\pr \mbk} ^{ba} 
\right]
}
\right]
\\
&+
\text{Re}
\Braket{
U_{\mbk \mbk\pr}^{ab} 
\left[
\left( \pd_{\mbk}^{\mu} + \pd_{\mbkpr}^{\mu} \right)
\left( \pd_{\mbk}^{\nu} + \pd_{\mbkpr}^{\nu} \right) U_{\mbk\pr \mbk} ^{ba} 
\right]
}
\biggr\},
\end{split}
\end{equation}
\begin{equation}
\begin{split}
\mathcal{N}^{(\text{qsj})}_{3,\mbk a}
=&
- \frac{\pi}{\hbar^2}  e^2 E^{\mu} E^{\nu}\tau_{\mbk}^a
\frac{\pd f_{0\mbk}^a}{\pd \veps_{\mbk}^a}
\sum_{\mbkpr b}
\delta ( \veps_{\mbk}^a - \veps_{\mbkpr}^b )
\biggl\{
\pd_{\mbkpr}^{\rho}
\left(
\frac{v_{\rho \mbkpr}^b}{\lvert \bs{v}_{\mbkpr}^b \rvert^2}
\pd_{\mbkpr}^{\sigma}
\left[
\frac{v_{\sigma \mbkpr}^b}{\lvert \bs{v}_{\mbkpr}^b \rvert^2}
\left(
v_{\mu \mbk}^a - v_{\mu \mbkpr}^b 
\right)
\text{Re}
\Braket{
U_{\mbk \mbk\pr}^{ab} 
\left[
\left( \pd_{\mbk}^{\nu} + \pd_{\mbkpr}^{\nu} \right) U_{\mbk\pr \mbk} ^{ba} 
\right]
}
\right]
\right)
\\
&+
2 \pd_{\mbkpr}^{\rho}
\left(
\frac{v_{\rho \mbkpr}^b}{\lvert \bs{v}_{\mbkpr}^b \rvert^2}
\text{Re}
\Braket{
U_{\mbk \mbk\pr}^{ab} 
\left[
\left( \pd_{\mbk}^{\mu} + \pd_{\mbkpr}^{\mu} \right)
\left( \pd_{\mbk}^{\nu} + \pd_{\mbkpr}^{\nu} \right) U_{\mbk\pr \mbk} ^{ba} 
\right]
}
\right)
\biggr\},
\end{split}
\end{equation}
\begin{equation}
\begin{split}
\mathcal{N}^{(\text{qsj})}_{4,\mbk a}
=&
\frac{4 \pi}{\hbar^2}  e^2 E^{\mu} E^{\nu}\tau_{\mbk}^a
\frac{\pd f_{0\mbk}^a}{\pd \veps_{\mbk}^a}
\sum_{\mbkpr b}
\delta ( \veps_{\mbk}^a - \veps_{\mbkpr}^b )
\biggl\{
\pd_{\mbkpr}^{\rho}
\left[
\frac{v_{\rho \mbkpr}^b}{\lvert \bs{v}_{\mbkpr}^b \rvert^2}
\left(
\mathcal{A}_{\mu \mbk}^a - \mathcal{A}_{\mu \mbkpr}^b 
\right)
\text{Im}
\Braket{
U_{\mbk \mbk\pr}^{ab} 
\left[
\left( \pd_{\mbk}^{\nu} + \pd_{\mbkpr}^{\nu} \right) U_{\mbk\pr \mbk} ^{ba} 
\right]
}
\right]
\\
&+
\frac{1}{2} \pd_{\mbkpr}^{\rho}
\left(
\frac{v_{\rho \mbkpr}^b}{\lvert \bs{v}_{\mbkpr}^b \rvert^2}
\left(
\mathcal{A}_{\mu \mbk}^a - \mathcal{A}_{\mu \mbkpr}^b 
\right)
\left(
\mathcal{A}_{\nu \mbk}^a - \mathcal{A}_{\nu \mbkpr}^b 
\right)
\Braket{U_{\mbk \mbk\pr}^{ab} U_{\mbk\pr \mbk}^{ba}}
\right)
\biggr\}
\\
&+
\frac{2 \pi}{\hbar}  e^2 E^{\mu} E^{\nu}\tau_{\mbk}^a
\frac{\pd^2 f_{0\mbk}^a}{\pd (\veps_{\mbk}^a)^2}
\sum_{\mbkpr b}
\delta ( \veps_{\mbk}^a - \veps_{\mbkpr}^b )
\biggl\{
\left(
\mathcal{A}_{\mu \mbk}^a - \mathcal{A}_{\mu \mbkpr}^b 
\right)
\text{Im}
\Braket{
U_{\mbk \mbk\pr}^{ab} 
\left[
\left( \pd_{\mbk}^{\nu} + \pd_{\mbkpr}^{\nu} \right) U_{\mbk\pr \mbk} ^{ba} 
\right]
}
\\
&+
\frac{1}{2}
\left(
\mathcal{A}_{\mu \mbk}^a - \mathcal{A}_{\mu \mbkpr}^b 
\right)
\left(
\mathcal{A}_{\nu \mbk}^a - \mathcal{A}_{\nu \mbkpr}^b 
\right)
\Braket{U_{\mbk \mbk\pr}^{ab} U_{\mbk\pr \mbk}^{ba}}
\biggr\},
\end{split}
\end{equation}
\begin{equation}
\begin{split}
\mathcal{N}^{(\text{qsj})}_{5,\mbk a}
=&
\pi  e^2 E^{\mu} E^{\nu} \tau_{\mbk}^a
\sum_{\mbkpr b c}
\frac{\pd^2 f_{0\mbkpr}^b}{\pd (\veps_{\mbkpr}^b)^2}
\frac{
\text{Im}
\left[
\Braket{U_{\mbk \mbk\pr}^{ab} U_{\mbk\pr \mbk}^{bc}} \mathcal{A}^{\prime ca}_{\mu \mbk}
\right]
}
{\veps_{\mbk}^a - \veps_{\mbk}^c}
\left[
\left(
v_{\mu \mbk}^a +v_{\mu \mbkpr}^b 
\right)
\delta ( \veps_{\mbk}^a - \veps_{\mbkpr}^b )
-
\left(
v_{\mu \mbk}^c +v_{\mu \mbkpr}^b 
\right)
\delta ( \veps_{\mbk}^c - \veps_{\mbkpr}^b )
\right]
\\
&+
\frac{2 \pi}{\hbar}  e^2 E^{\mu} E^{\nu} \tau_{\mbk}^a
\frac{\pd f_{0\mbk}^a}{\pd \veps_{\mbk}^a}
\sum_{\mbkpr b c}
\delta ( \veps_{\mbk}^a - \veps_{\mbkpr}^b )
\frac{
\text{Im}
\Braket{
U_{\mbk \mbk\pr}^{ab} 
\left[
\left( \pd_{\mbk}^{\nu} + \pd_{\mbkpr}^{\nu} \right) \left( U_{\mbk\pr \mbk} ^{bc} 
\mathcal{A}^{\prime ca}_{\mu \mbk} \right)
\right]
} 
}
{\veps_{\mbk}^a - \veps_{\mbk}^c}
\\
&+
2 \pi  e^2 E^{\mu} E^{\nu} \tau_{\mbk}^a
\sum_{\mbkpr b c}
\delta ( \veps_{\mbk}^a - \veps_{\mbkpr}^b )
\frac{
\text{Im}
\left[
\Braket{U_{\mbk \mbk\pr}^{ab} U_{\mbk\pr \mbk}^{bc}} \mathcal{A}^{\prime ca}_{\mu \mbk}
\right]
}
{\left( \veps_{\mbk}^a - \veps_{\mbk}^c \right)^2}
v_{\nu \mbk}^c
\left(
\frac{\pd f_{0\mbk}^a}{\pd \veps_{\mbk}^a}
-
\frac{\pd f_{0\mbk}^c}{\pd \veps_{\mbk}^c}\right)
\\
&-
\frac{2 \pi}{\hbar}  e^2 E^{\mu} E^{\nu} \tau_{\mbk}^a
\sum_{\mbkpr b c}
\delta ( \veps_{\mbk}^c- \veps_{\mbkpr}^b )
\frac{\pd f_{0\mbk}^c}{\pd \veps_{\mbk}^c}
\frac{
\text{Im}
\Braket{
U_{\mbk \mbk\pr}^{ab} 
\left[
\left( \pd_{\mbk}^{\nu} + \pd_{\mbkpr}^{\nu} \right) \left( U_{\mbk\pr \mbk} ^{bc} 
\right)
\right]
} \mathcal{A}^{\prime ca}_{\mu \mbk} 
}
{\veps_{\mbk}^a - \veps_{\mbk}^c},
\end{split}
\end{equation}
\begin{equation}
\begin{split}
\mathcal{N}^{(\text{qsj})}_{6,\mbk a}
=&
\frac{2 \pi}{\hbar}  e^2 E^{\mu} E^{\nu} \tau_{\mbk}^a
\frac{\pd f_{0\mbk}^a}{\pd \veps_{\mbk}^a}
\sum_{\mbkpr b c}
\delta ( \veps_{\mbk}^a - \veps_{\mbkpr}^b )
\frac{
\text{Im}
\Braket{
U_{\mbk \mbk\pr}^{ab} 
\left[
\left( \pd_{\mbk}^{\nu} + \pd_{\mbkpr}^{\nu} \right) \left( U_{\mbk\pr \mbk} ^{ca} 
\mathcal{A}^{\prime bc}_{\mu \mbkpr} \right)
\right]
} 
}
{\veps_{\mbk}^a - \veps_{\mbkpr}^c}
\\
&+
2 \pi  e^2 E^{\mu} E^{\nu} \tau_{\mbk}^a
\frac{\pd f_{0\mbk}^a}{\pd \veps_{\mbk}^a}
\sum_{\mbkpr b c}
\delta ( \veps_{\mbk}^a - \veps_{\mbkpr}^b )
\frac{
\text{Im}
\left[
\Braket{U_{\mbk \mbk\pr}^{ab} U_{\mbk\pr \mbk}^{ca}} \mathcal{A}^{\prime bc}_{\mu \mbkpr}
\right]
}
{\left( \veps_{\mbk}^a - \veps_{\mbkpr}^c \right)^2}
\left[
v_{\nu \mbkpr}^c
-
\frac{\bs{v}_{\mbkpr}^b \cdot \bs{v}_{\mbkpr}^c}
{\lvert \bs{v}_{\mbkpr}^b \rvert^2}
\left(
v_{\nu \mbk}^a - v_{\nu \mbkpr}^b
\right)
\right]
\\
&-
2 \pi  e^2 E^{\mu} E^{\nu} \tau_{\mbk}^a
\sum_{\mbkpr b c}
\delta ( \veps_{\mbk}^a - \veps_{\mbkpr}^b )
\frac{\pd f_{0\mbkpr}^c}{\pd \veps_{\mbkpr}^c}
\frac{
\text{Im}
\left[
\Braket{U_{\mbk \mbk\pr}^{ab} U_{\mbk\pr \mbk}^{ca}} \mathcal{A}^{\prime bc}_{\mu \mbkpr}
\right]
}
{\left( \veps_{\mbk}^a - \veps_{\mbkpr}^c \right)^2}
\left[
v_{\nu \mbkpr}^c
+
\frac{\bs{v}_{\mbkpr}^b \cdot \bs{v}_{\mbkpr}^c}
{\lvert \bs{v}_{\mbkpr}^b \rvert^2}
\left(
v_{\nu \mbk}^a - v_{\nu \mbkpr}^b
\right)
\right]
\\
&+
4 \pi  e^2 E^{\mu} E^{\nu} \tau_{\mbk}^a
\sum_{\mbkpr b c}
\delta ( \veps_{\mbk}^a - \veps_{\mbkpr}^b )
\text{Im}
\left[
\Braket{U_{\mbk \mbk\pr}^{ab} U_{\mbk\pr \mbk}^{ca}} \mathcal{A}^{\prime bc}_{\mu \mbkpr}
\right]
\frac{\bs{v}_{\mbkpr}^b \cdot \bs{v}_{\mbkpr}^c}
{\lvert \bs{v}_{\mbkpr}^b \rvert^2}
\left(
v_{\nu \mbk}^a - v_{\nu \mbkpr}^b 
\right)
\frac{
f_{0 \mbk}^a - f_{0 \mbkpr}^c
}
{\left( \veps_{\mbk}^a - \veps_{\mbkpr}^c \right)^3 } 
\\
&+
\frac{2 \pi}{\hbar}  e^2 E^{\mu} E^{\nu} \tau_{\mbk}^a
\frac{\pd f_{0\mbk}^a}{\pd \veps_{\mbk}^a}
\sum_{\mbkpr b c}
\delta ( \veps_{\mbk}^a- \veps_{\mbkpr}^c )
\frac{
\text{Im}
\Braket{
U_{\mbk \mbk\pr}^{ab} 
\left[
\left( \pd_{\mbk}^{\nu} + \pd_{\mbkpr}^{\nu} \right) \left( U_{\mbk\pr \mbk} ^{ca} 
\right)
\right]
} \mathcal{A}^{\prime bc}_{\mu \mbkpr} 
}
{\veps_{\mbk}^a - \veps_{\mbkpr}^b},
\end{split}
\end{equation}
\begin{equation}
\begin{split}
\mathcal{N}^{(\text{qsj})}_{7,\mbk a}
=&
\frac{2 \pi}{\hbar}  e^2 E^{\mu} E^{\nu} \tau_{\mbk}^a
\frac{\pd f_{0\mbk}^a}{\pd \veps_{\mbk}^a}
\sum_{\mbkpr b c}
\delta ( \veps_{\mbk}^a- \veps_{\mbkpr}^b )
\left(
\mathcal{A}_{\nu \mbk}^a - \mathcal{A}_{\nu \mbkpr}^b 
\right)
\frac{
\text{Im}
\Braket{
U_{\mbk \mbk\pr}^{ab} U_{\mbk\pr \mbk} ^{bc} 
} \mathcal{A}^{\prime ca}_{\mu \mbk} 
}
{\veps_{\mbk}^a - \veps_{\mbk}^c}
\\
&-
\frac{2 \pi}{\hbar}  e^2 E^{\mu} E^{\nu} \tau_{\mbk}^a
\sum_{\mbkpr b c}
\delta ( \veps_{\mbk}^c- \veps_{\mbkpr}^b )
\frac{\pd f_{0\mbk}^c}{\pd \veps_{\mbk}^c}
\left(
\mathcal{A}_{\nu \mbk}^c - \mathcal{A}_{\nu \mbkpr}^b 
\right)
\frac{
\text{Im}
\Braket{
U_{\mbk \mbk\pr}^{ab} U_{\mbk\pr \mbk} ^{bc} 
} \mathcal{A}^{\prime ca}_{\mu \mbk} 
}
{\veps_{\mbk}^a - \veps_{\mbk}^c}
\\
&+
\frac{4 \pi}{\hbar}  e^2 E^{\mu} E^{\nu} \tau_{\mbk}^a
\frac{\pd f_{0\mbk}^a}{\pd \veps_{\mbk}^a}
\sum_{\mbkpr b c}
\delta ( \veps_{\mbk}^a- \veps_{\mbkpr}^b )
\left(
\mathcal{A}_{\nu \mbk}^a - \mathcal{A}_{\nu \mbkpr}^b 
\right)
\frac{
\text{Im}
\Braket{
U_{\mbk \mbk\pr}^{ab} U_{\mbk\pr \mbk} ^{ca} 
} \mathcal{A}^{\prime bc}_{\mu \mbkpr} 
}
{\veps_{\mbk}^a - \veps_{\mbkpr}^c},
\end{split}
\end{equation}
\begin{equation}
\label{N_qsj_8}
\begin{split}
\mathcal{N}^{(\text{qsj})}_{8,\mbk a}
=&
\frac{2 \pi}{\hbar}  e^2 E^{\mu} E^{\nu} \tau_{\mbk}^a
\sum_{\mbkpr b c}
\delta ( \veps_{\mbk}^a- \veps_{\mbkpr}^b )
\Braket{
U_{\mbk \mbk\pr}^{ab} U_{\mbk\pr \mbk} ^{ba}}
\left[
\frac{g_{\mu\nu,\mbk}^{ac}}
{\veps_{\mbk}^a - \veps_{\mbk}^c}
\left(
\frac{\pd f_{0\mbk}^a}{\pd \veps_{\mbk}^a}
-
\frac{f_{0\mbk}^a - f_{0\mbk}^c}
{\veps_{\mbk}^a - \veps_{\mbk}^c}
\right)
-
\frac{g_{\mu\nu,\mbkpr}^{bc}}
{\veps_{\mbk}^a - \veps_{\mbkpr}^c}
\left(
\frac{\pd f_{0\mbk}^a}{\pd \veps_{\mbk}^a}
-
\frac{f_{0\mbk}^a - f_{0\mbkpr}^c}
{\veps_{\mbk}^a - \veps_{\mbkpr}^c}
\right)
\right].
\end{split}
\end{equation}

\subsection{Cubic skew scattering}

The density of cubic skew-scattered carriers arises from the lengthy off-diagonal element given by Eq.~(\ref{S_E2_Ipr_I-1}) and results in the expression
\begin{equation}
n_{E^2\mbk a}^{(\text{csk})}
=
\sum_{i=1}^{8}
\mathcal{N}^{(\text{csk})}_{i,\mbk a},
\end{equation}
with
\begin{equation}
\begin{split}
\mathcal{N}^{(\text{csk})}_{1,\mbk a}
=&
\frac{2\pi^3}{\hbar}
\tau_{\mbk}^a
\sum_{\mbkpr \mbkdpr \mbk_1 bcdef}
\delta(\veps_{\mbk}^a - \veps_{\mbkpr}^b )
\delta(\veps_{\mbk}^a - \veps_{\mbk_1}^d )
\frac{
\text{Re}
\left[
\Braket{U_{\mbk \mbkpr}^{ab} U_{\mbkpr \mbk}^{bc}}
\Braket{U_{\mbk \mbk_1}^{cd} U_{\mbk_1 \mbk}^{de}}
\Braket{U_{\mbk \mbkdpr}^{ef} U_{\mbkdpr \mbk}^{fa}}
\right]_{\substack{c,e \neq a}}
}
{(\veps_{\mbk}^a - \veps_{\mbk}^c)
(\veps_{\mbk}^a - \veps_{\mbk}^e)}
\\
&\times
\left[
( n_{E^2 \mbk a}^{(-2)} - n_{E^2 \mbkdpr f}^{(-2)}) 
\delta(\veps_{\mbk}^a -\veps_{\mbkdpr}^f)
+
( n_{E^2\mbk_1 d}^{(-2)} - n_{E^2\mbkdpr f}^{(-2)}) 
\delta(\veps_{\mbk_1}^d -\veps_{\mbkdpr}^f)
\right],
\end{split}
\end{equation}
\begin{equation}
\begin{split}
\mathcal{N}^{(\text{csk})}_{2,\mbk a}
=&
\frac{2\pi^3}{\hbar}
\tau_{\mbk}^a
\sum_{\mbkpr \mbkdpr \mbk_1 bcdef}
\delta(\veps_{\mbk}^a - \veps_{\mbkpr}^b )
\delta(\veps_{\mbk}^a - \veps_{\mbk_1}^d )
\frac{
\text{Re}
\left[
\Braket{U_{\mbk \mbkpr}^{ab} U_{\mbkpr \mbk}^{bc}}
\Braket{U_{\mbk \mbk_1}^{cd} U_{\mbk_1 \mbk}^{ea}}
\Braket{U_{\mbk_1 \mbkdpr}^{df} U_{\mbkdpr \mbk_1}^{fe}}
\right]_{\substack{c \neq a \\ e \neq d}}
}
{(\veps_{\mbk}^a - \veps_{\mbk}^c)
(\veps_{\mbk_1}^d - \veps_{\mbk_1}^e)}
\\
&\times
\left[
( n_{E^2 \mbk_1 e}^{(-2)} - n_{E^2 \mbkdpr f}^{(-2)}) 
\delta(\veps_{\mbk_1}^e -\veps_{\mbkdpr}^f)
+
( n_{E^2 \mbk_1 d}^{(-2)} - n_{E^2 \mbkdpr f}^{(-2)}) 
\delta(\veps_{\mbk_1}^d -\veps_{\mbkdpr}^f)
\right],
\end{split}
\end{equation}
\begin{equation}
\begin{split}
\mathcal{N}^{(\text{csk})}_{3,\mbk a}
=&
\frac{2\pi^3}{\hbar}
\tau_{\mbk}^a
\sum_{\mbkpr \mbkdpr \mbk_1 bcdef}
\delta(\veps_{\mbk}^a - \veps_{\mbkpr}^b )
\delta(\veps_{\mbk}^c - \veps_{\mbk_1}^d )
\frac{
\text{Re}
\left[
\Braket{U_{\mbk \mbkpr}^{ab} U_{\mbkpr \mbk}^{bc}}
\Braket{U_{\mbk \mbk_1}^{ce} U_{\mbk_1 \mbk}^{da}}
\Braket{U_{\mbk_1 \mbkdpr}^{ef} U_{\mbkdpr \mbk_1}^{fd}}
\right]_{\substack{c \neq a \\ e \neq d}}
}
{(\veps_{\mbk}^a - \veps_{\mbk}^c)
(\veps_{\mbk_1}^e - \veps_{\mbk_1}^d)}
\\
&\times
\left[
( n_{E^2 \mbk_1 e}^{(-2)} - n_{E^2 \mbkdpr f}^{(-2)}) 
\delta(\veps_{\mbk_1}^e -\veps_{\mbkdpr}^f)
+
( n_{E^2\mbk_1 d}^{(-2)} - n_{E^2\mbkdpr f}^{(-2)}) 
\delta(\veps_{\mbk_1}^d -\veps_{\mbkdpr}^f)
\right],
\end{split}
\end{equation}
\begin{equation}
\begin{split}
\mathcal{N}^{(\text{csk})}_{4,\mbk a}
=&
\frac{2\pi^3}{\hbar}
\tau_{\mbk}^a
\sum_{\mbkpr \mbkdpr \mbk_1 bcdef}
\delta(\veps_{\mbk}^a - \veps_{\mbkpr}^b )
\delta(\veps_{\mbk}^c - \veps_{\mbk_1}^d )
\frac{
\text{Re}
\left[
\Braket{U_{\mbk \mbkpr}^{ab} U_{\mbkpr \mbk}^{bc}}
\Braket{U_{\mbk \mbk_1}^{ed} U_{\mbk_1 \mbk}^{da}}
\Braket{U_{\mbk \mbkdpr}^{cf} U_{\mbkdpr \mbk}^{fe}}
\right]_{\substack{a,e \neq c}}
}
{(\veps_{\mbk}^a - \veps_{\mbk}^c)
(\veps_{\mbk}^e - \veps_{\mbk_1}^c)}
\\
&\times
\left[
( n_{E^2 \mbk e}^{(-2)} - n_{E^2 \mbkdpr f}^{(-2)}) 
\delta(\veps_{\mbk}^e -\veps_{\mbkdpr}^f)
+
( n_{E^2 \mbk c}^{(-2)} - n_{E^2 \mbkdpr f}^{(-2)}) 
\delta(\veps_{\mbk}^c -\veps_{\mbkdpr}^f)
\right],
\end{split}
\end{equation}
\begin{equation}
\begin{split}
\mathcal{N}^{(\text{csk})}_{5,\mbk a}
=&
\frac{2\pi^3}{\hbar}
\tau_{\mbk}^a
\sum_{\mbkpr \mbkdpr \mbk_1 bcdef}
\delta(\veps_{\mbk}^a - \veps_{\mbkpr}^b )
\delta(\veps_{\mbkpr}^c - \veps_{\mbk_1}^d )
\frac{
\text{Re}
\left[
\Braket{U_{\mbk \mbkpr}^{ab} U_{\mbkpr \mbk}^{ca}}
\Braket{U_{\mbkpr \mbk_1}^{bd} U_{\mbk_1 \mbkpr}^{de}}
\Braket{U_{\mbkpr \mbkdpr}^{ef} U_{\mbkdpr \mbkpr}^{fc}}
\right]_{\substack{b,e \neq c}}
}
{(\veps_{\mbk}^a - \veps_{\mbkpr}^c)
(\veps_{\mbkpr}^c - \veps_{\mbkpr}^e)}
\\
&\times
\left[
( n_{E^2 \mbkpr c}^{(-2)} - n_{E^2 \mbkdpr f}^{(-2)}) 
\delta(\veps_{\mbkpr}^c -\veps_{\mbkdpr}^f)
+
( n_{E^2\mbkpr e}^{(-2)} - n_{E^2\mbkdpr f}^{(-2)}) 
\delta(\veps_{\mbkpr}^e -\veps_{\mbkdpr}^f)
\right],
\end{split}
\end{equation}
\begin{equation}
\begin{split}
\mathcal{N}^{(\text{csk})}_{6,\mbk a}
=&
\frac{2\pi^3}{\hbar}
\tau_{\mbk}^a
\sum_{\mbkpr \mbkdpr \mbk_1 bcdef}
\delta(\veps_{\mbk}^a - \veps_{\mbkpr}^b )
\delta(\veps_{\mbkpr}^c - \veps_{\mbk_1}^d )
\frac{
\text{Re}
\left[
\Braket{U_{\mbk \mbkpr}^{ab} U_{\mbkpr \mbk}^{ca}}
\Braket{U_{\mbkpr \mbk_1}^{bd} U_{\mbk_1 \mbkpr}^{ec}}
\Braket{U_{\mbk_1 \mbkdpr}^{df} U_{\mbkdpr \mbk_1}^{fe}}
\right]_{\substack{c \neq b \\ e \neq d}}
}
{(\veps_{\mbk}^a - \veps_{\mbkpr}^c)
(\veps_{\mbk_1}^d - \veps_{\mbk_1}^e)}
\\
&\times
\left[
( n_{E^2 \mbk_1 e}^{(-2)} - n_{E^2 \mbkdpr f}^{(-2)}) 
\delta(\veps_{\mbk_1}^e -\veps_{\mbkdpr}^f)
+
( n_{E^2 \mbk_1 d}^{(-2)} - n_{E^2 \mbkdpr f}^{(-2)}) 
\delta(\veps_{\mbk_1}^d -\veps_{\mbkdpr}^f)
\right],
\end{split}
\end{equation}
\begin{equation}
\begin{split}
\mathcal{N}^{(\text{csk})}_{7,\mbk a}
=&
\frac{2\pi^3}{\hbar}
\tau_{\mbk}^a
\sum_{\mbkpr \mbkdpr \mbk_1 bcdef}
\delta(\veps_{\mbk}^a - \veps_{\mbkpr}^b )
\delta(\veps_{\mbkpr}^b - \veps_{\mbk_1}^d )
\frac{
\text{Re}
\left[
\Braket{U_{\mbk \mbkpr}^{ab} U_{\mbkpr \mbk}^{ca}}
\Braket{U_{\mbkpr \mbk_1}^{be} U_{\mbk_1 \mbkpr}^{dc}}
\Braket{U_{\mbk_1 \mbkdpr}^{ef} U_{\mbkdpr \mbk_1}^{fd}}
\right]_{\substack{c \neq b \\ e \neq d}}
}
{(\veps_{\mbk}^a - \veps_{\mbkpr}^c)
(\veps_{\mbk_1}^e - \veps_{\mbk_1}^d)}
\\
&\times
\left[
( n_{E^2 \mbk_1 e}^{(-2)} - n_{E^2 \mbkdpr f}^{(-2)}) 
\delta(\veps_{\mbk_1}^e -\veps_{\mbkdpr}^f)
+
( n_{E^2\mbk_1 d}^{(-2)} - n_{E^2\mbkdpr f}^{(-2)}) 
\delta(\veps_{\mbk_1}^d -\veps_{\mbkdpr}^f)
\right],
\end{split}
\end{equation}

\begin{equation}
\begin{split}
\mathcal{N}^{(\text{csk})}_{8,\mbk a}
=&
- \frac{2\pi^3}{\hbar}
\tau_{\mbk}^a
\sum_{\mbkpr \mbkdpr \mbk_1 bcdef}
\delta(\veps_{\mbk}^a - \veps_{\mbkpr}^b )
\delta(\veps_{\mbkpr}^b - \veps_{\mbk_1}^d )
\frac{
\text{Re}
\left[
\Braket{U_{\mbk \mbkpr}^{ab} U_{\mbkpr \mbk}^{ca}}
\Braket{U_{\mbkpr \mbk_1}^{ed} U_{\mbk_1 \mbkpr}^{dc}}
\Braket{U_{\mbkpr \mbkdpr}^{bf} U_{\mbkdpr \mbkpr}^{fe}}
\right]_{\substack{c,e \neq b}}
}
{(\veps_{\mbk}^a - \veps_{\mbkpr}^c )
(\veps_{\mbk}^a - \veps_{\mbkpr}^e )}
\\
&\times
\left[
( n_{E^2 \mbk e}^{(-2)} - n_{E^2 \mbkdpr f}^{(-2)}) 
\delta(\veps_{\mbk}^e -\veps_{\mbkdpr}^f)
+
( n_{E^2 \mbk c}^{(-2)} - n_{E^2 \mbkdpr f}^{(-2)}) 
\delta(\veps_{\mbk}^c -\veps_{\mbkdpr}^f)
\right].
\end{split}
\end{equation}

\subsection{Anomalous skew scattering}

The carrier density associated with anomalous skew scattering is given by
\begin{equation}
n_{E^2\mbk a}^{(\text{ask})}
=
\sum_{i=1}^{30}
\mathcal{N}^{(\text{ask})}_{i,\mbk a},
\end{equation}
\begin{equation}
\begin{split}
\mathcal{N}^{(\text{ask})}_{1,\mbk a}
=&
\frac{2 \pi^2}{\hbar^2}  e E^{\mu} \tau_{\mbk}^a
\sum_{\mbkpr \mbkdpr bcd}
\delta ( \veps_{\mbk}^a - \veps_{\mbkpr}^b )
\delta ( \veps_{\mbk}^a - \veps_{\mbkdpr}^d )
\\
&\times
\text{Re}
\Biggl\{
\frac{
\Braket{
U_{\mbk \mbk\pr}^{ab} U_{\mbk\pr \mbk} ^{bc} 
}
}
{\veps_{\mbk}^a - \veps_{\mbk}^c}
\pd_{\mbkdpr}^{\rho}
\left(
\frac{v_{\rho \mbkdpr}^d}{\lvert \bs{v}_{\mbkdpr}^d \rvert^2}
\Braket{
U_{\mbk \mbkdpr}^{cd} 
\left( \pd_{\mbk}^{\mu} + \pd_{\mbkdpr}^{\mu} \right) 
\left[
U_{\mbkdpr \mbk} ^{da}
\left(
n_{E \mbk a}^{(-1)} - n_{E \mbkdpr d}^{(-1)} 
\right) 
\right]
}
\right)
\\
&+
\frac{1}{2}
\frac{
\Braket{
U_{\mbk \mbk\pr}^{ab} U_{\mbk\pr \mbk} ^{bc} 
}
}
{\veps_{\mbk}^a - \veps_{\mbk}^c}
\pd_{\mbkdpr}^{\rho}
\left(
\frac{v_{\rho \mbkdpr}^d}{\lvert \bs{v}_{\mbkdpr}^d \rvert^2}
\pd_{\mbkdpr}^{\sigma}
\left[
\frac{v_{\sigma \mbkdpr}^d}{\lvert \bs{v}_{\mbkdpr}^d \rvert^2}
\left(
v_{\mu \mbk}^a - v_{\mu \mbkdpr}^d 
\right)
\Braket{
U_{\mbk \mbkdpr}^{cd}U_{\mbkdpr \mbk} ^{da} 
}
\left(
n_{E \mbk a}^{(-1)} - n_{E \mbkdpr d}^{(-1)} 
\right) 
\right]
\right)
\Biggr\}_{\substack{c \neq a}},
\end{split}
\end{equation}
\begin{equation}
\begin{split}
\mathcal{N}^{(\text{ask})}_{2,\mbk a}
=&
\frac{2 \pi^2}{\hbar^2}  e E^{\mu} \tau_{\mbk}^a
\sum_{\mbkpr \mbkdpr bcd}
\delta ( \veps_{\mbk}^a - \veps_{\mbkpr}^b )
\delta ( \veps_{\mbk}^a - \veps_{\mbkdpr}^d )
\\
&\times
\Biggl\{
\frac{1}{2}
\pd_{\mbkpr}^{\rho}
\left(
\frac{v_{\rho \mbkpr}^b}{\lvert \bs{v}_{\mbkpr}^b \rvert^2}
\pd_{\mbkpr}^{\sigma}
\left[
\frac{v_{\sigma \mbkpr}^b}{\lvert \bs{v}_{\mbkpr}^b \rvert^2}
\left(
v_{\mu \mbk}^a - v_{\mu \mbkpr}^b 
\right)
\frac{
\text{Re}
\left[
\Braket{
U_{\mbk \mbk\pr}^{ab} U_{\mbk\pr \mbk} ^{bc} 
}
\Braket{
U_{\mbk \mbkdpr}^{cd} U_{\mbkdpr \mbk} ^{da}
}
\right]
}
{\veps_{\mbk}^a - \veps_{\mbk}^c}
\left(
n_{E \mbk a}^{(-1)} - n_{E \mbkdpr d}^{(-1)} 
\right) 
\right]
\right)
\\
&+
\pd_{\mbkpr}^{\rho}
\left(
\frac{v_{\rho \mbkpr}^b}{\lvert \bs{v}_{\mbkpr}^b \rvert^2}
\text{Re}
\Braket{
U_{\mbk \mbk\pr}^{ab} 
\left( \pd_{\mbk}^{\mu} + \pd_{\mbkpr}^{\mu}\right)
\left[
U_{\mbk\pr \mbk} ^{bc}
\frac{
\Braket{
U_{\mbk \mbkdpr}^{cd} U_{\mbkdpr \mbk} ^{da} 
}
}
{\veps_{\mbk}^a - \veps_{\mbk}^c}
\left(
n_{E \mbk a}^{(-1)} - n_{E \mbkdpr d}^{(-1)} 
\right)
\right]  
}
\right)
\\
&+
\pd_{\mbkdpr}^{\rho}
\left(
\frac{v_{\rho \mbkdpr}^d}{\lvert \bs{v}_{\mbkdpr}^d \rvert^2}
\pd_{\mbkpr}^{\sigma}
\left[
\frac{v_{\sigma \mbkpr}^b}{\lvert \bs{v}_{\mbkpr}^b \rvert^2}
v_{\mu \mbk}^a
\frac{
\text{Re}
\left[
\Braket{
U_{\mbk \mbk\pr}^{ab} U_{\mbk\pr \mbk} ^{bc} 
}
\Braket{
U_{\mbk \mbkdpr}^{cd} U_{\mbkdpr \mbk} ^{da}
}
\right]
}
{\veps_{\mbk}^a - \veps_{\mbk}^c}
\left(
n_{E \mbk a}^{(-1)} - n_{E \mbkdpr d}^{(-1)} 
\right) 
\right]
\right)
\Biggr\}_{\substack{c \neq a}},
\end{split}
\end{equation}
\begin{equation}
\begin{split}
\mathcal{N}^{(\text{ask})}_{3,\mbk a}
=&
- \frac{2 \pi^2}{\hbar}  e E^{\mu} \tau_{\mbk}^a
\sum_{\mbkpr \mbkdpr bcd}
\delta ( \veps_{\mbk}^a - \veps_{\mbkpr}^b )
\delta ( \veps_{\mbk}^a - \veps_{\mbkdpr}^d )
\text{Re}
\Biggl\{
\frac{
\Braket{
U_{\mbk \mbk\pr}^{ab} U_{\mbk\pr \mbk} ^{bc} 
}
}
{\veps_{\mbk}^a - \veps_{\mbk}^c}
\pd_{\mbk}^{\mu}
\left[
\frac{v_{\rho \mbkdpr}^d}{\lvert \bs{v}_{\mbkdpr}^d \rvert^2}
\frac{
\Braket{
U_{\mbk \mbkdpr}^{cd} U_{\mbkdpr \mbk}^{da}
}
}
{\veps_{\mbk}^a - \veps_{\mbk}^c}
\left(
n_{E \mbk a}^{(-1)} - n_{E \mbkdpr d}^{(-1)} 
\right) 
\right]
\\
&+
\frac{
\Braket{
U_{\mbk \mbk\pr}^{ab} U_{\mbk\pr \mbk} ^{bc} 
}
}
{\veps_{\mbk}^a - \veps_{\mbk}^c}
\pd_{\mbkdpr}^{\rho}
\left[
\frac{v_{\rho \mbkdpr}^d}{\lvert \bs{v}_{\mbkdpr}^d \rvert^2}
v_{\mu \mbk}^a
\frac{
\Braket{
U_{\mbk \mbkdpr}^{cd} U_{\mbkdpr \mbk}^{da}
}
}
{\veps_{\mbk}^a - \veps_{\mbk}^c}
\left(
n_{E \mbk a}^{(-1)} - n_{E \mbkdpr d}^{(-1)} 
\right) 
\right]
\Biggr\}_{\substack{c \neq a}},
\end{split}
\end{equation}
\begin{equation}
\begin{split}
\mathcal{N}^{(\text{ask})}_{4,\mbk a}
=&
- \frac{2 \pi^2}{\hbar^2}  e E^{\mu} \tau_{\mbk}^a
\sum_{\mbkpr \mbkdpr bcd}
\delta ( \veps_{\mbk}^a - \veps_{\mbkpr}^b )
\delta ( \veps_{\mbk}^c - \veps_{\mbkdpr}^d )
\\
&\times
\text{Re}
\Biggl\{
\frac{
\Braket{
U_{\mbk \mbk\pr}^{ab} U_{\mbk\pr \mbk} ^{bc} 
}
}
{\veps_{\mbk}^a - \veps_{\mbk}^c}
\pd_{\mbkdpr}^{\rho}
\left(
\frac{v_{\rho \mbkdpr}^d}{\lvert \bs{v}_{\mbkdpr}^d \rvert^2}
\Braket{
\left( \pd_{\mbk}^{\mu} + \pd_{\mbkdpr}^{\mu} \right) 
\left[
U_{\mbk \mbkdpr}^{cd} 
\left(
n_{E \mbk c}^{(-1)} - n_{E \mbkdpr d}^{(-1)} 
\right) 
\right]
U_{\mbkdpr \mbk} ^{da}
}
\right)
\\
&+
\frac{1}{2}
\frac{
\Braket{
U_{\mbk \mbk\pr}^{ab} U_{\mbk\pr \mbk} ^{bc} 
}
}
{\veps_{\mbk}^a - \veps_{\mbk}^c}
\pd_{\mbkdpr}^{\rho}
\left(
\frac{v_{\rho \mbkdpr}^d}{\lvert \bs{v}_{\mbkdpr}^d \rvert^2}
\pd_{\mbkdpr}^{\sigma}
\left[
\frac{v_{\sigma \mbkdpr}^d}{\lvert \bs{v}_{\mbkdpr}^d \rvert^2}
\left(
v_{\mu \mbk}^c - v_{\mu \mbkdpr}^d 
\right)
\Braket{
U_{\mbk \mbkdpr}^{cd}U_{\mbkdpr \mbk} ^{da} 
}
\left(
n_{E \mbk c}^{(-1)} - n_{E \mbkdpr d}^{(-1)} 
\right) 
\right]
\right)
\Biggr\}_{\substack{c \neq a}},
\end{split}
\end{equation}
\begin{equation}
\begin{split}
\mathcal{N}^{(\text{ask})}_{5,\mbk a}
=&
\frac{2 \pi^2}{\hbar^2}  e E^{\mu} \tau_{\mbk}^a
\sum_{\mbkpr \mbkdpr bcd}
\delta ( \veps_{\mbk}^a - \veps_{\mbkpr}^b )
\delta ( \veps_{\mbk}^c - \veps_{\mbkdpr}^d )
\\
&\times
\Biggl\{
\frac{1}{2}
\pd_{\mbkpr}^{\rho}
\left(
\frac{v_{\rho \mbkpr}^b}{\lvert \bs{v}_{\mbkpr}^b \rvert^2}
\pd_{\mbkpr}^{\sigma}
\left[
\frac{v_{\sigma \mbkpr}^b}{\lvert \bs{v}_{\mbkpr}^b \rvert^2}
\left(
v_{\mu \mbk}^a - v_{\mu \mbkpr}^b 
\right)
\frac{
\text{Re}
\left[
\Braket{
U_{\mbk \mbk\pr}^{ab} U_{\mbk\pr \mbk} ^{bc} 
}
\Braket{
U_{\mbk \mbkdpr}^{cd} U_{\mbkdpr \mbk} ^{da}
}
\right]
}
{\veps_{\mbk}^a - \veps_{\mbk}^c}
\left(
n_{E \mbk c}^{(-1)} - n_{E \mbkdpr d}^{(-1)} 
\right) 
\right]
\right)
\\
&+
\pd_{\mbkpr}^{\rho}
\left(
\frac{v_{\rho \mbkpr}^b}{\lvert \bs{v}_{\mbkpr}^b \rvert^2}
\text{Re}
\Braket{
U_{\mbk \mbk\pr}^{ab} 
\left( \pd_{\mbk}^{\mu} + \pd_{\mbkpr}^{\mu}\right)
\left[
U_{\mbk\pr \mbk} ^{bc}
\frac{
\Braket{
U_{\mbk \mbkdpr}^{cd} U_{\mbkdpr \mbk} ^{da} 
}
}
{\veps_{\mbk}^a - \veps_{\mbk}^c}
\left(
n_{E \mbk c}^{(-1)} - n_{E \mbkdpr d}^{(-1)} 
\right)
\right]  
}
\right)
\\
&+
\pd_{\mbkdpr}^{\rho}
\left(
\frac{v_{\rho \mbkdpr}^d}{\lvert \bs{v}_{\mbkdpr}^d \rvert^2}
\pd_{\mbkpr}^{\sigma}
\left[
\frac{v_{\sigma \mbkpr}^b}{\lvert \bs{v}_{\mbkpr}^b \rvert^2}
v_{\mu \mbk}^c
\frac{
\text{Re}
\left[
\Braket{
U_{\mbk \mbk\pr}^{ab} U_{\mbk\pr \mbk} ^{bc} 
}
\Braket{
U_{\mbk \mbkdpr}^{cd} U_{\mbkdpr \mbk} ^{da}
}
\right]
}
{\veps_{\mbk}^a - \veps_{\mbk}^c}
\left(
n_{E \mbk c}^{(-1)} - n_{E \mbkdpr d}^{(-1)} 
\right) 
\right]
\right)
\Biggr\}_{\substack{c \neq a}},
\end{split}
\end{equation}
\begin{equation}
\begin{split}
\mathcal{N}^{(\text{ask})}_{6,\mbk a}
=&
- \frac{2 \pi^2}{\hbar}  e E^{\mu} \tau_{\mbk}^a
\sum_{\mbkpr \mbkdpr bcd}
\delta ( \veps_{\mbk}^a - \veps_{\mbkpr}^b )
\delta ( \veps_{\mbk}^c - \veps_{\mbkdpr}^d )
\text{Re}
\Biggl\{
\frac{
\Braket{
U_{\mbk \mbk\pr}^{ab} U_{\mbk\pr \mbk} ^{bc} 
}
}
{\veps_{\mbk}^a - \veps_{\mbk}^c}
\pd_{\mbk}^{\mu}
\left[
\frac{v_{\rho \mbkdpr}^d}{\lvert \bs{v}_{\mbkdpr}^d \rvert^2}
\frac{
\Braket{
U_{\mbk \mbkdpr}^{cd} U_{\mbkdpr \mbk}^{da}
}
}
{\veps_{\mbk}^a - \veps_{\mbk}^c}
\left(
n_{E \mbk c}^{(-1)} - n_{E \mbkdpr d}^{(-1)} 
\right) 
\right]
\\
&+
\frac{
\Braket{
U_{\mbk \mbk\pr}^{ab} U_{\mbk\pr \mbk} ^{bc} 
}
}
{\veps_{\mbk}^a - \veps_{\mbk}^c}
\pd_{\mbkdpr}^{\rho}
\left[
\frac{v_{\rho \mbkdpr}^d}{\lvert \bs{v}_{\mbkdpr}^d \rvert^2}
v_{\mu \mbk}^c
\frac{
\Braket{
U_{\mbk \mbkdpr}^{cd} U_{\mbkdpr \mbk}^{da}
}
}
{\veps_{\mbk}^a - \veps_{\mbk}^c}
\left(
n_{E \mbk c}^{(-1)} - n_{E \mbkdpr d}^{(-1)} 
\right) 
\right]
\Biggr\}_{\substack{c \neq a}},
\end{split}
\end{equation}
\begin{equation}
\begin{split}
\mathcal{N}^{(\text{ask})}_{7,\mbk a}
=&
- \frac{2 \pi^2}{\hbar^2}  e E^{\mu} \tau_{\mbk}^a
\sum_{\mbkpr \mbkdpr bcd}
\delta ( \veps_{\mbk}^a - \veps_{\mbkpr}^b )
\delta ( \veps_{\mbk}^a - \veps_{\mbkdpr}^d )
\\
&\times
\text{Re}
\Biggl\{
\frac{
\Braket{
U_{\mbk \mbk\pr}^{ab} U_{\mbk\pr \mbk} ^{ca} 
}
}
{\veps_{\mbk}^a - \veps_{\mbkpr}^c}
\pd_{\mbkdpr}^{\rho}
\left(
\frac{v_{\rho \mbkdpr}^d}{\lvert \bs{v}_{\mbkdpr}^d \rvert^2}
\Braket{
U_{\mbkdpr \mbkpr}^{dc}
\left( \pd_{\mbkpr}^{\mu} + \pd_{\mbkdpr}^{\mu} \right) 
\left[
U_{\mbkpr \mbkdpr} ^{bd}
\left(
n_{E \mbkpr b}^{(-1)} - n_{E \mbkdpr d}^{(-1)} 
\right) 
\right]
}
\right)
\\
&+
\frac{1}{2}
\frac{
\Braket{
U_{\mbk \mbk\pr}^{ab} U_{\mbk\pr \mbk} ^{ca} 
}
}
{\veps_{\mbk}^a - \veps_{\mbkpr}^c}
\pd_{\mbkdpr}^{\rho}
\left(
\frac{v_{\rho \mbkdpr}^d}{\lvert \bs{v}_{\mbkdpr}^d \rvert^2}
\pd_{\mbkdpr}^{\sigma}
\left[
\frac{v_{\sigma \mbkdpr}^d}{\lvert \bs{v}_{\mbkdpr}^d \rvert^2}
\left(
v_{\mu \mbkpr}^b - v_{\mu \mbkdpr}^d 
\right)
\Braket{
U_{\mbkpr \mbkdpr}^{bd}U_{\mbkdpr \mbkpr} ^{dc} 
}
\left(
n_{E \mbkpr b}^{(-1)} - n_{E \mbkdpr d}^{(-1)} 
\right) 
\right]
\right)
\Biggr\}_{\substack{c \neq b}},
\end{split}
\end{equation}
\begin{equation}
\begin{split}
\mathcal{N}^{(\text{ask})}_{8,\mbk a}
=&
\frac{2 \pi^2}{\hbar^2}  e E^{\mu} \tau_{\mbk}^a
\sum_{\mbkpr \mbkdpr bcd}
\delta ( \veps_{\mbk}^a - \veps_{\mbkpr}^b )
\delta ( \veps_{\mbk}^a - \veps_{\mbkdpr}^d )
\\
&\times
\Biggl\{
\frac{1}{2}
\pd_{\mbkpr}^{\rho}
\left(
\frac{v_{\rho \mbkpr}^b}{\lvert \bs{v}_{\mbkpr}^b \rvert^2}
\pd_{\mbkpr}^{\sigma}
\left[
\frac{v_{\sigma \mbkpr}^b}{\lvert \bs{v}_{\mbkpr}^b \rvert^2}
\left(
v_{\mu \mbk}^a - v_{\mu \mbkpr}^b 
\right)
\frac{
\text{Re}
\left[
\Braket{
U_{\mbk \mbk\pr}^{ab} U_{\mbk\pr \mbk} ^{ca} 
}
\Braket{
U_{\mbkpr \mbkdpr}^{bd} U_{\mbkdpr \mbkpr} ^{dc}
}
\right]
}
{\veps_{\mbkpr}^b - \veps_{\mbkpr}^c}
\left(
n_{E \mbkpr b}^{(-1)} - n_{E \mbkdpr d}^{(-1)} 
\right) 
\right]
\right)
\\
&+
\pd_{\mbkpr}^{\rho}
\left(
\frac{v_{\rho \mbkpr}^b}{\lvert \bs{v}_{\mbkpr}^b \rvert^2}
\pd_{\mbkdpr}^{\sigma}
\left[
\frac{v_{\sigma \mbkdpr}^d}{\lvert \bs{v}_{\mbkdpr}^d \rvert^2}
v_{\mu \mbk}^a
\frac{
\text{Re}
\left[
\Braket{
U_{\mbk \mbk\pr}^{ab} U_{\mbk\pr \mbk} ^{ca} 
}
\Braket{
U_{\mbkpr \mbkdpr}^{bd} U_{\mbkdpr \mbkpr} ^{dc}
}
\right]
}
{\veps_{\mbkpr}^b - \veps_{\mbkpr}^c}
\left(
n_{E \mbkpr b}^{(-1)} - n_{E \mbkdpr d}^{(-1)} 
\right) 
\right]
\right)
\\
&+
\frac{1}{2}
\pd_{\mbkdpr}^{\rho}
\left(
\frac{v_{\rho \mbkdpr}^d}{\lvert \bs{v}_{\mbkdpr}^d \rvert^2}
\pd_{\mbkdpr}^{\sigma}
\left[
\frac{v_{\sigma \mbkdpr}^d}{\lvert \bs{v}_{\mbkdpr}^d \rvert^2}
\left(
v_{\mu \mbk}^a + v_{\mu \mbkpr}^b 
\right)
\frac{
\text{Re}
\left[
\Braket{
U_{\mbk \mbk\pr}^{ab} U_{\mbk\pr \mbk} ^{ca} 
}
\Braket{
U_{\mbkpr \mbkdpr}^{bd} U_{\mbkdpr \mbkpr} ^{dc}
}
\right]
}
{\veps_{\mbkpr}^b - \veps_{\mbkpr}^c}
\left(
n_{E \mbkpr b}^{(-1)} - n_{E \mbkdpr d}^{(-1)} 
\right) 
\right]
\right)
\Biggr\}_{\substack{c \neq b}},
\end{split}
\end{equation}
\begin{equation}
\begin{split}
\mathcal{N}^{(\text{ask})}_{9,\mbk a}
=&
\frac{2 \pi^2}{\hbar^2}  e E^{\mu} \tau_{\mbk}^a
\sum_{\mbkpr \mbkdpr bcd}
\delta ( \veps_{\mbk}^a - \veps_{\mbkpr}^b )
\delta ( \veps_{\mbk}^a - \veps_{\mbkdpr}^d )
\\
&\times
\Biggl\{
\pd_{\mbkpr}^{\rho}
\left(
\frac{v_{\rho \mbkpr}^b}{\lvert \bs{v}_{\mbkpr}^b \rvert^2}
\text{Re}
\Braket{
U_{\mbk \mbk\pr}^{ab} 
\left( \pd_{\mbk}^{\mu} + \pd_{\mbkpr}^{\mu}\right)
\left[
U_{\mbk\pr \mbk} ^{ca}
\frac{
\Braket{
U_{\mbkpr \mbkdpr}^{bd} U_{\mbkdpr \mbkpr} ^{dc} 
}
}
{\veps_{\mbkpr}^b - \veps_{\mbkpr}^c}
\left(
n_{E \mbkpr b}^{(-1)} - n_{E \mbkdpr d}^{(-1)} 
\right)
\right]  
}
\right)
\\
&+
\pd_{\mbkdpr}^{\rho}
\left(
\frac{v_{\rho \mbkdpr}^d}{\lvert \bs{v}_{\mbkdpr}^d \rvert^2}
\text{Re}
\Braket{
U_{\mbk \mbk\pr}^{ab} 
\left( \pd_{\mbk}^{\mu} + \pd_{\mbkpr}^{\mu}\right)
\left[
U_{\mbk\pr \mbk} ^{ca}
\frac{
\Braket{
U_{\mbkpr \mbkdpr}^{bd} U_{\mbkdpr \mbkpr} ^{dc} 
}
}
{\veps_{\mbkpr}^b - \veps_{\mbkpr}^c}
\left(
n_{E \mbkpr b}^{(-1)} - n_{E \mbkdpr d}^{(-1)} 
\right)
\right]  
}
\right)
\Biggr\}_{\substack{c \neq b}},
\end{split}
\end{equation}
\begin{equation}
\begin{split}
\mathcal{N}^{(\text{ask})}_{10,\mbk a}
=&
\frac{2 \pi^2}{\hbar}  e E^{\mu} \tau_{\mbk}^a
\sum_{\mbkpr \mbkdpr bcd}
\delta ( \veps_{\mbk}^a - \veps_{\mbkpr}^b )
\delta ( \veps_{\mbk}^a - \veps_{\mbkdpr}^d )
\text{Re}
\Biggl\{
\frac{
\Braket{
U_{\mbk \mbk\pr}^{ab} U_{\mbk\pr \mbk} ^{ca} 
}
}
{\veps_{\mbk}^a - \veps_{\mbkpr}^c}
\pd_{\mbkpr}^{\mu}
\left[
\frac{
\Braket{
U_{\mbkpr \mbkdpr}^{bd} U_{\mbkdpr \mbkpr}^{dc}
}
}
{\veps_{\mbkpr}^b - \veps_{\mbkpr}^c}
\left(
n_{E \mbkpr b}^{(-1)} - n_{E \mbkdpr d}^{(-1)} 
\right) 
\right]
\\
&+
\frac{
\Braket{
U_{\mbk \mbk\pr}^{ab} U_{\mbk\pr \mbk} ^{ca} 
}
}
{\veps_{\mbk}^a - \veps_{\mbkpr}^c}
\pd_{\mbkdpr}^{\rho}
\left[
\frac{v_{\rho \mbkdpr}^d}{\lvert \bs{v}_{\mbkdpr}^d \rvert^2}
v_{\mu \mbkpr}^b
\frac{
\Braket{
U_{\mbkpr \mbkdpr}^{bd} U_{\mbkdpr \mbkpr}^{dc}
}
}
{\veps_{\mbkpr}^b - \veps_{\mbkpr}^c}
\left(
n_{E \mbkpr b}^{(-1)} - n_{E \mbkdpr d}^{(-1)} 
\right) 
\right]
\Biggr\}_{\substack{c \neq b}},
\end{split}
\end{equation}
\begin{equation}
\begin{split}
\mathcal{N}^{(\text{ask})}_{11,\mbk a}
=&
\frac{2 \pi^2}{\hbar^2}  e E^{\mu} \tau_{\mbk}^a
\sum_{\mbkpr \mbkdpr bcd}
\delta ( \veps_{\mbk}^a - \veps_{\mbkpr}^b )
\delta ( \veps_{\mbkpr}^c - \veps_{\mbkdpr}^d )
\\
&\times
\text{Re}
\Biggl\{
\frac{
\Braket{
U_{\mbk \mbk\pr}^{ab} U_{\mbk\pr \mbk} ^{ca} 
}
}
{\veps_{\mbk}^a - \veps_{\mbkpr}^c}
\pd_{\mbkdpr}^{\rho}
\left(
\frac{v_{\rho \mbkdpr}^d}{\lvert \bs{v}_{\mbkdpr}^d \rvert^2}
\Braket{
U_{\mbkpr \mbkdpr}^{bd}
\left( \pd_{\mbkpr}^{\mu} + \pd_{\mbkdpr}^{\mu} \right) 
\left[
U_{\mbkdpr \mbkpr} ^{dc}
\left(
n_{E \mbkpr c}^{(-1)} - n_{E \mbkdpr d}^{(-1)} 
\right) 
\right]
}
\right)
\\
&+
\frac{1}{2}
\frac{
\Braket{
U_{\mbk \mbk\pr}^{ab} U_{\mbk\pr \mbk} ^{ca} 
}
}
{\veps_{\mbk}^a - \veps_{\mbkpr}^c}
\pd_{\mbkdpr}^{\rho}
\left(
\frac{v_{\rho \mbkdpr}^d}{\lvert \bs{v}_{\mbkdpr}^d \rvert^2}
\pd_{\mbkdpr}^{\sigma}
\left[
\frac{v_{\sigma \mbkdpr}^d}{\lvert \bs{v}_{\mbkdpr}^d \rvert^2}
\left(
v_{\mu \mbkpr}^c - v_{\mu \mbkdpr}^d 
\right)
\Braket{
U_{\mbkpr \mbkdpr}^{bd}U_{\mbkdpr \mbkpr} ^{dc} 
}
\left(
n_{E \mbkpr c}^{(-1)} - n_{E \mbkdpr d}^{(-1)} 
\right) 
\right]
\right)
\Biggr\}_{\substack{c \neq b}},
\end{split}
\end{equation}
\begin{equation}
\begin{split}
\mathcal{N}^{(\text{ask})}_{12,\mbk a}
=&
\frac{2 \pi^2}{\hbar^2}  e E^{\mu} \tau_{\mbk}^a
\sum_{\mbkpr \mbkdpr bcd}
\delta ( \veps_{\mbk}^a - \veps_{\mbkpr}^b )
\delta ( \veps_{\mbkpr}^c - \veps_{\mbkdpr}^d )
\\
&\times
\Biggl\{
\frac{1}{2}
\pd_{\mbkpr}^{\rho}
\left(
\frac{v_{\rho \mbkpr}^b}{\lvert \bs{v}_{\mbkpr}^b \rvert^2}
\pd_{\mbkpr}^{\sigma}
\left[
\frac{v_{\sigma \mbkpr}^b}{\lvert \bs{v}_{\mbkpr}^b \rvert^2}
\left(
v_{\mu \mbk}^a - v_{\mu \mbkpr}^b 
\right)
\frac{
\text{Re}
\left[
\Braket{
U_{\mbk \mbk\pr}^{ab} U_{\mbk\pr \mbk} ^{ca} 
}
\Braket{
U_{\mbkpr \mbkdpr}^{bd} U_{\mbkdpr \mbkpr} ^{dc}
}
\right]
}
{\veps_{\mbkpr}^b - \veps_{\mbkpr}^c}
\left(
n_{E \mbkpr c}^{(-1)} - n_{E \mbkdpr d}^{(-1)} 
\right) 
\right]
\right)
\\
&+
\pd_{\mbkpr}^{\rho}
\left(
\frac{v_{\rho \mbkpr}^b}{\lvert \bs{v}_{\mbkpr}^b \rvert^2}
\pd_{\mbkdpr}^{\sigma}
\left[
\frac{v_{\sigma \mbkdpr}^d}{\lvert \bs{v}_{\mbkdpr}^d \rvert^2}
\left(
v_{\mu \mbk}^a - v_{\mu \mbkpr}^b + v_{\mu \mbkpr}^c
\right)
\frac{
\text{Re}
\left[
\Braket{
U_{\mbk \mbk\pr}^{ab} U_{\mbk\pr \mbk} ^{ca} 
}
\Braket{
U_{\mbkpr \mbkdpr}^{bd} U_{\mbkdpr \mbkpr} ^{dc}
}
\right]
}
{\veps_{\mbkpr}^b - \veps_{\mbkpr}^c}
\left(
n_{E \mbkpr c}^{(-1)} - n_{E \mbkdpr d}^{(-1)} 
\right) 
\right]
\right)
\\
&+
\frac{1}{2}
\pd_{\mbkdpr}^{\rho}
\left(
\frac{v_{\rho \mbkdpr}^d}{\lvert \bs{v}_{\mbkdpr}^d \rvert^2}
\pd_{\mbkdpr}^{\sigma}
\left[
\frac{v_{\sigma \mbkdpr}^d}{\lvert \bs{v}_{\mbkdpr}^d \rvert^2}
\left(
v_{\mu \mbk}^a - v_{\mu \mbkpr}^b + 2 v_{\mu \mbkpr}^c 
\right)
\frac{
\text{Re}
\left[
\Braket{
U_{\mbk \mbk\pr}^{ab} U_{\mbk\pr \mbk} ^{ca} 
}
\Braket{
U_{\mbkpr \mbkdpr}^{bd} U_{\mbkdpr \mbkpr} ^{dc}
}
\right]
}
{\veps_{\mbkpr}^b - \veps_{\mbkpr}^c}
\left(
n_{E \mbkpr c}^{(-1)} - n_{E \mbkdpr d}^{(-1)} 
\right) 
\right]
\right)
\Biggr\}_{\substack{c \neq b}},
\end{split}
\end{equation}
\begin{equation}
\begin{split}
\mathcal{N}^{(\text{ask})}_{13,\mbk a}
=&
\frac{2 \pi^2}{\hbar^2}  e E^{\mu} \tau_{\mbk}^a
\sum_{\mbkpr \mbkdpr bcd}
\delta ( \veps_{\mbk}^a - \veps_{\mbkpr}^b )
\delta ( \veps_{\mbkpr}^c - \veps_{\mbkdpr}^d )
\\
&\times
\Biggl\{
\pd_{\mbkpr}^{\rho}
\left(
\frac{v_{\rho \mbkpr}^b}{\lvert \bs{v}_{\mbkpr}^b \rvert^2}
\text{Re}
\Braket{
U_{\mbk \mbk\pr}^{ab} 
\left( \pd_{\mbk}^{\mu} + \pd_{\mbkpr}^{\mu}\right)
\left[
U_{\mbk\pr \mbk} ^{ca}
\frac{
\Braket{
U_{\mbkpr \mbkdpr}^{bd} U_{\mbkdpr \mbkpr} ^{dc} 
}
}
{\veps_{\mbkpr}^b - \veps_{\mbkpr}^c}
\left(
n_{E \mbkpr c}^{(-1)} - n_{E \mbkdpr d}^{(-1)} 
\right)
\right]  
}
\right)
\\
&+
\pd_{\mbkdpr}^{\rho}
\left(
\frac{v_{\rho \mbkdpr}^d}{\lvert \bs{v}_{\mbkdpr}^d \rvert^2}
\text{Re}
\Braket{
U_{\mbk \mbk\pr}^{ab} 
\left( \pd_{\mbk}^{\mu} + \pd_{\mbkpr}^{\mu}\right)
\left[
U_{\mbk\pr \mbk} ^{ca}
\frac{
\Braket{
U_{\mbkpr \mbkdpr}^{bd} U_{\mbkdpr \mbkpr} ^{dc} 
}
}
{\veps_{\mbkpr}^b - \veps_{\mbkpr}^c}
\left(
n_{E \mbkpr c}^{(-1)} - n_{E \mbkdpr d}^{(-1)} 
\right)
\right]  
}
\right)
\Biggr\}_{\substack{c \neq b}},
\end{split}
\end{equation}
\begin{equation}
\begin{split}
\mathcal{N}^{(\text{ask})}_{14,\mbk a}
=&
\frac{2 \pi^2}{\hbar}  e E^{\mu} \tau_{\mbk}^a
\sum_{\mbkpr \mbkdpr bcd}
\delta ( \veps_{\mbk}^a - \veps_{\mbkpr}^b )
\delta ( \veps_{\mbkpr}^c - \veps_{\mbkdpr}^d )
\text{Re}
\Biggl\{
\frac{
\Braket{
U_{\mbk \mbk\pr}^{ab} U_{\mbk\pr \mbk} ^{ca} 
}
}
{\veps_{\mbk}^a - \veps_{\mbkpr}^c}
\pd_{\mbkpr}^{\mu}
\left[
\frac{
\Braket{
U_{\mbkpr \mbkdpr}^{bd} U_{\mbkdpr \mbkpr}^{dc}
}
}
{\veps_{\mbkpr}^b - \veps_{\mbkpr}^c}
\left(
n_{E \mbkpr c}^{(-1)} - n_{E \mbkdpr d}^{(-1)} 
\right) 
\right]
\\
&+
\frac{
\Braket{
U_{\mbk \mbk\pr}^{ab} U_{\mbk\pr \mbk} ^{ca} 
}
}
{\veps_{\mbk}^a - \veps_{\mbkpr}^c}
\pd_{\mbkdpr}^{\rho}
\left[
\frac{v_{\rho \mbkdpr}^d}{\lvert \bs{v}_{\mbkdpr}^d \rvert^2}
v_{\mu \mbkpr}^c
\frac{
\Braket{
U_{\mbkpr \mbkdpr}^{bd} U_{\mbkdpr \mbkpr}^{dc}
}
}
{\veps_{\mbkpr}^b - \veps_{\mbkpr}^c}
\left(
n_{E \mbkpr c}^{(-1)} - n_{E \mbkdpr d}^{(-1)} 
\right) 
\right]
\Biggr\}_{\substack{c \neq b}},
\end{split}
\end{equation}
\begin{equation}
\begin{split}
\mathcal{N}^{(\text{ask})}_{15,\mbk a}
=&
- \frac{2 \pi^2}{\hbar^2}  e E^{\mu} \tau_{\mbk}^a
\sum_{\mbkpr \mbkdpr bcd}
\delta ( \veps_{\mbk}^a - \veps_{\mbkpr}^b )
\delta ( \veps_{\mbk}^a - \veps_{\mbkdpr}^d )
\\
&\times
\text{Im}
\Biggl\{
\frac{
\Braket{
U_{\mbk \mbk\pr}^{ab} U_{\mbk\pr \mbk} ^{bc} 
}
}
{\veps_{\mbk}^a - \veps_{\mbk}^c}
\pd_{\mbkdpr}^{\rho}
\left[
\frac{v_{\rho \mbkdpr}^d}{\lvert \bs{v}_{\mbkdpr}^d \rvert^2}
\Braket{
U_{\mbk \mbkdpr}^{cd} U_{\mbkdpr \mbk}^{da}}
\left( \mathcal{A}_{\mu \mbk}^a 
- 
\mathcal{A}_{\mu \mbkdpr}^d \right) 
\left(
n_{E \mbk a}^{(-1)} - n_{E \mbkdpr d}^{(-1)} 
\right) 
\right]
\Biggr\}_{\substack{c \neq a}}
\\
&-
\frac{2 \pi^2}{\hbar^2}  e E^{\mu} \tau_{\mbk}^a
\sum_{\mbkpr \mbkdpr bcd}
\delta ( \veps_{\mbk}^a - \veps_{\mbkpr}^b )
\delta ( \veps_{\mbk}^c - \veps_{\mbkdpr}^d )
\\
&\times
\text{Im}
\Biggl\{
\frac{
\Braket{
U_{\mbk \mbk\pr}^{ab} U_{\mbk\pr \mbk} ^{bc} 
}
}
{\veps_{\mbk}^a - \veps_{\mbk}^c}
\pd_{\mbkdpr}^{\rho}
\left[
\frac{v_{\rho \mbkdpr}^d}{\lvert \bs{v}_{\mbkdpr}^d \rvert^2}
\Braket{
U_{\mbk \mbkdpr}^{cd} U_{\mbkdpr \mbk}^{da}}
\left( \mathcal{A}_{\mu \mbk}^c 
- 
\mathcal{A}_{\mu \mbkdpr}^d \right) 
\left(
n_{E \mbk c}^{(-1)} - n_{E \mbkdpr d}^{(-1)} 
\right) 
\right]
\Biggr\}_{\substack{c \neq a}},
\end{split}
\end{equation}
\begin{equation}
\begin{split}
\mathcal{N}^{(\text{ask})}_{16,\mbk a}
=&
- \frac{2 \pi^2}{\hbar^2}  e E^{\mu} \tau_{\mbk}^a
\sum_{\mbkpr \mbkdpr bcd}
\delta ( \veps_{\mbk}^a - \veps_{\mbkpr}^b )
\delta ( \veps_{\mbk}^a - \veps_{\mbkdpr}^d )
\\
&\times
\pd_{\mbkpr}^{\rho}
\Biggl\{
\frac{v_{\rho \mbkpr}^b}{\lvert \bs{v}_{\mbkpr}^b \rvert^2}
\frac{
\text{Im}
\left[
\Braket{
U_{\mbk \mbk\pr}^{ab} U_{\mbk\pr \mbk} ^{bc} 
}
\Braket{
U_{\mbk \mbkdpr}^{cd} U_{\mbkdpr \mbk} ^{da} 
}
\right]
}
{\veps_{\mbk}^a - \veps_{\mbk}^c}
\left( \mathcal{A}_{\mu \mbk}^a 
- 
\mathcal{A}_{\mu \mbkpr}^b \right) 
\left(
n_{E \mbk a}^{(-1)} - n_{E \mbkdpr d}^{(-1)} 
\right) 
\Biggr\}_{\substack{c \neq a}}
\\
&-
\frac{2 \pi^2}{\hbar^2}  e E^{\mu} \tau_{\mbk}^a
\sum_{\mbkpr \mbkdpr bcd}
\delta ( \veps_{\mbk}^a - \veps_{\mbkpr}^b )
\delta ( \veps_{\mbk}^c - \veps_{\mbkdpr}^d )
\\
&\times
\pd_{\mbkpr}^{\rho}
\Biggl\{
\frac{v_{\rho \mbkpr}^b}{\lvert \bs{v}_{\mbkpr}^b \rvert^2}
\frac{
\text{Im}
\left[
\Braket{
U_{\mbk \mbk\pr}^{ab} U_{\mbk\pr \mbk} ^{bc} 
}
\Braket{
U_{\mbk \mbkdpr}^{cd} U_{\mbkdpr \mbk} ^{da} 
}
\right]
}
{\veps_{\mbk}^a - \veps_{\mbk}^c}
\left( \mathcal{A}_{\mu \mbk}^a 
- 
\mathcal{A}_{\mu \mbkpr}^b \right) 
\left(
n_{E \mbk c}^{(-1)} - n_{E \mbkdpr d}^{(-1)} 
\right) 
\Biggr\}_{\substack{c \neq a}},
\end{split}
\end{equation}
\begin{equation}
\begin{split}
\mathcal{N}^{(\text{ask})}_{17,\mbk a}
=&
\frac{2 \pi^2}{\hbar}  e E^{\mu} \tau_{\mbk}^a
\sum_{\mbkpr \mbkdpr bcd}
\delta ( \veps_{\mbk}^a - \veps_{\mbkpr}^b )
\delta ( \veps_{\mbk}^a - \veps_{\mbkdpr}^d )
\frac{
\text{Im}
\left[
\Braket{
U_{\mbk \mbk\pr}^{ab} U_{\mbk\pr \mbk} ^{bc} 
}
\Braket{
U_{\mbk \mbkdpr}^{cd} U_{\mbkdpr \mbk} ^{da} 
}
\right]_{\substack{c \neq a}}
}
{(\veps_{\mbk}^a - \veps_{\mbk}^c)^2}
\left( \mathcal{A}_{\mu \mbk}^a 
- 
\mathcal{A}_{\mu \mbk}^c \right) 
\left(
n_{E \mbk a}^{(-1)} - n_{E \mbkdpr d}^{(-1)} 
\right)
\\
&+
\frac{2 \pi^2}{\hbar}  e E^{\mu} \tau_{\mbk}^a
\sum_{\mbkpr \mbkdpr bcd}
\delta ( \veps_{\mbk}^a - \veps_{\mbkpr}^b )
\delta ( \veps_{\mbk}^c - \veps_{\mbkdpr}^d )
\frac{
\text{Im}
\left[
\Braket{
U_{\mbk \mbk\pr}^{ab} U_{\mbk\pr \mbk} ^{bc} 
}
\Braket{
U_{\mbk \mbkdpr}^{cd} U_{\mbkdpr \mbk} ^{da} 
}
\right]_{\substack{c \neq a}}
}
{(\veps_{\mbk}^a - \veps_{\mbk}^c)^2}
\left( \mathcal{A}_{\mu \mbk}^a 
- 
\mathcal{A}_{\mu \mbk}^c \right) 
\left(
n_{E \mbk c}^{(-1)} - n_{E \mbkdpr d}^{(-1)} 
\right),
\end{split}
\end{equation}
\begin{equation}
\begin{split}
\mathcal{N}^{(\text{ask})}_{18,\mbk a}
=&
- \frac{2 \pi^2}{\hbar^2}  e E^{\mu} \tau_{\mbk}^a
\sum_{\mbkpr \mbkdpr bcd}
\delta ( \veps_{\mbk}^a - \veps_{\mbkpr}^b )
\delta ( \veps_{\mbk}^a - \veps_{\mbkdpr}^d )
\\
&\times
\text{Im}
\Biggl\{
\frac{
\Braket{
U_{\mbk \mbk\pr}^{ab} U_{\mbk\pr \mbk} ^{ca} 
}
}
{\veps_{\mbk}^a - \veps_{\mbkpr}^c}
\pd_{\mbkdpr}^{\rho}
\left[
\frac{v_{\rho \mbkdpr}^d}{\lvert \bs{v}_{\mbkdpr}^d \rvert^2}
\Braket{
U_{\mbkpr \mbkdpr}^{bd} U_{\mbkdpr \mbkpr}^{dc}}
\left( \mathcal{A}_{\mu \mbkpr}^b 
- 
\mathcal{A}_{\mu \mbkdpr}^d \right) 
\left(
n_{E \mbkpr b}^{(-1)} - n_{E \mbkdpr d}^{(-1)} 
\right) 
\right]
\Biggr\}_{\substack{c \neq b}}
\\
&-
\frac{2 \pi^2}{\hbar^2}  e E^{\mu} \tau_{\mbk}^a
\sum_{\mbkpr \mbkdpr bcd}
\delta ( \veps_{\mbk}^a - \veps_{\mbkpr}^b )
\delta ( \veps_{\mbkpr}^c - \veps_{\mbkdpr}^d )
\\
&\times
\text{Im}
\Biggl\{
\frac{
\Braket{
U_{\mbk \mbk\pr}^{ab} U_{\mbk\pr \mbk} ^{ca} 
}
}
{\veps_{\mbk}^a - \veps_{\mbkpr}^c}
\pd_{\mbkdpr}^{\rho}
\left[
\frac{v_{\rho \mbkdpr}^d}{\lvert \bs{v}_{\mbkdpr}^d \rvert^2}
\Braket{
U_{\mbkpr \mbkdpr}^{bd} U_{\mbkdpr \mbkpr}^{dc}}
\left( \mathcal{A}_{\mu \mbkpr}^c 
- 
\mathcal{A}_{\mu \mbkdpr}^d \right) 
\left(
n_{E \mbkpr c}^{(-1)} - n_{E \mbkdpr d}^{(-1)} 
\right) 
\right]
\Biggr\}_{\substack{c \neq b}},
\end{split}
\end{equation}
\begin{equation}
\begin{split}
\mathcal{N}^{(\text{ask})}_{19,\mbk a}
=&
- \frac{2 \pi^2}{\hbar^2}  e E^{\mu} \tau_{\mbk}^a
\sum_{\mbkpr \mbkdpr bcd}
\delta ( \veps_{\mbk}^a - \veps_{\mbkpr}^b )
\delta ( \veps_{\mbk}^a - \veps_{\mbkdpr}^d )
\\
&\times
\pd_{\mbkpr}^{\rho}
\Biggl\{
\frac{v_{\rho \mbkpr}^b}{\lvert \bs{v}_{\mbkpr}^b \rvert^2}
\frac{
\text{Im}
\left[
\Braket{
U_{\mbk \mbk\pr}^{ab} U_{\mbk\pr \mbk} ^{ca} 
}
\Braket{
U_{\mbkpr \mbkdpr}^{bd} U_{\mbkdpr \mbkpr} ^{dc} 
}
\right]
}
{\veps_{\mbkpr}^b - \veps_{\mbkpr}^c}
\left( \mathcal{A}_{\mu \mbk}^a 
- 
\mathcal{A}_{\mu \mbkpr}^b \right) 
\left(
n_{E \mbkpr b}^{(-1)} - n_{E \mbkdpr d}^{(-1)} 
\right) 
\Biggr\}_{\substack{c \neq b}}
\\
&-
\frac{2 \pi^2}{\hbar^2}  e E^{\mu} \tau_{\mbk}^a
\sum_{\mbkpr \mbkdpr bcd}
\delta ( \veps_{\mbk}^a - \veps_{\mbkpr}^b )
\delta ( \veps_{\mbk}^c - \veps_{\mbkdpr}^d )
\\
&\times
\pd_{\mbkdpr}^{\rho}
\Biggl\{
\frac{v_{\rho \mbkdpr}^d}{\lvert \bs{v}_{\mbkdpr}^d \rvert^2}
\frac{
\text{Im}
\left[
\Braket{
U_{\mbk \mbk\pr}^{ab} U_{\mbk\pr \mbk} ^{ca} 
}
\Braket{
U_{\mbkpr \mbkdpr}^{bd} U_{\mbkdpr \mbkpr} ^{dc} 
}
\right]
}
{\veps_{\mbkpr}^b - \veps_{\mbkpr}^c}
\left( \mathcal{A}_{\mu \mbk}^a 
- 
\mathcal{A}_{\mu \mbkpr}^b \right) 
\left(
n_{E \mbkpr b}^{(-1)} - n_{E \mbkdpr d}^{(-1)} 
\right) 
\Biggr\}_{\substack{c \neq b}},
\end{split}
\end{equation}
\begin{equation}
\begin{split}
\mathcal{N}^{(\text{ask})}_{20,\mbk a}
=&
- \frac{2 \pi^2}{\hbar^2}  e E^{\mu} \tau_{\mbk}^a
\sum_{\mbkpr \mbkdpr bcd}
\delta ( \veps_{\mbk}^a - \veps_{\mbkpr}^b )
\delta ( \veps_{\mbkpr}^c - \veps_{\mbkdpr}^d )
\\
&\times
\pd_{\mbkpr}^{\rho}
\Biggl\{
\frac{v_{\rho \mbkpr}^b}{\lvert \bs{v}_{\mbkpr}^b \rvert^2}
\frac{
\text{Im}
\left[
\Braket{
U_{\mbk \mbk\pr}^{ab} U_{\mbk\pr \mbk} ^{ca} 
}
\Braket{
U_{\mbkpr \mbkdpr}^{bd} U_{\mbkdpr \mbkpr} ^{dc} 
}
\right]
}
{\veps_{\mbkpr}^b - \veps_{\mbkpr}^c}
\left( \mathcal{A}_{\mu \mbk}^a 
- 
\mathcal{A}_{\mu \mbkpr}^b \right) 
\left(
n_{E \mbkpr c}^{(-1)} - n_{E \mbkdpr d}^{(-1)} 
\right) 
\Biggr\}_{\substack{c \neq b}}
\\
&-
\frac{2 \pi^2}{\hbar^2}  e E^{\mu} \tau_{\mbk}^a
\sum_{\mbkpr \mbkdpr bcd}
\delta ( \veps_{\mbk}^a - \veps_{\mbkpr}^b )
\delta ( \veps_{\mbkpr}^c - \veps_{\mbkdpr}^d )
\\
&\times
\pd_{\mbkdpr}^{\rho}
\Biggl\{
\frac{v_{\rho \mbkdpr}^d}{\lvert \bs{v}_{\mbkdpr}^d \rvert^2}
\frac{
\text{Im}
\left[
\Braket{
U_{\mbk \mbk\pr}^{ab} U_{\mbk\pr \mbk} ^{ca} 
}
\Braket{
U_{\mbkpr \mbkdpr}^{bd} U_{\mbkdpr \mbkpr} ^{dc} 
}
\right]
}
{\veps_{\mbkpr}^b - \veps_{\mbkpr}^c}
\left( \mathcal{A}_{\mu \mbk}^a 
- 
\mathcal{A}_{\mu \mbkpr}^b \right) 
\left(
n_{E \mbkpr c}^{(-1)} - n_{E \mbkdpr d}^{(-1)} 
\right) 
\Biggr\}_{\substack{c \neq b}},
\end{split}
\end{equation}
\begin{equation}
\begin{split}
\mathcal{N}^{(\text{ask})}_{21,\mbk a}
=&
\frac{2 \pi^2}{\hbar}  e E^{\mu} \tau_{\mbk}^a
\sum_{\mbkpr \mbkdpr bcd}
\delta ( \veps_{\mbk}^a - \veps_{\mbkpr}^b )
\delta ( \veps_{\mbk}^a - \veps_{\mbkdpr}^d )
\frac{
\text{Im}
\left[
\Braket{
U_{\mbk \mbk\pr}^{ab} U_{\mbk\pr \mbk} ^{ca} 
}
\Braket{
U_{\mbkpr \mbkdpr}^{bd} U_{\mbkdpr \mbkpr} ^{dc} 
}
\right]_{\substack{c \neq b}}
}
{(\veps_{\mbk}^a - \veps_{\mbk}^c)^2}
\left( \mathcal{A}_{\mu \mbkpr}^b 
- 
\mathcal{A}_{\mu \mbkpr}^c \right) 
\left(
n_{E \mbkpr b}^{(-1)} - n_{E \mbkdpr d}^{(-1)} 
\right)
\\
&+
\frac{2 \pi^2}{\hbar}  e E^{\mu} \tau_{\mbk}^a
\sum_{\mbkpr \mbkdpr bcd}
\delta ( \veps_{\mbk}^a - \veps_{\mbkpr}^b )
\delta ( \veps_{\mbkpr}^c - \veps_{\mbkdpr}^d )
\frac{
\text{Im}
\left[
\Braket{
U_{\mbk \mbk\pr}^{ab} U_{\mbk\pr \mbk} ^{ca} 
}
\Braket{
U_{\mbkpr \mbkdpr}^{bd} U_{\mbkdpr \mbkpr} ^{dc} 
}
\right]_{\substack{c \neq b}}
}
{(\veps_{\mbk}^a - \veps_{\mbkpr}^c)^2}
\left( \mathcal{A}_{\mu \mbkpr}^b 
- 
\mathcal{A}_{\mu \mbkpr}^c \right) 
\left(
n_{E \mbkpr c}^{(-1)} - n_{E \mbkdpr d}^{(-1)} 
\right),
\end{split}
\end{equation}
\begin{equation}
\begin{split}
\mathcal{N}^{(\text{ask})}_{22,\mbk a}
=&
- \frac{2 \pi^2}{\hbar}  e E^{\mu} \tau_{\mbk}^a
\sum_{\mbkpr \mbkdpr bcde}
\delta ( \veps_{\mbk}^a - \veps_{\mbkpr}^b )
\frac{
\text{Im}
\left[
\Braket{
U_{\mbk \mbk\pr}^{ab} U_{\mbk\pr \mbk} ^{bc} 
}
\Braket{
U_{\mbk \mbkdpr}^{cd} U_{\mbkdpr \mbk} ^{ea} 
}
\mathcal{A}_{\mu \mbkdpr}^{\prime de}
\right]_{\substack{c \neq a}}
}
{\veps_{\mbk}^a - \veps_{\mbk}^c}
\left[
\frac{
n_{E \mbk a}^{(-1)} - n_{E \mbkdpr d}^{(-1)}
}
{\veps_{\mbk}^a - \veps_{\mbkdpr}^e}
\delta ( \veps_{\mbk}^a - \veps_{\mbkdpr}^d ) 
\right.
\\
&+
\left.
\frac{
n_{E \mbk c}^{(-1)} - n_{E \mbkdpr e}^{(-1)}
}
{\veps_{\mbk}^c - \veps_{\mbkdpr}^d}
\delta ( \veps_{\mbk}^c - \veps_{\mbkdpr}^e ) 
+
\frac{
n_{E \mbk a}^{(-1)} - n_{E \mbkdpr e}^{(-1)}
}
{\veps_{\mbk}^a - \veps_{\mbkdpr}^d}
\delta ( \veps_{\mbk}^a - \veps_{\mbkdpr}^e )
+
\frac{
n_{E \mbk c}^{(-1)} - n_{E \mbkdpr d}^{(-1)}
}
{\veps_{\mbk}^c - \veps_{\mbkdpr}^e}
\delta ( \veps_{\mbk}^c - \veps_{\mbkdpr}^d )
\right],
\end{split}
\end{equation}
\begin{equation}
\begin{split}
\mathcal{N}^{(\text{ask})}_{23,\mbk a}
=&
- \frac{2 \pi^2}{\hbar}  e E^{\mu} \tau_{\mbk}^a
\sum_{\mbkpr \mbkdpr bcde}
\delta ( \veps_{\mbk}^a - \veps_{\mbkpr}^b )
\frac{
\text{Im}
\left[
\Braket{
U_{\mbk \mbk\pr}^{ab} U_{\mbk\pr \mbk} ^{bc} 
}
\Braket{
U_{\mbk \mbkdpr}^{cd} U_{\mbkdpr \mbk} ^{de} 
}
\mathcal{A}_{\mu \mbk}^{\prime ea}
\right]_{\substack{c \neq a}}
}
{(\veps_{\mbk}^a - \veps_{\mbk}^c)
(\veps_{\mbk}^a - \veps_{\mbk}^e)}
\\
&\hspace{.27\linewidth}
\times
\left[
\left(
n_{E \mbk a}^{(-1)} - n_{E \mbkdpr d}^{(-1)}
\right)
\delta ( \veps_{\mbk}^a - \veps_{\mbkdpr}^d ) 
\right.
-
\left.
\left(
n_{E \mbk e}^{(-1)} - n_{E \mbkdpr d}^{(-1)}
\right)
\delta ( \veps_{\mbk}^e - \veps_{\mbkdpr}^d ) 
\right]
\\
&+
\frac{2 \pi^2}{\hbar}  e E^{\mu} \tau_{\mbk}^a
\sum_{\mbkpr \mbkdpr bcde}
\left[
\delta ( \veps_{\mbk}^a - \veps_{\mbkpr}^b )
-
\delta ( \veps_{\mbk}^e - \veps_{\mbkpr}^b )
\right]
\frac{
\text{Im}
\left[
\Braket{
U_{\mbk \mbk\pr}^{ab} U_{\mbk\pr \mbk} ^{bc} 
}
\Braket{
U_{\mbk \mbkdpr}^{cd} U_{\mbkdpr \mbk} ^{de} 
}
\mathcal{A}_{\mu \mbk}^{\prime ea}
\right]_{\substack{e \neq c}}
}
{(\veps_{\mbk}^a - \veps_{\mbk}^e)
(\veps_{\mbk}^c - \veps_{\mbk}^e)}
\\
&\hspace{.27\linewidth}
\times
\left[
\left(
n_{E \mbk c}^{(-1)} - n_{E \mbkdpr d}^{(-1)}
\right)
\delta ( \veps_{\mbk}^c - \veps_{\mbkdpr}^d ) 
\right.
+
\left.
\left(
n_{E \mbk e}^{(-1)} - n_{E \mbkdpr d}^{(-1)}
\right)
\delta ( \veps_{\mbk}^e - \veps_{\mbkdpr}^d ) 
\right],
\end{split}
\end{equation}
\begin{equation}
\begin{split}
\mathcal{N}^{(\text{ask})}_{24,\mbk a}
=&
- \frac{2 \pi^2}{\hbar}  e E^{\mu} \tau_{\mbk}^a
\sum_{\mbkpr \mbkdpr bcde}
\delta ( \veps_{\mbk}^a - \veps_{\mbkpr}^b )
\frac{
\text{Im}
\left[
\Braket{
U_{\mbk \mbk\pr}^{ab} U_{\mbk\pr \mbk} ^{bc} 
}
\Braket{
U_{\mbk \mbkdpr}^{ed} U_{\mbkdpr \mbk} ^{da} 
}
\mathcal{A}_{\mu \mbk}^{\prime ce}
\right]_{\substack{c \neq a}}
}
{(\veps_{\mbk}^a - \veps_{\mbk}^c)
(\veps_{\mbk}^c - \veps_{\mbk}^e)}
\\
&\hspace{.27\linewidth}
\times
\left[
\left(
n_{E \mbk c}^{(-1)} - n_{E \mbkdpr d}^{(-1)}
\right)
\delta ( \veps_{\mbk}^c - \veps_{\mbkdpr}^d ) 
\right.
-
\left.
\left(
n_{E \mbk e}^{(-1)} - n_{E \mbkdpr d}^{(-1)}
\right)
\delta ( \veps_{\mbk}^e - \veps_{\mbkdpr}^d ) 
\right]
\\
&-
\frac{2 \pi^2}{\hbar}  e E^{\mu} \tau_{\mbk}^a
\sum_{\mbkpr \mbkdpr bcde}
\delta ( \veps_{\mbk}^a - \veps_{\mbkpr}^b )
\frac{
\text{Im}
\left[
\Braket{
U_{\mbk \mbk\pr}^{ab} U_{\mbk\pr \mbk} ^{bc} 
}
\Braket{
U_{\mbk \mbkdpr}^{ed} U_{\mbkdpr \mbk} ^{da} 
}
\mathcal{A}_{\mu \mbk}^{\prime ce}
\right]_{\substack{c,e \neq a}}
}
{(\veps_{\mbk}^a - \veps_{\mbk}^c)
(\veps_{\mbk}^a - \veps_{\mbk}^e)}
\\
&\hspace{.27\linewidth}
\times
\left[
\left(
n_{E \mbk a}^{(-1)} - n_{E \mbkdpr d}^{(-1)}
\right)
\delta ( \veps_{\mbk}^a - \veps_{\mbkdpr}^d ) 
\right.
+
\left.
\left(
n_{E \mbk e}^{(-1)} - n_{E \mbkdpr d}^{(-1)}
\right)
\delta ( \veps_{\mbk}^e - \veps_{\mbkdpr}^d ) 
\right],
\end{split}
\end{equation}
\begin{equation}
\begin{split}
\mathcal{N}^{(\text{ask})}_{25,\mbk a}
=&
- \frac{2 \pi^2}{\hbar}  e E^{\mu} \tau_{\mbk}^a
\sum_{\mbkpr \mbkdpr bcde}
\delta ( \veps_{\mbk}^a - \veps_{\mbkpr}^b )
\frac{
\text{Im}
\left[
\Braket{
U_{\mbk \mbk\pr}^{ab} U_{\mbk\pr \mbk} ^{ca} 
}
\Braket{
U_{\mbkpr \mbkdpr}^{bd} U_{\mbkdpr \mbkpr} ^{ec} 
}
\mathcal{A}_{\mu \mbkdpr}^{\prime de}
\right]_{\substack{c \neq b}}
}
{\veps_{\mbk}^a - \veps_{\mbkpr}^c}
\left[
\frac{
n_{E \mbkpr b}^{(-1)} - n_{E \mbkdpr d}^{(-1)}
}
{\veps_{\mbk}^a - \veps_{\mbkdpr}^e}
\delta ( \veps_{\mbk}^a - \veps_{\mbkdpr}^d ) 
\right.
\\
&+
\left.
\frac{
n_{E \mbkpr c}^{(-1)} - n_{E \mbkdpr e}^{(-1)}
}
{\veps_{\mbkpr}^c - \veps_{\mbkdpr}^d}
\delta ( \veps_{\mbkpr}^c - \veps_{\mbkdpr}^e ) 
+
\frac{
n_{E \mbkpr b}^{(-1)} - n_{E \mbkdpr e}^{(-1)}
}
{\veps_{\mbk}^a - \veps_{\mbkdpr}^d}
\delta ( \veps_{\mbk}^a - \veps_{\mbkdpr}^e )
+
\frac{
n_{E \mbkpr c}^{(-1)} - n_{E \mbkdpr d}^{(-1)}
}
{\veps_{\mbkpr}^c - \veps_{\mbkdpr}^e}
\delta ( \veps_{\mbkpr}^c - \veps_{\mbkdpr}^d )
\right],
\end{split}
\end{equation}
\begin{equation}
\begin{split}
\mathcal{N}^{(\text{ask})}_{26,\mbk a}
=&
- \frac{2 \pi^2}{\hbar}  e E^{\mu} \tau_{\mbk}^a
\sum_{\mbkpr \mbkdpr bcde}
\delta ( \veps_{\mbk}^a - \veps_{\mbkpr}^b )
\frac{
\text{Im}
\left[
\Braket{
U_{\mbk \mbk\pr}^{ab} U_{\mbk\pr \mbk} ^{ca} 
}
\Braket{
U_{\mbkpr \mbkdpr}^{bd} U_{\mbkdpr \mbkpr} ^{de} 
}
\mathcal{A}_{\mu \mbkpr}^{\prime ec}
\right]_{\substack{c \neq b}}
}
{(\veps_{\mbk}^a - \veps_{\mbkpr}^c)
(\veps_{\mbkpr}^c - \veps_{\mbkpr}^e)}
\\
&\hspace{.27\linewidth}
\times
\left[
\left(
n_{E \mbkpr c}^{(-1)} - n_{E \mbkdpr d}^{(-1)}
\right)
\delta ( \veps_{\mbkpr}^c - \veps_{\mbkdpr}^d ) 
\right.
-
\left.
\left(
n_{E \mbkpr e}^{(-1)} - n_{E \mbkdpr d}^{(-1)}
\right)
\delta ( \veps_{\mbkpr}^e - \veps_{\mbkdpr}^d ) 
\right]
\\
&-
\frac{2 \pi^2}{\hbar}  e E^{\mu} \tau_{\mbk}^a
\sum_{\mbkpr \mbkdpr bcde}
\delta ( \veps_{\mbk}^a - \veps_{\mbkpr}^b )
\frac{
\text{Im}
\left[
\Braket{
U_{\mbk \mbk\pr}^{ab} U_{\mbk\pr \mbk} ^{ca} 
}
\Braket{
U_{\mbkpr \mbkdpr}^{bd} U_{\mbkdpr \mbkpr} ^{de} 
}
\mathcal{A}_{\mu \mbkpr}^{\prime ec}
\right]_{\substack{c,e \neq b}}
}
{(\veps_{\mbk}^a - \veps_{\mbkpr}^c)
(\veps_{\mbk}^a - \veps_{\mbkpr}^e)}
\\
&\hspace{.27\linewidth}
\times
\left[
\left(
n_{E \mbkpr b}^{(-1)} - n_{E \mbkdpr d}^{(-1)}
\right)
\delta ( \veps_{\mbk}^a - \veps_{\mbkdpr}^d ) 
\right.
+
\left.
\left(
n_{E \mbkpr e}^{(-1)} - n_{E \mbkdpr d}^{(-1)}
\right)
\delta ( \veps_{\mbkpr}^e - \veps_{\mbkdpr}^d ) 
\right],
\end{split}
\end{equation}
\begin{equation}
\begin{split}
\mathcal{N}^{(\text{ask})}_{27,\mbk a}
=&
- \frac{2 \pi^2}{\hbar}  e E^{\mu} \tau_{\mbk}^a
\sum_{\mbkpr \mbkdpr bcde}
\delta ( \veps_{\mbk}^a - \veps_{\mbkpr}^b )
\frac{
\text{Im}
\left[
\Braket{
U_{\mbk \mbk\pr}^{ab} U_{\mbk\pr \mbk} ^{ca} 
}
\Braket{
U_{\mbkpr \mbkdpr}^{ed} U_{\mbkdpr \mbkpr} ^{dc} 
}
\mathcal{A}_{\mu \mbkpr}^{\prime be}
\right]_{\substack{c \neq b}}
}
{(\veps_{\mbk}^a - \veps_{\mbkpr}^c)
(\veps_{\mbk}^a - \veps_{\mbkpr}^e)}
\\
&\hspace{.27\linewidth}
\times
\left[
\left(
n_{E \mbkpr b}^{(-1)} - n_{E \mbkdpr d}^{(-1)}
\right)
\delta ( \veps_{\mbkpr}^b - \veps_{\mbkdpr}^d ) 
\right.
-
\left.
\left(
n_{E \mbkpr e}^{(-1)} - n_{E \mbkdpr d}^{(-1)}
\right)
\delta ( \veps_{\mbkpr}^e - \veps_{\mbkdpr}^d ) 
\right]
\\
&-
\frac{2 \pi^2}{\hbar}  e E^{\mu} \tau_{\mbk}^a
\sum_{\mbkpr \mbkdpr bcde}
\delta ( \veps_{\mbk}^a - \veps_{\mbkpr}^b )
\frac{
\text{Im}
\left[
\Braket{
U_{\mbk \mbk\pr}^{ab} U_{\mbk\pr \mbk} ^{ca} 
}
\Braket{
U_{\mbkpr \mbkdpr}^{ed} U_{\mbkdpr \mbkpr} ^{dc} 
}
\mathcal{A}_{\mu \mbkpr}^{\prime be}
\right]_{\substack{e \neq c}}
}
{(\veps_{\mbk}^a - \veps_{\mbkpr}^e)
(\veps_{\mbkpr}^e - \veps_{\mbkpr}^c)}
\\
&\hspace{.27\linewidth}
\times
\left[
\left(
n_{E \mbkpr c}^{(-1)} - n_{E \mbkdpr d}^{(-1)}
\right)
\delta ( \veps_{\mbkpr}^c - \veps_{\mbkdpr}^d ) 
\right.
+
\left.
\left(
n_{E \mbkpr e}^{(-1)} - n_{E \mbkdpr d}^{(-1)}
\right)
\delta ( \veps_{\mbkpr}^e - \veps_{\mbkdpr}^d ) 
\right],
\end{split}
\end{equation}
\begin{equation}
\begin{split}
\mathcal{N}^{(\text{ask})}_{28,\mbk a}
=&
- \frac{2 \pi^2}{\hbar}  e E^{\mu} \tau_{\mbk}^a
\sum_{\mbkpr \mbkdpr bcde}
\delta ( \veps_{\mbk}^a - \veps_{\mbkpr}^e )
\frac{
\text{Im}
\left[
\Braket{
U_{\mbk \mbk\pr}^{ab} U_{\mbk\pr \mbk} ^{ca} 
}
\Braket{
U_{\mbkpr \mbkdpr}^{ed} U_{\mbkdpr \mbkpr} ^{dc} 
}
\mathcal{A}_{\mu \mbkpr}^{\prime be}
\right]_{\substack{e \neq c}}
}
{(\veps_{\mbk}^a - \veps_{\mbkpr}^b)
(\veps_{\mbk}^a - \veps_{\mbkpr}^c)}
\\
&\hspace{.27\linewidth}
\times
\left[
\left(
n_{E \mbkpr c}^{(-1)} - n_{E \mbkdpr d}^{(-1)}
\right)
\delta ( \veps_{\mbkpr}^c - \veps_{\mbkdpr}^d ) 
\right.
+
\left.
\left(
n_{E \mbkpr e}^{(-1)} - n_{E \mbkdpr d}^{(-1)}
\right)
\delta ( \veps_{\mbkpr}^e - \veps_{\mbkdpr}^d ) 
\right]
\\
&-
\frac{2 \pi^2}{\hbar}  e E^{\mu} \tau_{\mbk}^a
\sum_{\mbkpr \mbkdpr bcde}
\delta ( \veps_{\mbk}^a - \veps_{\mbkpr}^b )
\frac{
\text{Im}
\left[
\Braket{
U_{\mbk \mbk\pr}^{ab} U_{\mbk\pr \mbk} ^{ca} 
}
\Braket{
U_{\mbkpr \mbkdpr}^{ed} U_{\mbkdpr \mbkpr} ^{dc} 
}
\mathcal{A}_{\mu \mbkpr}^{\prime be}
\right]_{\substack{c \neq b,e}}
}
{(\veps_{\mbk}^a - \veps_{\mbkpr}^c)
(\veps_{\mbkpr}^c - \veps_{\mbkpr}^e)}
\\
&\hspace{.27\linewidth}
\times
\left[
\left(
n_{E \mbkpr c}^{(-1)} - n_{E \mbkdpr d}^{(-1)}
\right)
\delta ( \veps_{\mbkpr}^c - \veps_{\mbkdpr}^d ) 
\right.
+
\left.
\left(
n_{E \mbkpr e}^{(-1)} - n_{E \mbkdpr d}^{(-1)}
\right)
\delta ( \veps_{\mbkpr}^e - \veps_{\mbkdpr}^d ) 
\right],
\end{split}
\end{equation}
\begin{equation}
\begin{split}
\mathcal{N}^{(\text{ask})}_{29,\mbk a}
=&
- \frac{2 \pi^2}{\hbar}  e E^{\mu} \tau_{\mbk}^a
\sum_{\mbkpr \mbkdpr bcde}
\left[
\delta ( \veps_{\mbk}^a - \veps_{\mbkpr}^b )
-
\delta ( \veps_{\mbk}^a - \veps_{\mbkpr}^c )
\right]
\frac{
\text{Im}
\left[
\Braket{
U_{\mbk \mbk\pr}^{ab} U_{\mbk\pr \mbk} ^{cd} 
}
\Braket{
U_{\mbk \mbkdpr}^{de} U_{\mbkdpr \mbk} ^{ea} 
}
\mathcal{A}_{\mu \mbkpr}^{\prime bc}
\right]_{\substack{d \neq a}}
}
{(\veps_{\mbk}^a - \veps_{\mbk}^d)(\veps_{\mbkpr}^b - \veps_{\mbkpr}^c)}
\\
&\hspace{.27\linewidth}
\times
\left[
\left(
n_{E \mbk a}^{(-1)} - n_{E \mbkdpr e}^{(-1)}
\right)
\delta ( \veps_{\mbk}^a - \veps_{\mbkdpr}^e ) 
\right.
+
\left.
\left(
n_{E \mbk d}^{(-1)} - n_{E \mbkdpr e}^{(-1)}
\right)
\delta ( \veps_{\mbk}^d - \veps_{\mbkdpr}^e ) 
\right],
\end{split}
\end{equation}
\begin{equation}
\begin{split}
\mathcal{N}^{(\text{ask})}_{30,\mbk a}
=&
- \frac{2 \pi^2}{\hbar}  e E^{\mu} \tau_{\mbk}^a
\sum_{\mbkpr \mbkdpr bcde}
\left[
\delta ( \veps_{\mbk}^a - \veps_{\mbkpr}^b )
-
\delta ( \veps_{\mbkpr}^b - \veps_{\mbk}^d )
\right]
\frac{
\text{Im}
\left[
\Braket{
U_{\mbk \mbk\pr}^{ab} U_{\mbk\pr \mbk} ^{cd} 
}
\Braket{
U_{\mbkpr \mbkdpr}^{be} U_{\mbkdpr \mbkpr} ^{ec} 
}
\mathcal{A}_{\mu \mbk}^{\prime da}
\right]_{\substack{c \neq b}}
}
{(\veps_{\mbk}^a - \veps_{\mbk}^d)(\veps_{\mbkpr}^b - \veps_{\mbkpr}^c)}
\\
&\hspace{.27\linewidth}
\times
\left[
\left(
n_{E \mbkpr b}^{(-1)} - n_{E \mbkdpr e}^{(-1)}
\right)
\delta ( \veps_{\mbkpr}^b - \veps_{\mbkdpr}^e ) 
\right.
+
\left.
\left(
n_{E \mbkpr c}^{(-1)} - n_{E \mbkdpr e}^{(-1)}
\right)
\delta ( \veps_{\mbkpr}^c - \veps_{\mbkdpr}^e ) 
\right].
\end{split}
\end{equation}

\section{Carrier densities and conductivities of Dirac fermions}
\label{app_dirac}

\subsection{Carrier densities}
For the model of Dirac fermions given by Eq.~(\ref{H_Dirac}), to first order in tilting, the linear-response carrier densities associated with special scattering are obtained as
\begin{equation}
\label{n_Ek_sj_dirac}
n_{E \mbk}^{(\text{sj}) +}
=
\frac{e}{2 \hbar} \frac{n_I U_0^2}{v} E_x
\tau_{\mbk}^+ 
\frac{\pd f_{0 \mbk}^+}{\pd \veps_{\mbk}^+}
\cos \theta
\left[ \sin \theta \sin \phi
+
\frac{t_x}{4 v} \sin^2 \theta \sin 2\phi
-
\frac{t_y}{4 v} (3 + \cos 2\theta + \sin^2 \theta \cos 2\phi)
\right],
\end{equation}
\begin{equation}
\begin{split}
n_{E \mbk}^{(\text{sk}) +}
&=
\frac{3 e}{64 \hbar^2}
( n_I U_0^2 )^2
\frac{h_k}{v^3} E_x
(\tau_{\mbk}^+)^2 
\frac{\pd f_{0 \mbk}^+}{\pd \veps_{\mbk}^+}
\sin \theta \sin 2\theta
\biggl\{
\sin \theta \sin \phi
-
\frac{t_x}{6v} (1 + 5 \cos^2 \theta) \sin 2\phi
\\
&\hspace{.5\linewidth} -
\frac{t_y}{v} \left[1 + \cos^2 \theta
-
\frac{1}{6} (1 + 5 \cos^2 \theta) \cos 2\phi \right]
\biggr\},
\end{split}
\end{equation}
\begin{equation}
\begin{split}
n_{E \mbk}^{(\text{sk}) -}
&=
- \frac{3 e}{64 \hbar^2}
( n_I U_0^2 )^2
\frac{h_k}{v^3} E_x
\tau_{\mbk}^+ \tau_{\mbk}^-
\frac{\pd f_{0 \mbk}^+}{\pd \veps_{\mbk}^+}
\sin \theta \sin 2\theta
\biggl\{
\sin \theta \sin \phi
-
\frac{5 t_x}{6v} (1 + \cos^2 \theta) \sin 2\phi
\\
&\hspace{.5\linewidth} +
\frac{t_y}{v} \left[\frac{1}{3} - \cos^2 \theta
+
\frac{5}{6} (1 + \cos^2 \theta) \cos 2\phi \right]
\biggr\},
\end{split}
\end{equation}
where we note that $n_{E \mbk}^{(\text{sj}) -}=0$ when the Fermi level lies in the upper band. Note also that in the limit of vanishing tilting, these reduce to the forms obtained in Ref.~\cite{atencia2022semiclassical}.

Moving on to the mixed scattering densities in the quadratic responses, recalling Eq.~(\ref{n_E2k_sjo}), we find
\begin{equation}
\mathcal{N}^{(\text{sj,o})}_{1,\mbk +}
=
\frac{e^2}{4 \hbar} \frac{n_I U_0^2}{v h_k}
E_x^2 
\left(\tau_{\mbk}^+\right)^2 v_{x \mbk}^+
\frac{\pd f_{0 \mbk}^+}{\pd \veps_{\mbk}^+} \cos \theta
\left[
\sin \phi \sin \theta
+
\frac{t_x}{v} \sin 2\phi \sin^2 \theta
-
\frac{t_y}{v} \left( \cos 2 \theta + \sin^2 \theta \cos 2\phi \right)
\right],
\end{equation}
and
\begin{equation}
\mathcal{N}^{(\text{sj,o})}_{2,\mbk +}
=
- \frac{e^2}{4 \hbar} \frac{v^2}{h_k^2}
E_x^2 
\tau_{\mbk}^+
\frac{\pd f_{0 \mbk}^+}{\pd \veps_{\mbk}^+}
\frac{
\sin 2\theta \sin^2 \theta}
{5 + 3 \cos 2\theta}
\left[
\frac{t_x}{v} 
\left(3 \sin \phi + \sin 3\phi \right)
+
\frac{t_y}{v} 
\left(3 \cos \phi - \cos 3 \phi \right)
\right].
\end{equation}
The skew scattering contribution given by Eq.~(\ref{n_E2_sko}) may also be decomposed as
\begin{equation}
n_{E^2 \mbk +}^{(\text{sk,o})}
=
\mathcal{N}_{1, \mbk +}^{(\text{sk,o})}
+
\mathcal{N}_{2, \mbk +}^{(\text{sk,o})},
\end{equation}
where
\begin{equation}
\begin{split}
\mathcal{N}_{1, \mbk +}^{(\text{sk,o})}
=&
\frac{e^2}{128 \hbar} E_x^2 \tau_{\mbk}^+
\frac{\pd f_{0 \mbk}^+}{\pd \veps_{\mbk}^+}
\frac{v^2}{h_k^2}
\frac{ \sin 2\theta }
{ ( 5 + 3 \cos 2\theta)^4 }
\bigg\{
\frac{3 t_x}{v} \sin \phi (3487
+
6504 \cos 2\theta 
+
2188 \cos 4\theta
+
88 \cos 6\theta
+
21 \cos 8\theta
)
\\
&+
\frac{2 t_y}{v} \cos \phi (9463
+
9912 \cos 2\theta 
+
1100 \cos 4\theta
+
8 \cos 6\theta
-
3 \cos 8\theta
)
\biggr\},
\end{split}
\end{equation}

\begin{equation}
\begin{split}
\mathcal{N}_{2, \mbk +}^{(\text{sk,o})}
=&
\frac{3 e^2}{8 \hbar} E_x^2 \tau_{\mbk}^+
\frac{\pd^2 f_{0 \mbk}^+}{\pd (\veps_{\mbk}^+)^2}
\frac{v^2}{h_k}
\frac{ \sin^2\theta \sin 2\theta }
{ ( 5 + 3 \cos 2\theta)^3 }
\bigg\{
\frac{t_x}{v} \sin \phi (197
+
148 \cos 2\theta 
+
7 \cos 4\theta
)
+
\frac{t_y}{v} \cos \phi (29
+
4 \cos 2\theta 
-
\cos 4\theta
)
\biggr\},
\end{split}
\end{equation}
and
\begin{equation}
n_{E^2 \mbk -}^{(\text{sk,o})}
=
\mathcal{N}_{1, \mbk -}^{(\text{sk,o})}
+
\mathcal{N}_{2, \mbk -}^{(\text{sk,o})}
+
\mathcal{N}_{3, \mbk -}^{(\text{sk,o})},
\end{equation}
with
\begin{equation}
\begin{split}
\mathcal{N}_{1, \mbk -}^{(\text{sk,o})}
=&
- \frac{e^2}{128 \hbar} E_x^2 \tau_{\mbk}^-
\frac{\pd f_{0 \mbk}^+}{\pd \veps_{\mbk}^+}
\frac{v^2}{h_k^2}
\frac{ \sin 2\theta }
{ ( 5 + 3 \cos 2\theta)^4 }
\bigg\{
\frac{t_x}{v} \sin \phi (45677
+
68896 \cos 2\theta 
+
20628 \cos 4\theta
-
96 \cos 6\theta
+
63 \cos 8\theta
)
\\
&-
\frac{t_y}{v} \cos \phi (16523
+
32240 \cos 2\theta 
+
13100 \cos 4\theta
-
432 \cos 6\theta
+
9 \cos 8\theta
)
\biggr\},
\end{split}
\end{equation}

\begin{equation}
\begin{split}
\mathcal{N}_{2, \mbk -}^{(\text{sk,o})}
=&
- \frac{e^2}{16 \hbar} E_x^2 \tau_{\mbk}^-
\frac{\pd^2 f_{0 \mbk}^+}{\pd (\veps_{\mbk}^+)^2}
\frac{v^2}{h_k}
\frac{ \sin^2\theta \sin 2\theta }
{ ( 5 + 3 \cos 2\theta)^3 }
\bigg\{
\frac{t_x}{v} \sin \phi (1819
+
1516 \cos 2\theta 
+
57 \cos 4\theta
)
\\
&-
\frac{t_y}{v} \cos \phi (477
+
596 \cos 2\theta 
+
15 \cos 4\theta
)
\biggr\},
\end{split}
\end{equation}

\begin{equation}
\label{N_3k_sko_dirac}
\begin{split}
\mathcal{N}_{3, \mbk -}^{(\text{sk,o})}
=&
\frac{e^2}{16 \hbar^2} (n_I U_0^2)^2 E_x^2 \tau_{\mbk}^+ \tau_{\mbk}^-
\left[
\pd_{\mbk}^x (\tau_{\mbk}^+ v_{x \mbk}^+)
\frac{\pd f_{0 \mbk}^+}{\pd \veps_{\mbk}^+}
+
\hbar \tau_{\mbk}^+ (v_{x \mbk}^+)^2
\frac{\pd^2 f_{0 \mbk}^+}{\pd (\veps_{\mbk}^+)^2}
\right]
\frac{h_k}{v^4} \sin 2\theta
\left(
\frac{t_x}{v} \sin \phi
-
\frac{t_y}{v} \cos \phi
\right).
\end{split}
\end{equation}
\end{widetext}
\vspace{-.3cm}
And the remaining mixed scattering terms are readily evaluated by taking the derivatives of the linear response densities.
\vspace{-1cm}

\begin{table}
\caption{Transport coefficients of the lonigitudinal and transverse nonlinear conductivities.}

\begin{ruledtabular}
\begin{tabular}{l c c c c c}
{} & $n=0$ & $n=1$ & $n=2$ & $n=3$ & \multicolumn{1}{c}{$n=4$}
\\
\hline
$a_{2 n}^{(\text{o})}$ & 6207 & 6328 & -788 & 520 & 21
\\
$b_{2 n}^{(\text{o})}$ & 2665 & 2168 & -684 & -56 & 3
\\
$a_{2 n}^{(\text{sj,o})}$ & 109 & 139 & 15 & 0 & 0
\\
$b_{2 n}^{(\text{sj,o})}$ & 229 & 84 & 7 & 0 & 0
\\
$a_{2 n+1}^{(\text{o,sj})}$ & 9823 & 5301 & 1215 & 45 & 0 
\\
$b_{2 n}^{(\text{o,sj})}$ & 105 & 148 & 3 & 0 & 0
\\
$a_{2 n+1}^{(\text{sk,o})}$ & 3662 & 7856 & -1624 & -1717 & 15
\\
$b_{2 n+1}^{(\text{sk,o})}$ & 114852 & 59382 & 13258 & 987 & -63 
\\
$a_{2 n}^{(\text{o,sk})}$ & 7422 & 10159 & 2770 & 129 & 0 
\\
$b_{2 n}^{(\text{o,sk})}$ & 389 & 428 & 15 & 0 & 0
\\
$a_{2 n+1}^{(\text{ex})}$ & 6583 & 3057 & 555 & 45 & 0 
\\
$b_{2 n+1}^{(\text{ex})}$ & 15833 & 7375 & 1269 & 99 & 0 
\\
\end{tabular}
\end{ruledtabular}\label{tab1}
\end{table}

\subsection{Conductivities}
\vspace{-.2cm}
Here, we present the explicit forms of the functions appearing in the nonlinear conductivities in Section~(\ref{application}). The first functions read
\vspace{-.2cm}
\begin{subequations}
\label{S_muxx^o}
\begin{align}
\mathcal{S}_{xxx}^{(\text{o})}(\theta)
&=
\sum_{n=0}^{4}
\frac{a_{2 n}^{(\text{o})} \cos 2 n\theta}
{( 5 + 3 \cos 2 \theta )^4},
\\
\mathcal{S}_{yxx}^{(\text{o})}(\theta)
&=
\sum_{n=0}^{4}
\frac{b_{2 n}^{(\text{o})} \cos 2 n\theta}
{( 5 + 3 \cos 2 \theta )^4},
\end{align}
\end{subequations}
where $a_{2 n}^{(\text{o})}$ and $b_{2 n}^{(\text{o})}$ are numerical factors that emerge from the analytical integrations, the values of which are presented in Table~\ref{tab1}. The remaining functions read
\vspace{-.35cm}
\begin{subequations}
\begin{align}
\mathcal{S}_{xxx}^{(\text{sj,o})}(\theta)
&=
\sum_{n=0}^{2}
\frac{a_{2 n}^{(\text{sj,o})} \cos 2 n\theta \sin \theta \sin 2\theta} 
{( 5 + 3 \cos 2 \theta )^3},
\\
\mathcal{S}_{yxx}^{(\text{sj,o})}(\theta)
&=
\sum_{n=0}^{2}
\frac{b_{2 n}^{(\text{sj,o})} \cos 2 n\theta \sin \theta \sin 2\theta} 
{( 5 + 3 \cos 2 \theta )^3},
\end{align}
\end{subequations}
\begin{subequations}
\begin{align}
\mathcal{S}_{xxx}^{(\text{o,sj})}(\theta)
&=
\sum_{n=0}^{3}
\frac{a_{2n+1}^{(\text{o,sj})} \cos (2n+1) \theta} 
{( 5 + 3 \cos 2 \theta )^4},
\\
\mathcal{S}_{yxx}^{(\text{o,sj})}(\theta)
&=
\sum_{n=0}^{2}
\frac{b_{2n}^{(\text{o,sj})} \cos 2n \theta \sin \theta \sin 2\theta} 
{( 5 + 3 \cos 2 \theta )^4},
\end{align}
\end{subequations}
\begin{subequations}
\begin{align}
\mathcal{S}_{xxx}^{(\text{sk,o})}(\theta)
&=
\sum_{n=0}^{4}
\frac{a_{2n+1}^{(\text{sk,o})} \cos (2n+1)\theta \sin^2 \theta} 
{( 5 + 3 \cos 2 \theta )^5},
\\
\mathcal{S}_{yxx}^{(\text{sk,o})}(\theta)
&=
\sum_{n=0}^{4}
\frac{b_{2n+1}^{(\text{sk,o})} \cos (2n+1)\theta \sin^2 \theta} 
{( 5 + 3 \cos 2 \theta )^5},
\end{align}
\end{subequations}
\begin{subequations}
\begin{align}
\mathcal{S}_{xxx}^{(\text{o,sk})}(\theta)
&=
\sum_{n=0}^{3}
\frac{a_{2n}^{(\text{o,sk})} \cos 2n\theta \sin \theta \sin 2 \theta} 
{( 5 + 3 \cos 2 \theta )^5},
\\
\mathcal{S}_{yxx}^{(\text{o,sk})}(\theta)
&=
\sum_{n=0}^{2}
\frac{b_{2n}^{(\text{o,sk})} \cos 2n \theta \sin^3 \theta \sin 2 \theta} 
{( 5 + 3 \cos 2 \theta )^5},
\end{align}
\end{subequations}
\begin{subequations}
\begin{align}
\mathcal{S}_{xxx}^{(\text{ex})}(\theta)
&=
\sum_{n=0}^{3}
\frac{a_{2n+1}^{(\text{ex})} \cos (2n+1)\theta \sin^2 \theta} 
{( 5 + 3 \cos 2 \theta )^4},
\\
\mathcal{S}_{yxx}^{(\text{ex})}(\theta)
&=
\sum_{n=0}^{3}
\frac{b_{2n+1}^{(\text{ex})} \cos (2n+1)\theta \sin^2 \theta} 
{( 5 + 3 \cos 2 \theta )^4},
\end{align}
\end{subequations}
\begin{equation}
\label{S_yxx^BC}
\mathcal{S}_{yxx}^{(\text{BC})}(\theta)
=
\frac{ (7 + 3 \cos 2\theta) \sin \theta \sin 2\theta} 
{( 5 + 3 \cos 2 \theta )^2}.
\end{equation}

\bibliography{qkt}

\begin{thebibliography}{138}%
\makeatletter
\providecommand \@ifxundefined [1]{%
 \@ifx{#1\undefined}
}%
\providecommand \@ifnum [1]{%
 \ifnum #1\expandafter \@firstoftwo
 \else \expandafter \@secondoftwo
 \fi
}%
\providecommand \@ifx [1]{%
 \ifx #1\expandafter \@firstoftwo
 \else \expandafter \@secondoftwo
 \fi
}%
\providecommand \natexlab [1]{#1}%
\providecommand \enquote  [1]{``#1''}%
\providecommand \bibnamefont  [1]{#1}%
\providecommand \bibfnamefont [1]{#1}%
\providecommand \citenamefont [1]{#1}%
\providecommand \href@noop [0]{\@secondoftwo}%
\providecommand \href [0]{\begingroup \@sanitize@url \@href}%
\providecommand \@href[1]{\@@startlink{#1}\@@href}%
\providecommand \@@href[1]{\endgroup#1\@@endlink}%
\providecommand \@sanitize@url [0]{\catcode `\\12\catcode `\$12\catcode
  `\&12\catcode `\#12\catcode `\^12\catcode `\_12\catcode `\%12\relax}%
\providecommand \@@startlink[1]{}%
\providecommand \@@endlink[0]{}%
\providecommand \url  [0]{\begingroup\@sanitize@url \@url }%
\providecommand \@url [1]{\endgroup\@href {#1}{\urlprefix }}%
\providecommand \urlprefix  [0]{URL }%
\providecommand \Eprint [0]{\href }%
\providecommand \doibase [0]{https://doi.org/}%
\providecommand \selectlanguage [0]{\@gobble}%
\providecommand \bibinfo  [0]{\@secondoftwo}%
\providecommand \bibfield  [0]{\@secondoftwo}%
\providecommand \translation [1]{[#1]}%
\providecommand \BibitemOpen [0]{}%
\providecommand \bibitemStop [0]{}%
\providecommand \bibitemNoStop [0]{.\EOS\space}%
\providecommand \EOS [0]{\spacefactor3000\relax}%
\providecommand \BibitemShut  [1]{\csname bibitem#1\endcsname}%
\let\auto@bib@innerbib\@empty
\bibitem [{\citenamefont {Kubo}(1957)}]{kubo1957statistical}%
  \BibitemOpen
  \bibfield  {author} {\bibinfo {author} {\bibfnamefont {R.}~\bibnamefont
  {Kubo}},\ }\bibfield  {title} {\bibinfo {title} {Statistical-mechanical
  theory of irreversible processes. i. general theory and simple applications
  to magnetic and conduction problems},\ }\href
  {https://doi.org/10.1143/JPSJ.12.570} {\bibfield  {journal} {\bibinfo
  {journal} {J. Phys. Soc. Jpn.}\ }\textbf {\bibinfo {volume} {12}},\ \bibinfo
  {pages} {570} (\bibinfo {year} {1957})}\BibitemShut {NoStop}%
\bibitem [{\citenamefont {Keldysh}(2024)}]{keldysh2024diagram}%
  \BibitemOpen
  \bibfield  {author} {\bibinfo {author} {\bibfnamefont {L.~V.}\ \bibnamefont
  {Keldysh}},\ }\bibfield  {title} {\bibinfo {title} {{Diagram technique for
  nonequilibrium processes}},\ }in\ \href@noop {} {\emph {\bibinfo {booktitle}
  {{Selected Papers of Leonid V Keldysh}}}}\ (\bibinfo  {publisher} {World
  Scientific},\ \bibinfo {year} {2024})\ pp.\ \bibinfo {pages}
  {47--55}\BibitemShut {NoStop}%
\bibitem [{\citenamefont {Karplus}\ and\ \citenamefont
  {Luttinger}(1954)}]{karplus1954hall}%
  \BibitemOpen
  \bibfield  {author} {\bibinfo {author} {\bibfnamefont {R.}~\bibnamefont
  {Karplus}}\ and\ \bibinfo {author} {\bibfnamefont {J.~M.}\ \bibnamefont
  {Luttinger}},\ }\bibfield  {title} {\bibinfo {title} {{Hall Effect in
  Ferromagnetics}},\ }\href {https://link.aps.org/doi/10.1103/PhysRev.95.1154}
  {\bibfield  {journal} {\bibinfo  {journal} {Phys. Rev.}\ }\textbf {\bibinfo
  {volume} {95}},\ \bibinfo {pages} {1154} (\bibinfo {year}
  {1954})}\BibitemShut {NoStop}%
\bibitem [{\citenamefont {Kohn}\ and\ \citenamefont
  {Luttinger}(1957)}]{kohn1957quantum}%
  \BibitemOpen
  \bibfield  {author} {\bibinfo {author} {\bibfnamefont {W.}~\bibnamefont
  {Kohn}}\ and\ \bibinfo {author} {\bibfnamefont {J.~M.}\ \bibnamefont
  {Luttinger}},\ }\bibfield  {title} {\bibinfo {title} {{Quantum Theory of
  Electrical Transport Phenomena}},\ }\href
  {https://link.aps.org/doi/10.1103/PhysRev.108.590} {\bibfield  {journal}
  {\bibinfo  {journal} {Phys. Rev.}\ }\textbf {\bibinfo {volume} {108}},\
  \bibinfo {pages} {590} (\bibinfo {year} {1957})}\BibitemShut {NoStop}%
\bibitem [{\citenamefont {Luttinger}\ and\ \citenamefont
  {Kohn}(1958)}]{luttinger1958quantum}%
  \BibitemOpen
  \bibfield  {author} {\bibinfo {author} {\bibfnamefont {J.~M.}\ \bibnamefont
  {Luttinger}}\ and\ \bibinfo {author} {\bibfnamefont {W.}~\bibnamefont
  {Kohn}},\ }\bibfield  {title} {\bibinfo {title} {{Quantum Theory of
  Electrical Transport Phenomena. II}},\ }\href
  {https://link.aps.org/doi/10.1103/PhysRev.109.1892} {\bibfield  {journal}
  {\bibinfo  {journal} {Phys. Rev.}\ }\textbf {\bibinfo {volume} {109}},\
  \bibinfo {pages} {1892} (\bibinfo {year} {1958})}\BibitemShut {NoStop}%
\bibitem [{\citenamefont {Luttinger}(1958)}]{luttinger1958theory}%
  \BibitemOpen
  \bibfield  {author} {\bibinfo {author} {\bibfnamefont {J.~M.}\ \bibnamefont
  {Luttinger}},\ }\bibfield  {title} {\bibinfo {title} {Theory of the
  \text{H}all effect in ferromagnetic substances},\ }\href
  {https://link.aps.org/doi/10.1103/PhysRev.112.739} {\bibfield  {journal}
  {\bibinfo  {journal} {Phys. Rev.}\ }\textbf {\bibinfo {volume} {112}},\
  \bibinfo {pages} {739} (\bibinfo {year} {1958})}\BibitemShut {NoStop}%
\bibitem [{\citenamefont {Sinitsyn}\ \emph {et~al.}(2005)\citenamefont
  {Sinitsyn}, \citenamefont {Niu}, \citenamefont {Sinova},\ and\ \citenamefont
  {Nomura}}]{sinitsyn2005disorder}%
  \BibitemOpen
  \bibfield  {author} {\bibinfo {author} {\bibfnamefont {N.~A.}\ \bibnamefont
  {Sinitsyn}}, \bibinfo {author} {\bibfnamefont {Q.}~\bibnamefont {Niu}},
  \bibinfo {author} {\bibfnamefont {J.}~\bibnamefont {Sinova}},\ and\ \bibinfo
  {author} {\bibfnamefont {K.}~\bibnamefont {Nomura}},\ }\bibfield  {title}
  {\bibinfo {title} {{Disorder effects in the anomalous Hall effect induced by
  Berry curvature}},\ }\href
  {https://link.aps.org/doi/10.1103/PhysRevB.72.045346} {\bibfield  {journal}
  {\bibinfo  {journal} {Phys. Rev. B}\ }\textbf {\bibinfo {volume} {72}},\
  \bibinfo {pages} {045346} (\bibinfo {year} {2005})}\BibitemShut {NoStop}%
\bibitem [{\citenamefont {Sinitsyn}\ \emph {et~al.}(2007)\citenamefont
  {Sinitsyn}, \citenamefont {MacDonald}, \citenamefont {Jungwirth},
  \citenamefont {Dugaev},\ and\ \citenamefont
  {Sinova}}]{sinitsyn2007anomalous}%
  \BibitemOpen
  \bibfield  {author} {\bibinfo {author} {\bibfnamefont {N.~A.}\ \bibnamefont
  {Sinitsyn}}, \bibinfo {author} {\bibfnamefont {A.~H.}\ \bibnamefont
  {MacDonald}}, \bibinfo {author} {\bibfnamefont {T.}~\bibnamefont
  {Jungwirth}}, \bibinfo {author} {\bibfnamefont {V.~K.}\ \bibnamefont
  {Dugaev}},\ and\ \bibinfo {author} {\bibfnamefont {J.}~\bibnamefont
  {Sinova}},\ }\bibfield  {title} {\bibinfo {title} {{Anomalous Hall effect in
  a two-dimensional Dirac band: The link between the Kubo-Streda formula and
  the semiclassical Boltzmann equation approach}},\ }\href
  {https://link.aps.org/doi/10.1103/PhysRevB.75.045315} {\bibfield  {journal}
  {\bibinfo  {journal} {Phys. Rev. B}\ }\textbf {\bibinfo {volume} {75}},\
  \bibinfo {pages} {045315} (\bibinfo {year} {2007})}\BibitemShut {NoStop}%
\bibitem [{\citenamefont {Sinitsyn}(2007)}]{sinitsyn2007semiclassical}%
  \BibitemOpen
  \bibfield  {author} {\bibinfo {author} {\bibfnamefont {N.}~\bibnamefont
  {Sinitsyn}},\ }\bibfield  {title} {\bibinfo {title} {Semiclassical theories
  of the anomalous \text{H}all effect},\ }\href
  {https://iopscience.iop.org/article/10.1088/0953-8984/20/02/023201/meta?casa_token=1NQD_xlIEYQAAAAA:RwHn1TnjQG64Xi4VGds8xzxrnJluOiSQDF2aMhI_9oCdxe6bUsv4VlJhSUVtHHUIjd5kfjb_Vpw}
  {\bibfield  {journal} {\bibinfo  {journal} {J. Phys.: Condens. Matter}\
  }\textbf {\bibinfo {volume} {20}},\ \bibinfo {pages} {023201} (\bibinfo
  {year} {2007})}\BibitemShut {NoStop}%
\bibitem [{\citenamefont {Culcer}\ \emph {et~al.}(2017)\citenamefont {Culcer},
  \citenamefont {Sekine},\ and\ \citenamefont
  {MacDonald}}]{culcer2017interband}%
  \BibitemOpen
  \bibfield  {author} {\bibinfo {author} {\bibfnamefont {D.}~\bibnamefont
  {Culcer}}, \bibinfo {author} {\bibfnamefont {A.}~\bibnamefont {Sekine}},\
  and\ \bibinfo {author} {\bibfnamefont {A.~H.}\ \bibnamefont {MacDonald}},\
  }\bibfield  {title} {\bibinfo {title} {Interband coherence response to
  electric fields in crystals: \text{B}erry-phase contributions and disorder
  effects},\ }\href {https://link.aps.org/doi/10.1103/PhysRevB.96.035106}
  {\bibfield  {journal} {\bibinfo  {journal} {Phys. Rev. B}\ }\textbf {\bibinfo
  {volume} {96}},\ \bibinfo {pages} {035106} (\bibinfo {year}
  {2017})}\BibitemShut {NoStop}%
\bibitem [{\citenamefont {Sekine}\ \emph {et~al.}(2017)\citenamefont {Sekine},
  \citenamefont {Culcer},\ and\ \citenamefont {MacDonald}}]{sekine2017quantum}%
  \BibitemOpen
  \bibfield  {author} {\bibinfo {author} {\bibfnamefont {A.}~\bibnamefont
  {Sekine}}, \bibinfo {author} {\bibfnamefont {D.}~\bibnamefont {Culcer}},\
  and\ \bibinfo {author} {\bibfnamefont {A.~H.}\ \bibnamefont {MacDonald}},\
  }\bibfield  {title} {\bibinfo {title} {Quantum kinetic theory of the chiral
  anomaly},\ }\href {https://link.aps.org/doi/10.1103/PhysRevB.96.235134}
  {\bibfield  {journal} {\bibinfo  {journal} {Phys. Rev. B}\ }\textbf {\bibinfo
  {volume} {96}},\ \bibinfo {pages} {235134} (\bibinfo {year}
  {2017})}\BibitemShut {NoStop}%
\bibitem [{\citenamefont {Xiao}\ and\ \citenamefont
  {Niu}(2017)}]{xiao2017semiclassical}%
  \BibitemOpen
  \bibfield  {author} {\bibinfo {author} {\bibfnamefont {C.}~\bibnamefont
  {Xiao}}\ and\ \bibinfo {author} {\bibfnamefont {Q.}~\bibnamefont {Niu}},\
  }\bibfield  {title} {\bibinfo {title} {Semiclassical theory of spin-orbit
  torques in disordered multiband electron systems},\ }\href
  {https://link.aps.org/doi/10.1103/PhysRevB.96.045428} {\bibfield  {journal}
  {\bibinfo  {journal} {Phys. Rev. B}\ }\textbf {\bibinfo {volume} {96}},\
  \bibinfo {pages} {045428} (\bibinfo {year} {2017})}\BibitemShut {NoStop}%
\bibitem [{\citenamefont {Xiao}\ \emph
  {et~al.}(2019{\natexlab{a}})\citenamefont {Xiao}, \citenamefont {Liu},
  \citenamefont {Yuan}, \citenamefont {Yang},\ and\ \citenamefont
  {Niu}}]{xiao2019temperature}%
  \BibitemOpen
  \bibfield  {author} {\bibinfo {author} {\bibfnamefont {C.}~\bibnamefont
  {Xiao}}, \bibinfo {author} {\bibfnamefont {Y.}~\bibnamefont {Liu}}, \bibinfo
  {author} {\bibfnamefont {Z.}~\bibnamefont {Yuan}}, \bibinfo {author}
  {\bibfnamefont {S.~A.}\ \bibnamefont {Yang}},\ and\ \bibinfo {author}
  {\bibfnamefont {Q.}~\bibnamefont {Niu}},\ }\bibfield  {title} {\bibinfo
  {title} {{Temperature dependence of the side-jump spin Hall conductivity}},\
  }\href {https://link.aps.org/doi/10.1103/PhysRevB.100.085425} {\bibfield
  {journal} {\bibinfo  {journal} {Phys. Rev. B}\ }\textbf {\bibinfo {volume}
  {100}},\ \bibinfo {pages} {085425} (\bibinfo {year}
  {2019}{\natexlab{a}})}\BibitemShut {NoStop}%
\bibitem [{\citenamefont {Atencia}\ \emph {et~al.}(2022)\citenamefont
  {Atencia}, \citenamefont {Niu},\ and\ \citenamefont
  {Culcer}}]{atencia2022semiclassical}%
  \BibitemOpen
  \bibfield  {author} {\bibinfo {author} {\bibfnamefont {R.~B.}\ \bibnamefont
  {Atencia}}, \bibinfo {author} {\bibfnamefont {Q.}~\bibnamefont {Niu}},\ and\
  \bibinfo {author} {\bibfnamefont {D.}~\bibnamefont {Culcer}},\ }\bibfield
  {title} {\bibinfo {title} {Semiclassical response of disordered conductors:
  Extrinsic carrier velocity and spin and field-corrected collision integral},\
  }\href {https://link.aps.org/doi/10.1103/PhysRevResearch.4.013001} {\bibfield
   {journal} {\bibinfo  {journal} {Phys. Rev. Res.}\ }\textbf {\bibinfo
  {volume} {4}},\ \bibinfo {pages} {013001} (\bibinfo {year}
  {2022})}\BibitemShut {NoStop}%
\bibitem [{\citenamefont {Mahan}(2000)}]{mahan2000many}%
  \BibitemOpen
  \bibfield  {author} {\bibinfo {author} {\bibfnamefont {G.~D.}\ \bibnamefont
  {Mahan}},\ }\href@noop {} {\emph {\bibinfo {title} {\textit{Many Particle
  Physics, Third Edition}}}}\ (\bibinfo  {publisher} {Plenum},\ \bibinfo
  {address} {New York},\ \bibinfo {year} {2000})\BibitemShut {NoStop}%
\bibitem [{\citenamefont {Bruus}\ and\ \citenamefont
  {Flensberg}(2004)}]{bruus2004many}%
  \BibitemOpen
  \bibfield  {author} {\bibinfo {author} {\bibfnamefont {H.}~\bibnamefont
  {Bruus}}\ and\ \bibinfo {author} {\bibfnamefont {K.}~\bibnamefont
  {Flensberg}},\ }\href@noop {} {\emph {\bibinfo {title} {{Many-body Quantum
  Theory in Condensed Matter Physics: An Introduction}}}}\ (\bibinfo
  {publisher} {OUP Oxford},\ \bibinfo {year} {2004})\BibitemShut {NoStop}%
\bibitem [{\citenamefont {Vasko}\ and\ \citenamefont
  {Raichev}(2006)}]{vasko2006quantum}%
  \BibitemOpen
  \bibfield  {author} {\bibinfo {author} {\bibfnamefont {F.~T.}\ \bibnamefont
  {Vasko}}\ and\ \bibinfo {author} {\bibfnamefont {O.~E.}\ \bibnamefont
  {Raichev}},\ }\href@noop {} {\emph {\bibinfo {title} {{Quantum kinetic theory
  and applications: Electrons, photons, phonons}}}}\ (\bibinfo  {publisher}
  {Springer Science \& Business Media},\ \bibinfo {year} {2006})\BibitemShut
  {NoStop}%
\bibitem [{\citenamefont {Xiao}\ \emph {et~al.}(2010)\citenamefont {Xiao},
  \citenamefont {Chang},\ and\ \citenamefont {Niu}}]{xiao2010berry}%
  \BibitemOpen
  \bibfield  {author} {\bibinfo {author} {\bibfnamefont {D.}~\bibnamefont
  {Xiao}}, \bibinfo {author} {\bibfnamefont {M.-C.}\ \bibnamefont {Chang}},\
  and\ \bibinfo {author} {\bibfnamefont {Q.}~\bibnamefont {Niu}},\ }\bibfield
  {title} {\bibinfo {title} {{Berry phase effects on electronic properties}},\
  }\href {https://link.aps.org/doi/10.1103/RevModPhys.82.1959} {\bibfield
  {journal} {\bibinfo  {journal} {Rev. Mod. Phys.}\ }\textbf {\bibinfo {volume}
  {82}},\ \bibinfo {pages} {1959} (\bibinfo {year} {2010})}\BibitemShut
  {NoStop}%
\bibitem [{\citenamefont {Bhalla}\ \emph {et~al.}(2020)\citenamefont {Bhalla},
  \citenamefont {MacDonald},\ and\ \citenamefont
  {Culcer}}]{bhalla2020resonant}%
  \BibitemOpen
  \bibfield  {author} {\bibinfo {author} {\bibfnamefont {P.}~\bibnamefont
  {Bhalla}}, \bibinfo {author} {\bibfnamefont {A.~H.}\ \bibnamefont
  {MacDonald}},\ and\ \bibinfo {author} {\bibfnamefont {D.}~\bibnamefont
  {Culcer}},\ }\bibfield  {title} {\bibinfo {title} {{Resonant Photovoltaic
  Effect in Doped Magnetic Semiconductors}},\ }\href
  {https://link.aps.org/doi/10.1103/PhysRevLett.124.087402} {\bibfield
  {journal} {\bibinfo  {journal} {Phys. Rev. Lett.}\ }\textbf {\bibinfo
  {volume} {124}},\ \bibinfo {pages} {087402} (\bibinfo {year}
  {2020})}\BibitemShut {NoStop}%
\bibitem [{\citenamefont {Bhalla}\ \emph {et~al.}(2022)\citenamefont {Bhalla},
  \citenamefont {Das}, \citenamefont {Culcer},\ and\ \citenamefont
  {Agarwal}}]{bhalla2022resonant}%
  \BibitemOpen
  \bibfield  {author} {\bibinfo {author} {\bibfnamefont {P.}~\bibnamefont
  {Bhalla}}, \bibinfo {author} {\bibfnamefont {K.}~\bibnamefont {Das}},
  \bibinfo {author} {\bibfnamefont {D.}~\bibnamefont {Culcer}},\ and\ \bibinfo
  {author} {\bibfnamefont {A.}~\bibnamefont {Agarwal}},\ }\bibfield  {title}
  {\bibinfo {title} {{Resonant Second-Harmonic Generation as a Probe of Quantum
  Geometry}},\ }\href {https://link.aps.org/doi/10.1103/PhysRevLett.129.227401}
  {\bibfield  {journal} {\bibinfo  {journal} {Phys. Rev. Lett.}\ }\textbf
  {\bibinfo {volume} {129}},\ \bibinfo {pages} {227401} (\bibinfo {year}
  {2022})}\BibitemShut {NoStop}%
\bibitem [{\citenamefont {Cullen}\ \emph {et~al.}(2021)\citenamefont {Cullen},
  \citenamefont {Bhalla}, \citenamefont {Marcellina}, \citenamefont
  {Hamilton},\ and\ \citenamefont {Culcer}}]{cullen2021generating}%
  \BibitemOpen
  \bibfield  {author} {\bibinfo {author} {\bibfnamefont {J.~H.}\ \bibnamefont
  {Cullen}}, \bibinfo {author} {\bibfnamefont {P.}~\bibnamefont {Bhalla}},
  \bibinfo {author} {\bibfnamefont {E.}~\bibnamefont {Marcellina}}, \bibinfo
  {author} {\bibfnamefont {A.~R.}\ \bibnamefont {Hamilton}},\ and\ \bibinfo
  {author} {\bibfnamefont {D.}~\bibnamefont {Culcer}},\ }\bibfield  {title}
  {\bibinfo {title} {{Generating a Topological Anomalous Hall Effect in a
  Nonmagnetic Conductor: An In-Plane Magnetic Field as a Direct Probe of the
  Berry Curvature}},\ }\href
  {https://link.aps.org/doi/10.1103/PhysRevLett.126.256601} {\bibfield
  {journal} {\bibinfo  {journal} {Phys. Rev. Lett.}\ }\textbf {\bibinfo
  {volume} {126}},\ \bibinfo {pages} {256601} (\bibinfo {year}
  {2021})}\BibitemShut {NoStop}%
\bibitem [{\citenamefont {Bhalla}\ \emph {et~al.}(2023)\citenamefont {Bhalla},
  \citenamefont {Das}, \citenamefont {Agarwal},\ and\ \citenamefont
  {Culcer}}]{bhalla2023quantum}%
  \BibitemOpen
  \bibfield  {author} {\bibinfo {author} {\bibfnamefont {P.}~\bibnamefont
  {Bhalla}}, \bibinfo {author} {\bibfnamefont {K.}~\bibnamefont {Das}},
  \bibinfo {author} {\bibfnamefont {A.}~\bibnamefont {Agarwal}},\ and\ \bibinfo
  {author} {\bibfnamefont {D.}~\bibnamefont {Culcer}},\ }\bibfield  {title}
  {\bibinfo {title} {{Quantum kinetic theory of nonlinear optical currents:
  Finite Fermi surface and Fermi sea contributions}},\ }\href
  {https://link.aps.org/doi/10.1103/PhysRevB.107.165131} {\bibfield  {journal}
  {\bibinfo  {journal} {Phys. Rev. B}\ }\textbf {\bibinfo {volume} {107}},\
  \bibinfo {pages} {165131} (\bibinfo {year} {2023})}\BibitemShut {NoStop}%
\bibitem [{\citenamefont {Atencia}\ \emph {et~al.}(2023)\citenamefont
  {Atencia}, \citenamefont {Xiao},\ and\ \citenamefont
  {Culcer}}]{atencia2023disorder}%
  \BibitemOpen
  \bibfield  {author} {\bibinfo {author} {\bibfnamefont {R.~B.}\ \bibnamefont
  {Atencia}}, \bibinfo {author} {\bibfnamefont {D.}~\bibnamefont {Xiao}},\ and\
  \bibinfo {author} {\bibfnamefont {D.}~\bibnamefont {Culcer}},\ }\bibfield
  {title} {\bibinfo {title} {{Disorder in the nonlinear anomalous Hall effect
  of $\mathcal{PT}$-symmetric Dirac fermions}},\ }\href
  {https://link.aps.org/doi/10.1103/PhysRevB.108.L201115} {\bibfield  {journal}
  {\bibinfo  {journal} {Phys. Rev. B}\ }\textbf {\bibinfo {volume} {108}},\
  \bibinfo {pages} {L201115} (\bibinfo {year} {2023})}\BibitemShut {NoStop}%
\bibitem [{\citenamefont {Varshney}\ \emph {et~al.}(2023)\citenamefont
  {Varshney}, \citenamefont {Das}, \citenamefont {Bhalla},\ and\ \citenamefont
  {Agarwal}}]{varshney2023quantum}%
  \BibitemOpen
  \bibfield  {author} {\bibinfo {author} {\bibfnamefont {H.}~\bibnamefont
  {Varshney}}, \bibinfo {author} {\bibfnamefont {K.}~\bibnamefont {Das}},
  \bibinfo {author} {\bibfnamefont {P.}~\bibnamefont {Bhalla}},\ and\ \bibinfo
  {author} {\bibfnamefont {A.}~\bibnamefont {Agarwal}},\ }\bibfield  {title}
  {\bibinfo {title} {{Quantum kinetic theory of nonlinear thermal current}},\
  }\href {https://link.aps.org/doi/10.1103/PhysRevB.107.235419} {\bibfield
  {journal} {\bibinfo  {journal} {Phys. Rev. B}\ }\textbf {\bibinfo {volume}
  {107}},\ \bibinfo {pages} {235419} (\bibinfo {year} {2023})}\BibitemShut
  {NoStop}%
\bibitem [{\citenamefont {Ashcroft}\ and\ \citenamefont
  {Mermin}(1976)}]{ashcroft1976solid}%
  \BibitemOpen
  \bibfield  {author} {\bibinfo {author} {\bibfnamefont {N.~W.}\ \bibnamefont
  {Ashcroft}}\ and\ \bibinfo {author} {\bibfnamefont {N.~D.}\ \bibnamefont
  {Mermin}},\ }\href@noop {} {\emph {\bibinfo {title} {{Solid State
  Physics}}}}\ (\bibinfo  {publisher} {Holt-Saunders, New York},\ \bibinfo
  {year} {1976})\BibitemShut {NoStop}%
\bibitem [{\citenamefont {Adams}\ and\ \citenamefont
  {Blount}(1959)}]{adams1959energy}%
  \BibitemOpen
  \bibfield  {author} {\bibinfo {author} {\bibfnamefont {E.}~\bibnamefont
  {Adams}}\ and\ \bibinfo {author} {\bibfnamefont {E.}~\bibnamefont {Blount}},\
  }\bibfield  {title} {\bibinfo {title} {{Energy bands in the presence of an
  external force field—II: Anomalous velocities}},\ }\href
  {https://doi.org/10.1016/0022-3697(59)90004-6} {\bibfield  {journal}
  {\bibinfo  {journal} {J. Phys. Chem. Solids}\ }\textbf {\bibinfo {volume}
  {10}},\ \bibinfo {pages} {286} (\bibinfo {year} {1959})}\BibitemShut
  {NoStop}%
\bibitem [{\citenamefont {Chang}\ and\ \citenamefont
  {Niu}(1995)}]{chang1995berry}%
  \BibitemOpen
  \bibfield  {author} {\bibinfo {author} {\bibfnamefont {M.-C.}\ \bibnamefont
  {Chang}}\ and\ \bibinfo {author} {\bibfnamefont {Q.}~\bibnamefont {Niu}},\
  }\bibfield  {title} {\bibinfo {title} {{Berry Phase, Hyperorbits, and the
  Hofstadter Spectrum}},\ }\href
  {https://link.aps.org/doi/10.1103/PhysRevLett.75.1348} {\bibfield  {journal}
  {\bibinfo  {journal} {Phys. Rev. Lett.}\ }\textbf {\bibinfo {volume} {75}},\
  \bibinfo {pages} {1348} (\bibinfo {year} {1995})}\BibitemShut {NoStop}%
\bibitem [{\citenamefont {Chang}\ and\ \citenamefont
  {Niu}(1996)}]{chang1996berry}%
  \BibitemOpen
  \bibfield  {author} {\bibinfo {author} {\bibfnamefont {M.-C.}\ \bibnamefont
  {Chang}}\ and\ \bibinfo {author} {\bibfnamefont {Q.}~\bibnamefont {Niu}},\
  }\bibfield  {title} {\bibinfo {title} {{Berry phase, hyperorbits, and the
  Hofstadter spectrum: Semiclassical dynamics in magnetic Bloch bands}},\
  }\href {https://link.aps.org/doi/10.1103/PhysRevB.53.7010} {\bibfield
  {journal} {\bibinfo  {journal} {Phys. Rev. B}\ }\textbf {\bibinfo {volume}
  {53}},\ \bibinfo {pages} {7010} (\bibinfo {year} {1996})}\BibitemShut
  {NoStop}%
\bibitem [{\citenamefont {Sundaram}\ and\ \citenamefont
  {Niu}(1999)}]{sundaram1999wave}%
  \BibitemOpen
  \bibfield  {author} {\bibinfo {author} {\bibfnamefont {G.}~\bibnamefont
  {Sundaram}}\ and\ \bibinfo {author} {\bibfnamefont {Q.}~\bibnamefont {Niu}},\
  }\bibfield  {title} {\bibinfo {title} {{Wave-packet dynamics in slowly
  perturbed crystals: Gradient corrections and Berry-phase effects}},\ }\href
  {https://link.aps.org/doi/10.1103/PhysRevB.59.14915} {\bibfield  {journal}
  {\bibinfo  {journal} {Phys. Rev. B}\ }\textbf {\bibinfo {volume} {59}},\
  \bibinfo {pages} {14915} (\bibinfo {year} {1999})}\BibitemShut {NoStop}%
\bibitem [{\citenamefont {Sinitsyn}\ \emph {et~al.}(2006)\citenamefont
  {Sinitsyn}, \citenamefont {Niu},\ and\ \citenamefont
  {MacDonald}}]{sinitsyn2006coordinate}%
  \BibitemOpen
  \bibfield  {author} {\bibinfo {author} {\bibfnamefont {N.~A.}\ \bibnamefont
  {Sinitsyn}}, \bibinfo {author} {\bibfnamefont {Q.}~\bibnamefont {Niu}},\ and\
  \bibinfo {author} {\bibfnamefont {A.~H.}\ \bibnamefont {MacDonald}},\
  }\bibfield  {title} {\bibinfo {title} {{Coordinate shift in the semiclassical
  Boltzmann equation and the anomalous Hall effect}},\ }\href
  {https://link.aps.org/doi/10.1103/PhysRevB.73.075318} {\bibfield  {journal}
  {\bibinfo  {journal} {Phys. Rev. B}\ }\textbf {\bibinfo {volume} {73}},\
  \bibinfo {pages} {075318} (\bibinfo {year} {2006})}\BibitemShut {NoStop}%
\bibitem [{\citenamefont {Smit}(1958)}]{smit1958spontaneous}%
  \BibitemOpen
  \bibfield  {author} {\bibinfo {author} {\bibfnamefont {J.}~\bibnamefont
  {Smit}},\ }\bibfield  {title} {\bibinfo {title} {{The spontaneous Hall effect
  in ferromagnetics II}},\ }\href
  {https://doi.org/10.1016/S0031-8914(58)93541-9} {\bibfield  {journal}
  {\bibinfo  {journal} {Physica}\ }\textbf {\bibinfo {volume} {24}},\ \bibinfo
  {pages} {39} (\bibinfo {year} {1958})}\BibitemShut {NoStop}%
\bibitem [{\citenamefont {Berger}(1970)}]{berger1970side}%
  \BibitemOpen
  \bibfield  {author} {\bibinfo {author} {\bibfnamefont {L.}~\bibnamefont
  {Berger}},\ }\bibfield  {title} {\bibinfo {title} {{Side-Jump Mechanism for
  the Hall Effect of Ferromagnets}},\ }\href
  {https://link.aps.org/doi/10.1103/PhysRevB.2.4559} {\bibfield  {journal}
  {\bibinfo  {journal} {Phys. Rev. B}\ }\textbf {\bibinfo {volume} {2}},\
  \bibinfo {pages} {4559} (\bibinfo {year} {1970})}\BibitemShut {NoStop}%
\bibitem [{\citenamefont {Nozi{\`e}res}\ and\ \citenamefont
  {Lewiner}(1973)}]{nozieres1973simple}%
  \BibitemOpen
  \bibfield  {author} {\bibinfo {author} {\bibfnamefont {P.}~\bibnamefont
  {Nozi{\`e}res}}\ and\ \bibinfo {author} {\bibfnamefont {C.}~\bibnamefont
  {Lewiner}},\ }\bibfield  {title} {\bibinfo {title} {{A simple theory of the
  anomalous Hall effect in semiconductors}},\ }\href
  {https://doi.org/10.1051/jphys:019730034010090100} {\bibfield  {journal}
  {\bibinfo  {journal} {J. Phys. (Paris)}\ }\textbf {\bibinfo {volume} {34}},\
  \bibinfo {pages} {901} (\bibinfo {year} {1973})}\BibitemShut {NoStop}%
\bibitem [{\citenamefont {Wang}\ \emph {et~al.}(2006)\citenamefont {Wang},
  \citenamefont {Yates}, \citenamefont {Souza},\ and\ \citenamefont
  {Vanderbilt}}]{wang2006abinitio}%
  \BibitemOpen
  \bibfield  {author} {\bibinfo {author} {\bibfnamefont {X.}~\bibnamefont
  {Wang}}, \bibinfo {author} {\bibfnamefont {J.~R.}\ \bibnamefont {Yates}},
  \bibinfo {author} {\bibfnamefont {I.}~\bibnamefont {Souza}},\ and\ \bibinfo
  {author} {\bibfnamefont {D.}~\bibnamefont {Vanderbilt}},\ }\bibfield  {title}
  {\bibinfo {title} {{Ab initio calculation of the anomalous Hall conductivity
  by Wannier interpolation}},\ }\href
  {https://link.aps.org/doi/10.1103/PhysRevB.74.195118} {\bibfield  {journal}
  {\bibinfo  {journal} {Phys. Rev. B}\ }\textbf {\bibinfo {volume} {74}},\
  \bibinfo {pages} {195118} (\bibinfo {year} {2006})}\BibitemShut {NoStop}%
\bibitem [{\citenamefont {Wang}\ \emph {et~al.}(2007)\citenamefont {Wang},
  \citenamefont {Vanderbilt}, \citenamefont {Yates},\ and\ \citenamefont
  {Souza}}]{wang2007fermi}%
  \BibitemOpen
  \bibfield  {author} {\bibinfo {author} {\bibfnamefont {X.}~\bibnamefont
  {Wang}}, \bibinfo {author} {\bibfnamefont {D.}~\bibnamefont {Vanderbilt}},
  \bibinfo {author} {\bibfnamefont {J.~R.}\ \bibnamefont {Yates}},\ and\
  \bibinfo {author} {\bibfnamefont {I.}~\bibnamefont {Souza}},\ }\bibfield
  {title} {\bibinfo {title} {{Fermi-surface calculation of the anomalous Hall
  conductivity}},\ }\href {https://link.aps.org/doi/10.1103/PhysRevB.76.195109}
  {\bibfield  {journal} {\bibinfo  {journal} {Phys. Rev. B}\ }\textbf {\bibinfo
  {volume} {76}},\ \bibinfo {pages} {195109} (\bibinfo {year}
  {2007})}\BibitemShut {NoStop}%
\bibitem [{\citenamefont {Gradhand}\ \emph {et~al.}(2012)\citenamefont
  {Gradhand}, \citenamefont {Fedorov}, \citenamefont {Pientka}, \citenamefont
  {Zahn}, \citenamefont {Mertig},\ and\ \citenamefont
  {Gy{\"o}rffy}}]{gradhand2012first}%
  \BibitemOpen
  \bibfield  {author} {\bibinfo {author} {\bibfnamefont {M.}~\bibnamefont
  {Gradhand}}, \bibinfo {author} {\bibfnamefont {D.}~\bibnamefont {Fedorov}},
  \bibinfo {author} {\bibfnamefont {F.}~\bibnamefont {Pientka}}, \bibinfo
  {author} {\bibfnamefont {P.}~\bibnamefont {Zahn}}, \bibinfo {author}
  {\bibfnamefont {I.}~\bibnamefont {Mertig}},\ and\ \bibinfo {author}
  {\bibfnamefont {B.}~\bibnamefont {Gy{\"o}rffy}},\ }\bibfield  {title}
  {\bibinfo {title} {{First-principle calculations of the Berry curvature of
  Bloch states for charge and spin transport of electrons}},\ }\href
  {https://iopscience.iop.org/article/10.1088/0953-8984/24/21/213202}
  {\bibfield  {journal} {\bibinfo  {journal} {J. Phys. Condens. Matter}\
  }\textbf {\bibinfo {volume} {24}},\ \bibinfo {pages} {213202} (\bibinfo
  {year} {2012})}\BibitemShut {NoStop}%
\bibitem [{\citenamefont {He}\ \emph {et~al.}(2012)\citenamefont {He},
  \citenamefont {Moore},\ and\ \citenamefont {Varma}}]{he2012berry}%
  \BibitemOpen
  \bibfield  {author} {\bibinfo {author} {\bibfnamefont {Y.}~\bibnamefont
  {He}}, \bibinfo {author} {\bibfnamefont {J.}~\bibnamefont {Moore}},\ and\
  \bibinfo {author} {\bibfnamefont {C.~M.}\ \bibnamefont {Varma}},\ }\bibfield
  {title} {\bibinfo {title} {{Berry phase and anomalous Hall effect in a
  three-orbital tight-binding Hamiltonian}},\ }\href
  {https://link.aps.org/doi/10.1103/PhysRevB.85.155106} {\bibfield  {journal}
  {\bibinfo  {journal} {Phys. Rev. B}\ }\textbf {\bibinfo {volume} {85}},\
  \bibinfo {pages} {155106} (\bibinfo {year} {2012})}\BibitemShut {NoStop}%
\bibitem [{\citenamefont {Chen}\ \emph {et~al.}(2013)\citenamefont {Chen},
  \citenamefont {Bergman},\ and\ \citenamefont {Burkov}}]{chen2013weyl}%
  \BibitemOpen
  \bibfield  {author} {\bibinfo {author} {\bibfnamefont {Y.}~\bibnamefont
  {Chen}}, \bibinfo {author} {\bibfnamefont {D.~L.}\ \bibnamefont {Bergman}},\
  and\ \bibinfo {author} {\bibfnamefont {A.~A.}\ \bibnamefont {Burkov}},\
  }\bibfield  {title} {\bibinfo {title} {{Weyl fermions and the anomalous Hall
  effect in metallic ferromagnets}},\ }\href
  {https://link.aps.org/doi/10.1103/PhysRevB.88.125110} {\bibfield  {journal}
  {\bibinfo  {journal} {Phys. Rev. B}\ }\textbf {\bibinfo {volume} {88}},\
  \bibinfo {pages} {125110} (\bibinfo {year} {2013})}\BibitemShut {NoStop}%
\bibitem [{\citenamefont {Bianco}\ \emph {et~al.}(2014)\citenamefont {Bianco},
  \citenamefont {Resta},\ and\ \citenamefont {Souza}}]{bianco2014how}%
  \BibitemOpen
  \bibfield  {author} {\bibinfo {author} {\bibfnamefont {R.}~\bibnamefont
  {Bianco}}, \bibinfo {author} {\bibfnamefont {R.}~\bibnamefont {Resta}},\ and\
  \bibinfo {author} {\bibfnamefont {I.}~\bibnamefont {Souza}},\ }\bibfield
  {title} {\bibinfo {title} {{How disorder affects the Berry-phase anomalous
  Hall conductivity: A reciprocal-space analysis}},\ }\href
  {https://link.aps.org/doi/10.1103/PhysRevB.90.125153} {\bibfield  {journal}
  {\bibinfo  {journal} {Phys. Rev. B}\ }\textbf {\bibinfo {volume} {90}},\
  \bibinfo {pages} {125153} (\bibinfo {year} {2014})}\BibitemShut {NoStop}%
\bibitem [{\citenamefont {Chen}\ \emph {et~al.}(2014)\citenamefont {Chen},
  \citenamefont {Niu},\ and\ \citenamefont {MacDonald}}]{chen2014anomalous}%
  \BibitemOpen
  \bibfield  {author} {\bibinfo {author} {\bibfnamefont {H.}~\bibnamefont
  {Chen}}, \bibinfo {author} {\bibfnamefont {Q.}~\bibnamefont {Niu}},\ and\
  \bibinfo {author} {\bibfnamefont {A.~H.}\ \bibnamefont {MacDonald}},\
  }\bibfield  {title} {\bibinfo {title} {{Anomalous Hall Effect Arising from
  Noncollinear Antiferromagnetism}},\ }\href
  {https://link.aps.org/doi/10.1103/PhysRevLett.112.017205} {\bibfield
  {journal} {\bibinfo  {journal} {Phys. Rev. Lett.}\ }\textbf {\bibinfo
  {volume} {112}},\ \bibinfo {pages} {017205} (\bibinfo {year}
  {2014})}\BibitemShut {NoStop}%
\bibitem [{\citenamefont {Olsen}\ and\ \citenamefont
  {Souza}(2015)}]{olsen2015valley}%
  \BibitemOpen
  \bibfield  {author} {\bibinfo {author} {\bibfnamefont {T.}~\bibnamefont
  {Olsen}}\ and\ \bibinfo {author} {\bibfnamefont {I.}~\bibnamefont {Souza}},\
  }\bibfield  {title} {\bibinfo {title} {{Valley Hall effect in disordered
  monolayer ${\mathrm{MoS}}_{2}$ from first principles}},\ }\href
  {https://link.aps.org/doi/10.1103/PhysRevB.92.125146} {\bibfield  {journal}
  {\bibinfo  {journal} {Phys. Rev. B}\ }\textbf {\bibinfo {volume} {92}},\
  \bibinfo {pages} {125146} (\bibinfo {year} {2015})}\BibitemShut {NoStop}%
\bibitem [{\citenamefont {Feng}\ \emph {et~al.}(2016)\citenamefont {Feng},
  \citenamefont {Liu}, \citenamefont {Liu}, \citenamefont {Zhou},\ and\
  \citenamefont {Yao}}]{feng2016first}%
  \BibitemOpen
  \bibfield  {author} {\bibinfo {author} {\bibfnamefont {W.}~\bibnamefont
  {Feng}}, \bibinfo {author} {\bibfnamefont {C.-C.}\ \bibnamefont {Liu}},
  \bibinfo {author} {\bibfnamefont {G.-B.}\ \bibnamefont {Liu}}, \bibinfo
  {author} {\bibfnamefont {J.-J.}\ \bibnamefont {Zhou}},\ and\ \bibinfo
  {author} {\bibfnamefont {Y.}~\bibnamefont {Yao}},\ }\bibfield  {title}
  {\bibinfo {title} {{First-principles investigations on the Berry phase effect
  in spin--orbit coupling materials}},\ }\href
  {https://doi.org/10.1016/j.commatsci.2015.09.020} {\bibfield  {journal}
  {\bibinfo  {journal} {Comput. Mater. Sci.}\ }\textbf {\bibinfo {volume}
  {112}},\ \bibinfo {pages} {428} (\bibinfo {year} {2016})}\BibitemShut
  {NoStop}%
\bibitem [{\citenamefont {Dai}\ \emph {et~al.}(2017)\citenamefont {Dai},
  \citenamefont {Du},\ and\ \citenamefont {Lu}}]{dai2017negative}%
  \BibitemOpen
  \bibfield  {author} {\bibinfo {author} {\bibfnamefont {X.}~\bibnamefont
  {Dai}}, \bibinfo {author} {\bibfnamefont {Z.~Z.}\ \bibnamefont {Du}},\ and\
  \bibinfo {author} {\bibfnamefont {H.-Z.}\ \bibnamefont {Lu}},\ }\bibfield
  {title} {\bibinfo {title} {{Negative Magnetoresistance without Chiral Anomaly
  in Topological Insulators}},\ }\href
  {https://link.aps.org/doi/10.1103/PhysRevLett.119.166601} {\bibfield
  {journal} {\bibinfo  {journal} {Phys. Rev. Lett.}\ }\textbf {\bibinfo
  {volume} {119}},\ \bibinfo {pages} {166601} (\bibinfo {year}
  {2017})}\BibitemShut {NoStop}%
\bibitem [{\citenamefont {Martiny}\ \emph {et~al.}(2019)\citenamefont
  {Martiny}, \citenamefont {Kaasbjerg},\ and\ \citenamefont
  {Jauho}}]{martiny2019tunable}%
  \BibitemOpen
  \bibfield  {author} {\bibinfo {author} {\bibfnamefont {J.~H.~J.}\
  \bibnamefont {Martiny}}, \bibinfo {author} {\bibfnamefont {K.}~\bibnamefont
  {Kaasbjerg}},\ and\ \bibinfo {author} {\bibfnamefont {A.-P.}\ \bibnamefont
  {Jauho}},\ }\bibfield  {title} {\bibinfo {title} {{Tunable valley Hall effect
  in gate-defined graphene superlattices}},\ }\href
  {https://link.aps.org/doi/10.1103/PhysRevB.100.155414} {\bibfield  {journal}
  {\bibinfo  {journal} {Phys. Rev. B}\ }\textbf {\bibinfo {volume} {100}},\
  \bibinfo {pages} {155414} (\bibinfo {year} {2019})}\BibitemShut {NoStop}%
\bibitem [{\citenamefont {Wuttke}\ \emph {et~al.}(2019)\citenamefont {Wuttke},
  \citenamefont {Caglieris}, \citenamefont {Sykora}, \citenamefont
  {Scaravaggi}, \citenamefont {Wolter}, \citenamefont {Manna}, \citenamefont
  {S\"uss}, \citenamefont {Shekhar}, \citenamefont {Felser}, \citenamefont
  {B\"uchner},\ and\ \citenamefont {Hess}}]{wuttke2019berry}%
  \BibitemOpen
  \bibfield  {author} {\bibinfo {author} {\bibfnamefont {C.}~\bibnamefont
  {Wuttke}}, \bibinfo {author} {\bibfnamefont {F.}~\bibnamefont {Caglieris}},
  \bibinfo {author} {\bibfnamefont {S.}~\bibnamefont {Sykora}}, \bibinfo
  {author} {\bibfnamefont {F.}~\bibnamefont {Scaravaggi}}, \bibinfo {author}
  {\bibfnamefont {A.~U.~B.}\ \bibnamefont {Wolter}}, \bibinfo {author}
  {\bibfnamefont {K.}~\bibnamefont {Manna}}, \bibinfo {author} {\bibfnamefont
  {V.}~\bibnamefont {S\"uss}}, \bibinfo {author} {\bibfnamefont
  {C.}~\bibnamefont {Shekhar}}, \bibinfo {author} {\bibfnamefont
  {C.}~\bibnamefont {Felser}}, \bibinfo {author} {\bibfnamefont
  {B.}~\bibnamefont {B\"uchner}},\ and\ \bibinfo {author} {\bibfnamefont
  {C.}~\bibnamefont {Hess}},\ }\bibfield  {title} {\bibinfo {title} {{Berry
  curvature unravelled by the anomalous Nernst effect in
  ${\mathrm{Mn}}_{3}\mathrm{Ge}$}},\ }\href
  {https://link.aps.org/doi/10.1103/PhysRevB.100.085111} {\bibfield  {journal}
  {\bibinfo  {journal} {Phys. Rev. B}\ }\textbf {\bibinfo {volume} {100}},\
  \bibinfo {pages} {085111} (\bibinfo {year} {2019})}\BibitemShut {NoStop}%
\bibitem [{\citenamefont {Du}\ \emph {et~al.}(2020)\citenamefont {Du},
  \citenamefont {Tang}, \citenamefont {Li}, \citenamefont {Lin}, \citenamefont
  {Xu}, \citenamefont {Duan},\ and\ \citenamefont {Rubio}}]{du2020berry}%
  \BibitemOpen
  \bibfield  {author} {\bibinfo {author} {\bibfnamefont {S.}~\bibnamefont
  {Du}}, \bibinfo {author} {\bibfnamefont {P.}~\bibnamefont {Tang}}, \bibinfo
  {author} {\bibfnamefont {J.}~\bibnamefont {Li}}, \bibinfo {author}
  {\bibfnamefont {Z.}~\bibnamefont {Lin}}, \bibinfo {author} {\bibfnamefont
  {Y.}~\bibnamefont {Xu}}, \bibinfo {author} {\bibfnamefont {W.}~\bibnamefont
  {Duan}},\ and\ \bibinfo {author} {\bibfnamefont {A.}~\bibnamefont {Rubio}},\
  }\bibfield  {title} {\bibinfo {title} {{Berry curvature engineering by gating
  two-dimensional antiferromagnets}},\ }\href
  {https://link.aps.org/doi/10.1103/PhysRevResearch.2.022025} {\bibfield
  {journal} {\bibinfo  {journal} {Phys. Rev. Res.}\ }\textbf {\bibinfo {volume}
  {2}},\ \bibinfo {pages} {022025} (\bibinfo {year} {2020})}\BibitemShut
  {NoStop}%
\bibitem [{\citenamefont {He}\ \emph {et~al.}(2020)\citenamefont {He},
  \citenamefont {Goldhaber-Gordon},\ and\ \citenamefont {Law}}]{he2020giant}%
  \BibitemOpen
  \bibfield  {author} {\bibinfo {author} {\bibfnamefont {W.-Y.}\ \bibnamefont
  {He}}, \bibinfo {author} {\bibfnamefont {D.}~\bibnamefont
  {Goldhaber-Gordon}},\ and\ \bibinfo {author} {\bibfnamefont {K.~T.}\
  \bibnamefont {Law}},\ }\bibfield  {title} {\bibinfo {title} {{Giant orbital
  magnetoelectric effect and current-induced magnetization switching in twisted
  bilayer graphene}},\ }\href {https://doi.org/10.1038/s41467-020-15473-9}
  {\bibfield  {journal} {\bibinfo  {journal} {Nat. Commun.}\ }\textbf {\bibinfo
  {volume} {11}},\ \bibinfo {pages} {1650} (\bibinfo {year}
  {2020})}\BibitemShut {NoStop}%
\bibitem [{\citenamefont {He}\ and\ \citenamefont
  {Law}(2021)}]{he2021superconducting}%
  \BibitemOpen
  \bibfield  {author} {\bibinfo {author} {\bibfnamefont {W.-Y.}\ \bibnamefont
  {He}}\ and\ \bibinfo {author} {\bibfnamefont {K.~T.}\ \bibnamefont {Law}},\
  }\bibfield  {title} {\bibinfo {title} {{Superconducting orbital
  magnetoelectric effect and its evolution across the superconductor-normal
  metal phase transition}},\ }\href
  {https://link.aps.org/doi/10.1103/PhysRevResearch.3.L032012} {\bibfield
  {journal} {\bibinfo  {journal} {Phys. Rev. Res.}\ }\textbf {\bibinfo {volume}
  {3}},\ \bibinfo {pages} {L032012} (\bibinfo {year} {2021})}\BibitemShut
  {NoStop}%
\bibitem [{\citenamefont {Du}\ \emph {et~al.}(2021{\natexlab{a}})\citenamefont
  {Du}, \citenamefont {Lu},\ and\ \citenamefont {Xie}}]{du2021nonlinear}%
  \BibitemOpen
  \bibfield  {author} {\bibinfo {author} {\bibfnamefont {Z.}~\bibnamefont
  {Du}}, \bibinfo {author} {\bibfnamefont {H.-Z.}\ \bibnamefont {Lu}},\ and\
  \bibinfo {author} {\bibfnamefont {X.}~\bibnamefont {Xie}},\ }\bibfield
  {title} {\bibinfo {title} {{Nonlinear Hall effects}},\ }\href
  {https://doi.org/10.1038/s42254-021-00359-6} {\bibfield  {journal} {\bibinfo
  {journal} {Nat. Rev. Phys.}\ }\textbf {\bibinfo {volume} {3}},\ \bibinfo
  {pages} {744} (\bibinfo {year} {2021}{\natexlab{a}})}\BibitemShut {NoStop}%
\bibitem [{\citenamefont {Ideue}\ and\ \citenamefont
  {Iwasa}(2021)}]{ideue2021symmetry}%
  \BibitemOpen
  \bibfield  {author} {\bibinfo {author} {\bibfnamefont {T.}~\bibnamefont
  {Ideue}}\ and\ \bibinfo {author} {\bibfnamefont {Y.}~\bibnamefont {Iwasa}},\
  }\bibfield  {title} {\bibinfo {title} {{Symmetry breaking and nonlinear
  electric transport in van der Waals nanostructures}},\ }\href
  {https://doi.org/10.1146/annurev-conmatphys-060220-100347} {\bibfield
  {journal} {\bibinfo  {journal} {Annu. Rev. Condens. Matter Phys.}\ }\textbf
  {\bibinfo {volume} {12}},\ \bibinfo {pages} {201} (\bibinfo {year}
  {2021})}\BibitemShut {NoStop}%
\bibitem [{\citenamefont {Ortix}(2021)}]{ortix2021nonlinear}%
  \BibitemOpen
  \bibfield  {author} {\bibinfo {author} {\bibfnamefont {C.}~\bibnamefont
  {Ortix}},\ }\bibfield  {title} {\bibinfo {title} {{Nonlinear Hall Effect with
  Time-Reversal Symmetry: Theory and Material Realizations}},\ }\href
  {https://doi.org/10.1002/qute.202100056} {\bibfield  {journal} {\bibinfo
  {journal} {Adv. Quantum Technol.}\ }\textbf {\bibinfo {volume} {4}},\
  \bibinfo {pages} {2100056} (\bibinfo {year} {2021})}\BibitemShut {NoStop}%
\bibitem [{\citenamefont {Nagaosa}\ and\ \citenamefont
  {Yanase}(2024)}]{nagaosa2024nonreciprocal}%
  \BibitemOpen
  \bibfield  {author} {\bibinfo {author} {\bibfnamefont {N.}~\bibnamefont
  {Nagaosa}}\ and\ \bibinfo {author} {\bibfnamefont {Y.}~\bibnamefont
  {Yanase}},\ }\bibfield  {title} {\bibinfo {title} {{Nonreciprocal transport
  and optical phenomena in quantum materials}},\ }\href
  {https://doi.org/10.1146/annurev-conmatphys-032822-033734} {\bibfield
  {journal} {\bibinfo  {journal} {Annu. Rev. Condens. Matter Phys.}\ }\textbf
  {\bibinfo {volume} {15}},\ \bibinfo {pages} {63} (\bibinfo {year}
  {2024})}\BibitemShut {NoStop}%
\bibitem [{\citenamefont {Shim}\ \emph {et~al.}()\citenamefont {Shim},
  \citenamefont {Mehraeen}, \citenamefont {Sklenar}, \citenamefont {Zhang},
  \citenamefont {Hoffmann},\ and\ \citenamefont {Mason}}]{shim2024spin}%
  \BibitemOpen
  \bibfield  {author} {\bibinfo {author} {\bibfnamefont {S.}~\bibnamefont
  {Shim}}, \bibinfo {author} {\bibfnamefont {M.}~\bibnamefont {Mehraeen}},
  \bibinfo {author} {\bibfnamefont {J.}~\bibnamefont {Sklenar}}, \bibinfo
  {author} {\bibfnamefont {S.~S.-L.}\ \bibnamefont {Zhang}}, \bibinfo {author}
  {\bibfnamefont {A.}~\bibnamefont {Hoffmann}},\ and\ \bibinfo {author}
  {\bibfnamefont {N.}~\bibnamefont {Mason}},\ }\bibfield  {title} {\bibinfo
  {title} {{Spin-polarized antiferromagnetic metals}},\ }\href
  {https://doi.org/10.1146/annurev-conmatphys-042924-123620} {\bibfield
  {journal} {\bibinfo  {journal} {Annu. Rev. Condens. Matter Phys.}\ }\textbf
  {\bibinfo {volume} {16}}}\BibitemShut {NoStop}%
\bibitem [{\citenamefont {Avci}\ \emph
  {et~al.}(2015{\natexlab{a}})\citenamefont {Avci}, \citenamefont {Garello},
  \citenamefont {Mendil}, \citenamefont {Ghosh}, \citenamefont {Blasakis},
  \citenamefont {Gabureac}, \citenamefont {Trassin}, \citenamefont {Fiebig},\
  and\ \citenamefont {Gambardella}}]{avci2015magnetoresistance}%
  \BibitemOpen
  \bibfield  {author} {\bibinfo {author} {\bibfnamefont {C.~O.}\ \bibnamefont
  {Avci}}, \bibinfo {author} {\bibfnamefont {K.}~\bibnamefont {Garello}},
  \bibinfo {author} {\bibfnamefont {J.}~\bibnamefont {Mendil}}, \bibinfo
  {author} {\bibfnamefont {A.}~\bibnamefont {Ghosh}}, \bibinfo {author}
  {\bibfnamefont {N.}~\bibnamefont {Blasakis}}, \bibinfo {author}
  {\bibfnamefont {M.}~\bibnamefont {Gabureac}}, \bibinfo {author}
  {\bibfnamefont {M.}~\bibnamefont {Trassin}}, \bibinfo {author} {\bibfnamefont
  {M.}~\bibnamefont {Fiebig}},\ and\ \bibinfo {author} {\bibfnamefont
  {P.}~\bibnamefont {Gambardella}},\ }\bibfield  {title} {\bibinfo {title}
  {{Magnetoresistance of heavy and light metal/ferromagnet bilayers}},\ }\href
  {https://doi.org/10.1063/1.4935497} {\bibfield  {journal} {\bibinfo
  {journal} {Appl. Phys. Lett.}\ }\textbf {\bibinfo {volume} {107}} (\bibinfo
  {year} {2015}{\natexlab{a}})}\BibitemShut {NoStop}%
\bibitem [{\citenamefont {Avci}\ \emph
  {et~al.}(2015{\natexlab{b}})\citenamefont {Avci}, \citenamefont {Garello},
  \citenamefont {Ghosh}, \citenamefont {Gabureac}, \citenamefont {Alvarado},\
  and\ \citenamefont {Gambardella}}]{avci2015unidirectional}%
  \BibitemOpen
  \bibfield  {author} {\bibinfo {author} {\bibfnamefont {C.~O.}\ \bibnamefont
  {Avci}}, \bibinfo {author} {\bibfnamefont {K.}~\bibnamefont {Garello}},
  \bibinfo {author} {\bibfnamefont {A.}~\bibnamefont {Ghosh}}, \bibinfo
  {author} {\bibfnamefont {M.}~\bibnamefont {Gabureac}}, \bibinfo {author}
  {\bibfnamefont {S.~F.}\ \bibnamefont {Alvarado}},\ and\ \bibinfo {author}
  {\bibfnamefont {P.}~\bibnamefont {Gambardella}},\ }\bibfield  {title}
  {\bibinfo {title} {{Unidirectional spin Hall magnetoresistance in
  ferromagnet/normal metal bilayers}},\ }\href
  {https://doi.org/10.1038/nphys3356} {\bibfield  {journal} {\bibinfo
  {journal} {Nat. Phys.}\ }\textbf {\bibinfo {volume} {11}},\ \bibinfo {pages}
  {570} (\bibinfo {year} {2015}{\natexlab{b}})}\BibitemShut {NoStop}%
\bibitem [{\citenamefont {Olejn\'{\i}k}\ \emph {et~al.}(2015)\citenamefont
  {Olejn\'{\i}k}, \citenamefont {Nov\'ak}, \citenamefont {Wunderlich},\ and\
  \citenamefont {Jungwirth}}]{olejnik2015electrical}%
  \BibitemOpen
  \bibfield  {author} {\bibinfo {author} {\bibfnamefont {K.}~\bibnamefont
  {Olejn\'{\i}k}}, \bibinfo {author} {\bibfnamefont {V.}~\bibnamefont
  {Nov\'ak}}, \bibinfo {author} {\bibfnamefont {J.}~\bibnamefont
  {Wunderlich}},\ and\ \bibinfo {author} {\bibfnamefont {T.}~\bibnamefont
  {Jungwirth}},\ }\bibfield  {title} {\bibinfo {title} {Electrical detection of
  magnetization reversal without auxiliary magnets},\ }\href
  {https://link.aps.org/doi/10.1103/PhysRevB.91.180402} {\bibfield  {journal}
  {\bibinfo  {journal} {Phys. Rev. B}\ }\textbf {\bibinfo {volume} {91}},\
  \bibinfo {pages} {180402} (\bibinfo {year} {2015})}\BibitemShut {NoStop}%
\bibitem [{\citenamefont {Zhang}\ and\ \citenamefont
  {Vignale}(2016)}]{zhang2016theory}%
  \BibitemOpen
  \bibfield  {author} {\bibinfo {author} {\bibfnamefont {S.~S.-L.}\
  \bibnamefont {Zhang}}\ and\ \bibinfo {author} {\bibfnamefont
  {G.}~\bibnamefont {Vignale}},\ }\bibfield  {title} {\bibinfo {title} {{Theory
  of unidirectional spin Hall magnetoresistance in
  heavy-metal/ferromagnetic-metal bilayers}},\ }\href
  {https://link.aps.org/doi/10.1103/PhysRevB.94.140411} {\bibfield  {journal}
  {\bibinfo  {journal} {Phys. Rev. B}\ }\textbf {\bibinfo {volume} {94}},\
  \bibinfo {pages} {140411} (\bibinfo {year} {2016})}\BibitemShut {NoStop}%
\bibitem [{\citenamefont {Yasuda}\ \emph {et~al.}(2016)\citenamefont {Yasuda},
  \citenamefont {Tsukazaki}, \citenamefont {Yoshimi}, \citenamefont
  {Takahashi}, \citenamefont {Kawasaki},\ and\ \citenamefont
  {Tokura}}]{yasuda2016large}%
  \BibitemOpen
  \bibfield  {author} {\bibinfo {author} {\bibfnamefont {K.}~\bibnamefont
  {Yasuda}}, \bibinfo {author} {\bibfnamefont {A.}~\bibnamefont {Tsukazaki}},
  \bibinfo {author} {\bibfnamefont {R.}~\bibnamefont {Yoshimi}}, \bibinfo
  {author} {\bibfnamefont {K.~S.}\ \bibnamefont {Takahashi}}, \bibinfo {author}
  {\bibfnamefont {M.}~\bibnamefont {Kawasaki}},\ and\ \bibinfo {author}
  {\bibfnamefont {Y.}~\bibnamefont {Tokura}},\ }\bibfield  {title} {\bibinfo
  {title} {Large unidirectional magnetoresistance in a magnetic topological
  insulator},\ }\href {https://link.aps.org/doi/10.1103/PhysRevLett.117.127202}
  {\bibfield  {journal} {\bibinfo  {journal} {Phys. Rev. Lett.}\ }\textbf
  {\bibinfo {volume} {117}},\ \bibinfo {pages} {127202} (\bibinfo {year}
  {2016})}\BibitemShut {NoStop}%
\bibitem [{\citenamefont {Avci}\ \emph {et~al.}(2018)\citenamefont {Avci},
  \citenamefont {Mendil}, \citenamefont {Beach},\ and\ \citenamefont
  {Gambardella}}]{avci2018origins}%
  \BibitemOpen
  \bibfield  {author} {\bibinfo {author} {\bibfnamefont {C.~O.}\ \bibnamefont
  {Avci}}, \bibinfo {author} {\bibfnamefont {J.}~\bibnamefont {Mendil}},
  \bibinfo {author} {\bibfnamefont {G.~S.~D.}\ \bibnamefont {Beach}},\ and\
  \bibinfo {author} {\bibfnamefont {P.}~\bibnamefont {Gambardella}},\
  }\bibfield  {title} {\bibinfo {title} {{Origins of the Unidirectional Spin
  Hall Magnetoresistance in Metallic Bilayers}},\ }\href
  {https://link.aps.org/doi/10.1103/PhysRevLett.121.087207} {\bibfield
  {journal} {\bibinfo  {journal} {Phys. Rev. Lett.}\ }\textbf {\bibinfo
  {volume} {121}},\ \bibinfo {pages} {087207} (\bibinfo {year}
  {2018})}\BibitemShut {NoStop}%
\bibitem [{\citenamefont {Lv}\ \emph {et~al.}(2018)\citenamefont {Lv},
  \citenamefont {Kally}, \citenamefont {Zhang}, \citenamefont {Lee},
  \citenamefont {Jamali}, \citenamefont {Samarth},\ and\ \citenamefont
  {Wang}}]{lv2018unidirectional}%
  \BibitemOpen
  \bibfield  {author} {\bibinfo {author} {\bibfnamefont {Y.}~\bibnamefont
  {Lv}}, \bibinfo {author} {\bibfnamefont {J.}~\bibnamefont {Kally}}, \bibinfo
  {author} {\bibfnamefont {D.}~\bibnamefont {Zhang}}, \bibinfo {author}
  {\bibfnamefont {J.~S.}\ \bibnamefont {Lee}}, \bibinfo {author} {\bibfnamefont
  {M.}~\bibnamefont {Jamali}}, \bibinfo {author} {\bibfnamefont
  {N.}~\bibnamefont {Samarth}},\ and\ \bibinfo {author} {\bibfnamefont {J.-P.}\
  \bibnamefont {Wang}},\ }\bibfield  {title} {\bibinfo {title} {{Unidirectional
  spin-Hall and Rashba- Edelstein magnetoresistance in topological
  insulator-ferromagnet layer heterostructures}},\ }\href
  {https://doi.org/10.1038/s41467-017-02491-3} {\bibfield  {journal} {\bibinfo
  {journal} {Nat. Commun.}\ }\textbf {\bibinfo {volume} {9}},\ \bibinfo {pages}
  {1} (\bibinfo {year} {2018})}\BibitemShut {NoStop}%
\bibitem [{\citenamefont {Duy~Khang}\ and\ \citenamefont
  {Hai}(2019)}]{duy2019giant}%
  \BibitemOpen
  \bibfield  {author} {\bibinfo {author} {\bibfnamefont {N.~H.}\ \bibnamefont
  {Duy~Khang}}\ and\ \bibinfo {author} {\bibfnamefont {P.~N.}\ \bibnamefont
  {Hai}},\ }\bibfield  {title} {\bibinfo {title} {{Giant unidirectional spin
  Hall magnetoresistance in topological insulator--ferromagnetic semiconductor
  heterostructures}},\ }\href {https://doi.org/10.1063/1.5134728} {\bibfield
  {journal} {\bibinfo  {journal} {J. Appl. Phys.}\ }\textbf {\bibinfo {volume}
  {126}} (\bibinfo {year} {2019})}\BibitemShut {NoStop}%
\bibitem [{\citenamefont {{Guillet}}\ \emph {et~al.}(2020)\citenamefont
  {{Guillet}}, \citenamefont {{Zucchetti}}, \citenamefont {{Barbedienne}},
  \citenamefont {{Marty}}, \citenamefont {{Isella}}, \citenamefont {{Cagnon}},
  \citenamefont {{Vergnaud}}, \citenamefont {{Jaffr{\`e}s}}, \citenamefont
  {{Reyren}}, \citenamefont {{George}} \emph
  {et~al.}}]{guillet2020observation}%
  \BibitemOpen
  \bibfield  {author} {\bibinfo {author} {\bibfnamefont {T.}~\bibnamefont
  {{Guillet}}}, \bibinfo {author} {\bibfnamefont {C.}~\bibnamefont
  {{Zucchetti}}}, \bibinfo {author} {\bibfnamefont {Q.}~\bibnamefont
  {{Barbedienne}}}, \bibinfo {author} {\bibfnamefont {A.}~\bibnamefont
  {{Marty}}}, \bibinfo {author} {\bibfnamefont {G.}~\bibnamefont {{Isella}}},
  \bibinfo {author} {\bibfnamefont {L.}~\bibnamefont {{Cagnon}}}, \bibinfo
  {author} {\bibfnamefont {C.}~\bibnamefont {{Vergnaud}}}, \bibinfo {author}
  {\bibfnamefont {H.}~\bibnamefont {{Jaffr{\`e}s}}}, \bibinfo {author}
  {\bibfnamefont {N.}~\bibnamefont {{Reyren}}}, \bibinfo {author}
  {\bibfnamefont {J.~M.}\ \bibnamefont {{George}}}, \emph {et~al.},\ }\bibfield
   {title} {\bibinfo {title} {Observation of large unidirectional \text{R}ashba
  magnetoresistance in \text{G}e(111)},\ }\href
  {https://link.aps.org/doi/10.1103/PhysRevLett.124.027201} {\bibfield
  {journal} {\bibinfo  {journal} {Phys. Rev. Lett.}\ }\textbf {\bibinfo
  {volume} {124}},\ \bibinfo {pages} {027201} (\bibinfo {year}
  {2020})}\BibitemShut {NoStop}%
\bibitem [{\citenamefont {\ifmmode~\check{Z}\else \v{Z}\fi{}elezn\'y}\ \emph
  {et~al.}(2021)\citenamefont {\ifmmode~\check{Z}\else \v{Z}\fi{}elezn\'y},
  \citenamefont {Fang}, \citenamefont {Olejn\'{\i}k}, \citenamefont {Patchett},
  \citenamefont {Gerhard}, \citenamefont {Gould}, \citenamefont {Molenkamp},
  \citenamefont {Gomez-Olivella}, \citenamefont {Zemen}, \citenamefont
  {Tich\'y} \emph {et~al.}}]{zelezny2021unidirectional}%
  \BibitemOpen
  \bibfield  {author} {\bibinfo {author} {\bibfnamefont {J.}~\bibnamefont
  {\ifmmode~\check{Z}\else \v{Z}\fi{}elezn\'y}}, \bibinfo {author}
  {\bibfnamefont {Z.}~\bibnamefont {Fang}}, \bibinfo {author} {\bibfnamefont
  {K.}~\bibnamefont {Olejn\'{\i}k}}, \bibinfo {author} {\bibfnamefont
  {J.}~\bibnamefont {Patchett}}, \bibinfo {author} {\bibfnamefont
  {F.}~\bibnamefont {Gerhard}}, \bibinfo {author} {\bibfnamefont
  {C.}~\bibnamefont {Gould}}, \bibinfo {author} {\bibfnamefont {L.~W.}\
  \bibnamefont {Molenkamp}}, \bibinfo {author} {\bibfnamefont {C.}~\bibnamefont
  {Gomez-Olivella}}, \bibinfo {author} {\bibfnamefont {J.}~\bibnamefont
  {Zemen}}, \bibinfo {author} {\bibfnamefont {T.}~\bibnamefont {Tich\'y}},
  \emph {et~al.},\ }\bibfield  {title} {\bibinfo {title} {{Unidirectional
  magnetoresistance and spin-orbit torque in NiMnSb}},\ }\href
  {https://link.aps.org/doi/10.1103/PhysRevB.104.054429} {\bibfield  {journal}
  {\bibinfo  {journal} {Phys. Rev. B}\ }\textbf {\bibinfo {volume} {104}},\
  \bibinfo {pages} {054429} (\bibinfo {year} {2021})}\BibitemShut {NoStop}%
\bibitem [{\citenamefont {{Guillet}}\ \emph {et~al.}(2021)\citenamefont
  {{Guillet}}, \citenamefont {{Marty}}, \citenamefont {{Vergnaud}},
  \citenamefont {{Jamet}}, \citenamefont {{Zucchetti}}, \citenamefont
  {{Isella}}, \citenamefont {{Barbedienne}}, \citenamefont {{Jaffr{\`e}s}},
  \citenamefont {{Reyren}}, \citenamefont {{George}} \emph
  {et~al.}}]{guillet2021large}%
  \BibitemOpen
  \bibfield  {author} {\bibinfo {author} {\bibfnamefont {T.}~\bibnamefont
  {{Guillet}}}, \bibinfo {author} {\bibfnamefont {A.}~\bibnamefont {{Marty}}},
  \bibinfo {author} {\bibfnamefont {C.}~\bibnamefont {{Vergnaud}}}, \bibinfo
  {author} {\bibfnamefont {M.}~\bibnamefont {{Jamet}}}, \bibinfo {author}
  {\bibfnamefont {C.}~\bibnamefont {{Zucchetti}}}, \bibinfo {author}
  {\bibfnamefont {G.}~\bibnamefont {{Isella}}}, \bibinfo {author}
  {\bibfnamefont {Q.}~\bibnamefont {{Barbedienne}}}, \bibinfo {author}
  {\bibfnamefont {H.}~\bibnamefont {{Jaffr{\`e}s}}}, \bibinfo {author}
  {\bibfnamefont {N.}~\bibnamefont {{Reyren}}}, \bibinfo {author}
  {\bibfnamefont {J.~M.}\ \bibnamefont {{George}}}, \emph {et~al.},\ }\bibfield
   {title} {\bibinfo {title} {{Large Rashba unidirectional magnetoresistance in
  the Fe/Ge(111) interface states}},\ }\href
  {https://doi.org/10.1103/PhysRevB.103.064411} {\bibfield  {journal} {\bibinfo
   {journal} {Phys. Rev. B}\ }\textbf {\bibinfo {volume} {103}},\ \bibinfo
  {pages} {064411} (\bibinfo {year} {2021})}\BibitemShut {NoStop}%
\bibitem [{\citenamefont {Hasegawa}\ \emph {et~al.}(2021)\citenamefont
  {Hasegawa}, \citenamefont {Koyama},\ and\ \citenamefont
  {Chiba}}]{hasegawa2021enhanced}%
  \BibitemOpen
  \bibfield  {author} {\bibinfo {author} {\bibfnamefont {K.}~\bibnamefont
  {Hasegawa}}, \bibinfo {author} {\bibfnamefont {T.}~\bibnamefont {Koyama}},\
  and\ \bibinfo {author} {\bibfnamefont {D.}~\bibnamefont {Chiba}},\ }\bibfield
   {title} {\bibinfo {title} {{Enhanced unidirectional spin Hall
  magnetoresistance in a Pt/Co system with a Cu interlayer}},\ }\href
  {https://link.aps.org/doi/10.1103/PhysRevB.103.L020411} {\bibfield  {journal}
  {\bibinfo  {journal} {Phys. Rev. B}\ }\textbf {\bibinfo {volume} {103}},\
  \bibinfo {pages} {L020411} (\bibinfo {year} {2021})}\BibitemShut {NoStop}%
\bibitem [{\citenamefont {Liu}\ \emph {et~al.}(2021{\natexlab{a}})\citenamefont
  {Liu}, \citenamefont {Wang}, \citenamefont {Luan}, \citenamefont {Zhou},
  \citenamefont {Xia}, \citenamefont {Yang}, \citenamefont {Tian},
  \citenamefont {Guo}, \citenamefont {Du},\ and\ \citenamefont
  {Wu}}]{liu2021magnonic}%
  \BibitemOpen
  \bibfield  {author} {\bibinfo {author} {\bibfnamefont {G.}~\bibnamefont
  {Liu}}, \bibinfo {author} {\bibfnamefont {X.-g.}\ \bibnamefont {Wang}},
  \bibinfo {author} {\bibfnamefont {Z.~Z.}\ \bibnamefont {Luan}}, \bibinfo
  {author} {\bibfnamefont {L.~F.}\ \bibnamefont {Zhou}}, \bibinfo {author}
  {\bibfnamefont {S.~Y.}\ \bibnamefont {Xia}}, \bibinfo {author} {\bibfnamefont
  {B.}~\bibnamefont {Yang}}, \bibinfo {author} {\bibfnamefont {Y.~Z.}\
  \bibnamefont {Tian}}, \bibinfo {author} {\bibfnamefont {G.-h.}\ \bibnamefont
  {Guo}}, \bibinfo {author} {\bibfnamefont {J.}~\bibnamefont {Du}},\ and\
  \bibinfo {author} {\bibfnamefont {D.}~\bibnamefont {Wu}},\ }\bibfield
  {title} {\bibinfo {title} {Magnonic unidirectional spin \text{H}all
  magnetoresistance in a heavy-metal--ferromagnetic-insulator bilayer},\ }\href
  {https://link.aps.org/doi/10.1103/PhysRevLett.127.207206} {\bibfield
  {journal} {\bibinfo  {journal} {Phys. Rev. Lett.}\ }\textbf {\bibinfo
  {volume} {127}},\ \bibinfo {pages} {207206} (\bibinfo {year}
  {2021}{\natexlab{a}})}\BibitemShut {NoStop}%
\bibitem [{\citenamefont {Liu}\ \emph {et~al.}(2021{\natexlab{b}})\citenamefont
  {Liu}, \citenamefont {Holder},\ and\ \citenamefont {Yan}}]{liu2021chirality}%
  \BibitemOpen
  \bibfield  {author} {\bibinfo {author} {\bibfnamefont {Y.}~\bibnamefont
  {Liu}}, \bibinfo {author} {\bibfnamefont {T.}~\bibnamefont {Holder}},\ and\
  \bibinfo {author} {\bibfnamefont {B.}~\bibnamefont {Yan}},\ }\bibfield
  {title} {\bibinfo {title} {Chirality-induced giant unidirectional
  magnetoresistance in twisted bilayer graphene},\ }\href
  {https://doi.org/10.1016/j.xinn.2021.100085} {\bibfield  {journal} {\bibinfo
  {journal} {The Innovation}\ }\textbf {\bibinfo {volume} {2}},\ \bibinfo
  {pages} {100085} (\bibinfo {year} {2021}{\natexlab{b}})}\BibitemShut
  {NoStop}%
\bibitem [{\citenamefont {Chang}\ \emph {et~al.}(2021)\citenamefont {Chang},
  \citenamefont {Cheng}, \citenamefont {Huang}, \citenamefont {Peng},
  \citenamefont {Huang}, \citenamefont {Chen}, \citenamefont {Liu},\ and\
  \citenamefont {Pai}}]{chang2021large}%
  \BibitemOpen
  \bibfield  {author} {\bibinfo {author} {\bibfnamefont {T.-Y.}\ \bibnamefont
  {Chang}}, \bibinfo {author} {\bibfnamefont {C.-L.}\ \bibnamefont {Cheng}},
  \bibinfo {author} {\bibfnamefont {C.-C.}\ \bibnamefont {Huang}}, \bibinfo
  {author} {\bibfnamefont {C.-W.}\ \bibnamefont {Peng}}, \bibinfo {author}
  {\bibfnamefont {Y.-H.}\ \bibnamefont {Huang}}, \bibinfo {author}
  {\bibfnamefont {T.-Y.}\ \bibnamefont {Chen}}, \bibinfo {author}
  {\bibfnamefont {Y.-T.}\ \bibnamefont {Liu}},\ and\ \bibinfo {author}
  {\bibfnamefont {C.-F.}\ \bibnamefont {Pai}},\ }\bibfield  {title} {\bibinfo
  {title} {{Large unidirectional magnetoresistance in metallic heterostructures
  in the spin transfer torque regime}},\ }\href
  {https://link.aps.org/doi/10.1103/PhysRevB.104.024432} {\bibfield  {journal}
  {\bibinfo  {journal} {Phys. Rev. B}\ }\textbf {\bibinfo {volume} {104}},\
  \bibinfo {pages} {024432} (\bibinfo {year} {2021})}\BibitemShut {NoStop}%
\bibitem [{\citenamefont {Shim}\ \emph {et~al.}(2022)\citenamefont {Shim},
  \citenamefont {Mehraeen}, \citenamefont {Sklenar}, \citenamefont {Oh},
  \citenamefont {Gibbons}, \citenamefont {Saglam}, \citenamefont {Hoffmann},
  \citenamefont {Zhang},\ and\ \citenamefont {Mason}}]{shim2022unidirectional}%
  \BibitemOpen
  \bibfield  {author} {\bibinfo {author} {\bibfnamefont {S.}~\bibnamefont
  {Shim}}, \bibinfo {author} {\bibfnamefont {M.}~\bibnamefont {Mehraeen}},
  \bibinfo {author} {\bibfnamefont {J.}~\bibnamefont {Sklenar}}, \bibinfo
  {author} {\bibfnamefont {J.}~\bibnamefont {Oh}}, \bibinfo {author}
  {\bibfnamefont {J.}~\bibnamefont {Gibbons}}, \bibinfo {author} {\bibfnamefont
  {H.}~\bibnamefont {Saglam}}, \bibinfo {author} {\bibfnamefont
  {A.}~\bibnamefont {Hoffmann}}, \bibinfo {author} {\bibfnamefont {S.~S.-L.}\
  \bibnamefont {Zhang}},\ and\ \bibinfo {author} {\bibfnamefont
  {N.}~\bibnamefont {Mason}},\ }\bibfield  {title} {\bibinfo {title}
  {Unidirectional magnetoresistance in antiferromagnet/heavy-metal bilayers},\
  }\href {https://link.aps.org/doi/10.1103/PhysRevX.12.021069} {\bibfield
  {journal} {\bibinfo  {journal} {Phys. Rev. X}\ }\textbf {\bibinfo {volume}
  {12}},\ \bibinfo {pages} {021069} (\bibinfo {year} {2022})}\BibitemShut
  {NoStop}%
\bibitem [{\citenamefont {Mehraeen}\ and\ \citenamefont
  {Zhang}(2022)}]{mehraeen2022spin}%
  \BibitemOpen
  \bibfield  {author} {\bibinfo {author} {\bibfnamefont {M.}~\bibnamefont
  {Mehraeen}}\ and\ \bibinfo {author} {\bibfnamefont {S.~S.-L.}\ \bibnamefont
  {Zhang}},\ }\bibfield  {title} {\bibinfo {title} {Spin anomalous-\text{H}all
  unidirectional magnetoresistance},\ }\href
  {https://link.aps.org/doi/10.1103/PhysRevB.105.184423} {\bibfield  {journal}
  {\bibinfo  {journal} {Phys. Rev. B}\ }\textbf {\bibinfo {volume} {105}},\
  \bibinfo {pages} {184423} (\bibinfo {year} {2022})}\BibitemShut {NoStop}%
\bibitem [{\citenamefont {Ding}\ \emph {et~al.}(2022)\citenamefont {Ding},
  \citenamefont {No\"el}, \citenamefont {Krishnaswamy},\ and\ \citenamefont
  {Gambardella}}]{ding2022unidirectional}%
  \BibitemOpen
  \bibfield  {author} {\bibinfo {author} {\bibfnamefont {S.}~\bibnamefont
  {Ding}}, \bibinfo {author} {\bibfnamefont {P.}~\bibnamefont {No\"el}},
  \bibinfo {author} {\bibfnamefont {G.~K.}\ \bibnamefont {Krishnaswamy}},\ and\
  \bibinfo {author} {\bibfnamefont {P.}~\bibnamefont {Gambardella}},\
  }\bibfield  {title} {\bibinfo {title} {Unidirectional orbital
  magnetoresistance in light-metal--ferromagnet bilayers},\ }\href
  {https://link.aps.org/doi/10.1103/PhysRevResearch.4.L032041} {\bibfield
  {journal} {\bibinfo  {journal} {Phys. Rev. Res.}\ }\textbf {\bibinfo {volume}
  {4}},\ \bibinfo {pages} {L032041} (\bibinfo {year} {2022})}\BibitemShut
  {NoStop}%
\bibitem [{\citenamefont {Lou}\ \emph {et~al.}(2022)\citenamefont {Lou},
  \citenamefont {Zhao}, \citenamefont {Jiang},\ and\ \citenamefont
  {Bi}}]{lou2022large}%
  \BibitemOpen
  \bibfield  {author} {\bibinfo {author} {\bibfnamefont {K.}~\bibnamefont
  {Lou}}, \bibinfo {author} {\bibfnamefont {Q.}~\bibnamefont {Zhao}}, \bibinfo
  {author} {\bibfnamefont {B.}~\bibnamefont {Jiang}},\ and\ \bibinfo {author}
  {\bibfnamefont {C.}~\bibnamefont {Bi}},\ }\bibfield  {title} {\bibinfo
  {title} {Large anomalous unidirectional magnetoresistance in a single
  ferromagnetic layer},\ }\href
  {https://link.aps.org/doi/10.1103/PhysRevApplied.17.064052} {\bibfield
  {journal} {\bibinfo  {journal} {Phys. Rev. Appl.}\ }\textbf {\bibinfo
  {volume} {17}},\ \bibinfo {pages} {064052} (\bibinfo {year}
  {2022})}\BibitemShut {NoStop}%
\bibitem [{\citenamefont {Mehraeen}\ \emph {et~al.}(2023)\citenamefont
  {Mehraeen}, \citenamefont {Shen},\ and\ \citenamefont
  {Zhang}}]{mehraeen2023quantum}%
  \BibitemOpen
  \bibfield  {author} {\bibinfo {author} {\bibfnamefont {M.}~\bibnamefont
  {Mehraeen}}, \bibinfo {author} {\bibfnamefont {P.}~\bibnamefont {Shen}},\
  and\ \bibinfo {author} {\bibfnamefont {S.~S.-L.}\ \bibnamefont {Zhang}},\
  }\bibfield  {title} {\bibinfo {title} {Quantum unidirectional
  magnetoresistance},\ }\href
  {https://link.aps.org/doi/10.1103/PhysRevB.108.014411} {\bibfield  {journal}
  {\bibinfo  {journal} {Phys. Rev. B}\ }\textbf {\bibinfo {volume} {108}},\
  \bibinfo {pages} {014411} (\bibinfo {year} {2023})}\BibitemShut {NoStop}%
\bibitem [{\citenamefont {Cheng}\ \emph {et~al.}(2023)\citenamefont {Cheng},
  \citenamefont {Tang}, \citenamefont {Michel}, \citenamefont {Chong},
  \citenamefont {Yang}, \citenamefont {Cheng},\ and\ \citenamefont
  {Wang}}]{cheng2023unidirectional}%
  \BibitemOpen
  \bibfield  {author} {\bibinfo {author} {\bibfnamefont {Y.}~\bibnamefont
  {Cheng}}, \bibinfo {author} {\bibfnamefont {J.}~\bibnamefont {Tang}},
  \bibinfo {author} {\bibfnamefont {J.~J.}\ \bibnamefont {Michel}}, \bibinfo
  {author} {\bibfnamefont {S.~K.}\ \bibnamefont {Chong}}, \bibinfo {author}
  {\bibfnamefont {F.}~\bibnamefont {Yang}}, \bibinfo {author} {\bibfnamefont
  {R.}~\bibnamefont {Cheng}},\ and\ \bibinfo {author} {\bibfnamefont {K.~L.}\
  \bibnamefont {Wang}},\ }\bibfield  {title} {\bibinfo {title} {Unidirectional
  spin \text{H}all magnetoresistance in antiferromagnetic heterostructures},\
  }\href {https://link.aps.org/doi/10.1103/PhysRevLett.130.086703} {\bibfield
  {journal} {\bibinfo  {journal} {Phys. Rev. Lett.}\ }\textbf {\bibinfo
  {volume} {130}},\ \bibinfo {pages} {086703} (\bibinfo {year}
  {2023})}\BibitemShut {NoStop}%
\bibitem [{\citenamefont {Fan}\ \emph {et~al.}(2023)\citenamefont {Fan},
  \citenamefont {Zhang}, \citenamefont {Han}, \citenamefont {Lv}, \citenamefont
  {Liu},\ and\ \citenamefont {Wang}}]{fan2023observation}%
  \BibitemOpen
  \bibfield  {author} {\bibinfo {author} {\bibfnamefont {Y.}~\bibnamefont
  {Fan}}, \bibinfo {author} {\bibfnamefont {P.}~\bibnamefont {Zhang}}, \bibinfo
  {author} {\bibfnamefont {J.}~\bibnamefont {Han}}, \bibinfo {author}
  {\bibfnamefont {Y.}~\bibnamefont {Lv}}, \bibinfo {author} {\bibfnamefont
  {L.}~\bibnamefont {Liu}},\ and\ \bibinfo {author} {\bibfnamefont {J.-P.}\
  \bibnamefont {Wang}},\ }\bibfield  {title} {\bibinfo {title} {{Observation of
  the unidirectional magnetoresistance in antiferromagnetic insulator Fe2O3/Pt
  bilayers}},\ }\href {https://doi.org/10.1002/aelm.202300232} {\bibfield
  {journal} {\bibinfo  {journal} {Adv. Electron. Mater.}\ ,\ \bibinfo {pages}
  {2300232}} (\bibinfo {year} {2023})}\BibitemShut {NoStop}%
\bibitem [{\citenamefont {Zheng}\ \emph {et~al.}(2023)\citenamefont {Zheng},
  \citenamefont {Gu}, \citenamefont {Zhang}, \citenamefont {Zhang},
  \citenamefont {Zhao}, \citenamefont {Li}, \citenamefont {Ren}, \citenamefont
  {Jia}, \citenamefont {Xiao}, \citenamefont {Zhou} \emph
  {et~al.}}]{zheng2023coexistence}%
  \BibitemOpen
  \bibfield  {author} {\bibinfo {author} {\bibfnamefont {Z.}~\bibnamefont
  {Zheng}}, \bibinfo {author} {\bibfnamefont {Y.}~\bibnamefont {Gu}}, \bibinfo
  {author} {\bibfnamefont {Z.}~\bibnamefont {Zhang}}, \bibinfo {author}
  {\bibfnamefont {X.}~\bibnamefont {Zhang}}, \bibinfo {author} {\bibfnamefont
  {T.}~\bibnamefont {Zhao}}, \bibinfo {author} {\bibfnamefont {H.}~\bibnamefont
  {Li}}, \bibinfo {author} {\bibfnamefont {L.}~\bibnamefont {Ren}}, \bibinfo
  {author} {\bibfnamefont {L.}~\bibnamefont {Jia}}, \bibinfo {author}
  {\bibfnamefont {R.}~\bibnamefont {Xiao}}, \bibinfo {author} {\bibfnamefont
  {H.-A.}\ \bibnamefont {Zhou}}, \emph {et~al.},\ }\bibfield  {title} {\bibinfo
  {title} {{Coexistence of Magnon-Induced and Rashba-Induced Unidirectional
  Magnetoresistance in Antiferromagnets}},\ }\href
  {https://doi.org/10.1021/acs.nanolett.3c01082} {\bibfield  {journal}
  {\bibinfo  {journal} {Nano Lett.}\ } (\bibinfo {year} {2023})}\BibitemShut
  {NoStop}%
\bibitem [{\citenamefont {Mehraeen}\ and\ \citenamefont
  {Zhang}(2024)}]{mehraeen2024proximity}%
  \BibitemOpen
  \bibfield  {author} {\bibinfo {author} {\bibfnamefont {M.}~\bibnamefont
  {Mehraeen}}\ and\ \bibinfo {author} {\bibfnamefont {S.~S.-L.}\ \bibnamefont
  {Zhang}},\ }\bibfield  {title} {\bibinfo {title} {Proximity-induced nonlinear
  magnetoresistances on topological insulators},\ }\href
  {https://link.aps.org/doi/10.1103/PhysRevB.109.024421} {\bibfield  {journal}
  {\bibinfo  {journal} {Phys. Rev. B}\ }\textbf {\bibinfo {volume} {109}},\
  \bibinfo {pages} {024421} (\bibinfo {year} {2024})}\BibitemShut {NoStop}%
\bibitem [{\citenamefont {Zou}\ \emph {et~al.}(2024)\citenamefont {Zou},
  \citenamefont {Geng}, \citenamefont {Ma}, \citenamefont {Chen}, \citenamefont
  {Sheng},\ and\ \citenamefont {Xing}}]{zou2024nonreciprocal}%
  \BibitemOpen
  \bibfield  {author} {\bibinfo {author} {\bibfnamefont {M.~H.}\ \bibnamefont
  {Zou}}, \bibinfo {author} {\bibfnamefont {H.}~\bibnamefont {Geng}}, \bibinfo
  {author} {\bibfnamefont {R.}~\bibnamefont {Ma}}, \bibinfo {author}
  {\bibfnamefont {W.}~\bibnamefont {Chen}}, \bibinfo {author} {\bibfnamefont
  {L.}~\bibnamefont {Sheng}},\ and\ \bibinfo {author} {\bibfnamefont {D.~Y.}\
  \bibnamefont {Xing}},\ }\bibfield  {title} {\bibinfo {title} {Nonreciprocal
  ballistic transport in asymmetric bands},\ }\href
  {https://link.aps.org/doi/10.1103/PhysRevB.109.155302} {\bibfield  {journal}
  {\bibinfo  {journal} {Phys. Rev. B}\ }\textbf {\bibinfo {volume} {109}},\
  \bibinfo {pages} {155302} (\bibinfo {year} {2024})}\BibitemShut {NoStop}%
\bibitem [{\citenamefont {Zhao}\ \emph {et~al.}(2024)\citenamefont {Zhao},
  \citenamefont {Li}, \citenamefont {Liu}, \citenamefont {Li}, \citenamefont
  {Wang}, \citenamefont {Liu}, \citenamefont {Yang}, \citenamefont {Jiang},\
  and\ \citenamefont {Gao}}]{zhao2024large}%
  \BibitemOpen
  \bibfield  {author} {\bibinfo {author} {\bibfnamefont {L.}~\bibnamefont
  {Zhao}}, \bibinfo {author} {\bibfnamefont {Y.}~\bibnamefont {Li}}, \bibinfo
  {author} {\bibfnamefont {F.}~\bibnamefont {Liu}}, \bibinfo {author}
  {\bibfnamefont {T.}~\bibnamefont {Li}}, \bibinfo {author} {\bibfnamefont
  {Y.}~\bibnamefont {Wang}}, \bibinfo {author} {\bibfnamefont {X.}~\bibnamefont
  {Liu}}, \bibinfo {author} {\bibfnamefont {D.}~\bibnamefont {Yang}}, \bibinfo
  {author} {\bibfnamefont {C.}~\bibnamefont {Jiang}},\ and\ \bibinfo {author}
  {\bibfnamefont {C.}~\bibnamefont {Gao}},\ }\bibfield  {title} {\bibinfo
  {title} {{Large unidirectional magnetoresistance from the dual functionality
  of copper oxide in naturally oxidized light-metal Al/Cu bilayer films}},\
  }\href {https://link.aps.org/doi/10.1103/PhysRevApplied.21.044020} {\bibfield
   {journal} {\bibinfo  {journal} {Phys. Rev. Appl.}\ }\textbf {\bibinfo
  {volume} {21}},\ \bibinfo {pages} {044020} (\bibinfo {year}
  {2024})}\BibitemShut {NoStop}%
\bibitem [{\citenamefont {Aoki}\ \emph {et~al.}(2024)\citenamefont {Aoki},
  \citenamefont {Ohshima}, \citenamefont {Shinjo}, \citenamefont {Shiraishi},\
  and\ \citenamefont {Ando}}]{aoki2024evaluation}%
  \BibitemOpen
  \bibfield  {author} {\bibinfo {author} {\bibfnamefont {M.}~\bibnamefont
  {Aoki}}, \bibinfo {author} {\bibfnamefont {R.}~\bibnamefont {Ohshima}},
  \bibinfo {author} {\bibfnamefont {T.}~\bibnamefont {Shinjo}}, \bibinfo
  {author} {\bibfnamefont {M.}~\bibnamefont {Shiraishi}},\ and\ \bibinfo
  {author} {\bibfnamefont {Y.}~\bibnamefont {Ando}},\ }\bibfield  {title}
  {\bibinfo {title} {{Evaluation of Spin Hall Effect in Ferromagnets by Means
  of Unidirectional Spin Hall Magnetoresistance in Ta/Co Bilayers}},\ }\href
  {https://doi.org/10.3379/msjmag.2403R003} {\bibfield  {journal} {\bibinfo
  {journal} {J. Magn. Soc. Jpn.}\ }\textbf {\bibinfo {volume} {48}},\ \bibinfo
  {pages} {28} (\bibinfo {year} {2024})}\BibitemShut {NoStop}%
\bibitem [{\citenamefont {Huang}\ \emph {et~al.}(2024)\citenamefont {Huang},
  \citenamefont {Cui}, \citenamefont {Wang}, \citenamefont {Xie}, \citenamefont
  {Bai}, \citenamefont {Tian}, \citenamefont {Cao},\ and\ \citenamefont
  {Yan}}]{huang2024spin}%
  \BibitemOpen
  \bibfield  {author} {\bibinfo {author} {\bibfnamefont {Q.}~\bibnamefont
  {Huang}}, \bibinfo {author} {\bibfnamefont {X.}~\bibnamefont {Cui}}, \bibinfo
  {author} {\bibfnamefont {S.}~\bibnamefont {Wang}}, \bibinfo {author}
  {\bibfnamefont {R.}~\bibnamefont {Xie}}, \bibinfo {author} {\bibfnamefont
  {L.}~\bibnamefont {Bai}}, \bibinfo {author} {\bibfnamefont {Y.}~\bibnamefont
  {Tian}}, \bibinfo {author} {\bibfnamefont {Q.}~\bibnamefont {Cao}},\ and\
  \bibinfo {author} {\bibfnamefont {S.}~\bibnamefont {Yan}},\ }\bibfield
  {title} {\bibinfo {title} {{Spin-anomalous-Hall unidirectional
  magnetoresistance in light-metal/ferromagnetic-metal bilayers}},\ }\href
  {https://doi.org/10.1063/5.0194720} {\bibfield  {journal} {\bibinfo
  {journal} {Appl. Phys. Rev.}\ }\textbf {\bibinfo {volume} {11}} (\bibinfo
  {year} {2024})}\BibitemShut {NoStop}%
\bibitem [{\citenamefont {Kao}\ \emph {et~al.}(2024)\citenamefont {Kao},
  \citenamefont {Tang}, \citenamefont {Ortiz}, \citenamefont {Zhu},
  \citenamefont {Yuan}, \citenamefont {Rao}, \citenamefont {Li}, \citenamefont
  {Edgar}, \citenamefont {Yan}, \citenamefont {Mandrus} \emph
  {et~al.}}]{kao2024unconventional}%
  \BibitemOpen
  \bibfield  {author} {\bibinfo {author} {\bibfnamefont {I.}~\bibnamefont
  {Kao}}, \bibinfo {author} {\bibfnamefont {J.}~\bibnamefont {Tang}}, \bibinfo
  {author} {\bibfnamefont {G.~C.}\ \bibnamefont {Ortiz}}, \bibinfo {author}
  {\bibfnamefont {M.}~\bibnamefont {Zhu}}, \bibinfo {author} {\bibfnamefont
  {S.}~\bibnamefont {Yuan}}, \bibinfo {author} {\bibfnamefont {R.}~\bibnamefont
  {Rao}}, \bibinfo {author} {\bibfnamefont {J.}~\bibnamefont {Li}}, \bibinfo
  {author} {\bibfnamefont {J.~H.}\ \bibnamefont {Edgar}}, \bibinfo {author}
  {\bibfnamefont {J.}~\bibnamefont {Yan}}, \bibinfo {author} {\bibfnamefont
  {D.~G.}\ \bibnamefont {Mandrus}}, \emph {et~al.},\ }\bibfield  {title}
  {\bibinfo {title} {{Unconventional Unidirectional Magnetoresistance in vdW
  Heterostructures}},\ }\href {https://arxiv.org/abs/2405.10889} {\bibfield
  {journal} {\bibinfo  {journal} {arXiv:2405.10889}\ } (\bibinfo {year}
  {2024})}\BibitemShut {NoStop}%
\bibitem [{\citenamefont {Rikken}\ \emph {et~al.}(2001)\citenamefont {Rikken},
  \citenamefont {F\"olling},\ and\ \citenamefont
  {Wyder}}]{rikken2001electrical}%
  \BibitemOpen
  \bibfield  {author} {\bibinfo {author} {\bibfnamefont {G.~L. J.~A.}\
  \bibnamefont {Rikken}}, \bibinfo {author} {\bibfnamefont {J.}~\bibnamefont
  {F\"olling}},\ and\ \bibinfo {author} {\bibfnamefont {P.}~\bibnamefont
  {Wyder}},\ }\bibfield  {title} {\bibinfo {title} {{Electrical Magnetochiral
  Anisotropy}},\ }\href
  {https://link.aps.org/doi/10.1103/PhysRevLett.87.236602} {\bibfield
  {journal} {\bibinfo  {journal} {Phys. Rev. Lett.}\ }\textbf {\bibinfo
  {volume} {87}},\ \bibinfo {pages} {236602} (\bibinfo {year}
  {2001})}\BibitemShut {NoStop}%
\bibitem [{\citenamefont {He}\ \emph {et~al.}(2018)\citenamefont {He},
  \citenamefont {Zhang}, \citenamefont {Zhu}, \citenamefont {Liu},
  \citenamefont {Wang}, \citenamefont {Yu}, \citenamefont {Vignale},\ and\
  \citenamefont {Yang}}]{he2018bilinear}%
  \BibitemOpen
  \bibfield  {author} {\bibinfo {author} {\bibfnamefont {P.}~\bibnamefont
  {He}}, \bibinfo {author} {\bibfnamefont {S.~S.-L.}\ \bibnamefont {Zhang}},
  \bibinfo {author} {\bibfnamefont {D.}~\bibnamefont {Zhu}}, \bibinfo {author}
  {\bibfnamefont {Y.}~\bibnamefont {Liu}}, \bibinfo {author} {\bibfnamefont
  {Y.}~\bibnamefont {Wang}}, \bibinfo {author} {\bibfnamefont {J.}~\bibnamefont
  {Yu}}, \bibinfo {author} {\bibfnamefont {G.}~\bibnamefont {Vignale}},\ and\
  \bibinfo {author} {\bibfnamefont {H.}~\bibnamefont {Yang}},\ }\bibfield
  {title} {\bibinfo {title} {Bilinear magnetoelectric resistance as a probe of
  three-dimensional spin texture in topological surface states},\ }\href
  {https://doi.org/10.1038/s41567-017-0039-y} {\bibfield  {journal} {\bibinfo
  {journal} {Nat. Phys.}\ }\textbf {\bibinfo {volume} {14}},\ \bibinfo {pages}
  {495} (\bibinfo {year} {2018})}\BibitemShut {NoStop}%
\bibitem [{\citenamefont {Dyrda\l{}}\ \emph {et~al.}(2020)\citenamefont
  {Dyrda\l{}}, \citenamefont {Barna\ifmmode~\acute{s}\else \'{s}\fi{}},\ and\
  \citenamefont {Fert}}]{dyrdal2020spin}%
  \BibitemOpen
  \bibfield  {author} {\bibinfo {author} {\bibfnamefont {A.}~\bibnamefont
  {Dyrda\l{}}}, \bibinfo {author} {\bibfnamefont {J.}~\bibnamefont
  {Barna\ifmmode~\acute{s}\else \'{s}\fi{}}},\ and\ \bibinfo {author}
  {\bibfnamefont {A.}~\bibnamefont {Fert}},\ }\bibfield  {title} {\bibinfo
  {title} {Spin-momentum-locking inhomogeneities as a source of bilinear
  magnetoresistance in topological insulators},\ }\href
  {https://link.aps.org/doi/10.1103/PhysRevLett.124.046802} {\bibfield
  {journal} {\bibinfo  {journal} {Phys. Rev. Lett.}\ }\textbf {\bibinfo
  {volume} {124}},\ \bibinfo {pages} {046802} (\bibinfo {year}
  {2020})}\BibitemShut {NoStop}%
\bibitem [{\citenamefont {Zhang}\ \emph {et~al.}(2022)\citenamefont {Zhang},
  \citenamefont {Kalappattil}, \citenamefont {Liu}, \citenamefont {Mehraeen},
  \citenamefont {Zhang}, \citenamefont {Ding}, \citenamefont {Erugu},
  \citenamefont {Chen}, \citenamefont {Tian}, \citenamefont {Liu} \emph
  {et~al.}}]{zhang2022large}%
  \BibitemOpen
  \bibfield  {author} {\bibinfo {author} {\bibfnamefont {Y.}~\bibnamefont
  {Zhang}}, \bibinfo {author} {\bibfnamefont {V.}~\bibnamefont {Kalappattil}},
  \bibinfo {author} {\bibfnamefont {C.}~\bibnamefont {Liu}}, \bibinfo {author}
  {\bibfnamefont {M.}~\bibnamefont {Mehraeen}}, \bibinfo {author}
  {\bibfnamefont {S.~S.-L.}\ \bibnamefont {Zhang}}, \bibinfo {author}
  {\bibfnamefont {J.}~\bibnamefont {Ding}}, \bibinfo {author} {\bibfnamefont
  {U.}~\bibnamefont {Erugu}}, \bibinfo {author} {\bibfnamefont
  {Z.}~\bibnamefont {Chen}}, \bibinfo {author} {\bibfnamefont {J.}~\bibnamefont
  {Tian}}, \bibinfo {author} {\bibfnamefont {K.}~\bibnamefont {Liu}}, \emph
  {et~al.},\ }\bibfield  {title} {\bibinfo {title} {Large magnetoelectric
  resistance in the topological \text{D}irac semimetal $\alpha$-\text{S}n},\
  }\href {https://doi.org/10.1126/sciadv.abo0052} {\bibfield  {journal}
  {\bibinfo  {journal} {Sci. Adv.}\ }\textbf {\bibinfo {volume} {8}},\ \bibinfo
  {pages} {eabo0052} (\bibinfo {year} {2022})}\BibitemShut {NoStop}%
\bibitem [{\citenamefont {Wang}\ \emph
  {et~al.}(2022{\natexlab{a}})\citenamefont {Wang}, \citenamefont {Liu},
  \citenamefont {Huang}, \citenamefont {Mambakkam}, \citenamefont {Wang},
  \citenamefont {Yang}, \citenamefont {Sheng}, \citenamefont {Law},\ and\
  \citenamefont {Xiao}}]{wang2022large}%
  \BibitemOpen
  \bibfield  {author} {\bibinfo {author} {\bibfnamefont {Y.}~\bibnamefont
  {Wang}}, \bibinfo {author} {\bibfnamefont {B.}~\bibnamefont {Liu}}, \bibinfo
  {author} {\bibfnamefont {Y.-X.}\ \bibnamefont {Huang}}, \bibinfo {author}
  {\bibfnamefont {S.~V.}\ \bibnamefont {Mambakkam}}, \bibinfo {author}
  {\bibfnamefont {Y.}~\bibnamefont {Wang}}, \bibinfo {author} {\bibfnamefont
  {S.~A.}\ \bibnamefont {Yang}}, \bibinfo {author} {\bibfnamefont {X.-L.}\
  \bibnamefont {Sheng}}, \bibinfo {author} {\bibfnamefont {S.~A.}\ \bibnamefont
  {Law}},\ and\ \bibinfo {author} {\bibfnamefont {J.~Q.}\ \bibnamefont
  {Xiao}},\ }\bibfield  {title} {\bibinfo {title} {Large bilinear
  magnetoresistance from \text{R}ashba spin-splitting on the surface of a
  topological insulator},\ }\href
  {https://link.aps.org/doi/10.1103/PhysRevB.106.L241401} {\bibfield  {journal}
  {\bibinfo  {journal} {Phys. Rev. B}\ }\textbf {\bibinfo {volume} {106}},\
  \bibinfo {pages} {L241401} (\bibinfo {year}
  {2022}{\natexlab{a}})}\BibitemShut {NoStop}%
\bibitem [{\citenamefont {Fu}\ \emph {et~al.}(2022)\citenamefont {Fu},
  \citenamefont {Li}, \citenamefont {Papin}, \citenamefont {Noel},
  \citenamefont {Teresi}, \citenamefont {Cosset-Ch{\'e}neau}, \citenamefont
  {Grezes}, \citenamefont {Guillet}, \citenamefont {Thomas}, \citenamefont
  {Niquet} \emph {et~al.}}]{fu2022bilinear}%
  \BibitemOpen
  \bibfield  {author} {\bibinfo {author} {\bibfnamefont {Y.}~\bibnamefont
  {Fu}}, \bibinfo {author} {\bibfnamefont {J.}~\bibnamefont {Li}}, \bibinfo
  {author} {\bibfnamefont {J.}~\bibnamefont {Papin}}, \bibinfo {author}
  {\bibfnamefont {P.}~\bibnamefont {Noel}}, \bibinfo {author} {\bibfnamefont
  {S.}~\bibnamefont {Teresi}}, \bibinfo {author} {\bibfnamefont
  {M.}~\bibnamefont {Cosset-Ch{\'e}neau}}, \bibinfo {author} {\bibfnamefont
  {C.}~\bibnamefont {Grezes}}, \bibinfo {author} {\bibfnamefont
  {T.}~\bibnamefont {Guillet}}, \bibinfo {author} {\bibfnamefont
  {C.}~\bibnamefont {Thomas}}, \bibinfo {author} {\bibfnamefont {Y.-M.}\
  \bibnamefont {Niquet}}, \emph {et~al.},\ }\bibfield  {title} {\bibinfo
  {title} {{Bilinear magnetoresistance in HgTe topological insulator: opposite
  signs at opposite surfaces demonstrated by gate control}},\ }\href
  {https://doi.org/10.1021/acs.nanolett.2c02585} {\bibfield  {journal}
  {\bibinfo  {journal} {Nano Lett.}\ }\textbf {\bibinfo {volume} {22}},\
  \bibinfo {pages} {7867} (\bibinfo {year} {2022})}\BibitemShut {NoStop}%
\bibitem [{\citenamefont {Golub}\ \emph {et~al.}(2023)\citenamefont {Golub},
  \citenamefont {Ivchenko},\ and\ \citenamefont
  {Spivak}}]{golub2023electrical}%
  \BibitemOpen
  \bibfield  {author} {\bibinfo {author} {\bibfnamefont {L.~E.}\ \bibnamefont
  {Golub}}, \bibinfo {author} {\bibfnamefont {E.~L.}\ \bibnamefont
  {Ivchenko}},\ and\ \bibinfo {author} {\bibfnamefont {B.}~\bibnamefont
  {Spivak}},\ }\bibfield  {title} {\bibinfo {title} {{Electrical magnetochiral
  current in tellurium}},\ }\href
  {https://link.aps.org/doi/10.1103/PhysRevB.108.245202} {\bibfield  {journal}
  {\bibinfo  {journal} {Phys. Rev. B}\ }\textbf {\bibinfo {volume} {108}},\
  \bibinfo {pages} {245202} (\bibinfo {year} {2023})}\BibitemShut {NoStop}%
\bibitem [{\citenamefont {Marx}\ \emph {et~al.}(2024)\citenamefont {Marx},
  \citenamefont {Jafari}, \citenamefont {Tekelenburg}, \citenamefont {Loi},
  \citenamefont {S\l{}awi\ifmmode~\acute{n}\else \'{n}\fi{}ska},\ and\
  \citenamefont {Guimar\~aes}}]{marx2024nonlinear}%
  \BibitemOpen
  \bibfield  {author} {\bibinfo {author} {\bibfnamefont {A.~C.}\ \bibnamefont
  {Marx}}, \bibinfo {author} {\bibfnamefont {H.}~\bibnamefont {Jafari}},
  \bibinfo {author} {\bibfnamefont {E.~K.}\ \bibnamefont {Tekelenburg}},
  \bibinfo {author} {\bibfnamefont {M.~A.}\ \bibnamefont {Loi}}, \bibinfo
  {author} {\bibfnamefont {J.}~\bibnamefont {S\l{}awi\ifmmode~\acute{n}\else
  \'{n}\fi{}ska}},\ and\ \bibinfo {author} {\bibfnamefont {M.~H.~D.}\
  \bibnamefont {Guimar\~aes}},\ }\bibfield  {title} {\bibinfo {title}
  {{Nonlinear magnetotransport in ${\mathrm{MoTe}}_{2}$}},\ }\href
  {https://link.aps.org/doi/10.1103/PhysRevB.109.125408} {\bibfield  {journal}
  {\bibinfo  {journal} {Phys. Rev. B}\ }\textbf {\bibinfo {volume} {109}},\
  \bibinfo {pages} {125408} (\bibinfo {year} {2024})}\BibitemShut {NoStop}%
\bibitem [{\citenamefont {Boboshko}\ and\ \citenamefont
  {Dyrda\l{}}(2024)}]{boboshko2024bilinear}%
  \BibitemOpen
  \bibfield  {author} {\bibinfo {author} {\bibfnamefont {K.}~\bibnamefont
  {Boboshko}}\ and\ \bibinfo {author} {\bibfnamefont {A.}~\bibnamefont
  {Dyrda\l{}}},\ }\bibfield  {title} {\bibinfo {title} {{Bilinear
  magnetoresistance and planar Hall effect in topological insulators: Interplay
  of scattering on spin-orbital impurities and nonequilibrium spin
  polarization}},\ }\href
  {https://link.aps.org/doi/10.1103/PhysRevB.109.155420} {\bibfield  {journal}
  {\bibinfo  {journal} {Phys. Rev. B}\ }\textbf {\bibinfo {volume} {109}},\
  \bibinfo {pages} {155420} (\bibinfo {year} {2024})}\BibitemShut {NoStop}%
\bibitem [{\citenamefont {Kim}\ \emph {et~al.}(2024)\citenamefont {Kim},
  \citenamefont {Kim}, \citenamefont {Lee}, \citenamefont {Oh}, \citenamefont
  {Chen}, \citenamefont {Yang}, \citenamefont {Pu}, \citenamefont {Liu},
  \citenamefont {Hu}, \citenamefont {Cao~Van} \emph {et~al.}}]{kim2024spin}%
  \BibitemOpen
  \bibfield  {author} {\bibinfo {author} {\bibfnamefont {D.-J.}\ \bibnamefont
  {Kim}}, \bibinfo {author} {\bibfnamefont {K.-W.}\ \bibnamefont {Kim}},
  \bibinfo {author} {\bibfnamefont {K.}~\bibnamefont {Lee}}, \bibinfo {author}
  {\bibfnamefont {J.~H.}\ \bibnamefont {Oh}}, \bibinfo {author} {\bibfnamefont
  {X.}~\bibnamefont {Chen}}, \bibinfo {author} {\bibfnamefont {S.}~\bibnamefont
  {Yang}}, \bibinfo {author} {\bibfnamefont {Y.}~\bibnamefont {Pu}}, \bibinfo
  {author} {\bibfnamefont {Y.}~\bibnamefont {Liu}}, \bibinfo {author}
  {\bibfnamefont {F.}~\bibnamefont {Hu}}, \bibinfo {author} {\bibfnamefont
  {P.}~\bibnamefont {Cao~Van}}, \emph {et~al.},\ }\bibfield  {title} {\bibinfo
  {title} {{Spin Hall-induced bilinear magnetoelectric resistance}},\ }\href
  {https://doi.org/10.1038/s41563-024-02000-0} {\bibfield  {journal} {\bibinfo
  {journal} {Nat. Mater.}\ ,\ \bibinfo {pages} {1}} (\bibinfo {year}
  {2024})}\BibitemShut {NoStop}%
\bibitem [{\citenamefont {Sodemann}\ and\ \citenamefont
  {Fu}(2015)}]{sodemann2015quantum}%
  \BibitemOpen
  \bibfield  {author} {\bibinfo {author} {\bibfnamefont {I.}~\bibnamefont
  {Sodemann}}\ and\ \bibinfo {author} {\bibfnamefont {L.}~\bibnamefont {Fu}},\
  }\bibfield  {title} {\bibinfo {title} {{Quantum Nonlinear Hall Effect Induced
  by Berry Curvature Dipole in Time-Reversal Invariant Materials}},\ }\href
  {https://link.aps.org/doi/10.1103/PhysRevLett.115.216806} {\bibfield
  {journal} {\bibinfo  {journal} {Phys. Rev. Lett.}\ }\textbf {\bibinfo
  {volume} {115}},\ \bibinfo {pages} {216806} (\bibinfo {year}
  {2015})}\BibitemShut {NoStop}%
\bibitem [{\citenamefont {Low}\ \emph {et~al.}(2015)\citenamefont {Low},
  \citenamefont {Jiang},\ and\ \citenamefont {Guinea}}]{low2015topological}%
  \BibitemOpen
  \bibfield  {author} {\bibinfo {author} {\bibfnamefont {T.}~\bibnamefont
  {Low}}, \bibinfo {author} {\bibfnamefont {Y.}~\bibnamefont {Jiang}},\ and\
  \bibinfo {author} {\bibfnamefont {F.}~\bibnamefont {Guinea}},\ }\bibfield
  {title} {\bibinfo {title} {{Topological currents in black phosphorus with
  broken inversion symmetry}},\ }\href
  {https://link.aps.org/doi/10.1103/PhysRevB.92.235447} {\bibfield  {journal}
  {\bibinfo  {journal} {Phys. Rev. B}\ }\textbf {\bibinfo {volume} {92}},\
  \bibinfo {pages} {235447} (\bibinfo {year} {2015})}\BibitemShut {NoStop}%
\bibitem [{\citenamefont {Yasuda}\ \emph {et~al.}(2017)\citenamefont {Yasuda},
  \citenamefont {Tsukazaki}, \citenamefont {Yoshimi}, \citenamefont {Kondou},
  \citenamefont {Takahashi}, \citenamefont {Otani}, \citenamefont {Kawasaki},\
  and\ \citenamefont {Tokura}}]{yasuda2017current}%
  \BibitemOpen
  \bibfield  {author} {\bibinfo {author} {\bibfnamefont {K.}~\bibnamefont
  {Yasuda}}, \bibinfo {author} {\bibfnamefont {A.}~\bibnamefont {Tsukazaki}},
  \bibinfo {author} {\bibfnamefont {R.}~\bibnamefont {Yoshimi}}, \bibinfo
  {author} {\bibfnamefont {K.}~\bibnamefont {Kondou}}, \bibinfo {author}
  {\bibfnamefont {K.~S.}\ \bibnamefont {Takahashi}}, \bibinfo {author}
  {\bibfnamefont {Y.}~\bibnamefont {Otani}}, \bibinfo {author} {\bibfnamefont
  {M.}~\bibnamefont {Kawasaki}},\ and\ \bibinfo {author} {\bibfnamefont
  {Y.}~\bibnamefont {Tokura}},\ }\bibfield  {title} {\bibinfo {title}
  {{Current-Nonlinear Hall Effect and Spin-Orbit Torque Magnetization Switching
  in a Magnetic Topological Insulator}},\ }\href
  {https://link.aps.org/doi/10.1103/PhysRevLett.119.137204} {\bibfield
  {journal} {\bibinfo  {journal} {Phys. Rev. Lett.}\ }\textbf {\bibinfo
  {volume} {119}},\ \bibinfo {pages} {137204} (\bibinfo {year}
  {2017})}\BibitemShut {NoStop}%
\bibitem [{\citenamefont {Du}\ \emph {et~al.}(2018)\citenamefont {Du},
  \citenamefont {Wang}, \citenamefont {Lu},\ and\ \citenamefont
  {Xie}}]{du2018band}%
  \BibitemOpen
  \bibfield  {author} {\bibinfo {author} {\bibfnamefont {Z.~Z.}\ \bibnamefont
  {Du}}, \bibinfo {author} {\bibfnamefont {C.~M.}\ \bibnamefont {Wang}},
  \bibinfo {author} {\bibfnamefont {H.-Z.}\ \bibnamefont {Lu}},\ and\ \bibinfo
  {author} {\bibfnamefont {X.~C.}\ \bibnamefont {Xie}},\ }\bibfield  {title}
  {\bibinfo {title} {{Band Signatures for Strong Nonlinear Hall Effect in
  Bilayer ${\mathrm{WTe}}_{2}$}},\ }\href
  {https://link.aps.org/doi/10.1103/PhysRevLett.121.266601} {\bibfield
  {journal} {\bibinfo  {journal} {Phys. Rev. Lett.}\ }\textbf {\bibinfo
  {volume} {121}},\ \bibinfo {pages} {266601} (\bibinfo {year}
  {2018})}\BibitemShut {NoStop}%
\bibitem [{\citenamefont {Facio}\ \emph {et~al.}(2018)\citenamefont {Facio},
  \citenamefont {Efremov}, \citenamefont {Koepernik}, \citenamefont {You},
  \citenamefont {Sodemann},\ and\ \citenamefont {van~den
  Brink}}]{facio2018strongly}%
  \BibitemOpen
  \bibfield  {author} {\bibinfo {author} {\bibfnamefont {J.~I.}\ \bibnamefont
  {Facio}}, \bibinfo {author} {\bibfnamefont {D.}~\bibnamefont {Efremov}},
  \bibinfo {author} {\bibfnamefont {K.}~\bibnamefont {Koepernik}}, \bibinfo
  {author} {\bibfnamefont {J.-S.}\ \bibnamefont {You}}, \bibinfo {author}
  {\bibfnamefont {I.}~\bibnamefont {Sodemann}},\ and\ \bibinfo {author}
  {\bibfnamefont {J.}~\bibnamefont {van~den Brink}},\ }\bibfield  {title}
  {\bibinfo {title} {{Strongly Enhanced Berry Dipole at Topological Phase
  Transitions in BiTeI}},\ }\href
  {https://link.aps.org/doi/10.1103/PhysRevLett.121.246403} {\bibfield
  {journal} {\bibinfo  {journal} {Phys. Rev. Lett.}\ }\textbf {\bibinfo
  {volume} {121}},\ \bibinfo {pages} {246403} (\bibinfo {year}
  {2018})}\BibitemShut {NoStop}%
\bibitem [{\citenamefont {You}\ \emph {et~al.}(2018)\citenamefont {You},
  \citenamefont {Fang}, \citenamefont {Xu}, \citenamefont {Kaxiras},\ and\
  \citenamefont {Low}}]{you2018berry}%
  \BibitemOpen
  \bibfield  {author} {\bibinfo {author} {\bibfnamefont {J.-S.}\ \bibnamefont
  {You}}, \bibinfo {author} {\bibfnamefont {S.}~\bibnamefont {Fang}}, \bibinfo
  {author} {\bibfnamefont {S.-Y.}\ \bibnamefont {Xu}}, \bibinfo {author}
  {\bibfnamefont {E.}~\bibnamefont {Kaxiras}},\ and\ \bibinfo {author}
  {\bibfnamefont {T.}~\bibnamefont {Low}},\ }\bibfield  {title} {\bibinfo
  {title} {{Berry curvature dipole current in the transition metal
  dichalcogenides family}},\ }\href
  {https://link.aps.org/doi/10.1103/PhysRevB.98.121109} {\bibfield  {journal}
  {\bibinfo  {journal} {Phys. Rev. B}\ }\textbf {\bibinfo {volume} {98}},\
  \bibinfo {pages} {121109} (\bibinfo {year} {2018})}\BibitemShut {NoStop}%
\bibitem [{\citenamefont {Zhang}\ \emph
  {et~al.}(2018{\natexlab{a}})\citenamefont {Zhang}, \citenamefont {Van
  Den~Brink}, \citenamefont {Felser},\ and\ \citenamefont
  {Yan}}]{zhang2018electrically}%
  \BibitemOpen
  \bibfield  {author} {\bibinfo {author} {\bibfnamefont {Y.}~\bibnamefont
  {Zhang}}, \bibinfo {author} {\bibfnamefont {J.}~\bibnamefont {Van
  Den~Brink}}, \bibinfo {author} {\bibfnamefont {C.}~\bibnamefont {Felser}},\
  and\ \bibinfo {author} {\bibfnamefont {B.}~\bibnamefont {Yan}},\ }\bibfield
  {title} {\bibinfo {title} {{Electrically tuneable nonlinear anomalous Hall
  effect in two-dimensional transition-metal dichalcogenides WTe2 and MoTe2}},\
  }\href
  {https://iopscience.iop.org/article/10.1088/2053-1583/aad1ae/meta?casa_token=6lPJBbfismEAAAAA:61V2Scxg7xVW8nSGLjqtf3Goo7fEs_54K-CQ_yEIRBwRJhUsCpqNetZacfTpQJ6OFqSx6bQu5q-cOr5KO0GE-Q7FOVh3}
  {\bibfield  {journal} {\bibinfo  {journal} {2D Mater.}\ }\textbf {\bibinfo
  {volume} {5}},\ \bibinfo {pages} {044001} (\bibinfo {year}
  {2018}{\natexlab{a}})}\BibitemShut {NoStop}%
\bibitem [{\citenamefont {Zhang}\ \emph
  {et~al.}(2018{\natexlab{b}})\citenamefont {Zhang}, \citenamefont {Sun},\ and\
  \citenamefont {Yan}}]{zhang2018berry}%
  \BibitemOpen
  \bibfield  {author} {\bibinfo {author} {\bibfnamefont {Y.}~\bibnamefont
  {Zhang}}, \bibinfo {author} {\bibfnamefont {Y.}~\bibnamefont {Sun}},\ and\
  \bibinfo {author} {\bibfnamefont {B.}~\bibnamefont {Yan}},\ }\bibfield
  {title} {\bibinfo {title} {{Berry curvature dipole in Weyl semimetal
  materials: An ab initio study}},\ }\href
  {https://link.aps.org/doi/10.1103/PhysRevB.97.041101} {\bibfield  {journal}
  {\bibinfo  {journal} {Phys. Rev. B}\ }\textbf {\bibinfo {volume} {97}},\
  \bibinfo {pages} {041101} (\bibinfo {year} {2018}{\natexlab{b}})}\BibitemShut
  {NoStop}%
\bibitem [{\citenamefont {He}\ \emph {et~al.}(2019)\citenamefont {He},
  \citenamefont {Zhang}, \citenamefont {Zhu}, \citenamefont {Shi},
  \citenamefont {Heinonen}, \citenamefont {Vignale},\ and\ \citenamefont
  {Yang}}]{he2019nonlinear}%
  \BibitemOpen
  \bibfield  {author} {\bibinfo {author} {\bibfnamefont {P.}~\bibnamefont
  {He}}, \bibinfo {author} {\bibfnamefont {S.~S.-L.}\ \bibnamefont {Zhang}},
  \bibinfo {author} {\bibfnamefont {D.}~\bibnamefont {Zhu}}, \bibinfo {author}
  {\bibfnamefont {S.}~\bibnamefont {Shi}}, \bibinfo {author} {\bibfnamefont
  {O.~G.}\ \bibnamefont {Heinonen}}, \bibinfo {author} {\bibfnamefont
  {G.}~\bibnamefont {Vignale}},\ and\ \bibinfo {author} {\bibfnamefont
  {H.}~\bibnamefont {Yang}},\ }\bibfield  {title} {\bibinfo {title} {Nonlinear
  planar \text{H}all effect},\ }\href
  {https://link.aps.org/doi/10.1103/PhysRevLett.123.016801} {\bibfield
  {journal} {\bibinfo  {journal} {Phys. Rev. Lett.}\ }\textbf {\bibinfo
  {volume} {123}},\ \bibinfo {pages} {016801} (\bibinfo {year}
  {2019})}\BibitemShut {NoStop}%
\bibitem [{\citenamefont {Ma}\ \emph {et~al.}(2019)\citenamefont {Ma},
  \citenamefont {Xu}, \citenamefont {Shen}, \citenamefont {MacNeill},
  \citenamefont {Fatemi}, \citenamefont {Chang}, \citenamefont {Mier~Valdivia},
  \citenamefont {Wu}, \citenamefont {Du}, \citenamefont {Hsu} \emph
  {et~al.}}]{ma2019observation}%
  \BibitemOpen
  \bibfield  {author} {\bibinfo {author} {\bibfnamefont {Q.}~\bibnamefont
  {Ma}}, \bibinfo {author} {\bibfnamefont {S.-Y.}\ \bibnamefont {Xu}}, \bibinfo
  {author} {\bibfnamefont {H.}~\bibnamefont {Shen}}, \bibinfo {author}
  {\bibfnamefont {D.}~\bibnamefont {MacNeill}}, \bibinfo {author}
  {\bibfnamefont {V.}~\bibnamefont {Fatemi}}, \bibinfo {author} {\bibfnamefont
  {T.-R.}\ \bibnamefont {Chang}}, \bibinfo {author} {\bibfnamefont {A.~M.}\
  \bibnamefont {Mier~Valdivia}}, \bibinfo {author} {\bibfnamefont
  {S.}~\bibnamefont {Wu}}, \bibinfo {author} {\bibfnamefont {Z.}~\bibnamefont
  {Du}}, \bibinfo {author} {\bibfnamefont {C.-H.}\ \bibnamefont {Hsu}}, \emph
  {et~al.},\ }\bibfield  {title} {\bibinfo {title} {{Observation of the
  nonlinear Hall effect under time-reversal-symmetric conditions}},\ }\href
  {https://doi.org/10.1038/s41586-018-0807-6} {\bibfield  {journal} {\bibinfo
  {journal} {Nature}\ }\textbf {\bibinfo {volume} {565}},\ \bibinfo {pages}
  {337} (\bibinfo {year} {2019})}\BibitemShut {NoStop}%
\bibitem [{\citenamefont {Kang}\ \emph {et~al.}(2019)\citenamefont {Kang},
  \citenamefont {Li}, \citenamefont {Sohn}, \citenamefont {Shan},\ and\
  \citenamefont {Mak}}]{kang2019nonlinear}%
  \BibitemOpen
  \bibfield  {author} {\bibinfo {author} {\bibfnamefont {K.}~\bibnamefont
  {Kang}}, \bibinfo {author} {\bibfnamefont {T.}~\bibnamefont {Li}}, \bibinfo
  {author} {\bibfnamefont {E.}~\bibnamefont {Sohn}}, \bibinfo {author}
  {\bibfnamefont {J.}~\bibnamefont {Shan}},\ and\ \bibinfo {author}
  {\bibfnamefont {K.~F.}\ \bibnamefont {Mak}},\ }\bibfield  {title} {\bibinfo
  {title} {{Nonlinear anomalous Hall effect in few-layer WTe2}},\ }\href
  {https://doi.org/10.1038/s41563-019-0294-7} {\bibfield  {journal} {\bibinfo
  {journal} {Nat. Mater.}\ }\textbf {\bibinfo {volume} {18}},\ \bibinfo {pages}
  {324} (\bibinfo {year} {2019})}\BibitemShut {NoStop}%
\bibitem [{\citenamefont {Du}\ \emph {et~al.}(2019)\citenamefont {Du},
  \citenamefont {Wang}, \citenamefont {Li}, \citenamefont {Lu},\ and\
  \citenamefont {Xie}}]{du2019disorder}%
  \BibitemOpen
  \bibfield  {author} {\bibinfo {author} {\bibfnamefont {Z.}~\bibnamefont
  {Du}}, \bibinfo {author} {\bibfnamefont {C.}~\bibnamefont {Wang}}, \bibinfo
  {author} {\bibfnamefont {S.}~\bibnamefont {Li}}, \bibinfo {author}
  {\bibfnamefont {H.-Z.}\ \bibnamefont {Lu}},\ and\ \bibinfo {author}
  {\bibfnamefont {X.}~\bibnamefont {Xie}},\ }\bibfield  {title} {\bibinfo
  {title} {{Disorder-induced nonlinear Hall effect with time-reversal
  symmetry}},\ }\href {https://doi.org/10.1038/s41467-019-10941-3} {\bibfield
  {journal} {\bibinfo  {journal} {Nat. Commun.}\ }\textbf {\bibinfo {volume}
  {10}},\ \bibinfo {pages} {3047} (\bibinfo {year} {2019})}\BibitemShut
  {NoStop}%
\bibitem [{\citenamefont {Wang}\ \emph {et~al.}(2021)\citenamefont {Wang},
  \citenamefont {Gao},\ and\ \citenamefont {Xiao}}]{wang2021intrinsic}%
  \BibitemOpen
  \bibfield  {author} {\bibinfo {author} {\bibfnamefont {C.}~\bibnamefont
  {Wang}}, \bibinfo {author} {\bibfnamefont {Y.}~\bibnamefont {Gao}},\ and\
  \bibinfo {author} {\bibfnamefont {D.}~\bibnamefont {Xiao}},\ }\bibfield
  {title} {\bibinfo {title} {{Intrinsic Nonlinear \text{H}all Effect in
  Antiferromagnetic Tetragonal CuMnAs}},\ }\href
  {https://link.aps.org/doi/10.1103/PhysRevLett.127.277201} {\bibfield
  {journal} {\bibinfo  {journal} {Phys. Rev. Lett.}\ }\textbf {\bibinfo
  {volume} {127}},\ \bibinfo {pages} {277201} (\bibinfo {year}
  {2021})}\BibitemShut {NoStop}%
\bibitem [{\citenamefont {Li}\ \emph {et~al.}(2021)\citenamefont {Li},
  \citenamefont {Heinonen}, \citenamefont {Burkov},\ and\ \citenamefont
  {Zhang}}]{li2021nonlinear}%
  \BibitemOpen
  \bibfield  {author} {\bibinfo {author} {\bibfnamefont {R.-H.}\ \bibnamefont
  {Li}}, \bibinfo {author} {\bibfnamefont {O.~G.}\ \bibnamefont {Heinonen}},
  \bibinfo {author} {\bibfnamefont {A.~A.}\ \bibnamefont {Burkov}},\ and\
  \bibinfo {author} {\bibfnamefont {S.~S.-L.}\ \bibnamefont {Zhang}},\
  }\bibfield  {title} {\bibinfo {title} {{Nonlinear Hall effect in Weyl
  semimetals induced by chiral anomaly}},\ }\href
  {https://link.aps.org/doi/10.1103/PhysRevB.103.045105} {\bibfield  {journal}
  {\bibinfo  {journal} {Phys. Rev. B}\ }\textbf {\bibinfo {volume} {103}},\
  \bibinfo {pages} {045105} (\bibinfo {year} {2021})}\BibitemShut {NoStop}%
\bibitem [{\citenamefont {Zeng}\ \emph {et~al.}(2021)\citenamefont {Zeng},
  \citenamefont {Nandy},\ and\ \citenamefont {Tewari}}]{zeng2021nonlinear}%
  \BibitemOpen
  \bibfield  {author} {\bibinfo {author} {\bibfnamefont {C.}~\bibnamefont
  {Zeng}}, \bibinfo {author} {\bibfnamefont {S.}~\bibnamefont {Nandy}},\ and\
  \bibinfo {author} {\bibfnamefont {S.}~\bibnamefont {Tewari}},\ }\bibfield
  {title} {\bibinfo {title} {{Nonlinear transport in Weyl semimetals induced by
  Berry curvature dipole}},\ }\href
  {https://link.aps.org/doi/10.1103/PhysRevB.103.245119} {\bibfield  {journal}
  {\bibinfo  {journal} {Phys. Rev. B}\ }\textbf {\bibinfo {volume} {103}},\
  \bibinfo {pages} {245119} (\bibinfo {year} {2021})}\BibitemShut {NoStop}%
\bibitem [{\citenamefont {Wang}\ \emph
  {et~al.}(2022{\natexlab{b}})\citenamefont {Wang}, \citenamefont {Mambakkam},
  \citenamefont {Huang}, \citenamefont {Wang}, \citenamefont {Ji},
  \citenamefont {Xiao}, \citenamefont {Yang}, \citenamefont {Law},\ and\
  \citenamefont {Xiao}}]{wang2022observation}%
  \BibitemOpen
  \bibfield  {author} {\bibinfo {author} {\bibfnamefont {Y.}~\bibnamefont
  {Wang}}, \bibinfo {author} {\bibfnamefont {S.~V.}\ \bibnamefont {Mambakkam}},
  \bibinfo {author} {\bibfnamefont {Y.-X.}\ \bibnamefont {Huang}}, \bibinfo
  {author} {\bibfnamefont {Y.}~\bibnamefont {Wang}}, \bibinfo {author}
  {\bibfnamefont {Y.}~\bibnamefont {Ji}}, \bibinfo {author} {\bibfnamefont
  {C.}~\bibnamefont {Xiao}}, \bibinfo {author} {\bibfnamefont {S.~A.}\
  \bibnamefont {Yang}}, \bibinfo {author} {\bibfnamefont {S.~A.}\ \bibnamefont
  {Law}},\ and\ \bibinfo {author} {\bibfnamefont {J.~Q.}\ \bibnamefont
  {Xiao}},\ }\bibfield  {title} {\bibinfo {title} {Observation of nonlinear
  planar \text{H}all effect in magnetic-insulator--topological-insulator
  heterostructures},\ }\href
  {https://link.aps.org/doi/10.1103/PhysRevB.106.155408} {\bibfield  {journal}
  {\bibinfo  {journal} {Phys. Rev. B}\ }\textbf {\bibinfo {volume} {106}},\
  \bibinfo {pages} {155408} (\bibinfo {year} {2022}{\natexlab{b}})}\BibitemShut
  {NoStop}%
\bibitem [{\citenamefont {Gao}\ \emph {et~al.}(2023)\citenamefont {Gao},
  \citenamefont {Liu}, \citenamefont {Qiu}, \citenamefont {Ghosh},
  \citenamefont {V.~Trevisan}, \citenamefont {Onishi}, \citenamefont {Hu},
  \citenamefont {Qian}, \citenamefont {Tien}, \citenamefont {Chen} \emph
  {et~al.}}]{gao2023quantum}%
  \BibitemOpen
  \bibfield  {author} {\bibinfo {author} {\bibfnamefont {A.}~\bibnamefont
  {Gao}}, \bibinfo {author} {\bibfnamefont {Y.-F.}\ \bibnamefont {Liu}},
  \bibinfo {author} {\bibfnamefont {J.-X.}\ \bibnamefont {Qiu}}, \bibinfo
  {author} {\bibfnamefont {B.}~\bibnamefont {Ghosh}}, \bibinfo {author}
  {\bibfnamefont {T.}~\bibnamefont {V.~Trevisan}}, \bibinfo {author}
  {\bibfnamefont {Y.}~\bibnamefont {Onishi}}, \bibinfo {author} {\bibfnamefont
  {C.}~\bibnamefont {Hu}}, \bibinfo {author} {\bibfnamefont {T.}~\bibnamefont
  {Qian}}, \bibinfo {author} {\bibfnamefont {H.-J.}\ \bibnamefont {Tien}},
  \bibinfo {author} {\bibfnamefont {S.-W.}\ \bibnamefont {Chen}}, \emph
  {et~al.},\ }\bibfield  {title} {\bibinfo {title} {{Quantum metric nonlinear
  Hall effect in a topological antiferromagnetic heterostructure}},\ }\href
  {https://doi.org/10.1126/science.adf1506} {\bibfield  {journal} {\bibinfo
  {journal} {Science}\ ,\ \bibinfo {pages} {eadf1506}} (\bibinfo {year}
  {2023})}\BibitemShut {NoStop}%
\bibitem [{\citenamefont {Wang}\ \emph {et~al.}(2023)\citenamefont {Wang},
  \citenamefont {Kaplan}, \citenamefont {Zhang}, \citenamefont {Holder},
  \citenamefont {Cao}, \citenamefont {Wang}, \citenamefont {Zhou},
  \citenamefont {Zhou}, \citenamefont {Jiang}, \citenamefont {Zhang} \emph
  {et~al.}}]{wang2023quantum}%
  \BibitemOpen
  \bibfield  {author} {\bibinfo {author} {\bibfnamefont {N.}~\bibnamefont
  {Wang}}, \bibinfo {author} {\bibfnamefont {D.}~\bibnamefont {Kaplan}},
  \bibinfo {author} {\bibfnamefont {Z.}~\bibnamefont {Zhang}}, \bibinfo
  {author} {\bibfnamefont {T.}~\bibnamefont {Holder}}, \bibinfo {author}
  {\bibfnamefont {N.}~\bibnamefont {Cao}}, \bibinfo {author} {\bibfnamefont
  {A.}~\bibnamefont {Wang}}, \bibinfo {author} {\bibfnamefont {X.}~\bibnamefont
  {Zhou}}, \bibinfo {author} {\bibfnamefont {F.}~\bibnamefont {Zhou}}, \bibinfo
  {author} {\bibfnamefont {Z.}~\bibnamefont {Jiang}}, \bibinfo {author}
  {\bibfnamefont {C.}~\bibnamefont {Zhang}}, \emph {et~al.},\ }\bibfield
  {title} {\bibinfo {title} {{Quantum-metric-induced nonlinear transport in a
  topological antiferromagnet}},\ }\href
  {https://doi.org/10.1038/s41586-023-06363-3} {\bibfield  {journal} {\bibinfo
  {journal} {Nature}\ }\textbf {\bibinfo {volume} {621}},\ \bibinfo {pages}
  {487} (\bibinfo {year} {2023})}\BibitemShut {NoStop}%
\bibitem [{\citenamefont {Kaplan}\ \emph
  {et~al.}(2023{\natexlab{a}})\citenamefont {Kaplan}, \citenamefont {Holder},\
  and\ \citenamefont {Yan}}]{kaplan2023general}%
  \BibitemOpen
  \bibfield  {author} {\bibinfo {author} {\bibfnamefont {D.}~\bibnamefont
  {Kaplan}}, \bibinfo {author} {\bibfnamefont {T.}~\bibnamefont {Holder}},\
  and\ \bibinfo {author} {\bibfnamefont {B.}~\bibnamefont {Yan}},\ }\bibfield
  {title} {\bibinfo {title} {{General nonlinear Hall current in magnetic
  insulators beyond the quantum anomalous Hall effect}},\ }\href
  {https://doi.org/10.1038/s41467-023-38734-9} {\bibfield  {journal} {\bibinfo
  {journal} {Nat. Commun.}\ }\textbf {\bibinfo {volume} {14}},\ \bibinfo
  {pages} {3053} (\bibinfo {year} {2023}{\natexlab{a}})}\BibitemShut {NoStop}%
\bibitem [{\citenamefont {Das}\ \emph {et~al.}(2023)\citenamefont {Das},
  \citenamefont {Lahiri}, \citenamefont {Atencia}, \citenamefont {Culcer},\
  and\ \citenamefont {Agarwal}}]{das2023intrinsic}%
  \BibitemOpen
  \bibfield  {author} {\bibinfo {author} {\bibfnamefont {K.}~\bibnamefont
  {Das}}, \bibinfo {author} {\bibfnamefont {S.}~\bibnamefont {Lahiri}},
  \bibinfo {author} {\bibfnamefont {R.~B.}\ \bibnamefont {Atencia}}, \bibinfo
  {author} {\bibfnamefont {D.}~\bibnamefont {Culcer}},\ and\ \bibinfo {author}
  {\bibfnamefont {A.}~\bibnamefont {Agarwal}},\ }\bibfield  {title} {\bibinfo
  {title} {{Intrinsic nonlinear conductivities induced by the quantum
  metric}},\ }\href {https://link.aps.org/doi/10.1103/PhysRevB.108.L201405}
  {\bibfield  {journal} {\bibinfo  {journal} {Phys. Rev. B}\ }\textbf {\bibinfo
  {volume} {108}},\ \bibinfo {pages} {L201405} (\bibinfo {year}
  {2023})}\BibitemShut {NoStop}%
\bibitem [{\citenamefont {Ma}\ \emph {et~al.}(2023)\citenamefont {Ma},
  \citenamefont {Arora}, \citenamefont {Vignale},\ and\ \citenamefont
  {Song}}]{ma2023anomalous}%
  \BibitemOpen
  \bibfield  {author} {\bibinfo {author} {\bibfnamefont {D.}~\bibnamefont
  {Ma}}, \bibinfo {author} {\bibfnamefont {A.}~\bibnamefont {Arora}}, \bibinfo
  {author} {\bibfnamefont {G.}~\bibnamefont {Vignale}},\ and\ \bibinfo {author}
  {\bibfnamefont {J.~C.~W.}\ \bibnamefont {Song}},\ }\bibfield  {title}
  {\bibinfo {title} {{Anomalous Skew-Scattering Nonlinear Hall Effect and
  Chiral Photocurrents in $\mathcal{PT}$-Symmetric Antiferromagnets}},\ }\href
  {https://link.aps.org/doi/10.1103/PhysRevLett.131.076601} {\bibfield
  {journal} {\bibinfo  {journal} {Phys. Rev. Lett.}\ }\textbf {\bibinfo
  {volume} {131}},\ \bibinfo {pages} {076601} (\bibinfo {year}
  {2023})}\BibitemShut {NoStop}%
\bibitem [{\citenamefont {Zhuang}\ and\ \citenamefont
  {Yan}(2024)}]{zhuang2024intrinsic}%
  \BibitemOpen
  \bibfield  {author} {\bibinfo {author} {\bibfnamefont {Z.-Y.}\ \bibnamefont
  {Zhuang}}\ and\ \bibinfo {author} {\bibfnamefont {Z.}~\bibnamefont {Yan}},\
  }\bibfield  {title} {\bibinfo {title} {{Intrinsic nonlinear Hall effect in
  two-dimensional honeycomb topological antiferromagnets}},\ }\href
  {https://link.aps.org/doi/10.1103/PhysRevB.109.174443} {\bibfield  {journal}
  {\bibinfo  {journal} {Phys. Rev. B}\ }\textbf {\bibinfo {volume} {109}},\
  \bibinfo {pages} {174443} (\bibinfo {year} {2024})}\BibitemShut {NoStop}%
\bibitem [{\citenamefont {Wang}\ \emph {et~al.}(2024)\citenamefont {Wang},
  \citenamefont {Zhang}, \citenamefont {Zhu},\ and\ \citenamefont
  {Su}}]{wang2024intrinsic}%
  \BibitemOpen
  \bibfield  {author} {\bibinfo {author} {\bibfnamefont {Y.}~\bibnamefont
  {Wang}}, \bibinfo {author} {\bibfnamefont {Z.}~\bibnamefont {Zhang}},
  \bibinfo {author} {\bibfnamefont {Z.-G.}\ \bibnamefont {Zhu}},\ and\ \bibinfo
  {author} {\bibfnamefont {G.}~\bibnamefont {Su}},\ }\bibfield  {title}
  {\bibinfo {title} {{Intrinsic nonlinear Ohmic current}},\ }\href
  {https://link.aps.org/doi/10.1103/PhysRevB.109.085419} {\bibfield  {journal}
  {\bibinfo  {journal} {Phys. Rev. B}\ }\textbf {\bibinfo {volume} {109}},\
  \bibinfo {pages} {085419} (\bibinfo {year} {2024})}\BibitemShut {NoStop}%
\bibitem [{\citenamefont {Liu}\ \emph {et~al.}(2024)\citenamefont {Liu},
  \citenamefont {Qiang}, \citenamefont {Lu},\ and\ \citenamefont
  {Xie}}]{liu2024quantum}%
  \BibitemOpen
  \bibfield  {author} {\bibinfo {author} {\bibfnamefont {T.}~\bibnamefont
  {Liu}}, \bibinfo {author} {\bibfnamefont {X.-B.}\ \bibnamefont {Qiang}},
  \bibinfo {author} {\bibfnamefont {H.-Z.}\ \bibnamefont {Lu}},\ and\ \bibinfo
  {author} {\bibfnamefont {X.}~\bibnamefont {Xie}},\ }\bibfield  {title}
  {\bibinfo {title} {{Quantum geometry in condensed matter}},\ }\href
  {https://doi.org/10.1093/nsr/nwae334} {\bibfield  {journal} {\bibinfo
  {journal} {Natl. Sci. Rev.}\ ,\ \bibinfo {pages} {nwae334}} (\bibinfo {year}
  {2024})}\BibitemShut {NoStop}%
\bibitem [{\citenamefont {Liu}\ \emph {et~al.}(2021{\natexlab{c}})\citenamefont
  {Liu}, \citenamefont {Zhao}, \citenamefont {Huang}, \citenamefont {Wu},
  \citenamefont {Sheng}, \citenamefont {Xiao},\ and\ \citenamefont
  {Yang}}]{liu2021intrinsic}%
  \BibitemOpen
  \bibfield  {author} {\bibinfo {author} {\bibfnamefont {H.}~\bibnamefont
  {Liu}}, \bibinfo {author} {\bibfnamefont {J.}~\bibnamefont {Zhao}}, \bibinfo
  {author} {\bibfnamefont {Y.-X.}\ \bibnamefont {Huang}}, \bibinfo {author}
  {\bibfnamefont {W.}~\bibnamefont {Wu}}, \bibinfo {author} {\bibfnamefont
  {X.-L.}\ \bibnamefont {Sheng}}, \bibinfo {author} {\bibfnamefont
  {C.}~\bibnamefont {Xiao}},\ and\ \bibinfo {author} {\bibfnamefont {S.~A.}\
  \bibnamefont {Yang}},\ }\bibfield  {title} {\bibinfo {title} {{Intrinsic
  Second-Order Anomalous Hall Effect and Its Application in Compensated
  Antiferromagnets}},\ }\href
  {https://link.aps.org/doi/10.1103/PhysRevLett.127.277202} {\bibfield
  {journal} {\bibinfo  {journal} {Phys. Rev. Lett.}\ }\textbf {\bibinfo
  {volume} {127}},\ \bibinfo {pages} {277202} (\bibinfo {year}
  {2021}{\natexlab{c}})}\BibitemShut {NoStop}%
\bibitem [{\citenamefont {Parker}\ \emph {et~al.}(2019)\citenamefont {Parker},
  \citenamefont {Morimoto}, \citenamefont {Orenstein},\ and\ \citenamefont
  {Moore}}]{parker2019diagrammatic}%
  \BibitemOpen
  \bibfield  {author} {\bibinfo {author} {\bibfnamefont {D.~E.}\ \bibnamefont
  {Parker}}, \bibinfo {author} {\bibfnamefont {T.}~\bibnamefont {Morimoto}},
  \bibinfo {author} {\bibfnamefont {J.}~\bibnamefont {Orenstein}},\ and\
  \bibinfo {author} {\bibfnamefont {J.~E.}\ \bibnamefont {Moore}},\ }\bibfield
  {title} {\bibinfo {title} {{Diagrammatic approach to nonlinear optical
  response with application to Weyl semimetals}},\ }\href
  {https://link.aps.org/doi/10.1103/PhysRevB.99.045121} {\bibfield  {journal}
  {\bibinfo  {journal} {Phys. Rev. B}\ }\textbf {\bibinfo {volume} {99}},\
  \bibinfo {pages} {045121} (\bibinfo {year} {2019})}\BibitemShut {NoStop}%
\bibitem [{\citenamefont {Du}\ \emph {et~al.}(2021{\natexlab{b}})\citenamefont
  {Du}, \citenamefont {Wang}, \citenamefont {Sun}, \citenamefont {Lu},\ and\
  \citenamefont {Xie}}]{du2021quantum}%
  \BibitemOpen
  \bibfield  {author} {\bibinfo {author} {\bibfnamefont {Z.}~\bibnamefont
  {Du}}, \bibinfo {author} {\bibfnamefont {C.}~\bibnamefont {Wang}}, \bibinfo
  {author} {\bibfnamefont {H.-P.}\ \bibnamefont {Sun}}, \bibinfo {author}
  {\bibfnamefont {H.-Z.}\ \bibnamefont {Lu}},\ and\ \bibinfo {author}
  {\bibfnamefont {X.}~\bibnamefont {Xie}},\ }\bibfield  {title} {\bibinfo
  {title} {Quantum theory of the nonlinear \text{H}all effect},\ }\href
  {https://doi.org/10.1038/s41467-021-25273-4} {\bibfield  {journal} {\bibinfo
  {journal} {Nat. Commun.}\ }\textbf {\bibinfo {volume} {12}},\ \bibinfo
  {pages} {1} (\bibinfo {year} {2021}{\natexlab{b}})}\BibitemShut {NoStop}%
\bibitem [{\citenamefont {Rostami}\ \emph {et~al.}(2021)\citenamefont
  {Rostami}, \citenamefont {Katsnelson}, \citenamefont {Vignale},\ and\
  \citenamefont {Polini}}]{rostami2021gauge}%
  \BibitemOpen
  \bibfield  {author} {\bibinfo {author} {\bibfnamefont {H.}~\bibnamefont
  {Rostami}}, \bibinfo {author} {\bibfnamefont {M.~I.}\ \bibnamefont
  {Katsnelson}}, \bibinfo {author} {\bibfnamefont {G.}~\bibnamefont
  {Vignale}},\ and\ \bibinfo {author} {\bibfnamefont {M.}~\bibnamefont
  {Polini}},\ }\bibfield  {title} {\bibinfo {title} {{Gauge invariance and Ward
  identities in nonlinear response theory}},\ }\href
  {https://doi.org/10.1016/j.aop.2021.168523} {\bibfield  {journal} {\bibinfo
  {journal} {Ann. Phys.}\ }\textbf {\bibinfo {volume} {431}},\ \bibinfo {pages}
  {168523} (\bibinfo {year} {2021})}\BibitemShut {NoStop}%
\bibitem [{\citenamefont {Kaplan}\ \emph
  {et~al.}(2023{\natexlab{b}})\citenamefont {Kaplan}, \citenamefont {Holder},\
  and\ \citenamefont {Yan}}]{kaplan2023unifying}%
  \BibitemOpen
  \bibfield  {author} {\bibinfo {author} {\bibfnamefont {D.}~\bibnamefont
  {Kaplan}}, \bibinfo {author} {\bibfnamefont {T.}~\bibnamefont {Holder}},\
  and\ \bibinfo {author} {\bibfnamefont {B.}~\bibnamefont {Yan}},\ }\bibfield
  {title} {\bibinfo {title} {{Unifying semiclassics and quantum perturbation
  theory at nonlinear order}},\ }\href
  {https://scipost.org/10.21468/SciPostPhys.14.4.082} {\bibfield  {journal}
  {\bibinfo  {journal} {SciPost Phys.}\ }\textbf {\bibinfo {volume} {14}},\
  \bibinfo {pages} {082} (\bibinfo {year} {2023}{\natexlab{b}})}\BibitemShut
  {NoStop}%
\bibitem [{\citenamefont {McKay}\ \emph {et~al.}(2024)\citenamefont {McKay},
  \citenamefont {Mahmood},\ and\ \citenamefont {Bradlyn}}]{mckay2024charge}%
  \BibitemOpen
  \bibfield  {author} {\bibinfo {author} {\bibfnamefont {R.~C.}\ \bibnamefont
  {McKay}}, \bibinfo {author} {\bibfnamefont {F.}~\bibnamefont {Mahmood}},\
  and\ \bibinfo {author} {\bibfnamefont {B.}~\bibnamefont {Bradlyn}},\
  }\bibfield  {title} {\bibinfo {title} {{Charge Conservation beyond
  Uniformity: Spatially Inhomogeneous Electromagnetic Response in Periodic
  Solids}},\ }\href {https://link.aps.org/doi/10.1103/PhysRevX.14.011058}
  {\bibfield  {journal} {\bibinfo  {journal} {Phys. Rev. X}\ }\textbf {\bibinfo
  {volume} {14}},\ \bibinfo {pages} {011058} (\bibinfo {year}
  {2024})}\BibitemShut {NoStop}%
\bibitem [{\citenamefont {Xiao}\ \emph
  {et~al.}(2019{\natexlab{b}})\citenamefont {Xiao}, \citenamefont {Du},\ and\
  \citenamefont {Niu}}]{xiao2019theory}%
  \BibitemOpen
  \bibfield  {author} {\bibinfo {author} {\bibfnamefont {C.}~\bibnamefont
  {Xiao}}, \bibinfo {author} {\bibfnamefont {Z.~Z.}\ \bibnamefont {Du}},\ and\
  \bibinfo {author} {\bibfnamefont {Q.}~\bibnamefont {Niu}},\ }\bibfield
  {title} {\bibinfo {title} {{Theory of nonlinear Hall effects: Modified
  semiclassics from quantum kinetics}},\ }\href
  {https://link.aps.org/doi/10.1103/PhysRevB.100.165422} {\bibfield  {journal}
  {\bibinfo  {journal} {Phys. Rev. B}\ }\textbf {\bibinfo {volume} {100}},\
  \bibinfo {pages} {165422} (\bibinfo {year} {2019}{\natexlab{b}})}\BibitemShut
  {NoStop}%
\bibitem [{\citenamefont {Nandy}\ and\ \citenamefont
  {Sodemann}(2019)}]{nandy2019symmetry}%
  \BibitemOpen
  \bibfield  {author} {\bibinfo {author} {\bibfnamefont {S.}~\bibnamefont
  {Nandy}}\ and\ \bibinfo {author} {\bibfnamefont {I.}~\bibnamefont
  {Sodemann}},\ }\bibfield  {title} {\bibinfo {title} {{Symmetry and quantum
  kinetics of the nonlinear Hall effect}},\ }\href
  {https://link.aps.org/doi/10.1103/PhysRevB.100.195117} {\bibfield  {journal}
  {\bibinfo  {journal} {Phys. Rev. B}\ }\textbf {\bibinfo {volume} {100}},\
  \bibinfo {pages} {195117} (\bibinfo {year} {2019})}\BibitemShut {NoStop}%
\bibitem [{\citenamefont {Ba}\ \emph {et~al.}(2023)\citenamefont {Ba},
  \citenamefont {Wang}, \citenamefont {Duan}, \citenamefont {Deng},\ and\
  \citenamefont {Wang}}]{ba2023nonlinear}%
  \BibitemOpen
  \bibfield  {author} {\bibinfo {author} {\bibfnamefont {J.-Y.}\ \bibnamefont
  {Ba}}, \bibinfo {author} {\bibfnamefont {Y.-M.}\ \bibnamefont {Wang}},
  \bibinfo {author} {\bibfnamefont {H.-J.}\ \bibnamefont {Duan}}, \bibinfo
  {author} {\bibfnamefont {M.-X.}\ \bibnamefont {Deng}},\ and\ \bibinfo
  {author} {\bibfnamefont {R.-Q.}\ \bibnamefont {Wang}},\ }\bibfield  {title}
  {\bibinfo {title} {{Nonlinear planar Hall effect induced by interband
  transitions: Application to surface states of topological insulators}},\
  }\href {https://link.aps.org/doi/10.1103/PhysRevB.108.L241104} {\bibfield
  {journal} {\bibinfo  {journal} {Phys. Rev. B}\ }\textbf {\bibinfo {volume}
  {108}},\ \bibinfo {pages} {L241104} (\bibinfo {year} {2023})}\BibitemShut
  {NoStop}%
\bibitem [{\citenamefont {Huang}\ \emph {et~al.}(2023)\citenamefont {Huang},
  \citenamefont {Xiao}, \citenamefont {Yang},\ and\ \citenamefont
  {Li}}]{huang2023scaling}%
  \BibitemOpen
  \bibfield  {author} {\bibinfo {author} {\bibfnamefont {Y.-X.}\ \bibnamefont
  {Huang}}, \bibinfo {author} {\bibfnamefont {C.}~\bibnamefont {Xiao}},
  \bibinfo {author} {\bibfnamefont {S.~A.}\ \bibnamefont {Yang}},\ and\
  \bibinfo {author} {\bibfnamefont {X.}~\bibnamefont {Li}},\ }\bibfield
  {title} {\bibinfo {title} {{Scaling law for time-reversal-odd nonlinear
  transport}},\ }\href {https://arxiv.org/abs/2311.01219} {\bibfield  {journal}
  {\bibinfo  {journal} {arXiv:2311.01219}\ } (\bibinfo {year}
  {2023})}\BibitemShut {NoStop}%
\bibitem [{\citenamefont {Freimuth}\ \emph {et~al.}(2021)\citenamefont
  {Freimuth}, \citenamefont {Bl{\"u}gel},\ and\ \citenamefont
  {Mokrousov}}]{freimuth2021theory}%
  \BibitemOpen
  \bibfield  {author} {\bibinfo {author} {\bibfnamefont {F.}~\bibnamefont
  {Freimuth}}, \bibinfo {author} {\bibfnamefont {S.}~\bibnamefont
  {Bl{\"u}gel}},\ and\ \bibinfo {author} {\bibfnamefont {Y.}~\bibnamefont
  {Mokrousov}},\ }\bibfield  {title} {\bibinfo {title} {{Theory of
  unidirectional magnetoresistance and nonlinear Hall effect}},\ }\href
  {https://iopscience.iop.org/article/10.1088/1361-648X/ac327f/meta?casa_token=kQEn_-yG0EMAAAAA:01Alo9_HDofnEUbQF5aftDg3cEvoSsGHJEBnt6WDz2-lrGmQGaB2oTTUOq7RbGn8q3TPVput5WjINphCZZ3giKjXRAX_}
  {\bibfield  {journal} {\bibinfo  {journal} {J. Phys. Condens. Matter}\
  }\textbf {\bibinfo {volume} {34}},\ \bibinfo {pages} {055301} (\bibinfo
  {year} {2021})}\BibitemShut {NoStop}%
\bibitem [{\citenamefont {Nagaosa}\ \emph {et~al.}(2010)\citenamefont
  {Nagaosa}, \citenamefont {Sinova}, \citenamefont {Onoda}, \citenamefont
  {MacDonald},\ and\ \citenamefont {Ong}}]{nagaosa2010anomalous}%
  \BibitemOpen
  \bibfield  {author} {\bibinfo {author} {\bibfnamefont {N.}~\bibnamefont
  {Nagaosa}}, \bibinfo {author} {\bibfnamefont {J.}~\bibnamefont {Sinova}},
  \bibinfo {author} {\bibfnamefont {S.}~\bibnamefont {Onoda}}, \bibinfo
  {author} {\bibfnamefont {A.~H.}\ \bibnamefont {MacDonald}},\ and\ \bibinfo
  {author} {\bibfnamefont {N.~P.}\ \bibnamefont {Ong}},\ }\bibfield  {title}
  {\bibinfo {title} {{Anomalous Hall effect}},\ }\href
  {https://link.aps.org/doi/10.1103/RevModPhys.82.1539} {\bibfield  {journal}
  {\bibinfo  {journal} {Rev. Mod. Phys.}\ }\textbf {\bibinfo {volume} {82}},\
  \bibinfo {pages} {1539} (\bibinfo {year} {2010})}\BibitemShut {NoStop}%
\bibitem [{\citenamefont {Provost}\ and\ \citenamefont
  {Vallee}(1980)}]{provost1980riemannian}%
  \BibitemOpen
  \bibfield  {author} {\bibinfo {author} {\bibfnamefont {J.}~\bibnamefont
  {Provost}}\ and\ \bibinfo {author} {\bibfnamefont {G.}~\bibnamefont
  {Vallee}},\ }\bibfield  {title} {\bibinfo {title} {{Riemannian structure on
  manifolds of quantum states}},\ }\href {https://doi.org/10.1007/BF02193559}
  {\bibfield  {journal} {\bibinfo  {journal} {Commun. Math. Phys.}\ }\textbf
  {\bibinfo {volume} {76}},\ \bibinfo {pages} {289} (\bibinfo {year}
  {1980})}\BibitemShut {NoStop}%
\bibitem [{\citenamefont {Cheng}(2010)}]{cheng2010quantum}%
  \BibitemOpen
  \bibfield  {author} {\bibinfo {author} {\bibfnamefont {R.}~\bibnamefont
  {Cheng}},\ }\bibfield  {title} {\bibinfo {title} {{Quantum geometric tensor
  (Fubini-Study metric) in simple quantum system: A pedagogical
  introduction}},\ }\href {https://arxiv.org/abs/1012.1337} {\bibfield
  {journal} {\bibinfo  {journal} {arXiv:1012.1337}\ } (\bibinfo {year}
  {2010})}\BibitemShut {NoStop}%
\bibitem [{Note1()}]{Note1}%
  \BibitemOpen
  \bibinfo {note} {It should be stressed that the choice of scalar disorder
  here is merely one of convenience and the theory is readily applicable to
  spin-dependent disorder profiles as well.}\BibitemShut {Stop}%
\bibitem [{Note2()}]{Note2}%
  \BibitemOpen
  \bibinfo {note} {Here, as in Ref.~\cite {atencia2022semiclassical}, we assume
  the Markovian approximation, whereby time derivatives of the fluctuations can
  be neglected.}\BibitemShut {Stop}%
\bibitem [{\citenamefont {Blount}(1962)}]{blount1962formalisms}%
  \BibitemOpen
  \bibfield  {author} {\bibinfo {author} {\bibfnamefont {E.}~\bibnamefont
  {Blount}},\ }\bibfield  {title} {\bibinfo {title} {Formalisms of band
  theory},\ }in\ \href@noop {} {\emph {\bibinfo {booktitle} {Solid state
  physics}}},\ Vol.~\bibinfo {volume} {13}\ (\bibinfo  {publisher} {Elsevier},\
  \bibinfo {year} {1962})\ pp.\ \bibinfo {pages} {305--373}\BibitemShut
  {NoStop}%
\bibitem [{\citenamefont {Gao}\ \emph {et~al.}(2014)\citenamefont {Gao},
  \citenamefont {Yang},\ and\ \citenamefont {Niu}}]{gao2014field}%
  \BibitemOpen
  \bibfield  {author} {\bibinfo {author} {\bibfnamefont {Y.}~\bibnamefont
  {Gao}}, \bibinfo {author} {\bibfnamefont {S.~A.}\ \bibnamefont {Yang}},\ and\
  \bibinfo {author} {\bibfnamefont {Q.}~\bibnamefont {Niu}},\ }\bibfield
  {title} {\bibinfo {title} {{Field Induced Positional Shift of Bloch Electrons
  and Its Dynamical Implications}},\ }\href
  {https://link.aps.org/doi/10.1103/PhysRevLett.112.166601} {\bibfield
  {journal} {\bibinfo  {journal} {Phys. Rev. Lett.}\ }\textbf {\bibinfo
  {volume} {112}},\ \bibinfo {pages} {166601} (\bibinfo {year}
  {2014})}\BibitemShut {NoStop}%
\bibitem [{\citenamefont {Het\'enyi}\ and\ \citenamefont
  {L\'evay}(2023)}]{hetenyi2023fluctuations}%
  \BibitemOpen
  \bibfield  {author} {\bibinfo {author} {\bibfnamefont {B.}~\bibnamefont
  {Het\'enyi}}\ and\ \bibinfo {author} {\bibfnamefont {P.}~\bibnamefont
  {L\'evay}},\ }\bibfield  {title} {\bibinfo {title} {{Fluctuations,
  uncertainty relations, and the geometry of quantum state manifolds}},\ }\href
  {https://link.aps.org/doi/10.1103/PhysRevA.108.032218} {\bibfield  {journal}
  {\bibinfo  {journal} {Phys. Rev. A}\ }\textbf {\bibinfo {volume} {108}},\
  \bibinfo {pages} {032218} (\bibinfo {year} {2023})}\BibitemShut {NoStop}%
\bibitem [{\citenamefont {Goerbig}\ \emph {et~al.}(2008)\citenamefont
  {Goerbig}, \citenamefont {Fuchs}, \citenamefont {Montambaux},\ and\
  \citenamefont {Pi\'echon}}]{goerbig2008tilted}%
  \BibitemOpen
  \bibfield  {author} {\bibinfo {author} {\bibfnamefont {M.~O.}\ \bibnamefont
  {Goerbig}}, \bibinfo {author} {\bibfnamefont {J.-N.}\ \bibnamefont {Fuchs}},
  \bibinfo {author} {\bibfnamefont {G.}~\bibnamefont {Montambaux}},\ and\
  \bibinfo {author} {\bibfnamefont {F.}~\bibnamefont {Pi\'echon}},\ }\bibfield
  {title} {\bibinfo {title} {{Tilted anisotropic Dirac cones in quinoid-type
  graphene and
  $\ensuremath{\alpha}\text{\ensuremath{-}}{(\text{BEDT-TTF})}_{2}{\text{I}}_{3}$}},\
  }\href {https://link.aps.org/doi/10.1103/PhysRevB.78.045415} {\bibfield
  {journal} {\bibinfo  {journal} {Phys. Rev. B}\ }\textbf {\bibinfo {volume}
  {78}},\ \bibinfo {pages} {045415} (\bibinfo {year} {2008})}\BibitemShut
  {NoStop}%
\bibitem [{\citenamefont {Marinescu}\ and\ \citenamefont
  {Tewari}(2023)}]{marinescu2023magnetochiral}%
  \BibitemOpen
  \bibfield  {author} {\bibinfo {author} {\bibfnamefont {D.~C.}\ \bibnamefont
  {Marinescu}}\ and\ \bibinfo {author} {\bibfnamefont {S.}~\bibnamefont
  {Tewari}},\ }\bibfield  {title} {\bibinfo {title} {{Magnetochiral
  anisotropy-induced nonlinear Hall effect in spin-orbit coupled Rashba
  conductors}},\ }\href {https://link.aps.org/doi/10.1103/PhysRevB.108.195303}
  {\bibfield  {journal} {\bibinfo  {journal} {Phys. Rev. B}\ }\textbf {\bibinfo
  {volume} {108}},\ \bibinfo {pages} {195303} (\bibinfo {year}
  {2023})}\BibitemShut {NoStop}%
\bibitem [{\citenamefont {Gong}\ \emph {et~al.}(2024)\citenamefont {Gong},
  \citenamefont {Du}, \citenamefont {Sun}, \citenamefont {Lu},\ and\
  \citenamefont {Xie}}]{gong2024nonlinear}%
  \BibitemOpen
  \bibfield  {author} {\bibinfo {author} {\bibfnamefont {Z.-H.}\ \bibnamefont
  {Gong}}, \bibinfo {author} {\bibfnamefont {Z.}~\bibnamefont {Du}}, \bibinfo
  {author} {\bibfnamefont {H.-P.}\ \bibnamefont {Sun}}, \bibinfo {author}
  {\bibfnamefont {H.-Z.}\ \bibnamefont {Lu}},\ and\ \bibinfo {author}
  {\bibfnamefont {X.}~\bibnamefont {Xie}},\ }\bibfield  {title} {\bibinfo
  {title} {{Nonlinear transport theory at the order of quantum metric}},\
  }\href {https://arxiv.org/abs/2410.04995} {\bibfield  {journal} {\bibinfo
  {journal} {arXiv:2410.04995}\ } (\bibinfo {year} {2024})}\BibitemShut
  {NoStop}%
\end{thebibliography}%

\end{document}